\documentclass[%
reprint,
superscriptaddress,
nofootinbib,
amsmath,amssymb,
aps,
prl,
]{revtex4-1}
\usepackage{xr-hyper}
\usepackage{hyperref}
\usepackage{graphicx} 
\usepackage{dcolumn}
\usepackage{bm}
\usepackage{braket}
\usepackage{dsfont}
\usepackage{color,colortbl}
\usepackage[table,xcdraw]{xcolor}
\usepackage{multirow}
\usepackage{subcaption}
\usepackage{mwe}
\usepackage{booktabs}
\usepackage{caption}
\usepackage{lipsum}
\usepackage{babel,blindtext}
\usepackage{amsmath}
\usepackage[toc,page]{appendix}
\usepackage[symbol*]{footmisc}
\usepackage{float}

\usepackage{natbib}
\usepackage{filecontents}

\makeatletter

\newcommand*{\addFileDependency}[1]{
\typeout{(#1)}
\@addtofilelist{#1}
\IfFileExists{#1}{}{\typeout{No file #1.}}
}\makeatother

\newcommand*{\myexternaldocument}[1]{%
\externaldocument{#1}%
\addFileDependency{#1.tex}%
\addFileDependency{#1.aux}%
}

\myexternaldocument{SI}

\begin{document}
\title{Complexity of Many-Body Interactions in Transition Metals via Machine-Learned Force Fields from the TM23 Data Set}

\author{Cameron J. Owen$^{*, \dagger}$}
\affiliation{Harvard University}

\author{Steven B. Torrisi$^{*,a}$}
\affiliation{Harvard University}

\author{\\Yu Xie}
\affiliation{Harvard University}

\author{Simon Batzner}
\affiliation{Harvard University}

\author{Kyle Bystrom}
\affiliation{Harvard University}

\author{Jennifer Coulter}
\affiliation{Harvard University}

\author{Albert Musaelian}
\affiliation{Harvard University}

\author{Lixin Sun}
\affiliation{Harvard University}

\author{Boris Kozinsky$^{\dagger}$}
\affiliation{Harvard University}
\affiliation{Robert Bosch LLC Research and Technology Center}

\def\thefootnote{*}\footnotetext{\textbf{Equal Contribution.}\\}\def\thefootnote{\arabic{footnote}}

\def\thefootnote{a}\footnotetext{Currently at Toyota Research Institute.\\}\def\thefootnote{\arabic{footnote}}

\def\thefootnote{$\dagger$}\footnotetext{Corresponding authors\\C.J.O., E-mail: \url{cowen@g.harvard.edu}\\B.K., E-mail: \url{bkoz@g.harvard.edu}\\ }\def\thefootnote{\arabic{footnote}}

\newcommand\bvec{\mathbf}
\newcommand{\mathsc}[1]{{\normalfont\textsc{#1}}}

\begin{abstract}
This work examines challenges associated with the accuracy of machine-learned force fields (MLFFs) for bulk solid and liquid phases of \textit{d}-block elements.
In exhaustive detail, we contrast the performance of force, energy, and stress predictions across the transition metals for two leading MLFF models: a kernel-based atomic cluster expansion method implemented using sparse Gaussian processes (FLARE), and an equivariant message-passing neural network (NequIP).
Early transition metals present higher relative errors and are more difficult to learn relative to late platinum- and coinage-group elements, and this trend persists across model architectures.
Trends in complexity of interatomic interactions for different metals are revealed via comparison of the performance of representations with different many-body order and angular resolution.
Using arguments based on perturbation theory on the occupied and unoccupied $d$ states near the Fermi level, we determine that the large, sharp $d$ density of states both above and below the Fermi level in early transition metals leads to a more complex, harder-to-learn potential energy surface for these metals. 
Increasing the fictitious electronic temperature (smearing) modifies the angular sensitivity of forces and makes the early transition metal forces easier to learn.
This work illustrates challenges in capturing intricate properties of metallic bonding with current leading MLFFs and provides a reference data set for transition metals, aimed at benchmarking the accuracy and improving the development of emerging machine-learned approximations.
\end{abstract}

\maketitle

\section{Introduction}
Molecular dynamics (MD) simulations can reveal atomistic mechanisms for a wide range of fundamental material, chemical, and biological processes. 
\textit{Ab initio} methods like density functional theory (DFT) can calculate atomic forces, energies, and stresses, but are too expensive for MD simulations at large time- and length-scales.
Approximate surrogate models referred to as interatomic potentials or `classical' force fields (FFs) have bridged this gap but can take months or even years to develop, where practitioners exhaustively fit FF parameters to properties like experimental quantities (\emph{e.g.}, melting point, bulk lattice constant, structure factors, etc.) \citep{Mendelev2010DevelopmentIron}.
While their fixed analytical forms make them efficient and interpretable \citep{Daw1984Embedded-atomMetals,meam_baskes_1987,finnis_sinclair_1984,VanDuin2001ReaxFF:Hydrocarbons,Chenoweth2008ReaxFFOxidation,Senftle2016TheDirections}, predictions from classical FFs are limited in transferability beyond their initial training targets even for the same chemical system.
Application to complex chemistries and phenomena like bond-breaking in reactive systems requires great care and close supervision, as \emph{e.g.} assumptions that go into capturing bonded interactions can make decisive differences in simulation outcomes \citep{vandermause2022active,Johansson2022Micron-scaleLearning}.

In response to these challenges, atomistic FF development has, over the past decade, been revolutionized by the advent of machine-learned force fields (MLFFs), where the FF construction task is reduced to fitting a surrogate model of flexible form to first-principles data. 
Simple analytical forms of traditional FFs are replaced with flexible universal approximators to achieve increased accuracy and transferability \citep{Vandermause2020On-the-flyEvents,Batzner2021E3-EquivariantPotentials}.
This has allowed practitioners to fit MLFFs on demand for any desired system that can be computed by \emph{ab initio} methods.
This approach has already yielded successes for simple single-element systems, where MLFFs have been used to reveal surprising long-range mechanical behavior using MD \citep{Deringer2017MachineCarbon,Bartok2018MachineSilicon} and yield highly accurate and expressive power for determining a variety of material properties. 

Whether studying materials systems of one element or many, reliable reference data sets are of incredible importance to the task of MLFF model training and benchmarking \citep{Ramakrishnan2014QuantumMolecules,Deng2010ImageNet:Database,Jain2013}, where model architectures can be compared using the same set of training and test labels.
Presently, no dedicated benchmark data set exists for the \textit{d}-block of the periodic table, which makes it difficult to compare model architectures across a common standard in this set of elements, important for a wide range of applications \emph{e.g.} heterogeneous catalysis on bulk and nanoparticle (NP) structures, (high-entropy) alloys, surface reconstructions, and metallurgy.  
Another major challenge, which reference benchmark databases can help address, is the need to broker a compromise between efficiency and accuracy in the choice of ML formalism and hyperparameters, which also depends on the complexity of the system to be modeled.
Simplifications in the representation fidelity of atomic environments or in model architecture typically come at a cost to accuracy of the predictions made by the resulting MLFF, which must be weighed against the demands of the target application.
To further complicate this task, users typically lack the means and data to gauge what level of model architecture and representation fidelity is needed when approaching new systems, \emph{i.e.}, it is hard to know in advance when a simpler model will suffice.

Benchmark data sets and subsequent studies have been curated before within the community for solid-state materials: previous work carefully benchmarked the performance of a wide variety of model forms (GAP, MTP, NNP, SNAP, and qSNAP), with Zuo \emph{et al.} \cite{Zuo2020PerformancePotentials} releasing the associated data set.
This data set focused on a variety of structural motifs across a set of six elements (Li, Mo, Cu, Ni, Si, and Ge) chosen to represent different electronic character (metallic vs. covalent/semiconducting).

The models trained in this work, and in a followup investigation in Ref. \cite{Chen2022}, highlight larger force and energy errors on early transition metals, like Mo, as opposed to later transition metals, like Cu.
By explicitly comparing these metallic systems across model architectures in terms of predictive accuracy on energies and forces, hints of the strong disparity in model performance appear, but were not commented on in further detail.
More comprehensive reviews of the performance of model architectures like GAP and MTP on body-centered-cubic transition metals and alloys have been performed \cite{PhysRevMaterials.4.093802,Rosenbrock2021}, but the discussion of predictive accuracy was, again, not at the forefront.
Despite the potential of high predictive errors, correlated with where the models are trained on the periodic table, these efforts have demonstrated that one can still obtain proper physical descriptions of the systems (\emph{e.g.} phonons, alloy compositions, etc.), but there has been no systematic investigation into these model performances across the transition metal, which would yield increased understanding of the problem elements, and push the field towards better model architectures.

Another common benchmark data set used to compare accuracy of MLFFs is MD17 \citep{chmiela2017machine, schutt2017quantum, sgdml, christensen2020role} which is comprised of small, organic molecules in vacuum containing main-group elements (\textit{i.e.}, C, O, N, and H), where bond topology is typically rigid and many models can achieve chemical accuracy ($<$ 1 kcal/mol) \citep{Batzner2021E3-EquivariantPotentials}. 
While the latter is useful for the molecular chemistry community, a reference data set for transition metals would provide enormous benefit for the heterogeneous catalysis and metallurgical communities, among others tasked with building MLFFs for these elements.

Thus, better understanding of the tradeoffs between efficiency and efficacy in leading MLFFs for targeted elements could help to drive new methods development, as well as accelerate future model training. 
Moreover, there are increasingly many options for practitioners: MLFF development is now well into its second decade of application \citep{Blank1995NeuralSurfaces,Handley2009OptimalLearning,Behler2007,Bartok2010c,Handley2009OptimalLearning}, and many improvements have been made to the fitting processes, with uncertainty-based active learning \citep{Vandermause2020On-the-flyEvents,Xie2021BayesianStanene,vanderoord2022,Linfeng2019_dpgen} followed by exact mapping onto low-dimensional surrogate models (\emph{e.g.}, polynomial and spline models) \citep{Xie2021BayesianStanene,Xie2022Uncertainty-awareSiC,Glielmo2018EfficientLearning}.
A plethora of MLFF architectures exist: such as MTP \citep{Shapeev2016MomentPotentials}, GAP \citep{Bartok2015GaussianIntroduction}, ACE \citep{Drautz2019AtomicPotentials}, MACE \citep{https://doi.org/10.48550/arxiv.2206.07697}, PAiNN \citep{painn} SNAP \citep{Thompson2015SpectralPotentials}, SchNet \citep{schnet}, DeepMD \citep{Wang2018DeePMD-kit:Dynamics}, with each model architecture exhibiting its own strengths and weaknesses.
More recently, equivariant neural network methods, \textit{e.g.} NequIP and Allegro \citep{Batzner2021E3-EquivariantPotentials,Musaelian2023} have been shown to accurately predict the behaviour of a diverse range of molecular and materials systems ranging from solid-state ionic diffusion and heterogeneous catalysis to small molecules and water.
The models considered in this work (FLARE and NequIP) represent two of the recent leading MLFF approaches with inherent differences in how the representations are constructed for atomic environments.

Consideration of inherently more challenging material and chemical systems, however, will inevitably prompt further MLFF development.
To this point in time, MLFFs have demonstrated near-chemical accuracy on available organic molecule benchmark data, but have shown mixed results on materials \cite{Zuo2020PerformancePotentials}. 
By extending the composition and structural space wherein these MLFFs operate, novel model architectures that achieve state-of-the-art accuracy on organic systems may have to evolve from their current form to accomplish this task.
In hinting to the results presented here, we find that high angular resolution of NequIP is required to improve the predictive accuracy for more difficult transition metals in TM23, albeit at significantly increased computational cost.
Hence, novel model architectures are required that combine efficiency with representation resolutions to accurately capture these difficult solid state systems.

\begin{figure*}
    \centering
    \includegraphics[width=\textwidth]{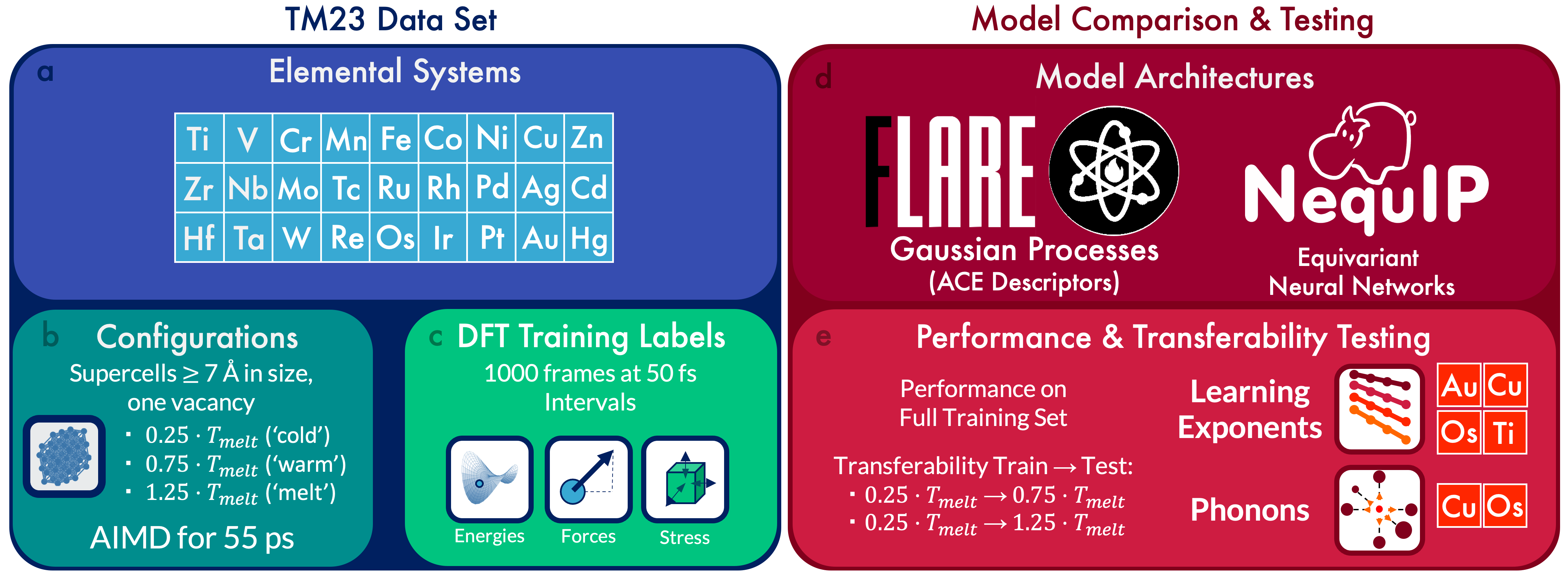}
\caption{Project Overview. (a) Complete table of transition metals studied, and the size requirement for super-cell creation.
(b) AIMD sampling of atomic environments. 
(c) Extraction procedure and high-fidelity DFT calculations. 
(d) Models were trained on high-fidelity labels via the FLARE and NequIP codes. 
(e) We gauged the model performance by using test sets from all temperatures, and by assessing transferability from `cool' to `warm' and `cool' to `melt' sets of data. 
We also compute learning exponents for Au, Cu, Ti, and Os and phonon dispersion curves for Cu and Os.}
    \label{fig:tableofcontents}
\end{figure*}

Such a data set curation task has the potential to benefit more than just MLFF developers.
Within the domains of surface science, heterogeneous catalysis, and alloys, crucial mechanisms such as surface restructuring \citep{Lim2020EvolutionDynamics}, active site dynamics, and cooperative reaction mechanisms on metal surfaces occur on long time- and length-scales \citep{Johansson2022Micron-scaleLearning}, making gains in MLFF efficiency and accuracy critical for the interpretation of experimental observables.
With important use-cases like this in mind, it behooves us to understand the role that varying atomic representations and model fidelity can play across machine-learning architectures, and to make the TM23 data set available for the MLFF development and broader scientific community. 
More available testing benchmarks will aid the overall project of developing and judging both purpose-built and general purpose FFs which can be flexibly applied to a wide variety of systems.

Finally, of central importance to this work is a demonstration of a new mode of use of MLFFs as a probe to extract fundamental physical and chemical trends from first-principles reference data, and to obtain a better understanding of the relationship of the predictive errors to the parameters employed in the underlying quantum mechanical method.
Specifically, by varying the DFT and MLFF model hyperparameters, such as the electronic temperature, or body-order and angular resolution of the representation, and comparing the accuracy with respect to first principles calculations, we can directly assess the complexity of the quantum many-body interatomic interactions and connect it to domain-knowledge intuition in terms of electronic structure of different metals.

\section{Results}
\subsection{Overview of the Benchmarking Task}
\label{sec:data set_task_overview}
We begin by presenting the TM23 data set, comprised of \emph{ab initio} molecular dynamics simulations of 27 \textit{d}-block metals, from which we sample a subset of training structures and associated energy, force, and stress labels from high-fidelity DFT calculations.
These benchmark data are then used to evaluate the performance of two different MLFF architectures employing different representations of the atomic environments.
To limit the breadth of comparison between the swath of available MLFF approaches, we only focus on two architectures: (1) Gaussian processes based on the atomic cluster expansion descriptors and (2) equivariant neural networks. 
Open-source implementations of these architectures are employed, namely FLARE \citep{Vandermause2020On-the-flyEvents} and NequIP \citep{Batzner2021E3-EquivariantPotentials}, respectively.
By evaluating a large set of relevant elements across two model architectures, it is demonstrated that even chemically simple systems -- mono-elemental bulk materials with a single vacancy in low-temperature crystalline, high-temperature crystalline, and molten phases -- present significant challenges for accurate learning by these leading models and a persistent trend of errors across the \textit{d}-block of the periodic table.

\begin{figure*}
  \centering
    \includegraphics[width=\textwidth]{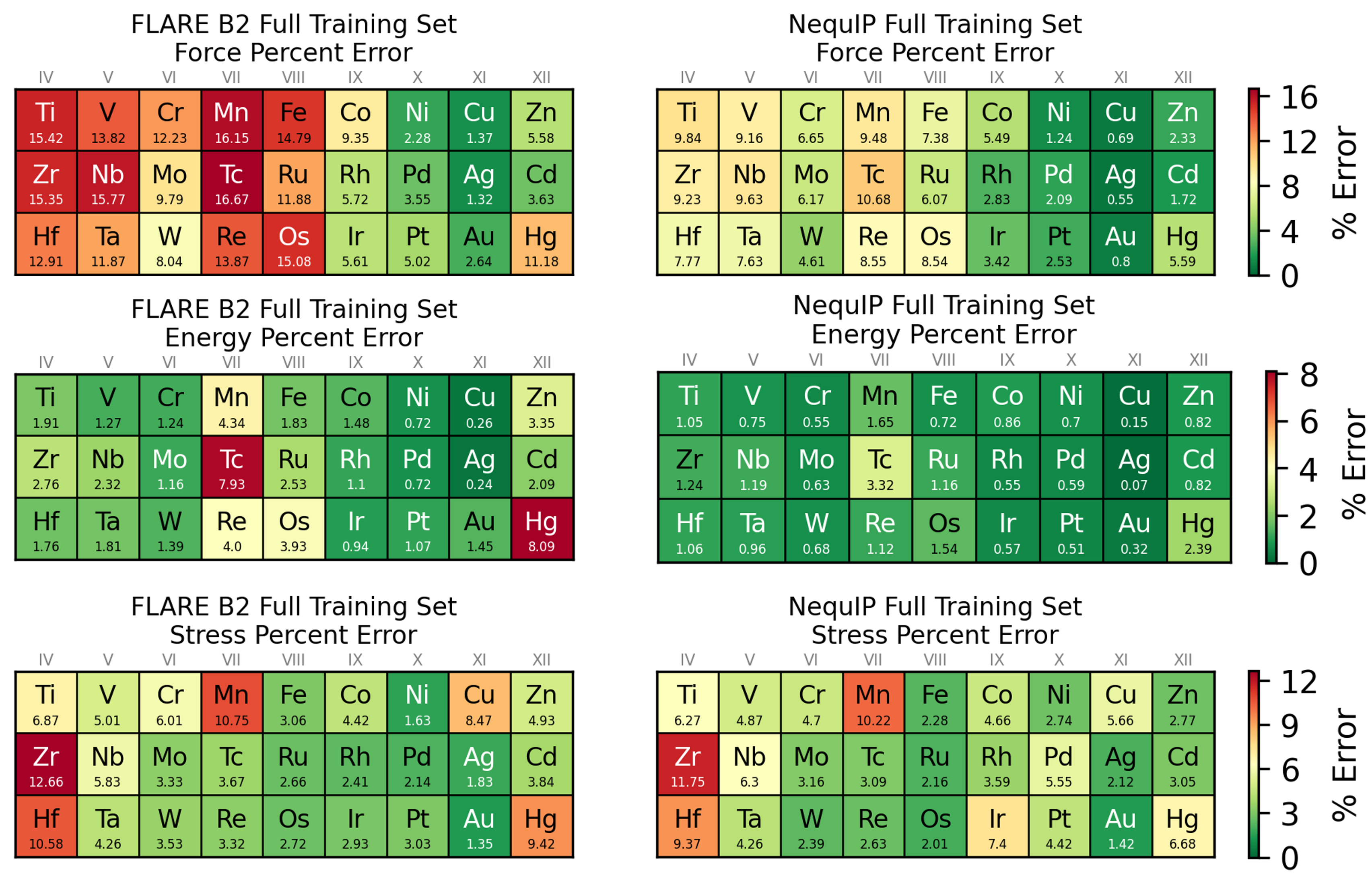}
\caption{Force, energy, and stress percent ($\%$) errors for the 27 TM systems for FLARE, using the B2 invariant ACE descriptors with $\zeta=2$ and NequIP.
Errors for NequIP are almost uniformly lower than for FLARE (with some exceptions in stress prediction). 
A version of this figure containing the MAEs is provided in Fig. \ref{fig:flare_v_nequip_main_mae}, and B1 results and MAEs for all other labels and models are provided in the Supplementary Information.
The layout for each panel reflects the \textit{d}-block of the periodic table, starting from Group IV (Ti, Zr, $\&$ Hf) and extending to Group XII (Zn, Cd, $\&$ Hg). 
A uniform color scheme is employed across all models and elements, where green represents the lowest $\%$ errors whereas red represents the highest $\%$ errors. }\label{fig:flare_v_nequip_main}
\end{figure*}

In benchmarking the accuracy of these methods, trade-offs that arise with differing model fidelities are explored and reasonable references in accuracy are provided for other practitioners approaching these systems.
A schematic of the benchmarking workflow is provided in Fig. \ref{fig:tableofcontents}, and the explicit details of each component are provided in Section IV.A. 
Importantly, clear trends in accuracy are found across the \textit{d}-block which hold regardless of model architecture, as quantified by relative errors of force, energy, and stress.
These benchmarks demonstrate the role that model architectures play across the same set of first principles reference data, where the choice of model parameters can influence their performance in a marked fashion for the more `difficult' metals as opposed to `easier' metals.
The TM23 data set is made freely available (upon publication) in order to facilitate direct comparison across the wide range of actively developed MLFF methods, and it is envisioned that this will act as a useful benchmark comparison target within the computational materials science community.

Lastly, we note that the data set used here was not intended to provide high-fidelity models for dynamic evolution of these systems. 
Rather, the primary task was to uncover the drastic differences in predictive accuracy across transition metals as well as the fundamental relationships between the DFT parameters employed, meaning that future work will consider the dynamic evolution of these systems and, importantly, if further augmentation of the data set is necessary.

\begin{figure*}
  \centering
    \includegraphics[width=\textwidth]{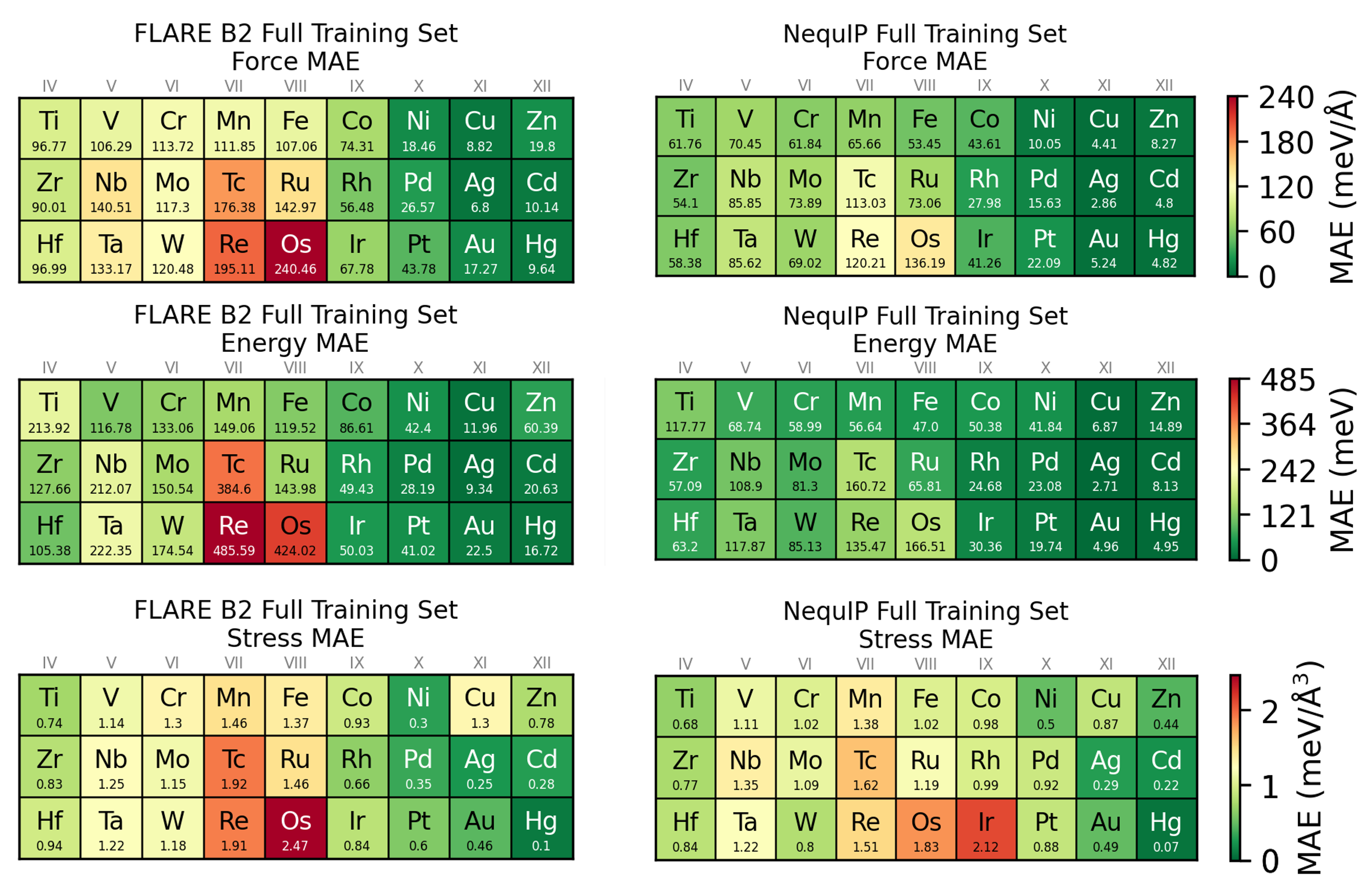}
\caption{Force, energy, and stress mean absolute errors (MAE) for the 27 TM systems for FLARE, using the B2 invariant ACE descriptors with $\zeta=2$ and NequIP.
Aside from the difference in error reporting, this Fig. is otherwise laid out identically to Fig. \ref{fig:flare_v_nequip_main}. Note that MAE label values scale with the magnitude of the forces, which in turn scale with the melting temperature $T_m$ of particular metals due to referencing AIMD temperatures against $T_m$ on a per-metal basis- in other words, higher melting point metals tend to have higher MAEs not only due to model error but to larger variation in the force values. See Section S2, Fig. S2, and Fig. S30-S56 in the SI for further details. 
}\label{fig:flare_v_nequip_main_mae}
\end{figure*}

\begin{table}[!htbp]
\resizebox{\columnwidth}{!}{\begin{tabular}{|c|c|c|c|c|c|}
\multicolumn{1}{c}{\bf }  \vline  & \multicolumn{2}{c}{\bf FLARE} \vline & \multicolumn{3}{c}{\bf   NequIP} \\
\multicolumn{1}{c}{\bf }  \vline  & \multicolumn{1}{c}{\bf ACE} \vline & \multicolumn{1}{c}{\bf   Total Train.} \vline & \multicolumn{1}{c}{\bf   } \vline & \multicolumn{1}{c}{\bf   Train. Speed} \vline & \multicolumn{1}{c}{\bf   GPU}\\
\multicolumn{1}{c}{\bf Elem. }  \vline  & \multicolumn{1}{c}{\bf Dim.} \vline   & \multicolumn{1}{c}{\bf   Time (hr)} \vline & \multicolumn{1}{c}{\bf   N$_{\text{wts}}$} \vline & \multicolumn{1}{c}{\bf   (hr/epoch)} \vline & \multicolumn{1}{c}{\bf   Arch.}\\
\hline
Ag & 180 & 18.799 & 265580 & 4.037 & A100 \\
\hline
Au & 220 & 39.308 & 612744 & 3.680 & A100 \\ 
\hline
Cd & 364 & 11.365 & 265580 & 3.375 & A100 \\ 
\hline
Co & 330 & 13.747 & 265580 & 2.760 & A100 \\
\hline
Cr & 270 & 9.904  & 265580 & 4.728 & A100  \\
\hline
Cu & 140 & 16.363 & 265580 & 7.271 & V100 \\
\hline
Fe & 462 & 10.499 & 612744 & 46.000& V100 \\
\hline
Hf & 225 & 8.348 & 265580 & 5.440  & V100 \\
\hline
Hg & 225 & 9.869 & 265580 & 1.780  & A100 \\
\hline
Ir & 765 & 18.367 & 612744 & 17.121& A100 \\
\hline
Mn & 462 & 13.497 & 265580 & 13.455& V100 \\
\hline
Mo & 330 & 11.357 & 265580 & 11.074& A100 \\
\hline
Nb & 225 & 10.711 & 265580 & 3.234 & A100 \\
\hline
Ni & 330 & 18.375 & 612744 & 10.897& V100 \\
\hline
Os & 225 & 11.833 & 265580 & 1.961 & A100 \\
\hline
Pd & 225 & 10.420 & 612744 & 25.407& V100 \\
\hline
Pt & 225 & 10.270 & 265580 & 1.486 & A100 \\
\hline
Re & 225 & 14.124 & 265580 & 3.484 & A100 \\
\hline
Rh & 330 & 13.770 & 265580 & 0.850 & V100 \\
\hline
Ru & 225 & 12.128 & 612744 & 6.453 & V100 \\
\hline
Ta & 225 & 11.543 & 265580 & 10.428& A100 \\
\hline
Tc & 225 & 15.695 & 265580 & 2.695 & V100 \\
\hline
Ti & 270 & 16.619 & 265580 & 17.391& V100 \\
\hline
V  & 270 & 12.413 & 265580 & 11.146& V100 \\
\hline
W  & 330 & 11.342 & 265580 & 8.543 & A100 \\
\hline
Zn & 364 & 13.429 & 265580 & 3.288 & A100 \\
\hline
Zr & 225 & 8.138 & 612744 & 2.564 & A100 \\
\hline
\end{tabular}}
\caption{FLARE ACE descriptor dimensions, total training times, and number of CPUs, as well as NequIP network weights, training speeds, and GPU architectures employed for the models used to generate the results in Fig. \ref{fig:flare_v_nequip_main}. All NequIP models were trained using a single GPU. \label{tab:size}}
\end{table}

\subsection{TM23 Data Set Description}
Super-cells for 27 transition metals were created from their experimentally verified ground state crystal structures, as provided by the Materials Project \citep{Jain2013} at their 0 K lattice constants predicted using the Perdew-Burke-Ernzerhof (PBE) exchange-correlation functional \citep{Perdew1996b}.
Super-cells were generated with a requirement that each lattice vector was at least 7 \AA{} in length. 
This value was chosen to balance the number of unique atomic environments observed within the finite radius cutoff of MLFF local structure representations against the number of atoms required for DFT calculation.
A single vacancy was introduced into each super-cell in order to diversify non-trivial atomistic configurations.
\textit{Ab initio} molecular dynamics (AIMD) simulations were then performed at three temperatures: 0.25$\cdot T_{melt}$, 0.75$\cdot T_{melt}$, and 1.25$\cdot T_{melt}$, where $T_{melt}$ is the experimental melting temperature, for a total of 55 ps.
Each system was evaluated at a single volume as defined by the ground state lattice parameters.
We henceforth refer to these temperatures as `cold', `warm', and `melt', respectively).
Melting temperatures for each system were extracted from \cite{macmillandata}.
We note that even though these are short time-scale AIMD simulations, the corresponding radial distribution functions for each temperature were analyzed to confirm the loss of crystalline-order at increased temperature.
Examples of these crystalline and amorphous radial distribution functions are provided in the SI in Fig. S57-S83.
This analysis is provided as a Jupyter notebook paired with the data set on Materials Cloud.

The first 5 ps of each trajectory was used for thermal equilibration at the desired temperature, and subsequently, representative frames (structure snapshots) were extracted from the remaining 50 ps of the trajectory at 50 fs intervals, in order to reduce their correlation.
Extracted frames were used as input for high-fidelity static DFT calculations, which produced the energy, force and stress labels used for model training and testing.
To accelerate each AIMD trajectory, \textbf{k}-point sampling was limited to only the $\Gamma$-point, whereas finer \textbf{k}-point grids were used for generating training labels for each extracted frame.
Spin-polarization was not included for all high-fidelity DFT frames, which is discussed in more detail in Section IV.A. 
This procedure yielded 1000 frames for each of the three AIMD temperatures and thus 3000 in total for each metal.
The entire collection of high-fidelity frames (81,000 in total) will be provided via the Materials Cloud upon publication.

\subsection{MLFF Training on TM23}
We then trained and tested both FLARE and NequIP models using energy, force and stress labels for each metal using the set of 3000 high-fidelity static DFT calculations.
For our first test, we obtained a holistic overview of model performance by using as wide a training/testing set as possible.
For each element, the training set drew from a combined set of 2700 frames using the first 900 extracted frames of each AIMD trajectory at the three temperatures.
The validation set for NequIP was chosen to be 10$\%$ of the training set, selected randomly.
The training weights in NequIP for forces and stress were set to 1, while the energies also employed a coefficient of 1, but used the `PerAtomMSELoss' as discussed in the NequIP github repository.
In the FLARE code, weights were not set for the energy, forces, and stresses, rather the noise hyperparameters were initially set according to the discussion in the Methods, and are optimized over the course of model training.
The test set contained a total of 300 frames taken to be the final 100 frames from each AIMD trajectory at the three temperatures considered.
Since prediction accuracy is influenced by the model parameters, \emph{e.g.}, representation cutoff, radial and angular bases for FLARE, or neural network depth and angular resolution in the case of NequIP, these model parameters were explicitly tested for each metal using a grid search on a smaller training set (200 frames) and test set (100 frames).
Results for each model architecture employing the best parameters are shown here in the main text, and the complete set of values is provided in Section S6. 
To put these results in the context of model size, we also provide the descriptor dimensions used for the FLARE models and number of weights used in the NequIP models in Table \ref{tab:size}. The descriptor dimension, $n_{\text{d}}$ is calculated using Eqn. \ref{eq:des}, given as
\begin{equation}\label{eq:des}
    n_{\text{d}} = (n_{\text{max}} \cdot n_{\text{species}} + 1) \cdot n_{\text{max}} \cdot \frac{n_{\text{species}}}{2} \cdot (\ell_{\text{max}} + 1).
\end{equation}

We then provide interpretation of the underlying trends across compositions in predictive accuracy, and examine the influence of model error on observable material properties.

\subsection{Accuracy Comparison of FLARE and NequIP} \label{comp}
Before proceeding, we comment upon the choice of error metric for each of the target labels.
Force errors are expressed as percentages, defined in Equation \ref{eqn:percs}, rather than mean absolute values (\textit{e.g.}, MAEs or RMSEs), since these quantities correlate with the average magnitude of the force and simultaneously the AIMD temperature, which naturally varies across metals.

\begin{equation}\label{eqn:percs}
    \% \textrm{Error} = \frac{\textrm{MAE}}{\textrm{MAV}}\cdot 100
\end{equation}

Full MAEs for forces, energies, and stresses are provided in Section S5. 
A comparison of force MAEs to the melting temperature are provided in Section S2 as a demonstration of this relationship.
Hence, test percent errors for the best FLARE and NequIP models on TM23 are provided in Fig. \ref{fig:flare_v_nequip_main} where trends in accuracy across the \textit{d}-block can be immediately observed.
Moving from left to right across Fig. \ref{fig:flare_v_nequip_main}, we see that early TMs exhibit higher test errors across forces, energies, and stresses relative to the late Pt-group and coinage metals (Groups IX, X, and XI, respectively), with the coinage metals producing the lowest errors of the entire set.

\begin{figure*}
  \centering
  \includegraphics[width=\textwidth]{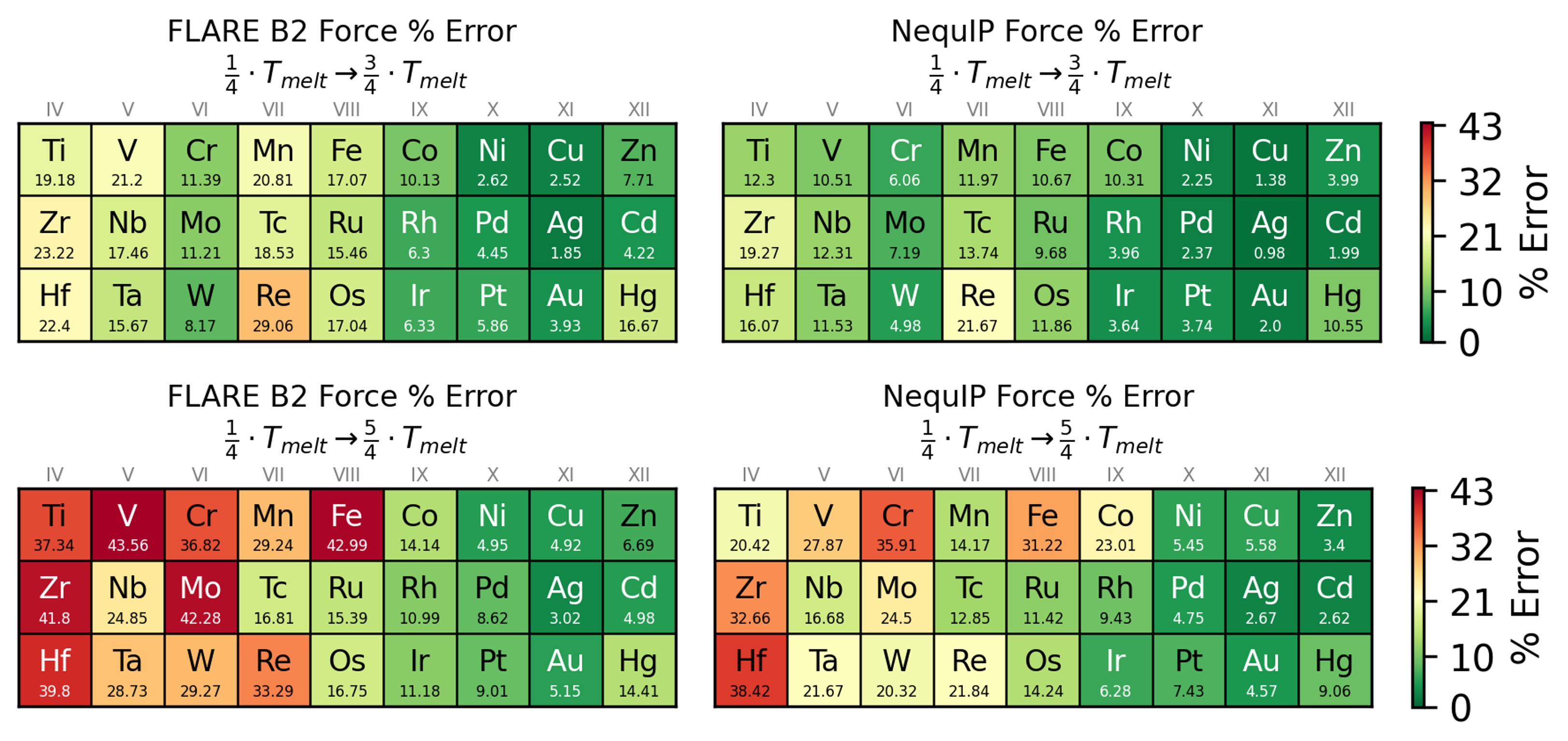}
    \caption{FLARE and NequIP force percent ($\%$) errors for the 27 TM systems by training on 1000 frames at 0.25$\cdot T_{melt}$ and testing on either 1000 frames at 0.75$\cdot T_{melt}$ (left panels) or 1.25$\cdot T_{melt}$ (right panels).
    FLARE values are obtained using the ACE B2 descriptor at $\zeta=2$.
    The formats again reflect the \textit{d}-block of the periodic table. 
    A uniform color scheme is employed across all models, where green represents the lowest $\%$ errors whereas red represents the highest $\%$ errors.}
    \label{fig:transferability}
\end{figure*}

\begin{figure*}
  \centering
  \includegraphics[width=\textwidth]{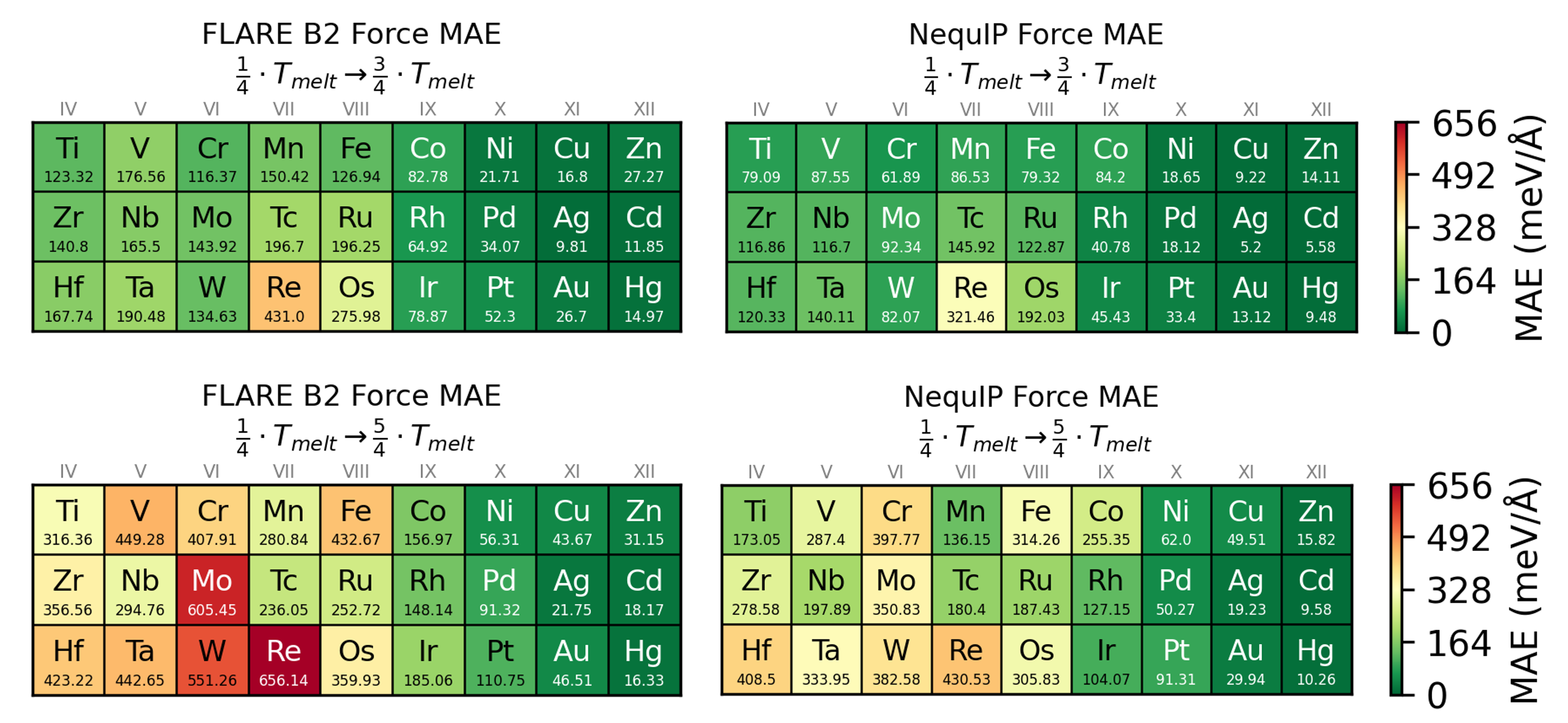}
    \caption{FLARE and NequIP mean absolute errors for the 27 TM systems by training on 1000 frames at 0.25$\cdot T_{melt}$ and testing on either 1000 frames at 0.75$\cdot T_{melt}$ (left panels) or 1.25$\cdot T_{melt}$ (right panels). Layout is otherwise identical to \ref{fig:transferability}.
}
    \label{fig:transferability_mae}
\end{figure*}

A trend across models can also be observed, where NequIP test errors are systematically lower than FLARE using ACE B2 descriptors.
FLARE test errors obtained using the ACE B1 descriptor are provided in the Supplementary Information, which are systematically higher than FLARE using ACE B2 and NequIP.
The observed trend between FLARE B2 and B1 ACE descriptors can be explained by an increase in effective body-order, where the 2-body B1 descriptor with a kernel power of 2 yields an effective but not complete 3-body interaction between environments, while the 3-body B2 descriptor at the same power yields 5-body terms.
Despite an overall improvement in the force $\%$ errors using NequIP compared to either of the FLARE models, the same trends remain persistent when comparing model performance across the \textit{d}-block metals, in that early transition metals exhibit noticeably higher errors than late transition metals across all model architectures.
Additionally, we note that we explicitly explored the relationship between observed model accuracy and \textbf{k}-point density, as is provided in Fig. S1.
We employ the minimum \textbf{k}-point density as a label for the inherent accuracy of the DFT calculations, and find no correlation to model test error.

Curiously, when moving from the coinage metals in Group XI to the Group XII transition metals (Zn, Cd, and Hg), an increase in the $\%$ error is also observed.
However, Group XII metals exhibit the lowest melting points of the elements considered in this study, meaning that the absolute magnitudes of the forces are small. 
Coupling this observation to the fact that the force MAEs for Group XII systems are on the order of only 10s of meV/\AA{}, it can be reasoned that these systems are sampling near the inherent noise of DFT given the convergence protocols, rather than reflecting difficulties in model learning.

The $\%$ error values in Fig. \ref{fig:flare_v_nequip_main} were determined using the entire training set of 2700 frames and testing on the remaining 300 frames.
Corresponding mean absolute error values for the same models and data are provided in Fig. \ref{fig:flare_v_nequip_main_mae}.
The training procedures for each model architecture are described in explicit detail in Section IV.C., and differ since FLARE implements a Gaussian process whereas NequIP is a neural network.
FLARE uses sparsification and MPI parallelization to circumvent the memory bottleneck present when considering the full training set of 2700 frames, where sparse representative atomic environments are selected from each training frame by the predictive uncertainty of the GP \cite{vandermause2022active}.
This yielded a total sparse set of 2700 atomic environments. 
On the other hand, the NequIP model trains using all atomic environments from each frame.

\subsection{Transferability of FLARE and NequIP Across Temperatures and Phases of TM23}
Measuring a model's ability to generalize beyond the training set distribution is of central interest for the development of MLFFs: a long simulation especially involving reactions and structural evolution may encounter new configurations not sampled in the training set. 
A valuable advantage of the TM23 data set is that the training data contains three distinct temperatures for each metal.
This provides an opportunity to explore the extrapolative ability of ML models between temperatures and structural phases, where thermal disorder can produce dramatically different atomic environments.

This latter statement is confirmed through the analysis of the radial distribution functions of the trajectories, where 0.25$\cdot T_{melt}$ retains crystalline order, reflected by the persistence of well-defined peaks which broaden and disappear at higher temperatures, indicating a transition to a molten phase.

We explore model transferability across temperatures in a similar fashion to an earlier work on motion of a single-molecule \cite{Kovacs2021LinearRMSE}. 
Models are thus trained using low temperature frames (0.25$\cdot T_{melt}$) and are tested using frames from higher temperatures (0.75$\cdot T_{melt}$ or 1.25$\cdot T_{melt}$).
Results for FLARE (using ACE B2) and NequIP models are provided in Fig. \ref{fig:transferability} and \ref{fig:transferability_mae}.
We note similar accuracy trends to the full multi-temperature training set results of Fig. \ref{fig:flare_v_nequip_main} and \ref{fig:flare_v_nequip_main_mae}.
With regards to metal-dependent performance, Group IX, X, and XI metals yield markedly lower errors than early transition metals (Group VIII and below) in forces, energies, and stresses for both transferability tasks. 
Trivially, both FLARE and NequIP yield lower predictive errors when tested on `warm' as opposed to the `melt' frames.
From the model architecture perspective, the global trend noted in the previous section remains consistent, namely that NequIP outperforms FLARE at the B2 descriptor, yielding lower errors across nearly all of the metals.

\begin{figure*}
    \centering
    \includegraphics[width=\textwidth]{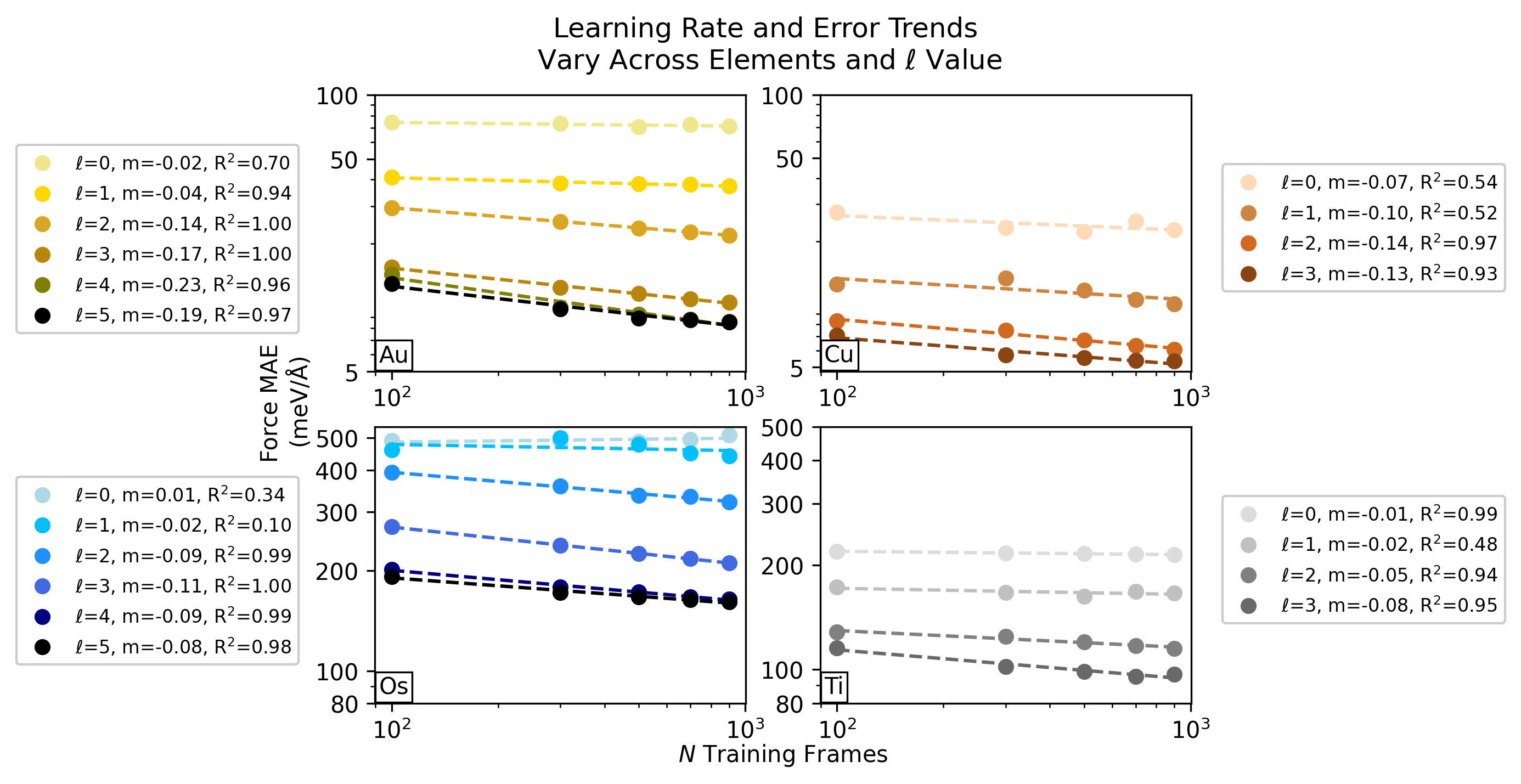}
    \caption{Model force MAE as a function of the number of training frames using varying $\ell_{max}$ values in NequIP for Au, Cu, Os, and Ti. 
    Power-law fits are made to the data in log-log space to obtain the training exponents. 
    To determine the robustness of each fit, $R^2$ values are also provided.}
    \label{fig:training_exponents}
\end{figure*}\

\subsection{Disparate Learning Behaviors of NequIP Across TM23} \label{exps}
In an effort to better understand the error as a function of the available training set size and model parameters, we investigated the effect of the NequIP architecture on the training exponent of a subset of these systems, a procedure discussed in the original NequIP paper \cite{Batzner2021E3-EquivariantPotentials}.   
It has been observed that the test error of deep learning systems follows a power law of the form $\epsilon = a N^b$, where $\epsilon$ refers to the predictive error, $N$ is the training set size, and $a$ and $b$ are constants. 
In \cite{Batzner2021E3-EquivariantPotentials}, it was shown that equivariant interatomic potentials, with the tensor rank, or angular rotation order of spherical harmonics, $\ell_{max} \geq 1$ exhibit higher values of $b$ as compared to invariant methods ($\ell_{max}=0$), meaning they learn faster with the number of data points. 
Here, we increase the value of $\ell_{max}$ (see Section IV.C. in for more detail) within the NequIP architecture, and demonstrate differences in the learning behavior.
The complete set of model weights, training speeds, irreducible feature coefficients, and computational architecture are provided in Table \ref{tab:el}.
To determine differences in the learning behavior across a subset of the metals, we vary $\ell_{max}$ from $0$ (limiting the model to invariant scalar features) to $\ell_{max} = 3$ (a fully equivariant model). 
For two metals (Au and Os), $\ell_{max}$ is increased further to 4 and 5 to explore the effect of even higher angular resolution of the equivariant representation.
Models are trained on data sets ranging from 100-900 frames taken from the 1.25$\cdot T_{melt}$ AIMD trajectory for each metal, in increments of 200, and force MAEs are employed to study learning dependencies on model architecture. 

\begin{table}[!htbp]
\resizebox{\columnwidth}{!}{\begin{tabular}{|c|c|c|c|}
\multicolumn{1}{c}{\bf }  & \multicolumn{1}{c}{\bf }  & \multicolumn{1}{c}{\bf   Train. Speed} & \multicolumn{1}{c}{\bf   Comp.} \\
\multicolumn{1}{c}{\bf Elem.$_{\ell_{max}}$} \vline   & \multicolumn{1}{c}{\bf  N$_{wts}$ } \vline & \multicolumn{1}{c}{\bf   (hr/epoch)} \vline & \multicolumn{1}{c}{\bf   Arch.} \\
\hline
Au$_{0}$ & 76346  & 0.019 & V100 \\
\hline
Au$_{1}$ & 137274 & 0.068 & V100 \\
\hline
Au$_{2}$ & 142138 & 0.223 & V100 \\
\hline
Au$_{3}$ & 134970 & 0.551 & V100 \\
\hline
Au$_{4}$ & 122234 & 0.818 & V100 \\
\hline
Au$_{5}$ & 106974 & 1.126 & V100 \\
\hline
Cu$_{0}$ & 75832  & 0.029 & A100 \\
\hline
Cu$_{1}$ & 136760 & 0.216 & A100 \\
\hline
Cu$_{2}$ & 141624 & 0.911 & A100 \\
\hline
Cu$_{3}$ & 134456 & 2.017 & A100 \\
\hline
Ti$_{0}$ & 75832 & 0.023 & A100 \\
\hline
Ti$_{1}$ & 136760 & 0.151 & A100 \\
\hline
Ti$_{2}$ & 141624 & 0.559 & A100 \\
\hline
Ti$_{3}$ & 134456 & 1.291 & A100 \\
\hline
Os$_{0}$ & 76346  & 0.011 & V100 \\
\hline
Os$_{1}$ & 137274 & 0.042 & V100 \\
\hline
Os$_{2}$ & 142138 & 0.126 & V100 \\
\hline
Os$_{3}$ & 134970 & 0.289 & V100 \\
\hline
Os$_{4}$ & 122234 & 0.462 & V100 \\
\hline
Os$_{5}$ & 106974 & 0.666 & V100 \\
\hline
\end{tabular}}
\caption{NequIP network weights, training speeds, and GPU architectures employed for the models at various values of $\ell_{max}$ used to generate the results in Fig. \ref{fig:training_exponents}. All NequIP models were trained using a single GPU. All training speeds were computed for training sets of 900 frames. The numerical coefficients of the irreducible feature representations were 64, 32, 16, 8, 4, and 2 for models with $\ell_{max}$ 5, 4, 3, 2, 1, and 0, respectively. \label{tab:el}}
\end{table}

Au and Cu are chosen as a representative subset of the `easy' metals, whereas Ti and Os represent the more `difficult' systems.
From Fig. \ref{fig:training_exponents}, we draw three immediate conclusions.
The first is that the error magnitude for all metals is markedly affected by increasing $\ell_{max}$.
Secondly, the learning exponent $m$ differentiates each model's ability to learn the forces as more training data is made available on a metal-basis: slopes are larger for the Au and Cu models using $\ell_{max}$ = 3, relative to Os and Ti, mean that Au and Cu models learn faster with new data than Os and Ti.
The Cu and Ti models are truncated at $\ell_{max} = 3$ given the results for Au and Os, where increasing the angular resolution to higher values increases computational cost of inference but does not bring about marked increases in predictive accuracies.

\begin{figure*}
    \centering
    \includegraphics[width=1.0\textwidth]{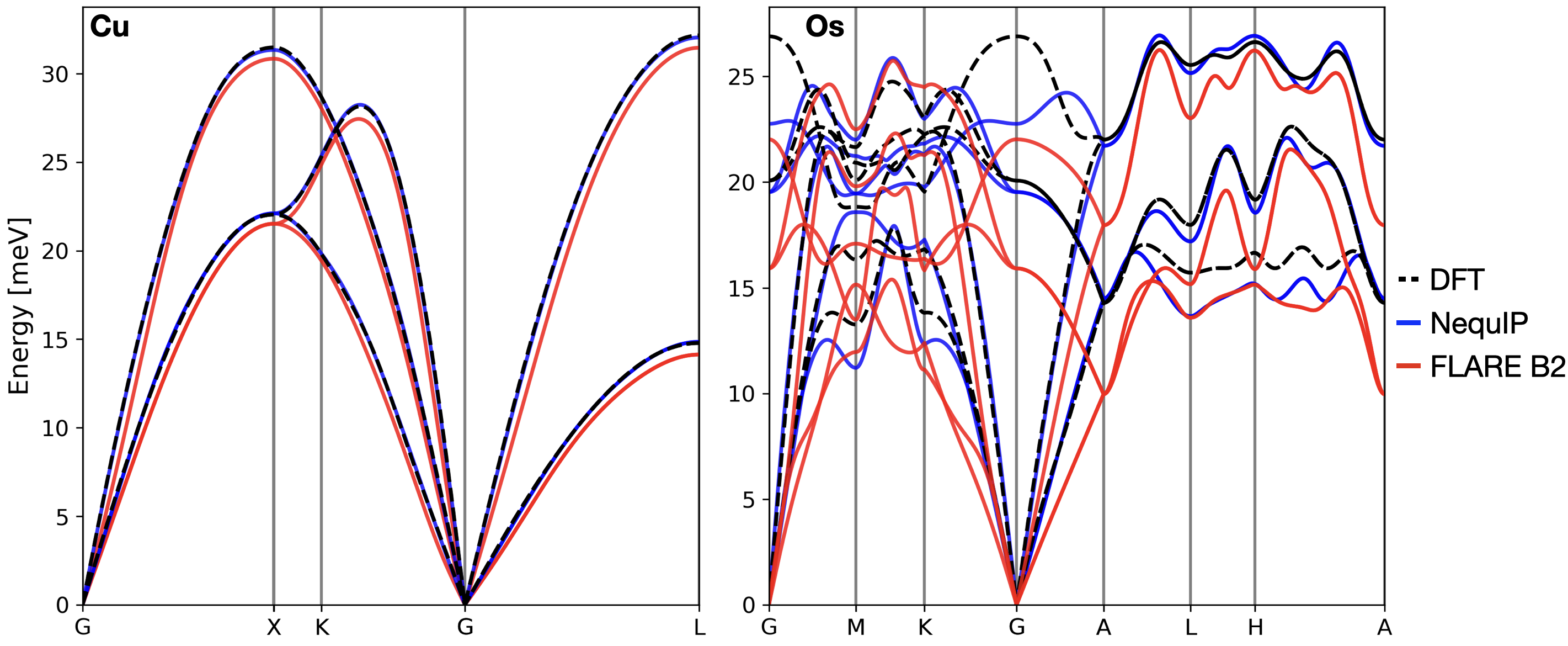}
    \caption{Phonon dispersion curves for Cu and Os where DFT is a dashed, black curve, NequIP is blue, and FLARE B2 is red. Each model curve is from the full training set, and employs optimal model parameters determined from grid test.}
    \label{fig:material_properties}
\end{figure*}\

Finally, a subtler and provoking conclusion from these data is that the observed increase in learning exponent, as a function of increasing $\ell_{max}$, revises previous understanding of the learning dynamics for equivariant models.
In the original work by \cite{Batzner2021E3-EquivariantPotentials}, power-law exponents were computed for the water data set of Cheng et al. \cite{cheng_water}
It was observed that the absolute value of the learning exponent increased when moving from $\ell_{max}=0$ to $\ell_{max}>0$ and the magnitude of the test error decreased, meaning that equivariant models with $\ell_{max}>0$ learn faster and attain lower overall error. 
That work also observed `diminishing returns', as the absolute value of the learning exponent increases the most from $\ell_{max}=0$ to $\ell_{max}=1$, significant, but smaller increases from $\ell_{max}=1$ to $\ell_{max}=2$, and then successive increases in $\ell$ presenting more modest changes to the learning exponent.
While we find that increasing $\ell_{max}$ tends to reduce the overall magnitude of error in all cases, we also find that the changes in learning rate differ across all metals.
However, the higher angular-resolution models do still present an overall lower error, with evidence of possible saturation for high $\ell_{max}$ values.
Furthermore, the biggest relative changes in the learning exponents do not occur between $\ell_{max}=0$ to $\ell_{max}=1$ in all four cases (Au, Cu, Os, and Ti), contrasting with the case for water in Reference \citep{Batzner2021E3-EquivariantPotentials}, where the influence of equivariance was largest from $\ell_{max} = 0$ to $\ell_{max} = 1$. 
Here, we see that the learning exponent steadily increases with increasing $\ell_{max}$ for Au until $\ell_{max}=4$ and Ti until the upper bound of $\ell_{max}=3$ is reached.
For Cu, the effect of $\ell_{max}$ seemingly saturates between $2$ and $3$ for Cu, which is similar to Os.
We note that the number of parameters (weights) and internal equivariant operations/tensor products in the neural network varies significantly with $\ell_{max}$, which may also confound these results.

\subsection{Finite Size Effects on Learning Difficulties}
\begin{figure}
    \centering
    \includegraphics[width=\columnwidth]{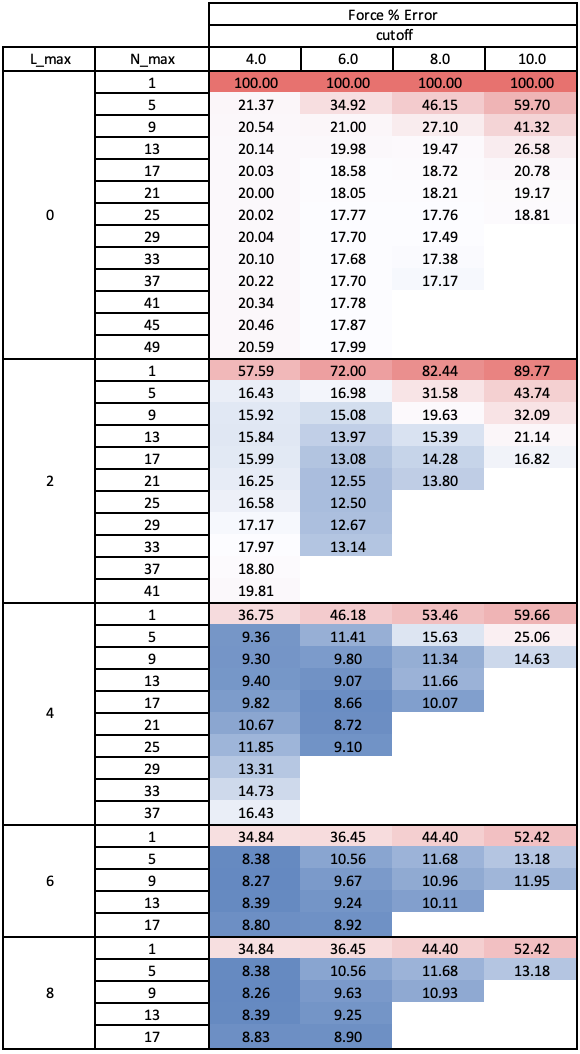}
    \caption{Force \% Error observed using FLARE B2, with power = 2 for a 640 atom unit cell of Os. The models are colored using a blue-white-red scale, with dark blue denoting lowest force \% error. The best models are employ $r_{cut}$ = 4.0 \AA{}, but with high values of $\ell_{max}$ of 6 and 8, and $n_{max}$ of 9. Blank cells denote models that hit a memory limit using a single, 48 CPU node for FLARE.}
    \label{fig:grid}
\end{figure}

We also considered the effect of unit-cell size on the ability of these MLFFs to adequately explore atomic representations up to longer cutoff radii.
This procedure is described in Section S7.B., where a 640 atom super-cell of Os was surveyed in AIMD at 1.25$\cdot T_{melt}$, and sequential frames were extracted to yield 200 train and 50 test frames.
A coarse grid test was then performed using FLARE B2, up to an r$_{cut}$ of 10 \AA{}, $\ell_{max}$ of 8, and $n_{max}$ of 49, the results of which are provided in Fig. \ref{fig:grid}. 
These results demonstrate that even if more unique atomic environments are made available to the model radially, within the cutoff of the representation, the `best' model still employs a short cutoff of 4 \AA{}, and high angular ($\ell_{max}$ = 6 or 8) and radial resolution (n$_{max}$ = 9 and larger).
This answers the question posed from the frames sampled within the TM23 dataset, which are shorter ranged, such that most atomic environments within the representation cutoff are periodic images.
These observations have direct implications in model design, where high angular and radial resolutions are required for early transition metals, which to this point, come with high computational cost to implement.

\subsection{Influence of Model Accuracy on 0 K Phonons}
To illustrate the influence of model error on material properties, a subset of the FLARE B2 and NequIP models trained across TM23 were used to calculate phonon dispersions.
The methods employed are provided in Section IV.D. 
Results for Cu and Os are shown in Fig. \ref{fig:material_properties}, and exhibit different levels of accuracy, consistent with the trends discussed previously on energy, force, stress labels.
Again, Cu represents an `easy' metal, whereas Os is markedly more difficult to learn as evidenced by lower MLFF accuracy on forces, energies, and stresses, as well as lower learning exponents.
Each model, using the full training set and the same as those presented in Fig. \ref{fig:flare_v_nequip_main}, is compared to ground-truth phonon dispersions obtained with DFT calculations.
In Fig. \ref{fig:material_properties}, both NequIP and FLARE (using ACE B2 with $\zeta = 2$) models for Cu perform very well in predicting the phonon band structure compared to DFT, whereas this task is more difficult for Os, with disagreement seen for various phonon band features. 
Acoustic phonon bands are better reproduced than optical phonon bands for NequIP and FLARE, with NequIP exhibiting better accuracy in both domains.
On the other hand, acoustic and optical bands for Os are systematically underestimated by FLARE, but the overall shapes of the bands are in qualitative agreement with DFT.
Systematic weakening of the phonon modes in this way can be explained by a slight overestimation of the Os \textit{hcp} $\hat{z}$-lattice vector of the primitive unit-cell predicted by FLARE (4.51 \AA{} versus 4.35 \AA{} predicted by NequIP).
A note must be made, however, that even though the NequIP model exhibits marginally more accurate vibrational spectra for both Cu and Os compared to FLARE using ACE B2, it is $\sim$ 200x slower than FLARE.
The trade-off between accuracy and efficiency can be made in the choice of models and architectures, and we provide the `best' model parameters for both FLARE and NequIP in Section S6 from minimization of the force MAE (and maximization of the model likelihood for FLARE) to aide in this decision.

\section{Discussion}
\subsection{Explanation of the Error Trends}
\subsubsection{Periodic Trends}
The observation of systematic variation in MLFF accuracy across the composition space in the TM23 data set points to potential underlying physical and chemical differences in the interatomic interactions for different metals. 
By comparing the performance of atomic representations with different many-body order and angular resolution we can extract insights into the fundamental complexity and character of bonding in transition metals. 
To generate human intuition, we interpret these test error trends in terms of both chemical and structural properties of metals.
Qualitatively, the test errors observed from Fig. \ref{fig:flare_v_nequip_main} loosely follow a trend with respect to \textit{d}-valence, where the metals with low numbers of valence \textit{d}-electrons exhibit higher errors compared to metals with full \textit{d}-shells.
In addition to \textit{d}-valence, the test errors also qualitatively follow a trend with respect to the initial crystal symmetry, as is shown in Fig. S3.
This is interesting, as the hexagonal crystals have two lattice parameters, which may provide a more difficult test-case with respect to the angular description of such materials.
However, this does not explain the observed errors for the body-centered-cubic metals.

\begin{figure*}
    \centering
    \includegraphics[width=\textwidth]{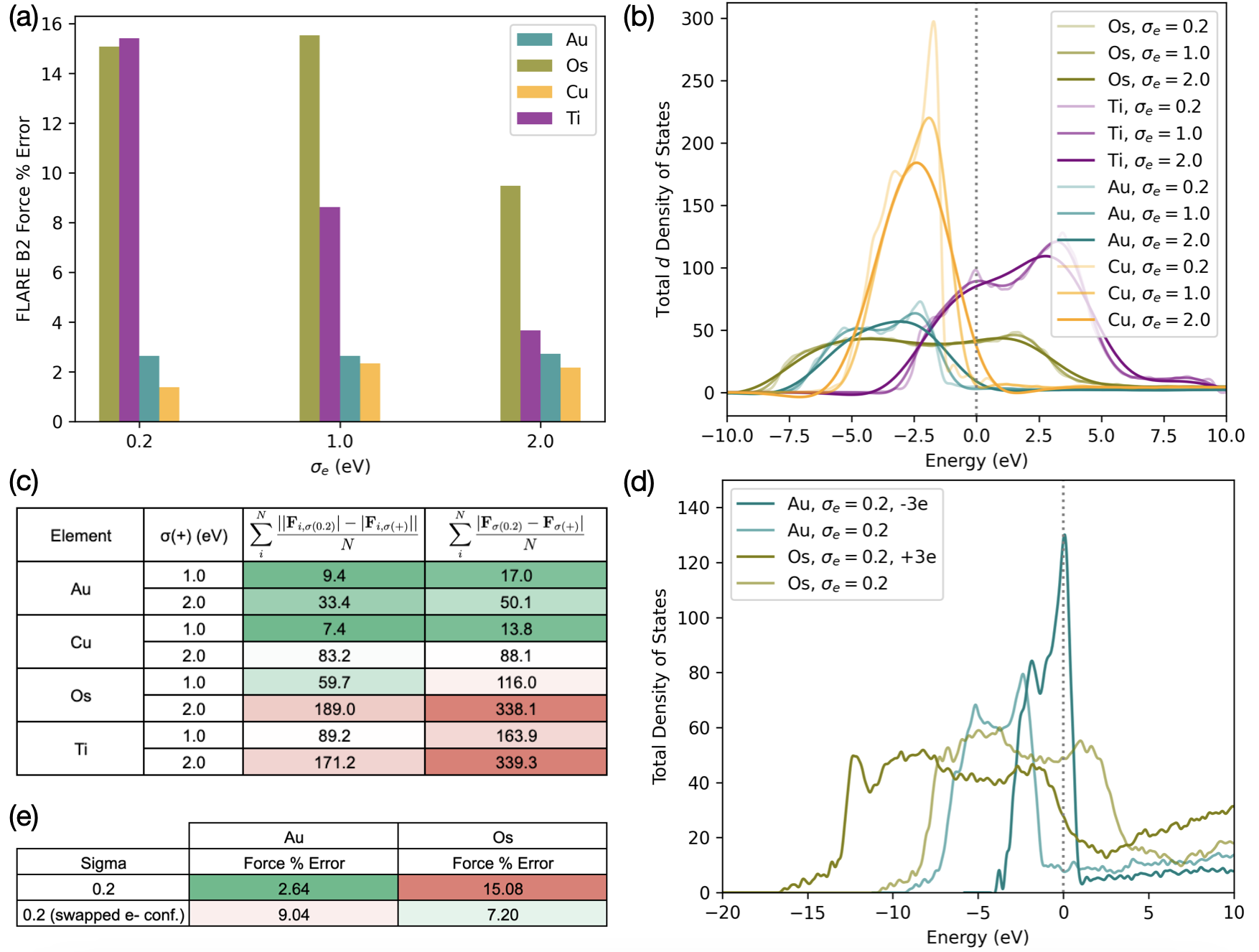}
    \caption{(a) FLARE (B2 with kernel power = 2) force percent test errors on TM23 data recomputed using higher values of the electronic smearing $\sigma_e$ for Au, Cu, Os, and Ti. (b) Total $d$ density of states for the four metals with different values of $\sigma_e$. (c) Mean absolute differences between force magnitudes (left column) and mean of the magnitude of the difference between the force vectors. (d) Total density of states for random frames from the `melt' test-set for Au and Os with and without modified total numbers of electrons. The values provided in the legend represent the total change in the number of electrons per atom. (e) FLARE (B2 with kernel power = 2) force percent test errors for Au and Os with unmodified, and swapped electronic configurations to illustrate the effect of artificially moving the Fermi level relative to the $d$-DOS on the resulting model errors.}
    \label{fig:final}
\end{figure*}

The higher relative errors of Group XII metals, disagreeing with this trend with \textit{d}-valence, are explained by the test errors for forces, energies, and stresses of these metals being on the same order of magnitude as the DFT noise, which are then compared to low-magnitude MAVs for these metals with relatively low melting points.
The trend with respect to \textit{d}-valence was first evaluated in the context of several previous studies noting the appearance of directional-bonding `behaviors' of early transition-metals \cite{PhysRevB.86.134106,PhysRevLett.106.246402,10.2307/53737}.
If substantiated, this correlation would partially explain why early transition metals require higher order angular resolution, as evidenced in the previous sections using both FLARE and NequIP. 
In \cite{PhysRevB.86.134106} and \cite{10.2307/53737}, the Cauchy pressure (defined by the relationship of elastic tensor components as $C_{12}-C_{44}$) is used as an indicator for directional bonding, which should be zero in pair-potentials.
To determine if this trend was present, using Cauchy pressure as a potential label, we extracted available elastic constants for each metal in TM23 from the Materials Project repository and compared to the observed NequIP force $\%$ error.
This comparison is provided in Section S4. 
Ultimately, no correlation is observed, but the quality of this label to determine directional bonding has also come under question \cite{GONG2020109174}.

\subsubsection{Electronic Structure and Density of States Influence MLFF Accuracy}
In this light, we additionally considered trends in another context, with respect to electronic structure, \textit{e.g.} \textit{d}-valence via \textit{d}-band center, which is commonly used as a label to differentiate chemical behaviors among the transition metals (\textit{e.g.} molecular adsorption and reactivity, especially in heterogeneous catalysis). 
The correlation of NequIP force $\%$ errors with the \textit{d}-band center of the metals is provided in the SI (Fig. S4), where a non-negligible linear trend is clearly observed between the force $\%$ test errors on the full set of data from NequIP and \textit{d}-band centers directly computed here using a small set (3) of random frames from the melted test set of each metal.
We also provide a complete representative set of the density of states plots of the TM23 metals in the SI (Fig. S5).
This result provides a strong indication that the discovered trends in test error across TM23 reveal a connection between the electronic structure and the many-body directional complexity of bonding present in these metals.

In seeking to quantitatively explain the test error trend among the transition metals with a measurable feature of the electronic structure, we also probed the correlation of test errors with DOS.
This choice was made because it gives us a lever to understand this relationship via the underlying DFT parameters, specifically how the smearing parameter $\sigma_e$ within the VASP calculation influences the electronic occupations and resulting forces. 
Efforts by Drautz et al. \cite{Drautz_2015,PhysRevB.74.174117} also established differences in angular characters of transition metals, which required special treatment (\emph{e.g.} explicit consideration of valence occupancy and moments of the density of states) for the construction of bond order potentials. 
Methfessel and Paxton \cite{PhysRevB.40.3616} showed that early transition metals have `more complicated' density of states profiles, with attention paid to metals like Zr exhibiting steep variations in the DOS at E$_{\text{fermi}}$ which, in the absence of smearing, results in `charge sloshing' during SCF calculations as the Fermi level position fluctuates. 
We extend this logic to the complexity of the PES as a function of atomic positions and the resulting difficulty of learning the PES.
Variations in local atomic configurations directly influence the electronic states and thus the energy of a configuration, and the occupancy changes of the bands (Kohn-Sham states at \emph{and near} E$_{\text{fermi}}$ are related to the potential energy surface).
Given this relationship, we then hypothesized that the labels computed using DFT for early transition metals, which have sharper DOS at and near E$_{\text{fermi}}$, would thus be more sensitive to slight perturbations in the atomic positions than metals with smoother states about the Fermi level.

We tested this hypothesis of the complexity of the DOS by systematically smoothing out the occupations for fixed configurations of Au, Cu, Os, and Ti by recalculating the force/energy/stress labels with increasing values of electronic smearing $\sigma_e$.
The intuition is that increasing the $\sigma_e$ value smears the electron occupations, producing a smoother DOS and thus letting us explicitly test the relationship between the complexity of the DOS and the observed test errors (even letting $\sigma_e$ vary to non-physical high values, \emph{e.g.} 1.0 and 2.0 eV). 
Moreover, since we have direct access to the atomic prediction targets, we can also determine the effect of changing this smearing on the force magnitudes, and importantly, angular distributions of force direction.
These results are provided in Fig. \ref{fig:final}, where panel (a) contains test errors for FLARE models (with B2 and kernel power 2) trained on the recomputed labels with different values of artificial smearing.
We can clearly see that the test error associated with models trained on increasingly smeared calculations decreases in Fig. \ref{fig:final}(a), with the high force \% errors for Os and Ti drastically reducing, to such an extent that they resemble the errors of Au and Cu in the case of Ti (despite the dramatic differences in their electron count) when a value of $\sigma_e$ = 2.0 eV is employed.
While we acknowledge that these values of $\sigma_e$ are extremely high, we find it striking that varying one parameter can so significantly reduce model fitting error with all else equal. 
This serves as a clue to what features of the elements are correlated with the regression difficulty.

In Fig. \ref{fig:final}(b), we directly show the effect of increasing $\sigma_e$ on the smoothness of the total $d$-DOS at and around E$_{fermi}$, with $\sigma_e$ = 2.0 eV smoothing out all of the steep DOS features, most especially for Ti.
We correlate this increase in smoothness, and thus decrease in complexity of the DOS with the sharp reduction in the test errors in panel (a).
Moreover, we look at the differences of the radial and angular components of the force labels computed at $\sigma_e$ = 1.0 and 2.0 eV relative to the original TM23 data at $\sigma_e$ = 0.2 eV in panel (c).
The middle-right column isolates any differences in the magnitudes of the forces for each atom across the entire training set, whereas the right-most column primarily isolates differences in the angular components, since the difference of the force vectors is first computed, then followed by the magnitude and mean. 
We note that the right-most column could also be shifted by a difference in the force magnitude.
However, to interpret these results, we have to look at the differences between the columns.
If the values in the `angular' (right-most) column are substantially larger than those in the `radial' (middle-right) column, then we can quantitatively assert that the radial distribution of the force vectors is sensitive to the value of $\sigma_e$, and this change then results in the earlier transition metals becoming `learn-able' using MLFFs with the same model architectures.

\subsubsection{Perturbation Theory Explains Relationship Between Density of States and Angular Sensitivity of Forces}
We explain the error trends further using arguments from perturbation theory. 
Consider a simple model system of an atom in a solid whose non-interacting electrons experience an external potential $V(r)$, which is produced by both the atom and the surrounding atoms in the solid. 
The electronic Hamiltonian is
\begin{equation}
    \mathcal{H} = \frac{p^2}{2m} + V(r).
\end{equation}
We assume for simplicity that this Hamiltonian produces eigenvalues and eigenvectors falling into one degenerate valence band (VB) and one degenerate conduction band (CB):
\begin{align}
    \text{VB}:\,\,\,&\braket{r|i}=\phi(r) Y_{L_i}(r),\,\,\,\mathcal{H}\ket{i}=0\\
    \text{CB}:\,\,\,&\braket{r|j}=\phi(r) Y_{L_j}(r),\,\,\,\mathcal{H}\ket{j}=\Delta\ket{j},
\end{align}
where $Y_L(r)$ are spherical harmonics, $L=l,m$ is a combined index for the principal and azimuthal quantum numbers, $\phi(r)$ is the radial component of the orbitals (with $\braket{\phi(r)|\phi(r)}=1$), and $\Delta$ is the energy gap between the valence and conduction states. 
$\phi(r)$ is assumed to be the same for each orbital for simplicity.
To match the state manifold of a typical transition metal, the combination of the VB and CB manifolds contains one $s$ shell, one $p$ shell, and one $d$ shell, analogous to transition metal valence and conduction states. 
While these `bands' do not account for bonding or the fact that the metal has no gap, they can be used as analogies to the manifold of states slightly above and below the Fermi level in a metal, with the key preserved feature being the different angular momentum character of each orbital. 
We assume the basic physical trends apply similarly well infinitesimally close to the gap and a small but finite energy difference $\Delta/2$ away from the gap.

The density of this system is
\begin{align}
    n(r) &= \sum_{i \in \text{VB}} \braket{r|i} \braket{i|r} \\
    &= \sum_{i \in \text{VB}} |\phi(r)|^2 |Y_{L_i}(r)|^2
\end{align}
and the energy of the system is 0 (because the valence bands all have zero energy).
Now consider a perturbing potential
\begin{equation}
    V^{(1)}(r)=-\sum_L V_L Y_L(r).
\end{equation}
This potential serves as a model for the effect of moving an ion in the solid, which changes the potential experienced by the electrons. 
By first order perturbation theory, the perturbation to the energy and valence orbitals is
\begin{align}
    \ket{i^{(1)}} &= \sum_{j \in \text{CB}} -\frac{1}{\Delta} \bra{j}V(r)\ket{i} \ket{j} \\
    &= \sum_L \sum_{j \in \text{CB}} \frac{C_{L_i L_j L} V_L}{\Delta} \ket{j},
\end{align}
where $C_{L_jL_iL} = \int d\Omega Y_{L_i}(\Omega) Y_{L_j}^*(\Omega) Y_L(\Omega)$ (same as the Gaunt coefficients except for the complex conjugation). The resulting perturbation to the density is
\begin{align}
    n^{(1)}(r) &= \frac{|\phi(r)|^2}{\Delta} \sum_{LL'} Y_{L}(r) V_{L'} P_{LL'} + \text{C.C.}\label{eq:pt_density_response}\\
    P_{LL'} &= \sum_{i \in VB} \sum_{j \in CB} C_{L_j L_i L} C_{L_i L_j L'},
\end{align}
where C.C. stands for the complex conjugate of the expression.
With non-integer occupation numbers $f_i$, the density response $P_{LL'}$ becomes
\begin{equation}
    P_{LL'} = \sum_{i \in VB} \sum_{j \in CB} (f_i - f_j) C_{L_j L_i L} C_{L_i L_j L'}. \label{eq:pt_angular_response}
\end{equation}
Similarly, the energy perturbation to first order is
\begin{equation}
    E^{(1)} = \sum_L V_L \left(\sum_{i \in \text{VB}} f_i C_{L_i L_i L} + \sum_{j \in \text{CB}} f_j C_{L_j L_j L}\right). \label{eq:e1}
\end{equation}

Because the forces on the nuclei are determined by the electronic charge distribution, we can interpret the trends in the MLFF force errors by analyzing the density response to a potential given by Equation \ref{eq:pt_density_response}, and in particular the angular dependence of the response given by Equation \ref{eq:pt_angular_response}. 
In general, we can expect that higher-order spherical harmonics in the density response will result in the forces being more difficult to predict, since the forces are determined by the interaction between the ions and electron density distribution.
Because $C_{L_1 L_2 L_3} = 0$ if $l_1 + l_2 < l_3$, $d$ to $d$ transitions both respond to higher-order $V_L$ terms (up to $\ell=4$) and produce higher-order density fluctuations in $n^{(1)}(r)$ via Equation \ref{eq:pt_density_response}. 
Therefore, we focus on the role of these $d$ to $d$ transitions for different metals to explain the complexity of the PES.

In the limiting case where the $d$ shell is mostly occupied (\emph{i.e.} the late transition metals in Groups X-XII), there are few or no conduction $d$ electrons, so the sum over $j$ in Equation \ref{eq:pt_angular_response} vanishes, resulting in a less complex density response and therefore a simpler PES. 
However, when there are $d$ orbitals in both the conduction and valence bands, the sum over $j$ has finite contributions, and $\ell=4$ responses to the potential arise, making the PES more complicated. 
In this simplified perturbation theory model, the magnitude of the $l=4$ density response should be roughly proportional to $(10 - N_d) N_d$, where $N_d$ is the number of $d$ electrons per atom in the system. 
This intuition is confirmed for the metals with full $d$ shells (Groups X-XII) and the roughly half-filled shells (Groups VI-IX), as the latter have much higher force errors than the former (Fig. \ref{fig:flare_v_nequip_main}).
We also show this to be true empirically, by artificially changing the number of electrons in the Au and Os systems, effectively swapping their electronic configurations, which changes the relative position of the Fermi level to the DOS, shown in Fig. \ref{fig:final}(d), and recomputing the FLARE force percent errors on the test set in Fig. \ref{fig:final}(e).
We see that by changing the number of electrons, and thus shifting the location of the Fermi energy, we can directly influence the difficulty of learning the force labels.

In spite of having fewer $d$ electrons than the metals with half-filled shells, the metals in Groups IV and V with mostly empty (but importantly, not completely empty) shells also have high force errors. 
We attribute this to the fact that these metals still have a large $d$-character DOS both above and below the Fermi level, similarly to the middle-group transition metals, whereas the later-group transition metals have very little unoccupied $d$-DOS near the Fermi level. 
This trend is illustrated in the SI (Fig. S5), which shows that transition metals in Groups IV-IX have a large $d$-DOS both above and below the Fermi level, whereas the $d$-DOS for Groups X-XII is mostly or entirely below the Fermi level.
Therefore, we expect the early transition metals to each have a complex PES like the middle transition metals, which is confirmed by Fig. \ref{fig:flare_v_nequip_main}.

The effects of smearing can also be considered within the perturbation theory model. 
Empirically, large, non-physical values of smearing significantly decrease the force errors for early and middle transition metals (Fig. \ref{fig:final}(a)). 
The case of the earlier transition metals is complicated, but there are a few reasons to expect the smearing to decrease the force errors. First, smearing out the occupations causes all the $m$ channels of the $d$ shell to become partly occupied (\emph{i.e.} $f_i\approx f_j$), which damps the density response in Equation \ref{eq:pt_angular_response}. 
Second, as shown in Fig. \ref{fig:final}(b), smearing evens out the DOS of the individual $d$ orbitals, which results in similar occupation numbers $f_i$ for different $m$ values, which also causes Eqn. \ref{eq:pt_angular_response} to vanish. 
Finally, larger smearing smooths out the total $d$ DOS near the Fermi level, especially in difficult transition metals like Ti and Os (Fig. \ref{fig:final}(b)), which might also smooth the PES.

Ultimately, early transition metals are found to possess more structurally sensitive many-body interactions, which we find to be correlated with their more complicated electronic structures, and thus require higher angular resolution of the MLFF tasked with learning forces, energies, and stresses. 
Confirming this is also the overall reduction in test error magnitude as the angular resolution of NequIP is increased, as was demonstrated in Section II.B. 
This finding motivates further equivariant MLFF development to access higher values of $\ell_{max}$ with increased computational efficiency, and will be the followup investigation to this work as these novel MLFF methods have recently become available (\emph{e.g.} Allegro \cite{Musaelian2023} and MACE \cite{https://doi.org/10.48550/arxiv.2206.07697}).

\subsection{Conclusions}
This work presents TM23, a benchmark data set for transition metals comprised of high-fidelity first-principles calculations, and provides a systematic comparison of two leading MLFF architectures.
FLARE and NequIP models were tasked with learning force, energy and stress labels from the same reference data, and marked differences in performance were observed.
These findings uncover persistent trends in MLFF accuracy across TM23, with interatomic interactions in early transition metals (Group VIII and below) being demonstrably more difficult to capture accurately than Group IX, X, XI, and XII metals. 
These trends were observed not only in final model accuracy but also in learning exponents, as a function of training set sizes using NequIP. 
Importantly, these trends were then explained using a detailed understanding of the electronic structures of these metals coupled with an explanation rooted in perturbation theory.
This work highlights the utility of using systematically controllable geometric representations, such as the atomic cluster expansion and equivariant neural networks, to uncover subtle and complex physical features of metallic interactions in a data-driven way from high-fidelity first-principles reference data. 
The accuracy trends for FLARE and NequIP across the space of transition metals point to the varying degree of importance of directionality and many-body character of interactions, the complexities of which can be artificially reduced via modification of the electronic structure in the underlying quantum mechanical calculation.
Ultimately, the TM23 data set provides some of the most challenging reference benchmarks for currently leading MLFF approaches, even if just for single-element bulk systems. 
Substantial improvements are therefore still needed and motivated by these results for future MLFF architectures, focusing on more efficient many-body representations that can expand to higher radial and angular resolutions without sacrificing computational efficiency.
We prove this empirically, by establishing the sensitivity of the radial distribution of the forces to the electronic bonding interactions within the metal.
Systematic work is needed on using this challenging  data set to benchmark different MLFF model architectures and regression methods.
Another extension of this work can target the augmentation of this data set, from which purpose-built models can be trained for property prediction (\emph{e.g.} melting point estimation, dislocation dynamics, etc).
We anticipate that our findings and reference data will help both to anticipate appropriate model parameters for practitioners studying transition metals and to advance model development for atomistic description of heterogeneous metal catalysts, multi-element alloys and many other applications where these elements are present.

\section{Methods}
\subsection{Training Data Acquisition: Density-Functional Theory and \textit{ab initio} Molecular Dynamics} \label{S-sec:data}
The complete computational workflow is detailed in Fig. \ref{fig:tableofcontents}.
Bulk configurations of all metals were extracted from the Materials Project (MP) repository \cite{Jain2013}, where the lowest-energy crystalline phase was selected for each system.
This workflow was employed for all systems, including those exhibit dynamic instabilities \cite{Qian_2018} in bcc phases at lower temperature (\emph{e.g.} Ti, Zr, and Hf).
Each phase was selected based on the convex hull energy, the lowest being chosen.
The selected phase was also confirmed to be experimentally observed, as is provided on the MP `Materials-Explorer' dashboard.
Reference MP identification numbers (MP-IDs) corresponding to each metal structure employed are provided in Table S1. 
Super-cells of each metal were then created, sized such that at least 7 \AA{} of spacing existed between an atom and its periodic images. 
This lattice requirement was chosen as a lower bound since previous FLARE force field training displayed sufficient accuracy in modelling 2-body behavior at this distance \cite{Lim2020EvolutionDynamics,Vandermause2020On-the-flyEvents}.
Computational efficiency in creating the TM23 data set was also considered, as super-cells of this size contained 32 to 71 atoms. 
The pymatgen library was employed to simplify super-cell creation and calculation setup \cite{Ong2013}.
To increase the diversity of atomic environments within each super-cell, a single vacancy defect was introduced by randomly deleting an atom.
Intuitively, introduction of a vacancy should increase the number of nontrivial atomic environments from the onset of each trajectory, and may facilitate more complicated dynamics (\emph{e.g.}, vacancy diffusion) throughout the course of the simulation.

Following creation of each super-cell and incorporation of a single vacancy, \textit{ab initio} molecular dynamics (AIMD) simulations were then performed in three thermal regimes: `cold' metals surveyed at 25\% of their experimental melting temperatures $T_{melt}$, `warm' metals at 0.75$\cdot T_{melt}$, and `melt' molten configurations from dynamics surveyed at 1.25$\cdot T_{melt}$. 
These relative temperatures were chosen such that diverse atomic orderings would be captured within the training set, ranging from fully crystalline to amorphous atomic environments. 
The experimental melting temperatures associated with all elements are listed in Table S2. 

AIMD trajectories and density functional theory (DFT) calculations were completed using the Vienna \textit{Ab initio} Simulation Package (VASP, version 5.4.1) \cite{Kresse1993AbMetals,Kresse1996EfficientSet,Kresse1996EfficiencySet,Kresse1999}.
The pseudopotentials employed for each metal are listed in Table S2.
AIMD trajectories were surveryed for a total of 55 picoseconds using a timestep of 5 femtoseconds for all metals.
The NVT ensemble with the Nos\'e-Hoover thermostat \cite{Hoover1985CanonicalDistributions,Martyna1998NoseHooverDynamics} and a Nos\'e--mass of 40 timesteps were employed.
To reduce the computational cost incurred by each AIMD trajectory, since sequential configurations of successive frames are strongly correlated, eigenvalues for the wavefunction were only sampled using the $\Gamma$ \textbf{k}-point. 
While the density of the \textbf{k}-point mesh could influence the dynamics that are observed throughout the course of a trajectory, one can rely on this effect being less pronounced at longer timescales.
This assumption is valid given the stochastic nature of MD trajectories, especially using a canonical ensemble like NVT with the Nose-Hoover thermostat.

An important discussion is provided here with respect to the exclusion of spin-polarization from both the AIMD trajectories and high-fidelity DFT calculations.
Spin-polarization can have a pronounced effect on the electronic structure for magnetic systems, and has been shown to influence the vacancy diffusion energy \cite{Angsten2014ElementalStructures} up to 300 K.
Moreover, we note that property predictions could be affected by magnetic effects, and use of these force fields would require caution under such conditions.
However, all metals in both AIMD trajectories at the `warm' and `melt' temperatures are well above their Curie temperatures, so in `real-world' dynamics at these elevated temperatures, spin ordering in the electronic degrees of freedom would not be present. 
Thus, there is an inherent issue with combining high-temperature AIMD for dynamics and static DFT for energy, force, and stress label calculation: even if the configuration itself has a high potential energy (\textit{i.e.} it is far from an equilibrium 0 K ordered structure) and was sampled from an AIMD trajectory above the Curie temperature, the configurations are ultimately endowed with labels calculated for a wavefunction solved at approximately 0 K electron temperature.
Finite temperature electronic smearing is supplied to each static DFT calculation via $\sigma_e$ in VASP, serving as a crude approximation to elevated electron temperature. 
The $\sigma_e$ value in VASP was initially not changed to approximate the electronic temperature of the systems as the temperature of the AIMD simulations was varied.
However, this value was varied to even non-physically high values of 1.0 and 2.0 eV for a subset of the metals to provide the results in Fig. 9 of the main text.
Mixing the `warm' and `melt' training labels with `cold' metals below the Curie temperature, without an MLFF designed to account for the inclusion of spin would likely cause the model to try to learn two different potential energy surfaces-- one with spin ordering and one without. 
Therefore, we compromised by neglecting spin entirely for both AIMD trajectories and the subsequent high-fidelity DFT calculations.
While this means that the sampled configurations and computed training labels for metals with significant spin interactions (\emph{e.g.} Fe, Co, Ni) are cruder approximations to their `real' dynamics and potential energy surfaces, we leave these questions to be answered in a future investigation. 

Individual frames were then extracted from each AIMD trajectory at intervals of 50 fs, excluding the first 5 ps of the trajectory.
Frames from the first 5 ps were excluded so that the extracted frames were given sufficient time to equilibrate with the applied thermostat.
This selection procedure resulted in 1000 frames from each trajectory, yielding a total of 3000 frames for each metal. 
In order to visualize the extent of the thermal disorder in our trajectories, radial distribution functions averaged across all trajectories were computed using the ASE package.
The methods and RDF plots are provided in the SI.

Higher-fidelity single point energy, force, and stress labels were then calculated using increased \textbf{k}-mesh densities converged on a per-element basis.
The \textbf{k}-point spacing was chosen such that the energy noise per atom was below 1 meV/atom and the force noise was below 5 meV/\AA{}.
Element specific \textbf{k}-point grids were employed to respect inherent symmetries of each system, as recommended by VASP. 
To further facilitate convergence, Methfessel-Paxton smearing at the Fermi-level was employed, with all metals using a value of 0.2 eV, but was eventually varied given to generate the labels corresponding to the models trained in Fig. 9.
Moreover, Fig. 9(d) employed the VASP `NELECT' parameter, which allowed us to artificially change the total number of electrons for the Au and Os labels, effectively testing the effect of swapping their electronic configurations.
All calculations employed a cutoff energy of 520 eV, which was sufficient for all ENMAX values provided by the pseudopotentials for all metals.

\subsection{Element Specific Semi-Core Corrections and \textbf{k}-point Densities}
\label{S-sec:dft}
In the interest of reproducibility, we also provide a complete description of the employed semi-core corrections used in each pseudopotential across the metals, as well as the minimum \textbf{k}-point densities applied for each system, provided in Table S1. 
Moreover, to address the pertinent question of DFT accuracy of this data set, we evaluate the correlation between NequIP predicted force, energy, and stress percent errors (using the full training set) against the minimum \textbf{k}-point density observed along the lattice vectors of each cell in Fig. S1. 
The pseudopotential naming convention maintains consistency with those presented in the VASP documentation.

\subsection{MLFF Architectures} 
\label{S-methods}
\subsubsection{Gaussian Processes in the Fast Learning of Atomistic Rare Events (FLARE) Architecture}
The Gaussian process machine-learning architecture implemented in FLARE incorporates the atomic cluster expansion (ACE) descriptors and the normalized dot-product kernel, described in detail elsewhere \cite{vandermause2022active}.
Unlike the 2+3 body atomic environment representations used in previous implementations of FLARE \cite{Vandermause2020On-the-flyEvents}, the representations for each individual atomic environment in the current iteration are computed using the atomic cluster expansion (ACE) of Drautz \cite{Drautz2019AtomicPotentials}. 
Briefly, ACE represents the local environment around each atom using a `fingerprint' which projects the distribution of neighboring atoms into a set of radial and angular basis functions.
The GP then compares the full set of atomic environments in the test frame to other descriptor vectors in the training set to perform inference, which provides the predictive energy, forces and stress, as well as quantitative uncertainties.
Here, we maintain consistency with the notation of \cite{vandermause2022active} and \cite{Drautz2019AtomicPotentials}. 
We use B1 and B2 descriptors from the equation (28) and (29) of the original ACE paper \cite{Drautz2019AtomicPotentials}, corresponding to rotation-invariant 2-body and 3-body representations. 
We note that using descriptors of higher body order can improve the accuracy, but is computation and memory intensive, especially considering the construction of SGP requires to store the descriptors and their gradients of the entire training data set. 
To develop a more scalable method for higher body order descriptors and benchmark their performance with the SGP will be the focus of our future work. 
For the kernel of SGP, we choose the normalized dot product raised to the power ($\zeta$) of 1 and 2, which lifted the body order of our model. 
Specifically, the 2-body B1 descriptors with $\zeta=2$ becomes effectively 3-body, and the 3-body B2 descriptors with $\zeta=2$ becomes effectively 5-body.
We provide `best' model parameters for both kernel powers, but only results for $\zeta=2,$ since these are shown to be systematically more accurate across all labels and metals.

Our GP model is trained by optimize hyperparameters to maximize the log likelihood function which describes the overall agreement of the model with training data and the complexity of the model.
For each model, the likelihood is optimized via gradient descent with respect to four hyperparameters: the signal variance ($\sigma$), and the three noise variances for each of energy, force, and stress ($\sigma_E, \sigma_F,$ and $\sigma_S$, respectively).
The initial value of signal variance is set to 2.0. 
Each noise parameter is initialized to the expected error for the corresponding quantity, specifically: $0.001$ eV/atom, $0.05 $ eV/\AA{}, and $0.005 $ ev/\AA{}$^3$, respectively. 
Additionally, depending upon the temperature of the AIMD data fed into the model, the initial $\sigma_F$ is varied accordingly, since the force MAE scales with the temperature of the AIMD trajectory (higher temperature configurations tend to have larger forces acting on the atoms).
For all FLARE training, hyperparameter optimization was completed periodically throughout training until the final frame addition, which varied depending on the size of the training set (from a total of 100 to 2700 frames), as hyperparameter optimization requires re-computing and inverting the covariance matrix at each step.
Furthermore, the gradient tolerance for the convergence criteria of L-BFGS-B optimization method was set to 1E-4 for the marginal log likelihood, and the maximum number of iterations was set to an appreciably large value (200) to ensure convergence, if necessary.

Lastly, an exhaustive grid-search through reasonable values of the descriptor parameters was also completed for each metal and B1 and B2 respectively.
These descriptor parameters are not optimized against the likelihood at training time like conventional hyperparameters, and can strongly influence model behavior.
The radial cutoff $r_{cut}$, radial basis length ($n_{max}$), and angular basis length ($l_{max}$) were all tested over a broad range of reasonable values.
Interested readers may find the full results in Tables S13-S16, whereas the main text only presents results for models that employ the best model parameters for each ACE descriptor, which were selected using a combination of maximum likelihood and minimum force MAE.
If a system presented the scenario where force MAE was minimized with different parameters when compared to maximum likelihood, the parameters were taken from the minimum force MAE calculation.
This scenario is plausible since FLARE training was done on forces, energies, and stresses.

\subsubsection{Equivariant Message-Passing Graph Neural Network (NequIP) Architecture} \label{nequip}
We also employ the equivariant message-passing interatomic potential NequIP \cite{Batzner2021E3-EquivariantPotentials}.
NequIP has recently been shown to be highly sample-efficient, and to be remarkably robust when compared to other existing ML methods in the MLFF literature on main-group benchmarks.
The NequIP architecture is described in detail elsewhere \cite{Batzner2021E3-EquivariantPotentials}, but the key idea lies in learning featurizations of the atomistic structure which are explicitly constructed to be equivariant under symmetry operations of the Euclidean group E(3). 
E(3) is comprised of of translations, rotations, and mirror operations, covering the physical symmetries present in atomistic systems.
Equivariance is distinct from invariance by the fact that invariant quantities do not vary under E(3) operations, whereas equivariant quantities transform appropriately with E(3) operations. 
In other words, an invariant scalar does not change under symmetry transformations, whereas \textit{e.g.} an equivariant vector transforms in a way that is commensurate with the group (see equation 3 of \cite{Musaelian2023}).
An example of an invariant MLFF is the SchNet potential \cite{schnet}, which only operates on invariant descriptions of the geometry (distances $r_{ij}$), whereas the equivariant NequIP potential additionally uses higher order tensor representations to encode more complex geometric information about atomic environments.
NequIP uses these features in a message-passing framework that represents the atomistic structure as a graph, and makes predictions by iteratively propagating information along that graph through a series of $N_{layer}$ update layers.

The training procedure for NequIP is similar to that of FLARE, where the train-test splits are equivalent, but NequIP additionally withholds a further percentage of the training data, 10$\%$ in this work, that is used to monitor the progress of the ongoing training procedure.
As was done for FLARE, an exhaustive grid search over the hyperparameters of the model was completed for each metal.
The hyperparameters scanned for NequIP were the: radial cutoff ($r_{cut}$), number of layers ($N_{layers}$), angular resolution of the network ($l_{max}$), number of features ($f_{irreducible}$), learning rate, and force/energy weights.
The best model parameters for each metal are provided in Section S6. 
The best model for each metal was chosen via minimization of the force error on the validation set.
To limit computational inefficiency in observing convergence for each NequIP model, a learning rate scheduler was employed, where the learning rate was reduced by 80$\%$ if the force MAE on the validation set did not improve within 100 epochs, and training is concluded if the learning rate reduced to be less than 1E-05.
In the loss function, energies were weighted using the per-atom-MSE, and the loss coefficients were set to 1 for both the forces and stresses, shown in Eqn.\ 30 of \cite{Musaelian2023}.
To limit the amount of computational resources required for this component, we set a hard-limit of 3 weeks of wall-time for training on a single A100 GPU.
Several models converged before this limit, but most were halted at this upper bound.

\subsection{Phonon Dispersion} \label{phonon}
Phonon dispersion curves were calculated for Cu and Os using the Phonopy \cite{phonopy2015} and Phoebe packages \cite{cepellotti2022phoebe}.
First, the ($1\times1\times1$) primitive unit-cell of each material was relaxed using LAMMPS, isotropically along each lattice vector using conjugate gradient descent until energy and force thresholds of $1\times10^{-12}$ were met.
To perform phonon calculations, Phonopy was used to generate displaced super-cell structures from which forces for each super-cell were calculated using FLARE and NequIP potentials. 
Then, Phonopy was used to collect the force constants from each super-cell calculation and Phoebe was applied to construct the dynamical matrix and plot the phonon dispersions. 

DFT phonon calculations were then completed for each material using the same workflow in order to provide `ground-truth' labels with which to compare the MLFF predictions. These DFT calculations were performed using the same VASP pseudopotentials, \textbf{k}-point densities, and INCAR parameters as the TM23 frames for Cu and Os.

\section{Data Availability}
The TM23 data set will be provided on the Materials Cloud upon publication.

\section{Code Availability}
The details about VASP, a proprietary code, can be found at https://www.vasp.at/.
The details about FLARE, and NequIP, which are open-source codes can be found at https://github.com/mir-group/flare and https://github.com/mir-group/nequip, respectively.

\subsubsection*{Author Contributions}
C.J.O.\ created the data set, compiled and analyzed the data, performed all model training and evaluation using FLARE and NequIP architectures, and co-wrote the manuscript. 
S.B.T.\ initiated the project, built Python scripts for data management and calculations, and co-wrote the manuscript. 
S.B.T.\ and C.J.O.\ jointly designed the figures. 
Y.X.\ aided in FLARE implementation and calculations, and provided detailed feedback on the FLARE methods section. 
S.B.\ and A.M.\ provided discussions regarding the NequIP architecture and feedback for the respective methods section. 
J.C.\ provided helpful discussion and analysis of the phonon dispersion results obtained via Phonopy and Phoebe codes and wrote the corresponding methods section. 
K.B.\ provided helpful discussions pertinent to the final interpretation of the error trends, and was responsible for the perturbation theory considered in this context.
L.S.\ advised components of this work. 
B.K.\ supervised all aspects of this work. 
All authors contributed to revision of the manuscript.

\subsubsection*{Acknowledgments}
We gratefully acknowledge Bill Curtain, Georg Kresse, Ralf Drautz, Gabor Csanyi, and Karsten Jacobsen for thoughtful discussions regarding interpretation of the data, methods, and training protocols.
We thank Dr. Jin Soo (David) Lim and Dr. Jonathan Vandermause for helpful conversations at the outset of this project.
This work was supported by the US Department of Energy, Office of Basic Energy Sciences Award No. DE-SC0022199 and No. DE-SC0020128, as well as by Robert Bosch LLC.
C.J.O. is supported by the National Science Foundation Graduate Research Fellowship Program under Grant No. DGE1745303. 
S.B.T. was supported by the Department of Energy Computational Science Graduate Fellowship under grant DE-FG02-97ER25308.
Computational resources were provided by the FAS Division of Science Research Computing Group at Harvard University. 

\section{Ethics Declarations}
\subsection{Competing interests}
The authors declare no competing interests.

\section{References}
\bibliography{bib}

\end{document}


\title{Supplementary Information for: Complexity of Many-Body Interactions in Transition Metals via Machine-Learned Force Fields from the TM23 Data Set}

\author{Cameron J. Owen$^{*, \dagger}$}
\affiliation{Harvard University}

\author{Steven B. Torrisi$^{*,a}$}
\affiliation{Harvard University}

\author{\\Yu Xie}
\affiliation{Harvard University}

\author{Simon Batzner}
\affiliation{Harvard University}

\author{Kyle Bystrom}
\affiliation{Harvard University}

\author{Jennifer Coulter}
\affiliation{Harvard University}

\author{Albert Musaelian}
\affiliation{Harvard University}

\author{Lixin Sun}
\affiliation{Harvard University}

\author{Boris Kozinsky$^{\dagger}$}
\affiliation{Harvard University}
\affiliation{Robert Bosch LLC Research and Technology Center}

\def\thefootnote{*}\footnotetext{\textbf{Equal Contribution.}\\}\def\thefootnote{\arabic{footnote}}

\def\thefootnote{a}\footnotetext{Currently at Toyota Research Institute.\\}\def\thefootnote{\arabic{footnote}}

\def\thefootnote{$\dagger$}\footnotetext{Corresponding authors\\C.J.O., E-mail: \url{cowen@g.harvard.edu}\\B.K., E-mail: \url{bkoz@seas.harvard.edu}\\ }\def\thefootnote{\arabic{footnote}}

\maketitle

\section{Crystal Symmetries and Melting Points}\label{S-sec:crystals}
In Table \ref{S-tab:crystal} we provide the complete list of MP-IDs, space group symbols, and melting temperatures that were employed to determine the temperatures during each AIMD trajectory.

\section{Temperature Correlation of Errors}\label{S-sec:temp}
Here, we provide evidence for our chosen error metric (percent error), instead of MAE, which is typically employed in the MLFF community.
This distinction between the two metrics is displayed in Figure \ref{S-fig:temp_corr}, where the correlation between NequIP Force MAE on the full training set and the melting temperature of each metal can be observed whereas this correlation disappears when force percent error is employed. 
We believe that this is a more appropriate metric by which to judge the success of our MLFF architectures on these systems, as the inherent relationship to the temperature at which frames were collected is removed.

\section{Correlation of Errors and Crystal Symmetries} \label{S-symm}
During the course of this investigation, several chemical property relationships were considered in order to try and understand the observed error trends. 
Based on the  distribution of ground-state crystal symmetries across the \textit{d}-block and the trends in error, we visualize and present a loose correlation in Figure \ref{S-fig:symm}.
This is purely qualitative, and is expanded upon in the following section with more quantitative proof of a periodic trend of electronic structure with error.

\section{Correlation of Errors with \textit{d}-Band Centers and Cauchy Pressures} \label{S-sec:cauchy_dband}
As discussed in the main text, several discussions with MLFF and DFT domain experts, as well as investigations in the literature, prompted additional analysis of the observed error trends with respect to electronic and material properties of the metals in TM23. 
We provide two comparisons here: NequIP force percent errors on the entire training set against available \textit{d}-band centers, and Cauchy pressures of the metals. 
These comparisons are provided in Figures \ref{S-fig:dband} and \ref{S-fig:cauchy}, respectively, where a reasonable ($R^2=0.6762$) trend is observed in the former, and no trend is observed in the latter, as evidenced by confidence of the linear regression fits.
Starting with Figure \ref{S-fig:cauchy}, this comparison was prompted by previous literature positing a correlation between the amount of directional-bonding seemingly present in early transition metals with Cauchy pressures\cite{PhysRevB.86.134106,PhysRevLett.106.246402,10.2307/53737}.
From these investigations, it was concluded that the amount of directional-bonding in early transition metals could influence the difficulty in building pair-potentials, as angular terms become more important.
This was then coupled with the apparent Cauchy pressure of the metals considered, defined as C$_{12}$-C$_{44}$ \cite{PhysRevB.86.134106,10.2307/53737}. 
However, when comparing the observed force $\%$ test errors from NequIP on the full training set to available Cauchy pressures, calculated from elastic constants obtained from the Materials Project\cite{Jain2013}, there is no correlation.
On the other hand, when the same NequIP test errors are compared to \textit{d}-band centers computed here, there is an obvious trend as evidenced by the linear fit, as evidenced in Figure \ref{S-fig:dband}.
This provides sufficient evidence that there exists a periodic trend of the observed MLFF accuracies with the electronic structure of the metals, which is extended upon in the main text.

\section{Mean Absolute Errors and Mean Absolute Values Across Labels} \label{S-section:maes}
In addition to the force percent errors presented in the main text, we also provide the corresponding MAEs and MAVs leading to the presented percent errors.
These values are given in Tables \ref{S-tab:mavs} - \ref{S-tab:nequip_cm}, respectively.
The MAE values are split between FLARE and NequIP, and among the various training scenarios for both MAEs and MAVs (full training set and each transferability task). 
MAEs and percentages are also provided in Figures \ref{S-fig:firsterror} - \ref{S-fig:lasterror}, using the same \textit{d}-block format as that employed in the main text.

We note that the MAEs of the forces in particular tend to correlate with the melting temperature of the metal. Some of this is due to some metals having more difficult to model, but some of it is also due to the fact that our AIMD simulations have temperatures chosen with reference to their melting points. To demonstrate this, we plot the distribution of force labels associated with each metal to show the natural variation in the force label that occurs across transition metals, as seen in Figures S\ref{S-fig:firsterrordist} - S\ref{S-fig:lasterrordist}.

\section{Radial Distribution Functions}

In order to visualize the extent of the thermal disorder in our trajectories, radial distribution functions averaged across all trajectories were computed using the ASE package \cite{larsen2017atomic}. 
We converted each frame into an ASE Atoms object and using the \verb|ase.geometry.Analysis| object and calling the \verb|get_rdf| function.
The RDFs are averaged over all cold, warm, and melt cells respectively, and plotted on the same axis.
The extent of melting can be visualized by the extent to which the trough between the first and second nearest-neighbor shells increases when moving from low to high temperature.
The maximum radius is kept small to avoid using periodic image atoms from neighboring cells.
The RDF plots are provided in Figures S\ref{S-fig:firstrdf} - S\ref{S-fig:lastrdf}.

\section{Best-Model Parameters}\label{S-sec:best_models}
Best-model parameters determined for both FLARE and NequIP models from grid-test are also provided.
We hope that these model parameters will reduce the overall time-to-solution required by practitioners tasked with building MLFFs for these transition metals, saving days or even weeks of compute resources.
We separate these results by architecture, FLARE and NequIP, as well as representation within FLARE (with ACE B1 and B2 with $\zeta$ = 1 or 2).
`Best' model parameters for both model architectures were determined from minimization of the force MAE (and maximization of the model likelihood for FLARE).

We start with the FLARE ACE B1 models, powers 1 \& 2, where the grid-search bounds were set accordingly: r$_{cutoff}$ (\AA{}) = [3.0,9.0,1.0], n$_{max}$ = [1,29,2], and l$_{max}$ = [0] and best models are provided in Table \ref{S-tab:flare_B1} and \ref{S-tab:flare_B1_p2}.
For the FLARE ACE B2 models, the grid-search bounds were set accordingly: r$_{cutoff}$ (\AA{}) = [3.0,9.0,1.0], n$_{max}$ = [1,19,2], and l$_{max}$ = [0,6,1] and best models are provided in Table \ref{S-tab:flare_b2} and \ref{S-tab:flare_b2_p2}.
All NequIP models employed 4 layers, with radial (number of) basis functions set equal to 8, and a batch size of 1.
The grid-search bounds were set accordingly: r$_{cutoff}$ (\AA{}) = [4.0,6.0,1.0], learning rate = (0.001,0.005,0.01), number of layers = [2,6,2], and feature expressions were sampled from the list (ranked in increasing order of angular complexity, $\ell=0$ to $\ell=5$) in Table \ref{S-tab:nequip_features} and best models are provided in Table \ref{S-tab:nequip_best}.

\section{Additional Tests}
\subsection{Additional Sparse Environments in FLARE}
An important choice for training the FLARE models was how many atomic environments should be selected per training frame.
This modeling choice was explicitly tested, and the results are provided in Table \ref{tab:test}.

\subsection{Forced Symmetries}
Given the results of Section \ref{S-symm}, we also considered the effect of forcing the crystal symmetries of a subset of the metals, e.g. Os from \textit{hcp} into \textit{fcc} and evaluating the corresponding test error.
These results are provided in Table \ref{S-tab:forced} using the FLARE ACE B1 descriptor on a previously obtained 0.9$\cdot T_{melt}$ dataset (800 training frames and 200 test frames ordered sequentially), since it is cheap to evaluate and reflected the same periodic trends as the FLARE ACE B2 and NequIP models, where forcing the symmetry does not effect the observed test error.
The forced symmetry training sets were collected using the same protocol employed for collection of the TM23 frames.
Hence, we determined that the cause of this observation was not directly correlated with the crystal symmetry of the metal, but something more inherent (i.e. electronic structure, \textit{d}-valence, etc.), which was shown in the case of \textit{d}-band center compared to NequIP force $\%$ errors.

\subsection{Large super-cells} \label{S-sec:big}
Lastly, a 640 atom super-cell of Os was also surveyed in AIMD at 1.25$\cdot T_{melt}$, and sequential frames were extracted to yield 200 train and 50 test frames.
This test was completed to understand the influence of cell dimension on the learning ability of MLFFs with finite descriptor length, as is employed in the FLARE ACE B2 model.
A coarse grid test was then performed, up to an r$_{cut}$ of 10 \AA{}, $\ell_{max}$ of 8, and $n_{max}$ of 40. 
The results are provided in Figure 8 in the main text, where the force percentages and model likelihoods are color coded to illustrate trends in model fidelity across these parameters.
Ultimately, the `best' model in terms of minimum force MAE and maximum likelihood employs a cutoff of 4 \AA{}, negating an argument for the use of larger super-cells to circumvent the observed error trends presented in the main text. 
The absolute magnitude of the error presented in Figure 8 in the main text cannot be compared to the values presented in the main text for Os, as these large super-cell frames are highly correlated since they are selected sequentially from the AIMD simulation, thus simplifying the prediction task.

\section*{References}
\bibliography{bib.bib}

\begin{table*}[t]
\caption{Semi-Core Corrected Pseudopotentials and \textbf{k}-point Densities Employed During High-Fidelity Label Creation}
\label{S-tab:kpoint}
\centering
\begin{tabular}{|c|c|c|c|c|}
\multicolumn{1}{c}{\bf Element}    &\multicolumn{1}{c}{\bf Pseudopot. Name} &\multicolumn{1}{c}{\bf Pseudopot. Valency}  &\multicolumn{1}{c}{\bf Min. \textbf{k}-point Density (\AA{}$^{-1}$)}   \\
\hline
Ag    	&	Ag	&	11	&	0.15	\\
\hline
Au    	&	Au	&	11	&	0.15	\\
\hline
Cd    	&	Cd	&	12	&	0.12	\\
\hline
Co    	&	Co	&	9	&	0.28	\\
\hline
Cr    	&	Cr\_pv	&	12	&	0.18	\\
\hline
Cu    	&	Cu\_pv	&	17	&	0.19	\\
\hline
Fe    	&	Fe\_pv	&	14	&	0.25	\\
\hline
Hf    	&	Hf\_pv	&	10	&	0.22	\\
\hline
Hg    	&	Hg	&	12	&	0.20	\\
\hline
Ir    	&	Ir	&	9	&	0.18	\\
\hline
Mn    	&	Mn\_pv	&	13	&	0.25	\\
\hline
Mo    	&	Mo\_pv	&	12	&	0.17	\\
\hline
Nb    	&	Nb\_sv	&	13	&	0.16	\\
\hline
Ni    	&	Ni\_pv	&	16	&	0.30	\\
\hline
Os    	&	Os\_pv	&	14	&	0.15	\\
\hline
Pd    	&	Pd	&	10	&	0.26	\\
\hline
Pt    	&	Pt	&	10	&	0.18	\\
\hline
Re    	&	Re\_pv	&	13	&	0.15	\\
\hline
Rh    	&	Rh\_pv	&	15	&	0.18	\\
\hline
Ru    	&	Ru\_pv	&	14	&	0.19	\\
\hline
Ta    	&	Ta\_pv	&	11	&	0.16	\\
\hline
Tc    	&	Tc\_pv	&	13	&	0.15	\\
\hline
Ti    	&	Ti\_pv	&	10	&	0.25	\\
\hline
V     	&	V\_pv	&	11	&	0.23	\\
\hline
W     	&	W\_sv	&	12	&	0.16	\\
\hline
Zn    	&	Zn	&	12	&	0.10	\\
\hline
Zr    	&	Zr\_sv	&	12	&	0.20	\\
\hline
\end{tabular}
\end{table*}

\begin{figure*}
    \centering
    \includegraphics[width=0.8\textwidth]{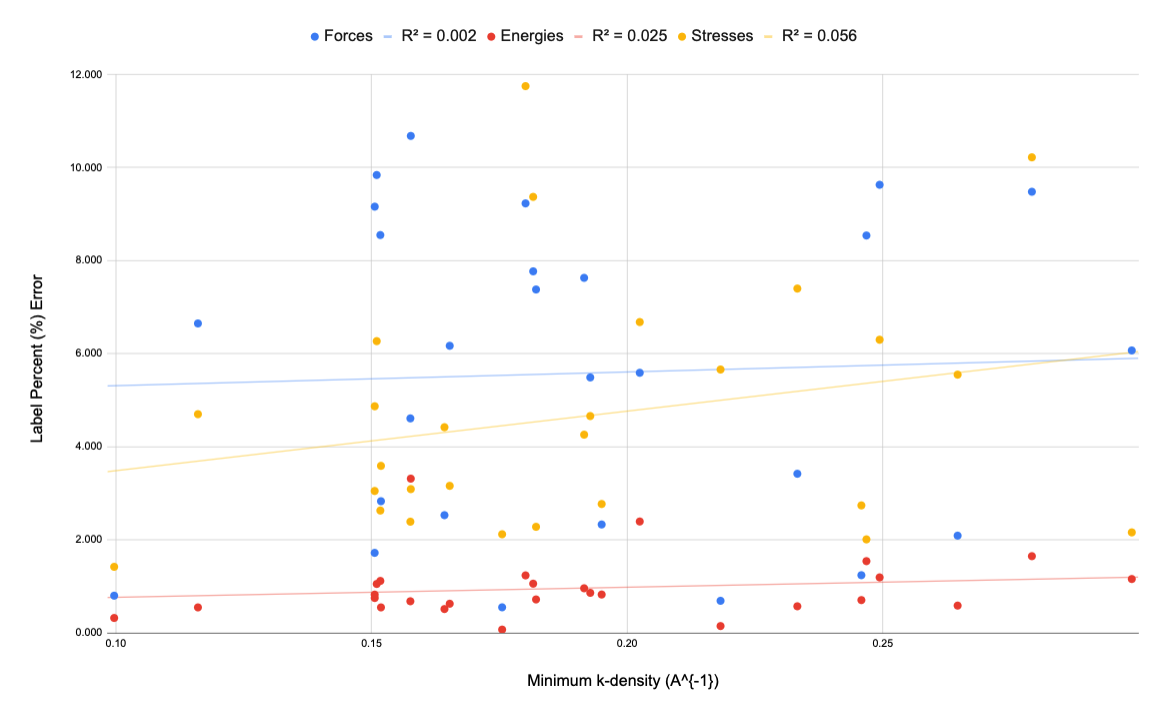}
    \caption{NequIP label percent errors across TM23 plotted against the minimum \textbf{k}-point density employed along the lattice vectors of each system. R$^{2}$ values are included for each set of labels, demonstrating the lack of correlation between the \textbf{k}-density and observed trends in predictive accuracy.}
    \label{S-fig:dft_acc_corr}
\end{figure*}\

\begin{table*}[t]
\caption{Crystal Symmetries, Materials Project IDs, and Melting Points}
\label{S-tab:crystal}
\centering
\begin{tabular}{|c|c|c|c|}
\multicolumn{1}{c}{\bf Element}    &\multicolumn{1}{c}{\bf Materials-Project ID}   &\multicolumn{1}{c}{\bf Space Group Symbol}   &\multicolumn{1}{c}{\bf Melting Point (K)}\\
\hline
Ag   & mp-124   & Fm$\hat{3}$m     & 1235 \\
\hline
Au   & mp-81    & Fm$\hat{3}$m     & 1337 \\
\hline
Cd   & mp-94    & P6$_3$/mmc       & 594 \\
\hline
Co   & mp-54    & P6$_3$/mmc       & 1768 \\
\hline
Cr   & mp-90    & Im$\hat{3}$m     & 2180 \\
\hline
Cu   & mp-30    & Fm$\hat{3}$m     & 1358 \\
\hline
Fe   & mp-13    & Im$\hat{3}$m     & 1811 \\
\hline
Hf   & mp-103   & P6$_3$/mmc       & 2506 \\
\hline
Hg   & mp-10861 & P6/mmm           & 234 \\
\hline
Ir   & mp-101   & Fm$\hat{3}$m     & 2739 \\
\hline
Mn   & mp-35    & I$\hat{4}$3m     & 1519 \\
\hline
Mo   & mp-129   & Im$\hat{3}$m     & 2896 \\
\hline
Nb   & mp-75    & Im$\hat{3}$m     & 2750 \\
\hline
Ni   & mp-23    & Fm$\hat{3}$m     & 1728 \\
\hline
Os   & mp-49    & P6$_3$/mmc       & 3306 \\
\hline
Pd   & mp-2     & Fm$\hat{3}$m     & 1828 \\
\hline
Pt   & mp-126   & Fm$\hat{3}$m     & 2041 \\
\hline
Re   & mp-8     & P6$_3$/mmc       & 3459 \\
\hline
Rh   & mp-74    & Fm$\hat{3}$m     & 2237 \\
\hline
Ru   & mp-33    & P6$_3$/mmc       & 2607 \\
\hline
Ta   & mp-50    & Im$\hat{3}$m     & 3290 \\
\hline
Tc   & mp-113   & P6$_3$/mmc       & 2430 \\
\hline
Ti   & mp-72    & P6/mmm       & 1941 \\
\hline
V    & mp-146   & Im$\hat{3}$m     & 2183 \\
\hline
W    & mp-91    & Im$\hat{3}$m     & 3695 \\
\hline
Zn   & mp-79    & P6$_3$/mmc       & 693 \\
\hline
Zr   & mp-131   & P6$_3$/mmc       & 2128 \\
\hline
\end{tabular}
\end{table*}

\begin{figure*}
    \centering
    \includegraphics[width=0.8\textwidth]{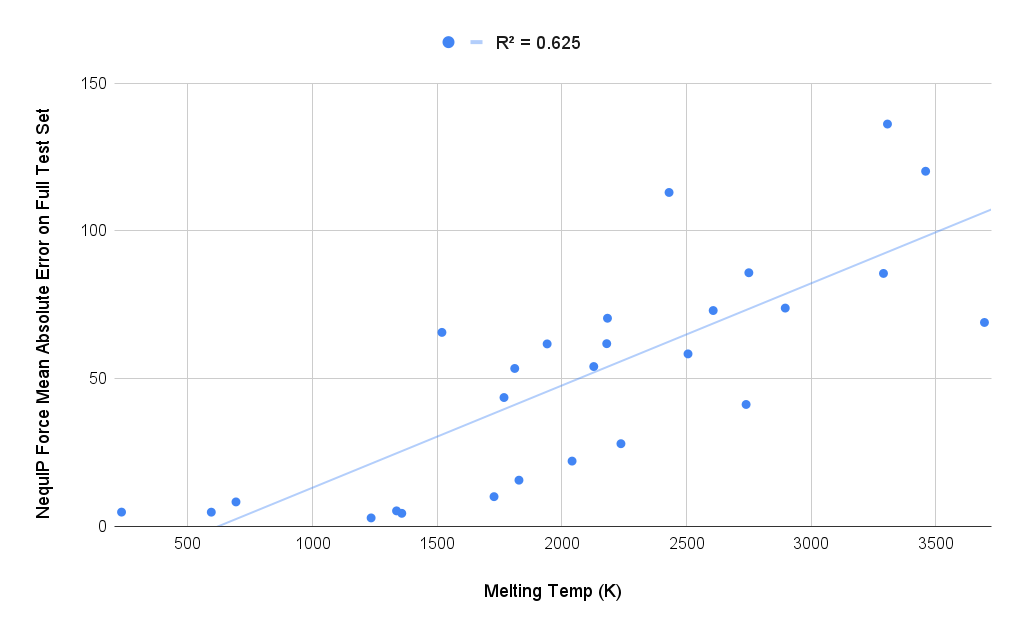}
    \includegraphics[width=0.8\textwidth]{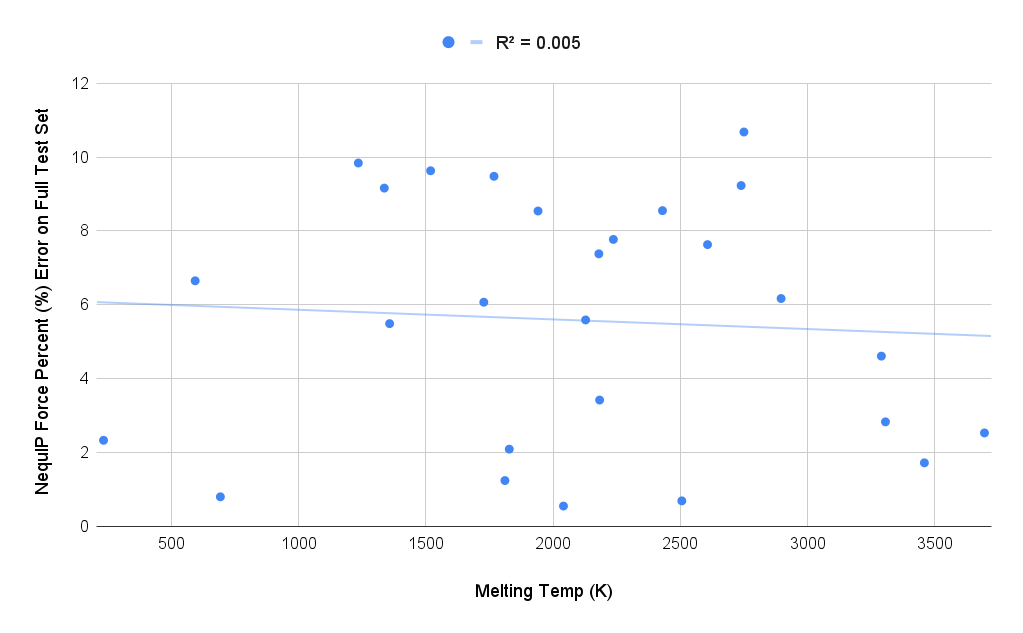}
    \caption{(Top Panel) NequIP Force MAE (meV/A) as a function of melting temperature (K) of each metal. The R$^2$ value of the linear fit is 0.625, designating a loose linear relationship between the two variables. (Bottom Panel) NequIP Force percent error as a function of melting temperature (K) of each metal. The R$^2$ value of the linear fit is 0.005, designating the absence of correlation between the two variables.}
    \label{S-fig:temp_corr}
\end{figure*}\

\begin{figure*}
    \centering
    \includegraphics[width=\textwidth]{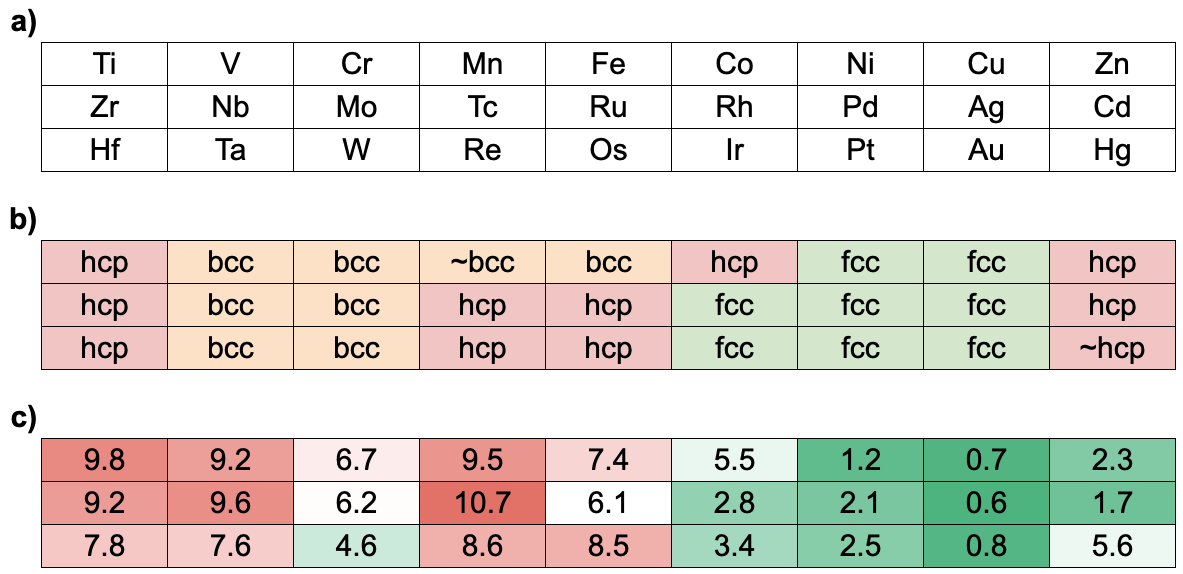}
    \caption{(a) TM23 viewed in the perspective of the \textit{d}-block. (b) TM23 metals colored by their respective crystal symmetries. (c) NequIP force percent error across the data set, presented in the same format as panels (a) and (b). }
    \label{S-fig:symm}
\end{figure*}\

\begin{figure*}
    \centering
    \includegraphics[width=0.7\textwidth]{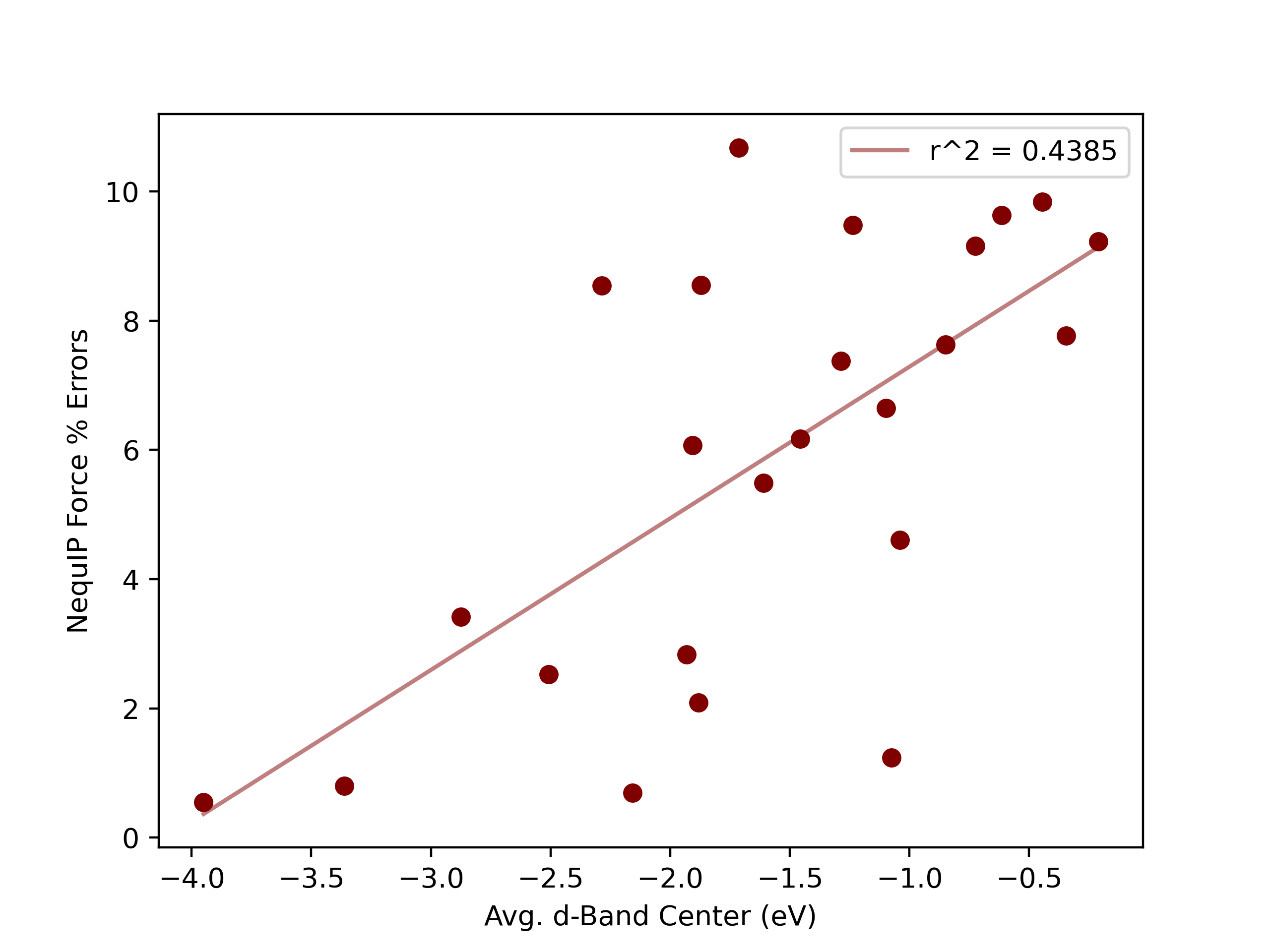}
    \caption{Correlation observed between \textit{d}-band center of the metals, as computed directly here on a random frame from the melted test set, using an average of the energies of the $d$-states weighted by their intensities, and the force \% error from NequIP on the full training set. This correlation is corroborated by the fit of the linear regression with r$^2$ = 0.4385.}
    \label{S-fig:dband}
\end{figure*}

\begin{figure*}
    \centering
    \includegraphics[width=\textwidth]{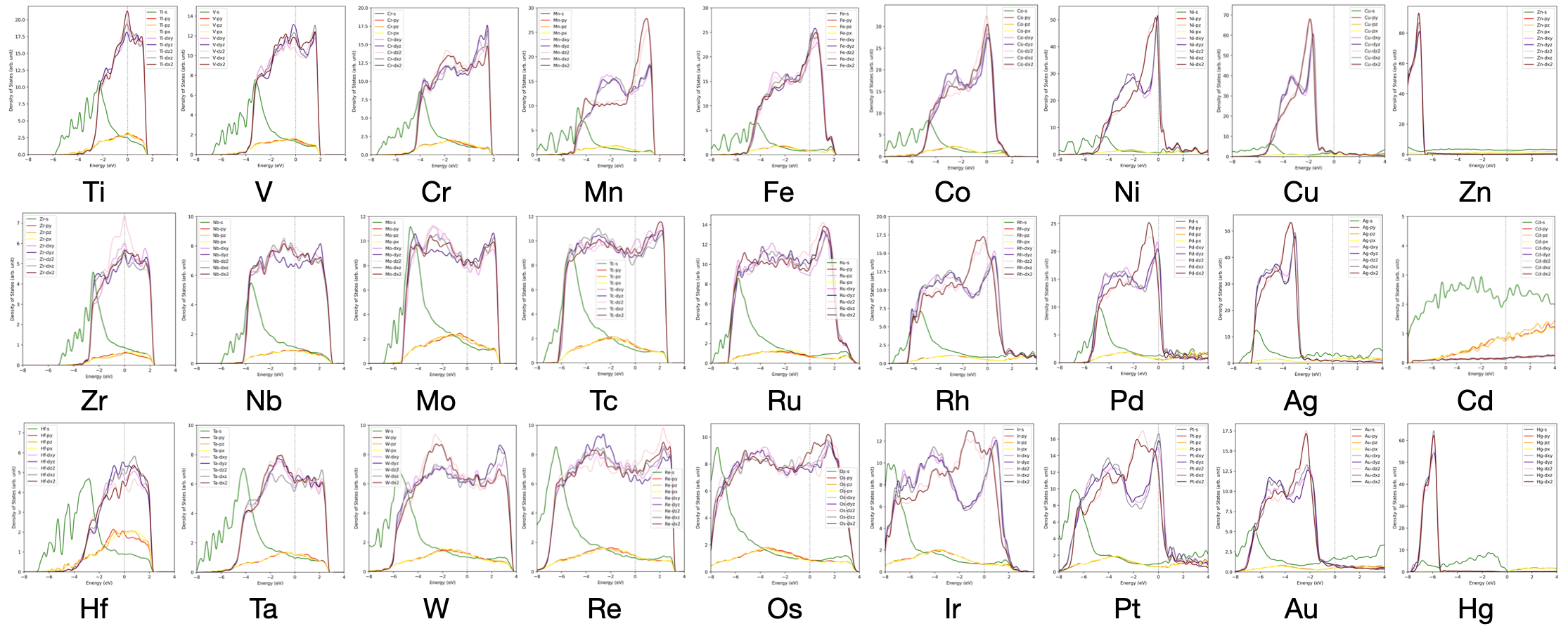}
    \caption{Complete set of angular momentum projected DOS for the TM23 set, using a randomly sampled frame from the melted test set.}
    \label{fig:DOS}
\end{figure*}

\begin{figure*}
    \centering
    \includegraphics[width=\textwidth]{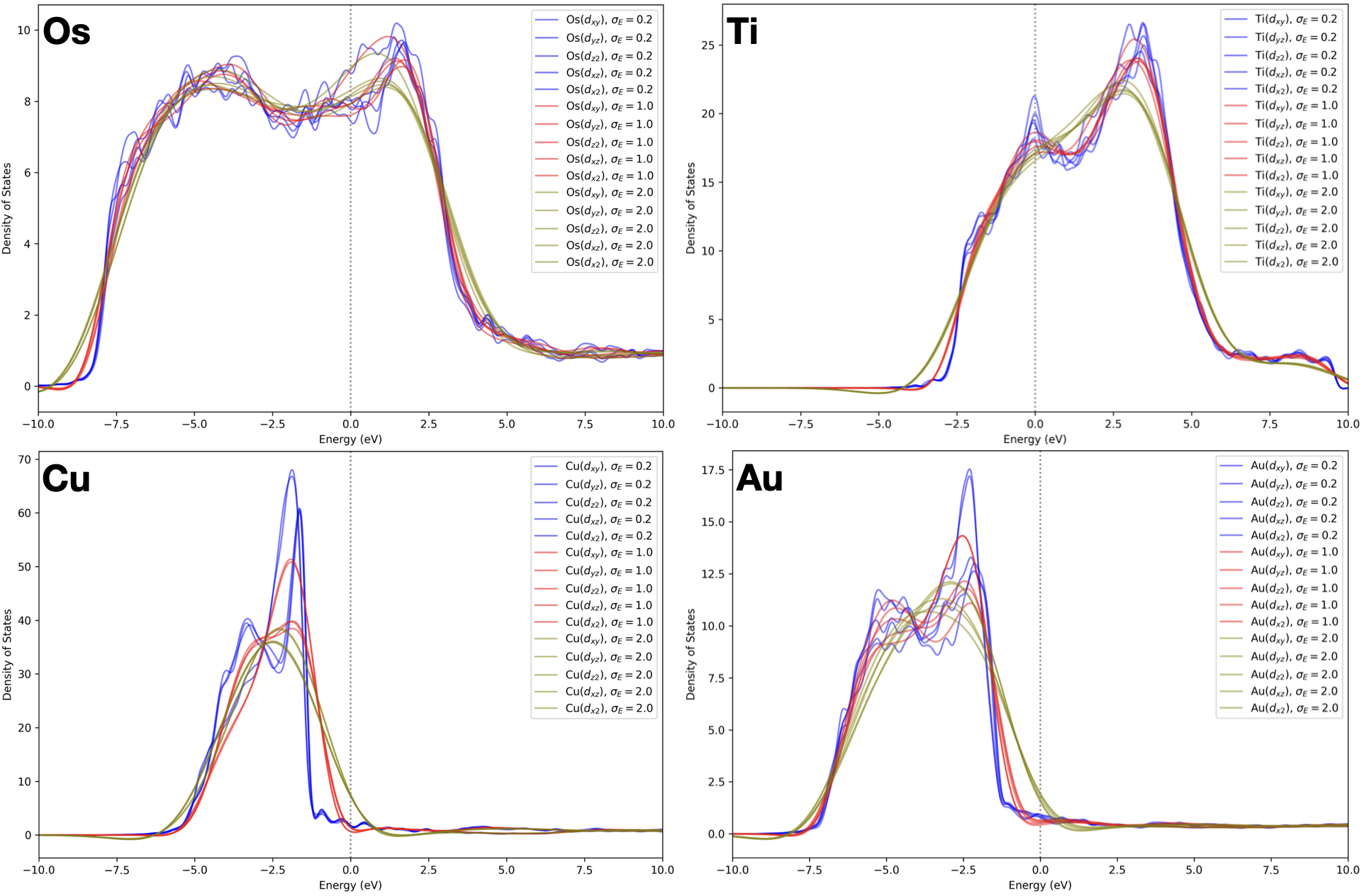}
    \caption{Angular momentum projected $d$-DOS for Au, Cu, Os, and Ti, using a randomly sampled frame from the melted test set with various values of $\sigma_e$.}
    \label{fig:dDOS}
\end{figure*}

\begin{figure*}
    \centering
    \includegraphics[width=0.6\textwidth]{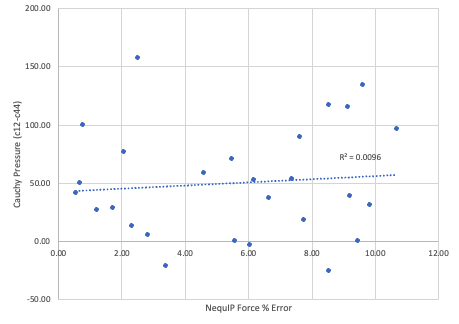}
    \caption{Correlation of NequIP force \% errors with Cauchy pressures obtained from elastic constants provided by the Materials Project.}
    \label{S-fig:cauchy}
\end{figure*}\

\begin{table*}[t]
\caption{Effect of Sparse Env. Addition on Test Error}
\label{tab:test}
\centering
\begin{tabular}{|c|c|c|c|c|c|}
\multicolumn{1}{c}{\bf Element}    &\multicolumn{1}{c}{\bf N$_{env}$} &\multicolumn{1}{c}{\bf Force \% Error}  &\multicolumn{1}{c}{\bf Energy \% Error}   \\
\hline
Au    	&	1	&	2.6118	&	0.0320	\\
Au    	&	2	&	2.5650	&	0.0320	\\
Au    	&	3	&	2.6089	&	0.0320	\\
\hline
\end{tabular}
\end{table*}

\begin{table*}[t]
\caption{Label Mean Absolute Values for Full Training Scenario Test Set (300 Frames, Final 100 from each AIMD temp)}
\label{S-tab:mavs}
\centering
\begin{tabular}{|c|c|c|c|}
\multicolumn{1}{c}{\bf }    &\multicolumn{1}{c}{\bf Force MAV}   &\multicolumn{1}{c}{\bf Energy MAV}   &\multicolumn{1}{c}{\bf Stress MAV} \\
\multicolumn{1}{c}{\bf Element}    &\multicolumn{1}{c}{\bf (meV/\AA{})}   &\multicolumn{1}{c}{\bf (meV)}   &\multicolumn{1}{c}{\bf (meV/\AA{}$^3$)} \\
\hline
Ag   & 515.91	&	3816.65	&	13.50 \\
\hline
Au   & 655.14	&	1548.15	&	34.20 \\
\hline
Cd   & 279.57	&	987.78	&	7.32 \\
\hline
Co   & 794.96	&	5850.09	&	20.98 \\
\hline
Cr   & 929.69	&	10755.07	&	21.64 \\
\hline
Cu   & 642.08	&	4669.96	&	15.35 \\
\hline
Fe   & 723.98	&	6536.83	&	44.82 \\
\hline
Hf   & 751.45	&	5975.32	&	8.94 \\
\hline
Hg   & 86.18	&	206.63	&	1.09 \\
\hline
Ir   & 1207.35	&	5303.52	&	28.68 \\
\hline
Mn   & 692.58	&	3435.07	&	13.54 \\
\hline
Mo   & 1197.63	&	12964.07	&	34.61 \\
\hline
Nb   & 891.14	&	9127.91	&	21.45 \\
\hline
Ni   & 810.41	&	5933.72	&	18.43 \\
\hline
Os   & 1594.23	&	10798.38	&	90.68 \\
\hline
Pd   & 748.09	&	3934.83	&	16.51 \\
\hline
Pt   & 872.66	&	3850.06	&	19.90 \\
\hline
Re   & 1406.68	&	12123.75	&	57.45 \\
\hline
Rh   & 987.17	&	4501.07	&	27.57 \\
\hline
Ru   & 1203.90	&	5682.51	&	55.01 \\
\hline
Ta   & 1122.21	&	12290.17	&	28.59 \\
\hline
Tc   & 1058.09	&	4847.76	&	52.36 \\
\hline
Ti   & 627.76	&	11190.78	&	10.78 \\
\hline
V    & 768.83	&	9169.87	&	22.80 \\
\hline
W    & 1497.51	&	12521.31	&	33.46 \\
\hline
Zn   & 354.61	&	1804.18	&	15.84 \\
\hline
Zr   & 586.41	&	4620.36	&	6.53 \\
\hline
\end{tabular}
\end{table*}

\begin{table*}[t]
\caption{Label Mean Absolute Values for Cold-to-Warm Transferability Test (MAVs for the 1000 Warm Frames)}
\label{S-flare_b1}
\centering
\begin{tabular}{|c|c|c|c|}
\multicolumn{1}{c}{\bf }    &\multicolumn{1}{c}{\bf Force MAV}   &\multicolumn{1}{c}{\bf Energy MAV}   &\multicolumn{1}{c}{\bf Stress MAV} \\
\multicolumn{1}{c}{\bf Element}    &\multicolumn{1}{c}{\bf (meV/\AA{})}   &\multicolumn{1}{c}{\bf (meV)}   &\multicolumn{1}{c}{\bf (meV/\AA{}$^3$)} \\
\hline
Ag&	530.35	&	5481.72	&	13.67	\\
\hline
Au	&655.14	&	1548.15	&	34.20	\\
\hline
Cd	&280.58	&	1093.64	&	6.77	\\
\hline
Co&	816.99	&	6313.36	&	8.78	\\
\hline
Cr	&1021.74	&	8691.27	&	6.35	\\
\hline
Cu	&666.74	&	6383.62	&	15.19	\\
\hline
Fe	&743.59	&	6107.72	&	49.36	\\
\hline
Hf	&748.89	&	6934.14	&	6.87	\\
\hline
Hg	&89.83	&	273.36	&	1.04	\\
\hline
Ir	&1246.43	&	7114.11	&	28.58	\\
\hline
Mn	&722.79	&	4853.81	&	5.25	\\
\hline
Mo	&1284.43	&	11565.13	&	27.92	\\
\hline
Nb	&947.92	&	7983.35	&	18.85	\\
\hline
Ni	&830.27	&	7506.52	&	17.32	\\
\hline
Os	&1619.29	&	11289.26	&	86.33	\\
\hline
Pd	&765.97	&	5143.21	&	16.01	\\
\hline
Pt&892.54	&	5161.68	&	19.36	\\
\hline
Re	&1483.27	&	20116.15	&	63.86	\\
\hline
Rh	&1030.30	&	5998.51	&	27.22	\\
\hline
Ru	&1269.30	&	3747.22	&	54.40	\\
\hline
Ta	&1215.39	&	15476.64	&	26.07	\\
\hline
Tc	&1061.78	&	3909.61	&	44.85	\\
\hline
Ti	&642.95	&	11827.47	&	9.20	\\
\hline
V	&832.71	&	7986.51	&	18.90	\\
\hline
W	&1647.33	&	14779.27	&	32.88	\\
\hline
Zn&	353.73	&	2028.22	&	13.89	\\
\hline
Zr	&606.50	&	4339.17	&	5.30	\\
\hline
\end{tabular}
\end{table*}

\begin{table*}[t]
\caption{Label Mean Absolute Values for Cold-to-Melt Transferability Test (MAVs for the 1000 Melt Frames)}
\label{S-flare_b1}
\centering
\begin{tabular}{|c|c|c|c|}
\multicolumn{1}{c}{\bf }    &\multicolumn{1}{c}{\bf Force MAV}   &\multicolumn{1}{c}{\bf Energy MAV}   &\multicolumn{1}{c}{\bf Stress MAV} \\
\multicolumn{1}{c}{\bf Element}    &\multicolumn{1}{c}{\bf (meV/\AA{})}   &\multicolumn{1}{c}{\bf (meV)}   &\multicolumn{1}{c}{\bf (meV/\AA{}$^3$)} \\
\hline
Ag	&720.58	&	11054.69	&	22.84	\\
\hline
Au	&655.14	&	1548.15	&	34.20	\\
\hline
Cd	&364.92	&	2351.20	&	10.33	\\
\hline
Co	&1109.81	&	16192.88	&	34.65	\\
\hline
Cr	&1107.74	&	28593.31	&	44.35	\\
\hline
Cu	&887.47	&	13020.66	&	28.48	\\
\hline
Fe	&1006.44	&	17263.18	&	14.66	\\
\hline
Hf	&1063.23	&	17818.92	&	18.04	\\
\hline
Hg	&113.28	&	550.76	&	1.72	\\
\hline
Ir	&1655.88	&	14675.42	&	52.23	\\
\hline
Mn	&960.57	&	9597.37	&	20.06	\\
\hline
Mo	&1432.04	&	34512.92	&	64.96	\\
\hline
Nb	&1186.20	&	24046.05	&	37.60	\\
\hline
Ni	&1136.77	&	16589.20	&	36.28	\\
\hline
Os	&2148.48	&	28305.11	&	133.14	\\
\hline
Pd	&1059.29	&	11068.59	&	32.73	\\
\hline
Pt	&1228.89	&	10795.12	&	38.36	\\
\hline
Re	&1970.88	&	34270.82	&	98.11	\\
\hline
Rh	&1348.04	&	12522.11	&	46.65	\\
\hline
Ru	&1641.95	&	15626.61	&	80.17	\\
\hline
Ta	&1540.81	&	33771.05	&	52.23	\\
\hline
Tc	&1403.84	&	14221.96	&	70.78	\\
\hline
Ti	&847.26	&	29558.41	&	18.81	\\
\hline
V	&1031.29	&	24605.74	&	43.97	\\
\hline
W	&1883.28	&	42208.89	&	77.09	\\
\hline
Zn	&465.78	&	4649.31	&	20.39	\\
\hline
Zr	&853.11	&	12641.14	&	11.35	\\
\hline
\end{tabular}
\end{table*}

\begin{table*}[t]
\caption{FLARE B1 Full Training Set (2700 Frames) MAEs on the Remaining Frames (300)}
\label{S-flare_b1}
\centering
\begin{tabular}{|c|c|c|c|}
\multicolumn{1}{c}{\bf }    &\multicolumn{1}{c}{\bf FLARE B1}   &\multicolumn{1}{c}{\bf FLARE B1}   &\multicolumn{1}{c}{\bf FLARE B1} \\
\multicolumn{1}{c}{\bf }    &\multicolumn{1}{c}{\bf Force MAE}   &\multicolumn{1}{c}{\bf Energy MAE}   &\multicolumn{1}{c}{\bf Stress MAE} \\
\multicolumn{1}{c}{\bf Element}    &\multicolumn{1}{c}{\bf (meV/\AA{})}   &\multicolumn{1}{c}{\bf (meV)}   &\multicolumn{1}{c}{\bf (meV/\AA{}$^3$)} \\
\hline
Ag   & 10.39  & 17.14   & 17.76  \\
\hline
Au   & 77.36  & 133.0   & 1.25   \\
\hline
Cd   & 26.21  & 66.75   & 10.69  \\
\hline
Co   & 132.75 & 265.61  & 3.10   \\
\hline
Cr   & 198.17 & 247.61  & 2.94   \\
\hline
Cu   & 18.32  & 45.00   & 35.01  \\
\hline
Fe   & 202.97 & 546.42  & 53.24  \\
\hline
Hf   & 187.05 & 314.37  & 1.90   \\
\hline
Hg   & 17.83  & 41.68   & 0.23   \\
\hline
Ir   & 131.01 & 156.60  & 2.06   \\
\hline
Mn   & 176.91 & 282.45  & 2.46   \\
\hline
Mo   & 226.08 & 392.02  & 2.45  \\
\hline
Nb   & 244.05 & 911.97  & 3.48   \\
\hline
Ni   & 33.38  & 47.96   & 47.74  \\
\hline
Os   & 447.91 & 2259.27 & 6.83   \\
\hline
Pd   & 41.11  & 45.41   & 30.77 \\
\hline
Pt   & 73.82  & 87.41   & 27.06  \\
\hline
Re   & 307.94 & 837.23  & 4.38   \\
\hline
Rh   & 91.24  & 115.32  & 1.56   \\
\hline
Ru   & 280.14 & 450.60  & 4.00  \\
\hline
Ta   & 223.10 & 440.00  & 2.55   \\
\hline
Tc   & 259.28 & 1239.27 & 4.08   \\
\hline
Ti   & 175.13 & 779.26  & 1.71   \\
\hline
V    & 211.47 & 512.06  & 2.64  \\
\hline
W    & 278.40 & 426.11  & 3.42   \\
\hline
Zn   & 41.18  & 221.25  & 9.38   \\
\hline
Zr   & 164.46 & 348.82  & 1.76   \\
\hline
\end{tabular}
\end{table*}

\begin{table*}[t]
\caption{FLARE B2 Full Training Set (2700 Frames) MAEs on the Remaining Frames (300)}
\label{S-flare_b1}
\centering
\begin{tabular}{|c|c|c|c|}
\multicolumn{1}{c}{\bf }     &\multicolumn{1}{c}{\bf FLARE B2}   &\multicolumn{1}{c}{\bf FLARE B2}   &\multicolumn{1}{c}{\bf FLARE B2}\\
\multicolumn{1}{c}{\bf }     &\multicolumn{1}{c}{\bf Force MAE}   &\multicolumn{1}{c}{\bf Energy MAE}   &\multicolumn{1}{c}{\bf Stress MAE}\\
\multicolumn{1}{c}{\bf Element}     &\multicolumn{1}{c}{\bf (meV/\AA{})}   &\multicolumn{1}{c}{\bf (meV)}   &\multicolumn{1}{c}{\bf (meV/\AA{}$^3$)}\\
\hline
Ag    & 6.80   & 9.34   & 0.25 \\
\hline
Au    & 17.27  & 22.50  & 0.46 \\
\hline
Cd  & 10.14  & 20.63  & 0.28 \\
\hline
Co    & 74.31  & 86.61  & 0.93 \\
\hline
Cr     & 113.72 & 133.07 & 1.30 \\
\hline
Cu    & 8.82   & 11.96  & 1.30 \\
\hline
Fe   & 107.06 & 119.52 & 1.37 \\
\hline
Hf    & 96.99  & 105.38 & 0.95 \\
\hline
Hg     & 9.64   & 16.72  & 0.10 \\
\hline
Ir     & 67.78  & 50.03  & 0.84 \\
\hline
Mn    & 111.85 & 149.06 & 1.46 \\
\hline
Mo    & 117.30 & 150.54 & 1.15 \\
\hline
Nb     & 140.52 & 212.07 & 1.25 \\
\hline
Ni   & 18.46  & 42.40  & 0.30 \\
\hline
Os   & 240.46 & 424.02 & 2.47 \\
\hline
Pd    & 26.57  & 28.19  & 0.35 \\
\hline
Pt    & 43.78  & 41.03  & 0.60 \\
\hline
Re    & 195.11 & 485.59 & 1.91 \\
\hline
Rh   & 56.48  & 49.43  & 0.66 \\
\hline
Ru    & 142.97 & 143.98 & 1.46 \\
\hline
Ta   & 133.17 & 222.35 & 1.22 \\
\hline
Tc     & 176.38 & 384.60 & 1.92 \\
\hline
Ti    & 96.77  & 213.92 & 0.74 \\
\hline
V   & 106.29 & 116.79 & 1.14 \\
\hline
W      & 120.48 & 174.54 & 1.18 \\
\hline
Zn   & 19.80  & 60.39  & 0.78 \\
\hline
Zr     & 90.00  & 127.66 & 0.83 \\
\hline
\end{tabular}
\end{table*}

\begin{table*}[t]
\caption{FLARE B2 Transferability `Cold' Training Set (1000 Frames) MAEs on the `Warm' Test Set (1000 Frames)}
\label{S-flare_b1}
\centering
\begin{tabular}{|c|c|c|c|}
\multicolumn{1}{c}{\bf }    &\multicolumn{1}{c}{\bf Force MAE}   &\multicolumn{1}{c}{\bf Energy MAE}   &\multicolumn{1}{c}{\bf Stress MAE}\\
\multicolumn{1}{c}{\bf Element}    &\multicolumn{1}{c}{\bf (meV/\AA{})}   &\multicolumn{1}{c}{\bf (meV)}   &\multicolumn{1}{c}{\bf (meV/\AA{}$^3$)}\\
\hline
Ag	&	9.81	&	15.00	&	0.67	\\
\hline
Au	&	26.70	&	45.30	&	0.83	\\
\hline
Cd	&	11.85	&	18.37	&	0.53	\\
\hline
Co	&	82.78	&	125.57	&	1.81	\\
\hline
Cr	&	116.37	&	145.81	&	2.29	\\
\hline
Cu	&	16.80	&	41.25	&	1.65	\\
\hline
Fe	&	126.94	&	162.44	&	2.13	\\
\hline
Hf	&	167.74	&	555.05	&	4.62	\\
\hline
Hg	&	14.97	&	42.93	&	0.29	\\
\hline
Ir	&	78.87	&	64.40	&	1.15	\\
\hline
Mn	&	150.42	&	177.21	&	2.18	\\
\hline
Mo	&	143.92	&	186.03	&	2.96	\\
\hline
Nb	&	165.50	&	762.10	&	1.94	\\
\hline
Ni	&	21.71	&	25.35	&	1.19	\\
\hline
Os	&	275.98	&	498.20	&	8.42	\\
\hline
Pd	&	34.07	&	30.39	&	0.88	\\
\hline
Pt	&	52.30	&	57.76	&	1.25	\\
\hline
Re	&	431.00	&	863.64	&	13.01	\\
\hline
Rh	&	64.92	&	51.78	&	1.31	\\
\hline
Ru	&	196.25	&	237.42	&	4.15	\\
\hline
Ta	&	190.48	&	374.26	&	2.94	\\
\hline
Tc	&	196.70	&	333.35	&	2.96	\\
\hline
Ti	&	123.32	&	192.84	&	1.27	\\
\hline
V	&	176.56	&	762.85	&	2.34	\\
\hline
W	&	134.63	&	144.63	&	3.13	\\
\hline
Zn	&	27.27	&	79.01	&	1.72	\\
\hline
Zr	&	140.80	&	276.34	&	1.98	\\
\hline
\end{tabular}
\end{table*}

\begin{table*}[t]
\caption{FLARE B2 Transferability `Cold' Training Set (1000 Frames) MAEs on the `Melt' Test Set (1000 Frames)}
\label{S-flare_b1}
\centering
\begin{tabular}{|c|c|c|c|}
\multicolumn{1}{c}{\bf }    &\multicolumn{1}{c}{\bf Force MAE}   &\multicolumn{1}{c}{\bf Energy MAE}   &\multicolumn{1}{c}{\bf Stress MAE}\\
\multicolumn{1}{c}{\bf Element}    &\multicolumn{1}{c}{\bf (meV/\AA{})}   &\multicolumn{1}{c}{\bf (meV)}   &\multicolumn{1}{c}{\bf (meV/\AA{}$^3$)}\\
\hline
Ag	&	21.75	&	43.78	&	1.70	\\
\hline
Au	&	46.51	&	126.04	&	1.97	\\
\hline
Cd	&	18.17	&	53.39	&	1.12	\\
\hline
Co	&	156.97	&	766.40	&	6.18	\\
\hline
Cr	&	407.91	&	2056.21	&	29.82	\\
\hline
Cu	&	43.67	&	223.30	&	2.68	\\
\hline
Fe	&	432.67	&	999.83	&	11.76	\\
\hline
Hf	&	423.22	&	688.58	&	10.38	\\
\hline
Hg	&	16.33	&	38.05	&	0.44	\\
\hline
Ir	&	185.06	&	376.53	&	2.51	\\
\hline
Mn	&	280.84	&	426.74	&	4.00	\\
\hline
Mo	&	605.45	&	4121.33	&	22.25	\\
\hline
Nb	&	294.76	&	1806.51	&	4.69	\\
\hline
Ni	&	56.31	&	139.10	&	5.00	\\
\hline
Os	&	359.93	&	1507.91	&	16.04	\\
\hline
Pd	&	91.32	&	263.56	&	3.29	\\
\hline
Pt	&	110.75	&	318.73	&	3.71	\\
\hline
Re	&	656.14	&	1342.49	&	12.15	\\
\hline
Rh	&	148.14	&	224.84	&	3.04	\\
\hline
Ru	&	252.72	&	636.88	&	6.64	\\
\hline
Ta	&	442.65	&	1286.94	&	5.41	\\
\hline
Tc	&	236.05	&	393.10	&	5.60	\\
\hline
Ti	&	316.36	&	967.36	&	8.33	\\
\hline
V	&	449.28	&	6690.84	&	16.04	\\
\hline
W	&	551.26	&	1111.75	&	18.34	\\
\hline
Zn	&	31.15	&	93.43	&	1.90	\\
\hline
Zr	&	356.56	&	2476.42	&	6.67	\\
\hline
\end{tabular}
\end{table*}

\begin{table*}[t]
\caption{NequIP Full Training Set (2700 Frames) MAEs on the Remaining Frames (300)}
\label{S-flare_b1}
\centering
\begin{tabular}{|c|c|c|c|}
\multicolumn{1}{c}{\bf }    &\multicolumn{1}{c}{\bf Force MAE}   &\multicolumn{1}{c}{\bf Energy MAE}   &\multicolumn{1}{c}{\bf Stress MAE} \\
\multicolumn{1}{c}{\bf Element}    &\multicolumn{1}{c}{\bf (meV/\AA{})}   &\multicolumn{1}{c}{\bf (meV)}   &\multicolumn{1}{c}{\bf (meV/\AA{}$^3$)} \\
\hline
Ag   & 2.86	&	2.71	&	0.29\\
\hline
Au   & 5.24	&	4.96	&	0.49\\
\hline
Cd   & 4.80	&	8.13	&	0.22\\
\hline
Co   & 43.61	&	50.38	&	0.98\\
\hline
Cr   & 61.84	&	58.99	&	1.02\\
\hline
Cu   & 4.41	&	6.87	&	0.87\\
\hline
Fe   & 53.45	&	47.00	&	1.02\\
\hline
Hf   & 58.38	&	63.20	&	0.84\\
\hline
Hg   & 4.82	&	4.95	&	0.07\\
\hline
Ir   & 41.26	&	30.36	&	2.12\\
\hline
Mn   & 65.66	&	56.64	&	1.38\\
\hline
Mo   & 73.89	&	81.30	&	1.09\\
\hline
Nb   & 85.85	&	108.90	&	1.35\\
\hline
Ni   & 10.05	&	41.84	&	0.50\\
\hline
Os   & 136.19	&	166.51	&	1.83\\
\hline
Pd   & 15.63	&	23.08	&	0.92\\
\hline
Pt   & 22.09	&	19.74	&	0.88\\
\hline
Re   & 120.21	&	135.47	&	1.51\\
\hline
Rh   & 27.98	&	24.68	&	0.99\\
\hline
Ru   & 73.06	&	65.81	&	1.19\\
\hline
Ta   & 85.62	&	117.87	&	1.22\\
\hline
Tc   & 113.03	&	160.72	&	1.62\\
\hline
Ti   & 61.76	&	117.77	&	0.68\\
\hline
V    & 70.45	&	68.74	&	1.11\\
\hline
W    & 69.02	&	85.13	&	0.80\\
\hline
Zn   & 8.27	&	14.89	&	0.44\\
\hline
Zr   & 54.10	&	57.09	&	0.77\\
\hline
\end{tabular}
\end{table*}

\begin{table*}[t]
\caption{NequIP Transferability `Cold' Training Set (1000 Frames) MAEs on the `Warm' Test Set (1000 Frames)}
\label{S-flare_b1}
\centering
\begin{tabular}{|c|c|c|c|}
\multicolumn{1}{c}{\bf }    &\multicolumn{1}{c}{\bf Force MAE}   &\multicolumn{1}{c}{\bf Energy MAE}   &\multicolumn{1}{c}{\bf Stress MAE} \\
\multicolumn{1}{c}{\bf Element}    &\multicolumn{1}{c}{\bf (meV/\AA{})}   &\multicolumn{1}{c}{\bf (meV)}   &\multicolumn{1}{c}{\bf (meV/\AA{}$^3$)} \\
\hline
Ag	&	5.20	&	6.16	&	0.22	\\
\hline
Au	&	13.13	&	66.68	&	0.77	\\
\hline
Cd	&	5.58	&	6.77	&	0.24	\\
\hline
Co	&	84.20	&	101.49	&	1.76	\\
\hline
Cr	&	61.89	&	79.54	&	1.42	\\
\hline
Cu	&	9.23	&	15.97	&	0.48	\\
\hline
Fe	&	79.32	&	101.25	&	1.88	\\
\hline
Hf	&	120.33	&	233.75	&	3.07	\\
\hline
Hg	&	9.48	&	14.17	&	0.09	\\
\hline
Ir	&	45.43	&	31.03	&	4.68	\\
\hline
Mn	&	86.53	&	99.18	&	2.67	\\
\hline
Mo	&	92.35	&	136.44	&	3.46	\\
\hline
Nb	&	116.70	&	174.67	&	3.61	\\
\hline
Ni	&	18.65	&	41.36	&	0.50	\\
\hline
Os	&	192.03	&	215.75	&	3.02	\\
\hline
Pd	&	18.12	&	24.87	&	2.36	\\
\hline
Pt	&	33.40	&	22.90	&	2.28	\\
\hline
Re	&	321.46	&	1473.16	&	10.37	\\
\hline
Rh	&	40.78	&	32.42	&	2.31	\\
\hline
Ru	&	122.87	&	694.33	&	3.41	\\
\hline
Ta	&	140.11	&	244.73	&	2.17	\\
\hline
Tc	&	145.92	&	186.39	&	1.77	\\
\hline
Ti	&	79.09	&	207.67	&	2.21	\\
\hline
V	&	87.55	&	91.38	&	1.54	\\
\hline
W	&	82.07	&	82.49	&	1.73	\\
\hline
Zn	&	14.11	&	39.54	&	0.70	\\
\hline
Zr	&	116.86	&	217.21	&	4.13	\\
\hline
\end{tabular}
\end{table*}

\begin{table*}[t]
\caption{NequIP Transferability `Cold' Training Set (1000 Frames) MAEs on the `Melt' Test Set (1000 Frames)}
\label{S-tab:nequip_cm}
\centering
\begin{tabular}{|c|c|c|c|}
\multicolumn{1}{c}{\bf }    &\multicolumn{1}{c}{\bf Force MAE}   &\multicolumn{1}{c}{\bf Energy MAE}   &\multicolumn{1}{c}{\bf Stress MAE} \\
\multicolumn{1}{c}{\bf Element}    &\multicolumn{1}{c}{\bf (meV/\AA{})}   &\multicolumn{1}{c}{\bf (meV)}   &\multicolumn{1}{c}{\bf (meV/\AA{}$^3$)} \\
\hline
Ag	&19.23	&	53.90	&	0.39	\\
\hline
Au	&29.94	&	60.93	&	1.21	\\
\hline
Cd	&9.58	&	20.94	&	0.30	\\
\hline
Co	&544.01	&	1709.43	&	15.30	\\
\hline
Cr	&397.77	&	1685.15	&	19.52	\\
\hline
Cu	&49.51	&	177.55	&	0.97	\\
\hline
Fe	&314.26	&	1902.89	&	3.85	\\
\hline
Hf	&408.50	&	1585.52	&	5.86	\\
\hline
Hg	&10.26	&	19.97	&	0.13	\\
\hline
Ir	&104.07	&	111.27	&	9.89	\\
\hline
Mn	&136.15	&	198.76	&	5.47	\\
\hline
Mo	&350.83	&	816.24	&	14.52	\\
\hline
Nb	&197.89	&	793.61	&	9.19	\\
\hline
Ni	&116.17	&	397.06	&	1.55	\\
\hline
Os	&305.83	&	522.27	&	5.76	\\
\hline
Pd	&50.27	&	56.15	&	4.85	\\
\hline
Pt	&242.19	&	260.68	&	8.45	\\
\hline
Re	&430.53	&	3277.55	&	12.89	\\
\hline
Rh	&867.98	&	558.19	&	28.70	\\
\hline
Ru	&187.43	&	319.94	&	2.73	\\
\hline
Ta	&333.95	&	1488.47	&	6.27	\\
\hline
Tc	&180.41	&	282.42	&	3.72	\\
\hline
Ti	&173.05	&	821.37	&	2.72	\\
\hline
V	&287.41	&	411.56	&	6.56	\\
\hline
W	&382.58	&	1071.21	&	10.07	\\
\hline
Zn	&15.82	&	37.82	&	0.75	\\
\hline
Zr	&383.11	&	1291.73	&	7.56	\\
\hline
\end{tabular}
\end{table*}

\begin{table*}[t]
\caption{FLARE B1 (Power = 1) Best Model Parameters}
\label{S-tab:flare_B1}
\centering
\begin{tabular}{|c|c|c|c|}
\multicolumn{1}{c}{\bf }    &\multicolumn{1}{c}{\bf Cutoff Radius }   &\multicolumn{1}{c}{\bf Radial Fidelity}   &\multicolumn{1}{c}{\bf Angular Fidelity }\\
\multicolumn{1}{c}{\bf Element}    &\multicolumn{1}{c}{\bf (r$_{cutoff}$, \AA{})}   &\multicolumn{1}{c}{\bf (n$_{max}$)}   &\multicolumn{1}{c}{\bf (l$_{max}$)}\\
\hline
Ag   & 9.0   & 23   & 0 \\
\hline
Au   & 9.0   & 16   & 0 \\
\hline
Cd   & 9.0   & 21   & 0 \\
\hline
Co   & 7.0   & 27   & 0 \\
\hline
Cr   & 6.0   & 29   & 0 \\
\hline
Cu   & 7.0   & 15   & 0 \\
\hline
Fe   & 8.0   & 15   & 0 \\
\hline
Hf   & 9.0   & 29   & 0 \\
\hline
Hg   & 9.0   & 21   & 0 \\
\hline
Ir   & 9.0   & 27   & 0 \\
\hline
Mn   & 6.0   & 27   & 0 \\
\hline
Mo   & 8.0   & 29   & 0 \\
\hline
Nb   & 8.0   & 25   & 0 \\
\hline
Ni   & 9.0   & 19   & 0 \\
\hline
Os   & 9.0   & 21   & 0 \\
\hline
Pd   & 9.0   & 19   & 0 \\
\hline
Pt   & 9.0   & 21   & 0 \\
\hline
Re   & 8.0   & 27   & 0 \\
\hline
Rh   & 9.0   & 19   & 0 \\
\hline
Ru   & 9.0   & 29   & 0 \\
\hline
Ta   & 9.0   & 25   & 0 \\
\hline
Tc   & 8.0   & 25   & 0 \\
\hline
Ti   & 8.0   & 29   & 0 \\
\hline
V    & 8.0   & 29   & 0 \\
\hline
W    & 8.0   & 29   & 0 \\
\hline
Zn   & 9.0   & 15   & 0 \\
\hline
Zr   & 9.0   & 25   & 0 \\
\hline
\end{tabular}
\end{table*}

\begin{table*}[t]
\caption{FLARE B1 (Power = 2) Best Model Parameters}
\label{S-tab:flare_B1_p2}
\centering
\begin{tabular}{|c|c|c|c|}
\multicolumn{1}{c}{\bf }    &\multicolumn{1}{c}{\bf Cutoff Radius }   &\multicolumn{1}{c}{\bf Radial Fidelity}   &\multicolumn{1}{c}{\bf Angular Fidelity }\\
\multicolumn{1}{c}{\bf Element}    &\multicolumn{1}{c}{\bf (r$_{cutoff}$, \AA{})}   &\multicolumn{1}{c}{\bf (n$_{max}$)}   &\multicolumn{1}{c}{\bf (l$_{max}$)}\\
\hline
Ag   & 9.0   & 25   & 0 \\
\hline
Au   & 9.0   & 16   & 0 \\
\hline
Cd   & 9.0   & 19   & 0 \\
\hline
Co   & 9.0   & 29   & 0 \\
\hline
Cr   & 7.0   & 29   & 0 \\
\hline
Cu   & 8.0   & 21   & 0 \\
\hline
Fe   & 7.0   & 29   & 0 \\
\hline
Hf   & 8.0   & 13   & 0 \\
\hline
Hg   & 9.0   & 23   & 0 \\
\hline
Ir   & 9.0   & 27   & 0 \\
\hline
Mn   & 8.0   & 27   & 0 \\
\hline
Mo   & 8.0   & 25   & 0 \\
\hline
Nb   & 9.0   & 25   & 0 \\
\hline
Ni   & 9.0   & 19   & 0 \\
\hline
Os   & 9.0   & 19   & 0 \\
\hline
Pd   & 9.0   & 25   & 0 \\
\hline
Pt   & 9.0   & 25   & 0 \\
\hline
Re   & 9.0   & 25   & 0 \\
\hline
Rh   & 9.0   & 27   & 0 \\
\hline
Ru   & 9.0   & 27   & 0 \\
\hline
Ta   & 9.0   & 29   & 0 \\
\hline
Tc   & 8.0   & 23   & 0 \\
\hline
Ti   & 8.0   & 29   & 0 \\
\hline
V    & 7.0   & 21   & 0 \\
\hline
W    & 7.0   & 19   & 0 \\
\hline
Zn   & 9.0   & 21   & 0 \\
\hline
Zr   & 8.0   & 17   & 0 \\
\hline
\end{tabular}
\end{table*}

\begin{table*}[t]
\caption{FLARE B2 (Power = 1) Best Model Parameters}
\label{S-tab:flare_b2}
\centering
\begin{tabular}{|c|c|c|c|}
\multicolumn{1}{c}{\bf }    &\multicolumn{1}{c}{\bf Cutoff Radius }   &\multicolumn{1}{c}{\bf Radial Fidelity}   &\multicolumn{1}{c}{\bf Angular Fidelity }\\
\multicolumn{1}{c}{\bf Element}    &\multicolumn{1}{c}{\bf (r$_{cutoff}$, \AA{})}   &\multicolumn{1}{c}{\bf (n$_{max}$)}   &\multicolumn{1}{c}{\bf (l$_{max}$)}\\
\hline
Ag   & 7.0   & 13   & 4 \\
\hline
Au   & 9.0   & 13   & 5 \\
\hline
Cd   & 9.0   & 15   & 5 \\
\hline
Co   & 5.0   & 15   & 4 \\
\hline
Cr   & 6.0   & 19   & 5 \\
\hline
Cu   & 6.0   & 13   & 4 \\
\hline
Fe   & 6.0   & 19   & 4 \\
\hline
Hf   & 8.0   & 17   & 4 \\
\hline
Hg   & 9.0   & 15   & 5 \\
\hline
Ir   & 7.0   & 19   & 5 \\
\hline
Mn   & 6.0   & 9    & 5 \\
\hline
Mo   & 6.0   & 17   & 4 \\
\hline
Nb   & 7.0   & 19   & 5 \\
\hline
Ni   & 8.0   & 19   & 5 \\
\hline
Os   & 7.0   & 19   & 4 \\
\hline
Pd   & 9.0   & 15   & 4 \\
\hline
Pt   & 8.0   & 19   & 5 \\
\hline
Re   & 7.0   & 19   & 4 \\
\hline
Rh   & 7.0   & 15   & 5 \\
\hline
Ru   & 6.0   & 15   & 3 \\
\hline
Ta   & 8.0   & 19   & 5 \\
\hline
Tc   & 7.0   & 19   & 4 \\
\hline
Ti   & 6.0   & 19   & 5 \\
\hline
V    & 6.0   & 19   & 5 \\
\hline
W    & 6.0   & 17   & 4 \\
\hline
Zn   & 9.0   & 15   & 5 \\
\hline
Zr   & 8.0   & 17   & 4 \\
\hline
\end{tabular}
\end{table*}

\begin{table*}[t]
\caption{FLARE B2 (Power = 2) Best Model Parameters}
\label{S-tab:flare_b2_p2}
\centering
\begin{tabular}{|c|c|c|c|}
\multicolumn{1}{c}{\bf }    &\multicolumn{1}{c}{\bf Cutoff Radius }   &\multicolumn{1}{c}{\bf Radial Fidelity}   &\multicolumn{1}{c}{\bf Angular Fidelity }\\
\multicolumn{1}{c}{\bf Element}    &\multicolumn{1}{c}{\bf (r$_{cutoff}$, \AA{})}   &\multicolumn{1}{c}{\bf (n$_{max}$)}   &\multicolumn{1}{c}{\bf (l$_{max}$)}\\
\hline
Ag   & 6.0   & 9    & 3 \\ 
\hline
Au   & 6.0   & 10   & 3 \\
\hline
Cd   & 7.0   & 13   & 3 \\
\hline
Co   & 5.0   & 11   & 4 \\
\hline
Cr   & 4.0   & 9    & 5 \\
\hline
Cu   & 5.0   & 7    & 4 \\
\hline
Fe   & 4.0   & 11   & 6 \\
\hline
Hf   & 6.0   & 9    & 4 \\
\hline
Hg   & 8.0   & 9    & 4 \\
\hline
Ir   & 6.0   & 17   & 4 \\
\hline
Mn   & 4.0   & 11   & 6 \\
\hline
Mo   & 5.0   & 11   & 4 \\
\hline
Nb   & 5.0   & 9    & 4 \\
\hline
Ni   & 5.0   & 11   & 4 \\
\hline
Os   & 5.0   & 9    & 4 \\
\hline
Pd   & 5.0   & 9    & 4 \\
\hline
Pt   & 5.0   & 9    & 4 \\
\hline
Re   & 5.0   & 9    & 4 \\
\hline
Rh   & 5.0   & 11   & 4 \\
\hline
Ru   & 5.0   & 9    & 4 \\
\hline
Ta   & 5.0   & 9    & 4 \\
\hline
Tc   & 5.0   & 9    & 4 \\
\hline
Ti   & 5.0   & 9    & 5 \\
\hline
V    & 5.0   & 9    & 5 \\
\hline
W    & 5.0   & 11   & 4 \\
\hline
Zn   & 6.0   & 13   & 3 \\
\hline
Zr   & 6.0   & 9    & 4 \\
\hline
\end{tabular}
\end{table*}

\begin{table*}[t]
\caption{NequIP Grid Test}
\label{S-tab:nequip_features}
\centering
\begin{tabular}{|c|c|c|c|}
\multicolumn{1}{c}{\bf Feature }\\
\multicolumn{1}{c}{\bf Representation}\\
\hline
32x0e \\
\hline
128x0e \\
\hline
256x0e \\
\hline
32x0e + 32x0o + 16x1e + 16x1o \\
\hline
32x0o + 32x0e + 32x1o + 32x1e \\
\hline
64x0o + 64x0e + 64x1o + 64x1e \\
\hline
16x0o + 16x0e + 16x1o + 16x1e + 16x2o + 16x2e  \\
\hline
32x0o + 32x0e + 32x1o + 32x1e + 32x2o + 32x2e \\
\hline
32x0o + 32x0e + 32x2o + 32x2e \\
\hline
64x0o + 64x0e + 64x2o + 64x2e \\
\hline
8x0o + 8x0e + 8x1o + 8x1e + 8x2o + 8x2e + 8x3o + 8x3e \\
\hline
16x0o + 16x0e + 16x1o + 16x1e + 16x2o + 16x2e + 16x3o + 16x3e \\
\hline
32x0o + 32x0e + 16x1o + 16x1e + 8x2o + 8x2e + 4x3o + 4x3e \\
\hline
32x0o + 32x0e + 32x1o + 32x1e + 32x2o + 32x2e + 32x3o + 32x3o \\
\hline
64x0o + 64x0e + 32x1o + 32x1e + 16x2o + 16x2e + 8x3o + 8x3e \\
\hline
32x0o + 32x0e + 32x3o + 32x3e \\
\hline
64x0o + 64x0e + 64x3o + 64x3e \\
\hline
4x0o + 4x0e + 4x1o + 4x1e + 4x2o + 4x2e + 4x3o + 4x3e + 4x4o + 4x4e \\
\hline
8x0o + 8x0e + 8x1o + 8x1e + 8x2o + 8x2e + 8x3o + 8x3e + 8x4o + 8x4e \\
\hline
32x0o + 32x0e + 16x1o + 16x1e + 8x2o + 8x2e + 4x3o + 4x3e + 2x4o + 2x4e \\
\hline
32x0o + 32x0e + 32x1o + 32x1e + 32x2o + 32x2e + 32x3o + 32x3o + 32x4o + 32x4o \\
\hline
64x0o + 64x0e + 32x1o + 32x1e + 16x2o + 16x2e + 8x3o + 8x3e + 4x4o + 4x4e \\
\hline
32x0o + 32x0e + 32x4o + 32x4e \\
\hline
64x0o + 64x0e + 64x4o + 64x4e\\
\hline
32x0o + 32x0e + 32x1o + 32x1e + 32x2o + 32x2e + 32x3o + 32x3o + 32x4o + 32x4o + 32x5o + 32x5o \\
\hline
\end{tabular}
\end{table*}

\begin{table*}[t]
\caption{NequIP Best Model Parameters}
\label{S-tab:nequip_best}
\centering
\begin{tabular}{|c|c|c|c|}
\multicolumn{1}{c}{\bf  }    & \multicolumn{1}{c}{\bf Cutoff Radius}   &  \multicolumn{1}{c}{\bf Learning } &  \multicolumn{1}{c}{\bf  Feature}\\
\multicolumn{1}{c}{\bf Element}    & \multicolumn{1}{c}{\bf (r$_{cutoff}$, \AA{})}   &  \multicolumn{1}{c}{\bf  Rate} &  \multicolumn{1}{c}{\bf  Expression}\\
\hline
Ag   & 6.0     & 0.01 &  32x(0e+0o) + 16x(1e+1o) + 8x(2e+2o) \\
     &            &         &  + 4x(3e+3o) + 2x(4e+4o) \\
\hline
Au   & 6.0     & 0.01 &  64x(0e+0o) + 32x(1e+1o) + 16x(2e+2o) \\
     &            &         &  + 8x(3e+3o) + 4x(4e+4o) \\
\hline
Cd   & 6.0     & 0.001 &  32x(0e+0o) + 16x(1e+1o) + 8x(2e+2o) \\
     &            &         &  + 4x(3e+3o) + 2x(4e+4o) \\
\hline
Co   & 4.0     & 0.01 &  32x(0e+0o) + 16x(1e+1o) + 8x(2e+2o) \\
     &            &         &  + 4x(3e+3o) + 2x(4e+4o) \\
\hline
Cr   & 6.0     & 0.001 &  32x(0e+0o) + 16x(1e+1o) + 8x(2e+2o) \\
     &            &         &  + 4x(3e+3o) + 2x(4e+4o) \\
\hline
Cu   & 6.0     & 0.01 &  32x(0e+0o) + 16x(1e+1o) + 8x(2e+2o) \\
     &            &         &  + 4x(3e+3o) + 2x(4e+4o) \\
\hline
Fe   & 4.0     & 0.001 &  64x(0e+0o) + 32x(1e+1o) + 16x(2e+2o) \\
     &            &         &  + 8x(3e+3o) + 4x(4e+4o) \\
\hline
Hf   & 6.0     & 0.001 &  32x(0e+0o) + 16x(1e+1o) + 8x(2e+2o) \\
     &            &         &  + 4x(3e+3o) + 2x(4e+4o) \\
\hline
Hg   & 6.0     & 0.001 &  32x(0e+0o) + 16x(1e+1o) + 8x(2e+2o) \\
     &            &         &  + 4x(3e+3o) + 2x(4e+4o) \\
\hline
Ir   & 6.0     & 0.001 &  64x(0e+0o) + 32x(1e+1o) + 16x(2e+2o) \\
     &            &         &  + 8x(3e+3o) + 4x(4e+4o) \\
\hline
Mn   & 6.0     & 0.001 &  32x(0e+0o) + 16x(1e+1o) + 8x(2e+2o) \\
     &            &         &  + 4x(3e+3o) + 2x(4e+4o) \\
\hline
Mo   & 6.0     & 0.001 &  32x(0e+0o) + 16x(1e+1o) + 8x(2e+2o) \\
     &            &         &  + 4x(3e+3o) + 2x(4e+4o) \\
\hline
Nb   & 6.0     & 0.001 &  32x(0e+0o) + 16x(1e+1o) + 8x(2e+2o) \\
     &            &         &  + 4x(3e+3o) + 2x(4e+4o) \\
\hline
Ni   & 4.0     & 0.01 &  64x(0e+0o) + 32x(1e+1o) + 16x(2e+2o) \\
     &            &         &  + 8x(3e+3o) + 4x(4e+4o) \\
\hline
Os   & 4.0     & 0.001 &  32x(0e+0o) + 16x(1e+1o) + 8x(2e+2o) \\
     &            &         &  + 4x(3e+3o) + 2x(4e+4o) \\
\hline
Pd   & 6.0     & 0.001 &  64x(0e+0o) + 32x(1e+1o) + 16x(2e+2o) \\
     &            &         &  + 8x(3e+3o) + 4x(4e+4o) \\
\hline
Pt   & 4.0     & 0.01 &  32x(0e+0o) + 16x(1e+1o) + 8x(2e+2o) \\
     &            &         &  + 4x(3e+3o) + 2x(4e+4o) \\
\hline
Re   & 6.0     & 0.001 &  32x(0e+0o) + 16x(1e+1o) + 8x(2e+2o) \\
     &            &         &  + 4x(3e+3o) + 2x(4e+4o) \\
\hline
Rh   & 4.0     & 0.01 &  32x(0e+0o) + 16x(1e+1o) + 8x(2e+2o) \\
     &            &         &  + 4x(3e+3o) + 2x(4e+4o) \\
\hline
Ru   & 4.0     & 0.001 &  64x(0e+0o) + 32x(1e+1o) + 16x(2e+2o) \\
     &            &         &  + 8x(3e+3o) + 4x(4e+4o) \\
\hline
Ta   & 6.0     & 0.001 &  32x(0e+0o) + 16x(1e+1o) + 8x(2e+2o) \\
     &            &         &  + 4x(3e+3o) + 2x(4e+4o) \\
\hline
Tc   & 6.0     & 0.001 &  32x(0e+0o) + 16x(1e+1o) + 8x(2e+2o) \\
     &            &         &  + 4x(3e+3o) + 2x(4e+4o) \\
\hline
Ti   & 6.0     & 0.001 &  32x(0e+0o) + 16x(1e+1o) + 8x(2e+2o) \\
     &            &         &  + 4x(3e+3o) + 2x(4e+4o) \\
\hline
V    & 6.0     & 0.001 &  32x(0e+0o) + 16x(1e+1o) + 8x(2e+2o) \\
     &            &         &  + 4x(3e+3o) + 2x(4e+4o) \\
\hline
W    & 6.0     & 0.001 &  32x(0e+0o) + 16x(1e+1o) + 8x(2e+2o) \\
     &            &         &  + 4x(3e+3o) + 2x(4e+4o) \\
\hline
Zn   & 6.0     & 0.001 &  32x(0e+0o) + 16x(1e+1o) + 8x(2e+2o) \\
     &            &         &  + 4x(3e+3o) + 2x(4e+4o) \\
\hline
Zr   & 6.0     & 0.001 &  64x(0e+0o) + 32x(1e+1o) + 16x(2e+2o) \\
     &            &         &  + 8x(3e+3o) + 4x(4e+4o) \\
\hline
\end{tabular}
\end{table*}

\begin{table*}[t]
\caption{Forced Symmetry Tests}
\label{S-tab:forced}
\centering
\begin{tabular}{|c|c|c|c|c|}
 \multicolumn{1}{c}{\bf   }   &   \multicolumn{1}{c}{\bf   }  &  \multicolumn{1}{c}{\bf  Force } & \multicolumn{1}{c}{\bf  Force } & \multicolumn{1}{c}{\bf  Force }\\
\multicolumn{1}{c}{\bf Element}    & \multicolumn{1}{c}{\bf Symmetry}   &  \multicolumn{1}{c}{\bf   MAE (meV/\AA{})} & \multicolumn{1}{c}{\bf   $\%$ Error} & \multicolumn{1}{c}{\bf   MAV (meV/\AA{})}\\
\hline
   &   \textit{fcc}   & 66 & 9 & 742 \\
Au   &   \textit{hcp}   & 51 & 8 & 637 \\
\hline
   &   \textit{hcp}   & 183 & 22 & 838 \\
Hf   &   \textit{fcc}   & 222 & 25 & 874 \\
\hline
   &   \textit{hcp}   & 457 & 26 & 1755 \\
Os   &   \textit{fcc}   & 329 & 18 & 1789 \\
\hline
   &   \textit{bcc}   & 256 & 20 & 1277 \\
Ta   &   \textit{fcc}   & 264 & 22 & 1181 \\
\hline
   &   \textit{bcc}   & 341 & 19 & 1781 \\
W   &   \textit{fcc}   & 349 & 23 & 1499 \\
\hline
\end{tabular}
\end{table*}

    \begin{figure*}
        \centering
        \includegraphics[width=.8\textwidth]{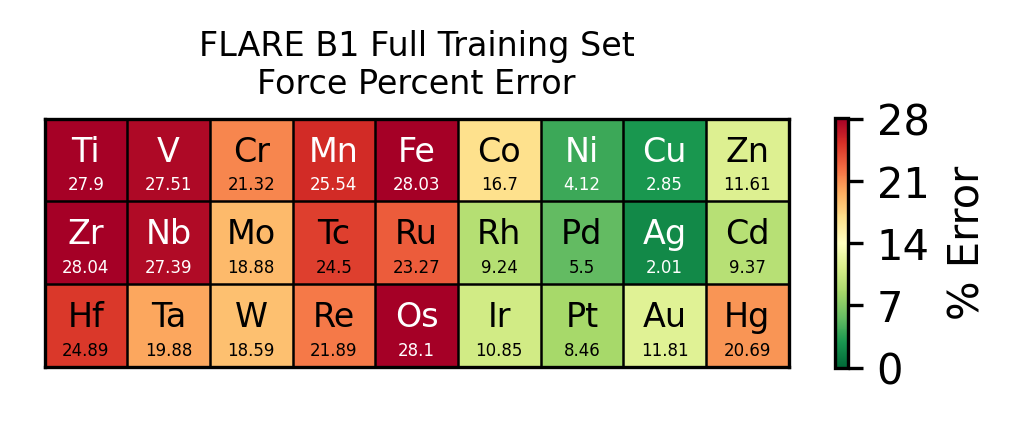}
        \caption{}
        \label{S-fig:firsterror}
    \end{figure*}

    \begin{figure*}
        \centering
        \includegraphics[width=.8\textwidth]{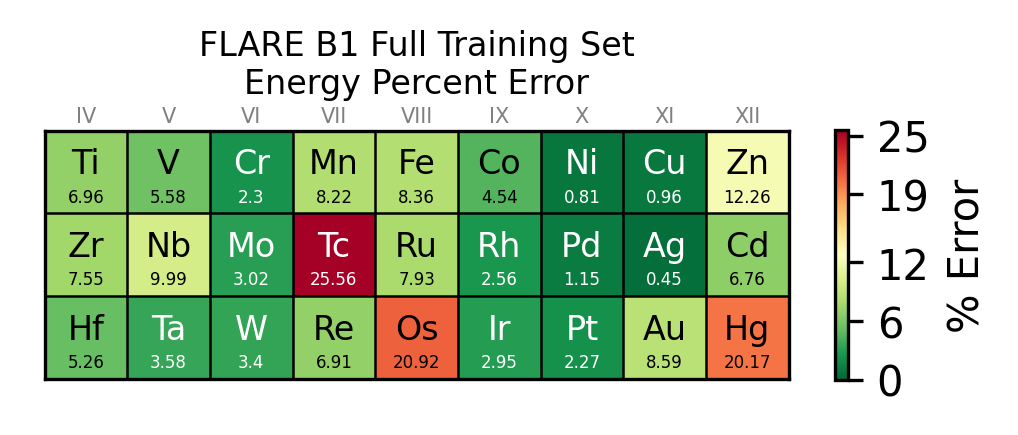}
        \caption{}
    \end{figure*}

    \begin{figure*}
        \centering
        \includegraphics[width=.8\textwidth]{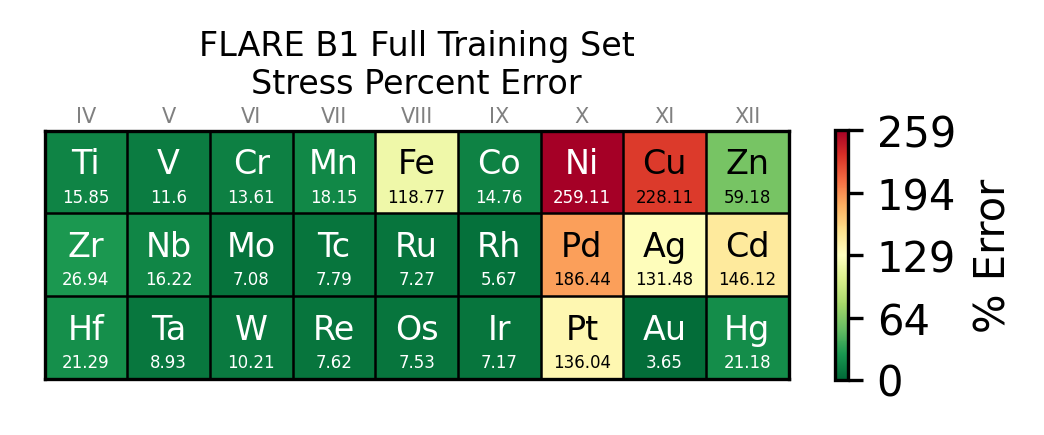}
        \caption{}
    \end{figure*}

    \begin{figure*}
        \centering
        \includegraphics[width=.8\textwidth]{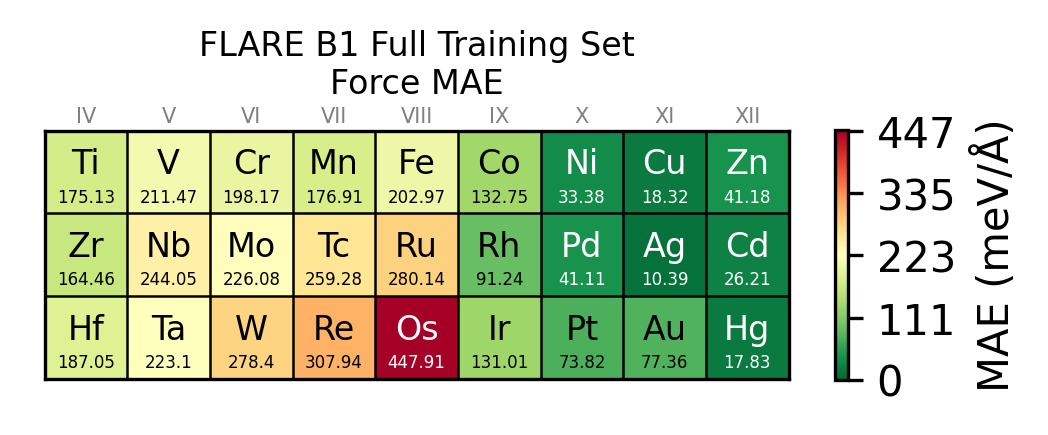}
        \caption{}
    \end{figure*}

    \begin{figure*}
        \centering
        \includegraphics[width=.8\textwidth]{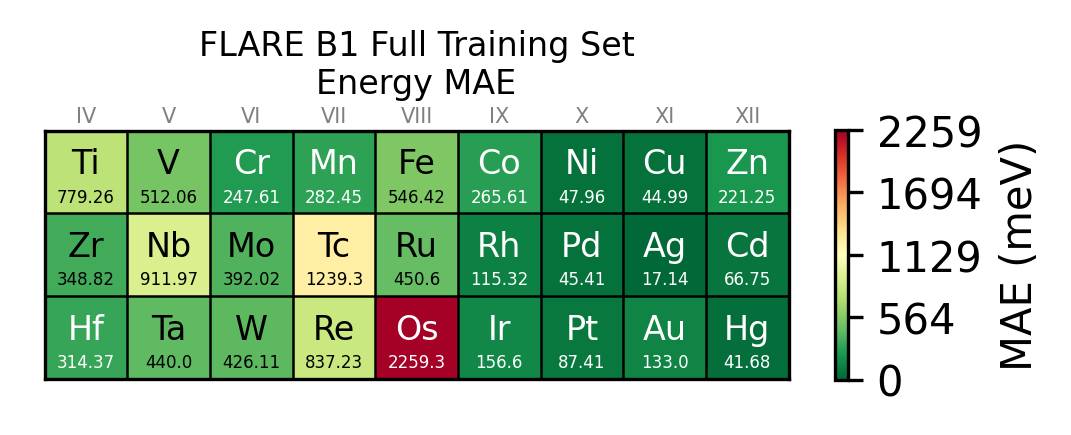}
        \caption{}
    \end{figure*}

    \begin{figure*}
        \centering
        \includegraphics[width=.8\textwidth]{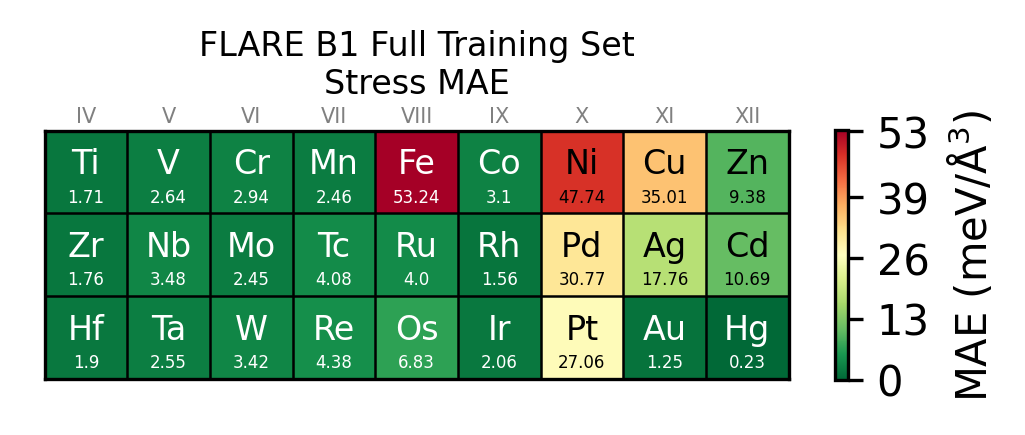}
        \caption{}
    \end{figure*}

    \begin{figure*}
        \centering
        \includegraphics[width=.8\textwidth]{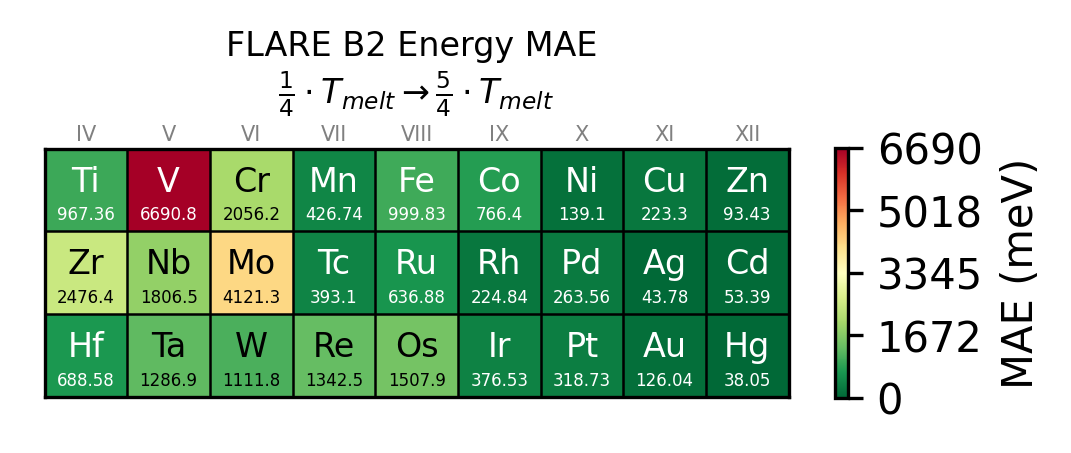}
        \caption{}
    \end{figure*}

    \begin{figure*}
        \centering
        \includegraphics[width=.8\textwidth]{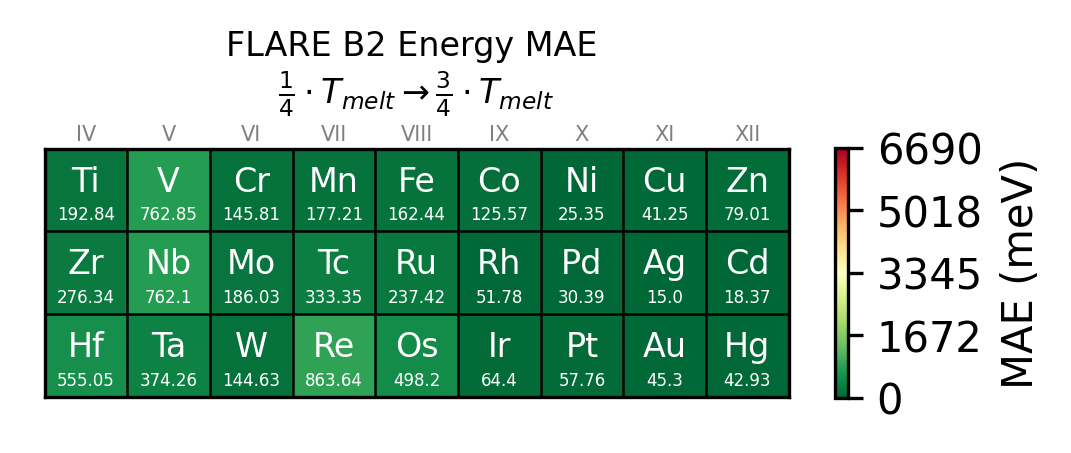}
        \caption{}
    \end{figure*}

    \begin{figure*}
        \centering
        \includegraphics[width=.8\textwidth]{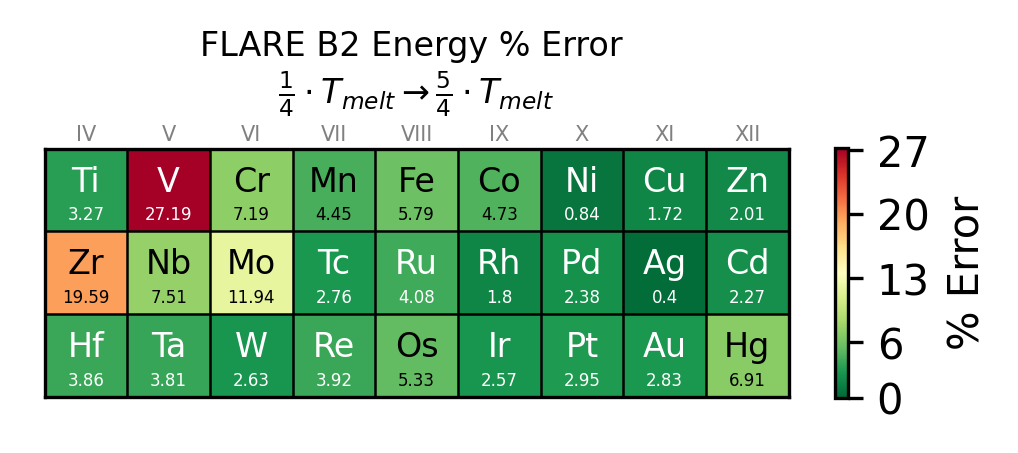}
        \caption{}
    \end{figure*}

    \begin{figure*}
        \centering
        \includegraphics[width=.8\textwidth]{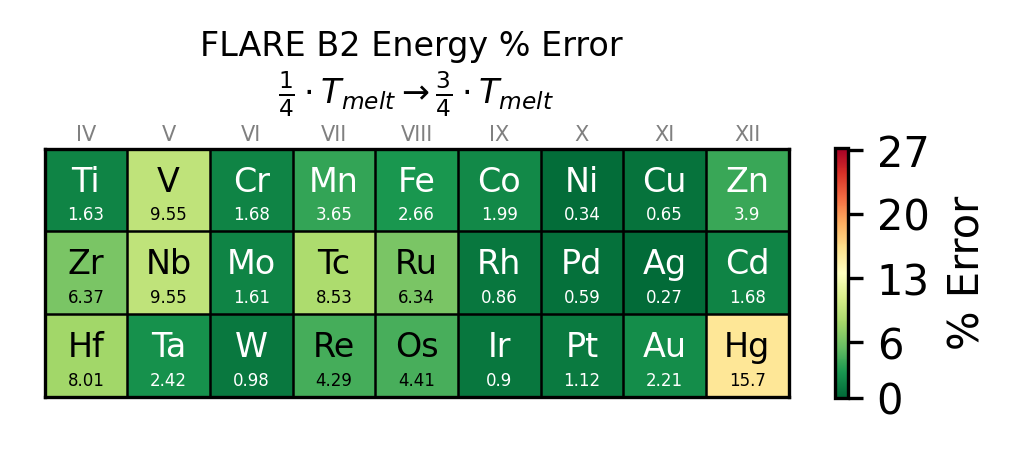}
        \caption{}
    \end{figure*}

\clearpage

    \begin{figure*}
        \centering
        \includegraphics[width=.8\textwidth]{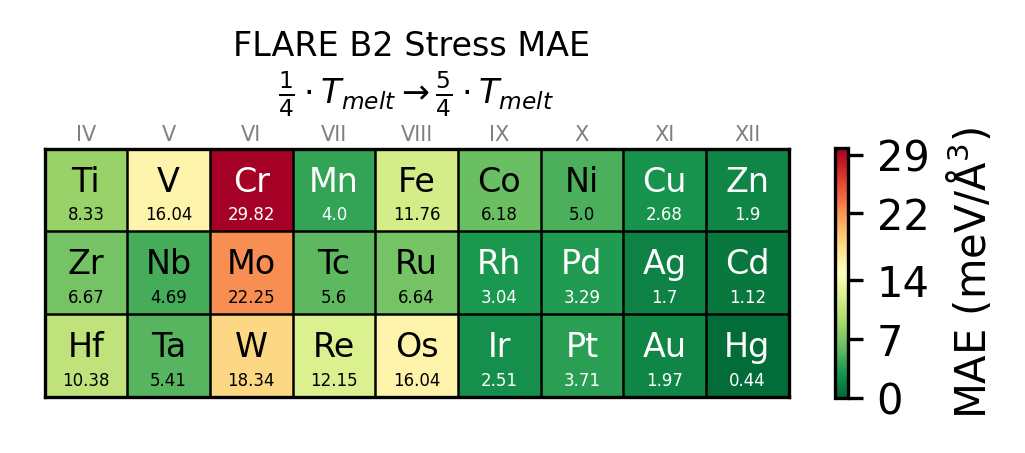}
        \caption{}
    \end{figure*}

    \begin{figure*}
        \centering
        \includegraphics[width=.8\textwidth]{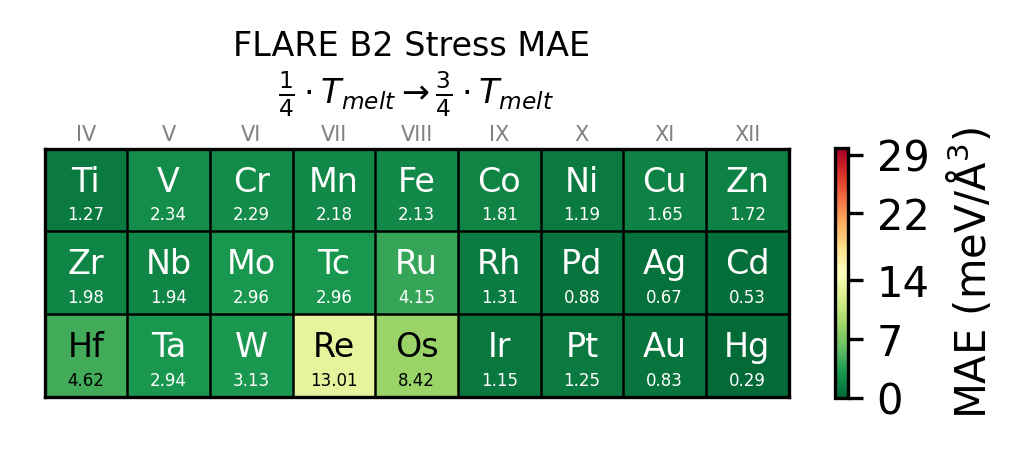}
        \caption{}
    \end{figure*}

    \begin{figure*}
        \centering
        \includegraphics[width=.8\textwidth]{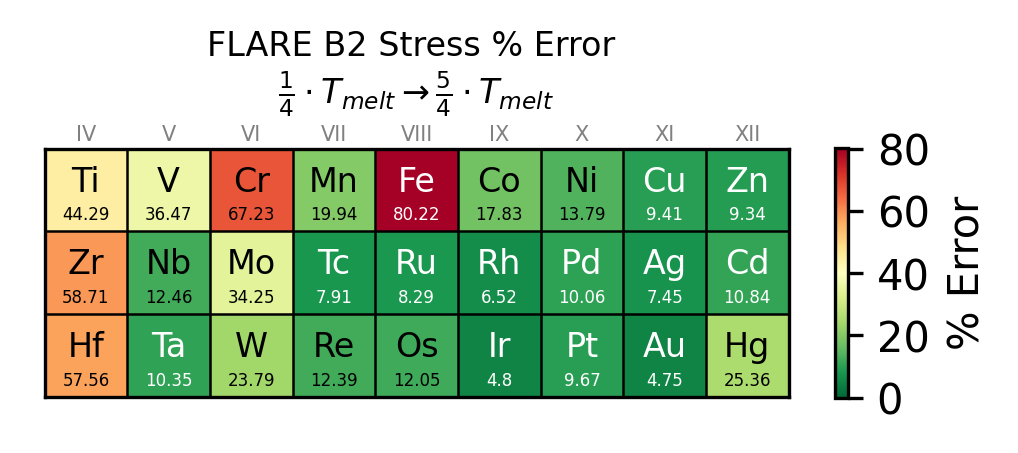}
        \caption{}
    \end{figure*}

    \begin{figure*}
        \centering
        \includegraphics[width=.8\textwidth]{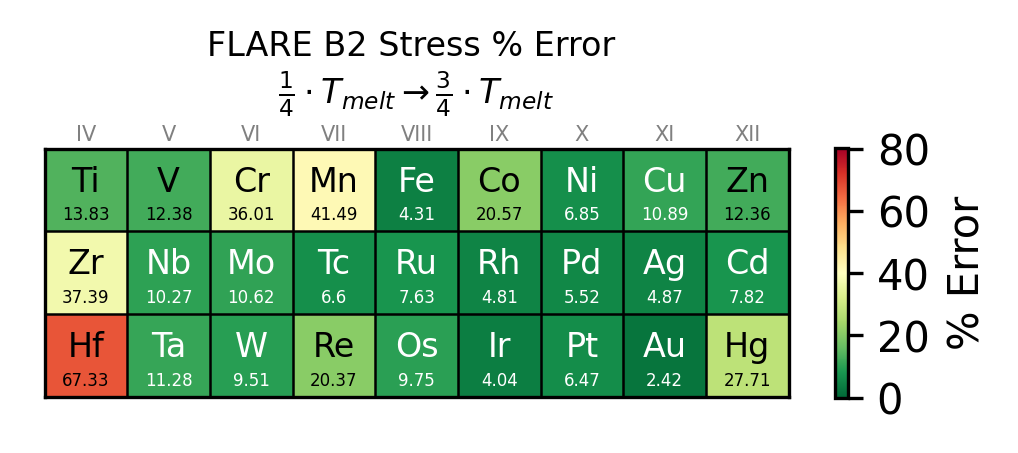}
        \caption{}
    \end{figure*}

    \begin{figure*}
        \centering
        \includegraphics[width=.8\textwidth]{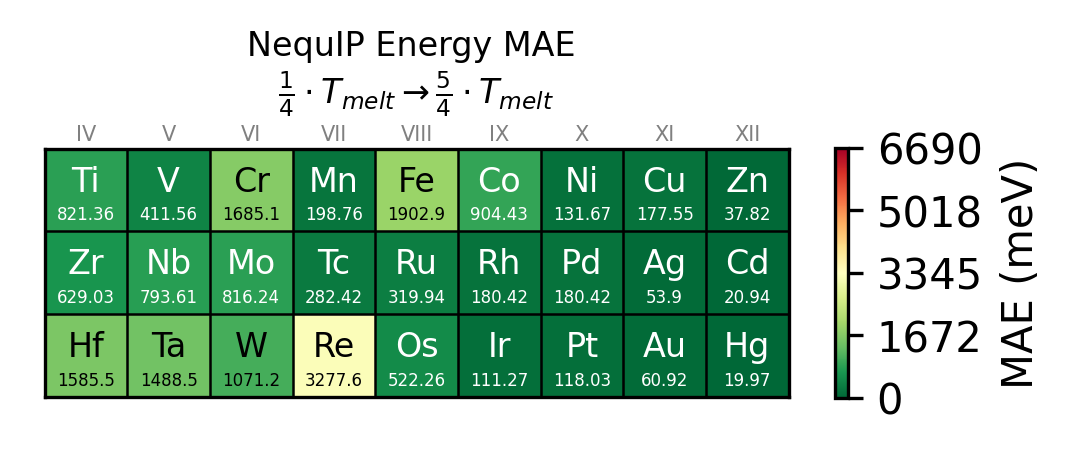}
        \caption{}
    \end{figure*}

    \begin{figure*}
        \centering
        \includegraphics[width=.8\textwidth]{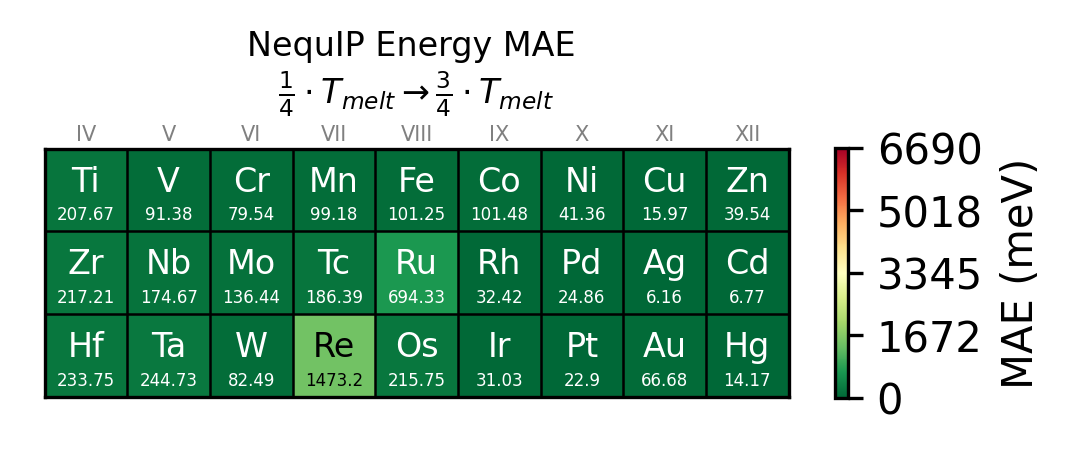}
        \caption{}
    \end{figure*}

    \begin{figure*}
        \centering
        \includegraphics[width=.8\textwidth]{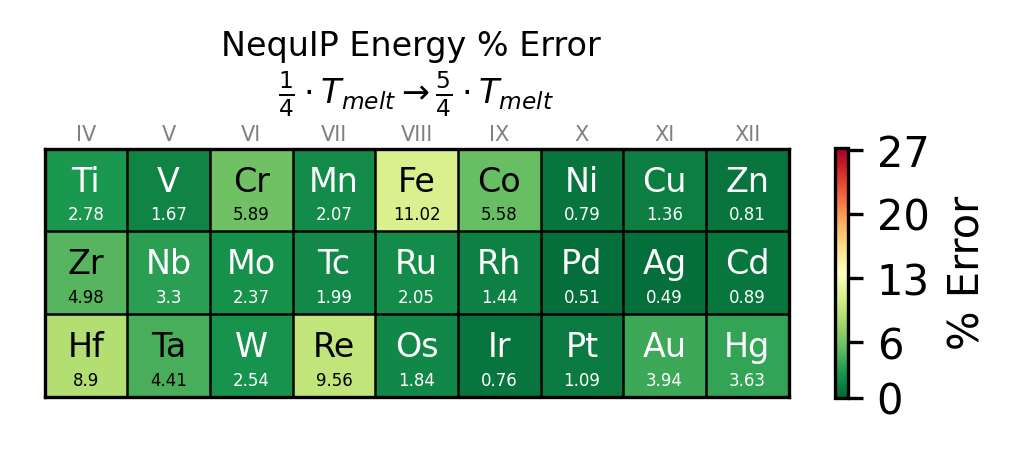}
        \caption{}
    \end{figure*}

    \begin{figure*}
        \centering
        \includegraphics[width=.8\textwidth]{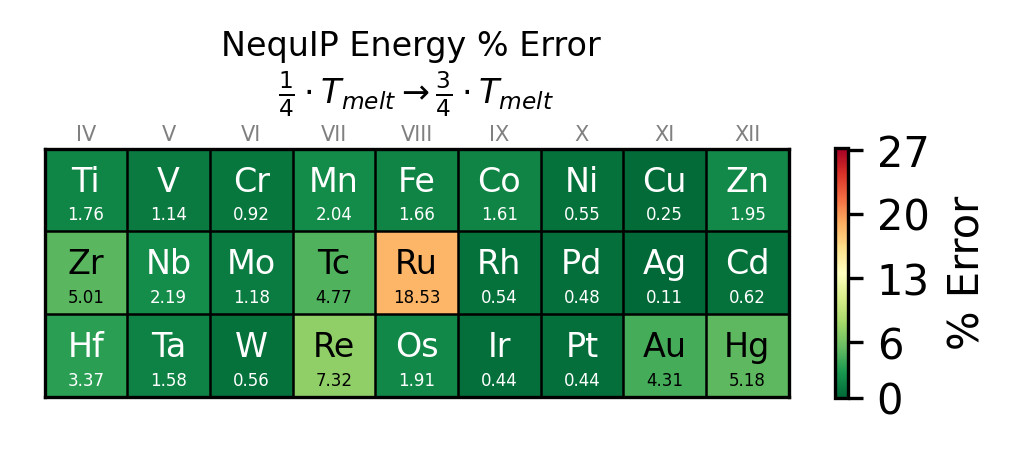}
        \caption{}
    \end{figure*}

    \begin{figure*}
        \centering
        \includegraphics[width=.8\textwidth]{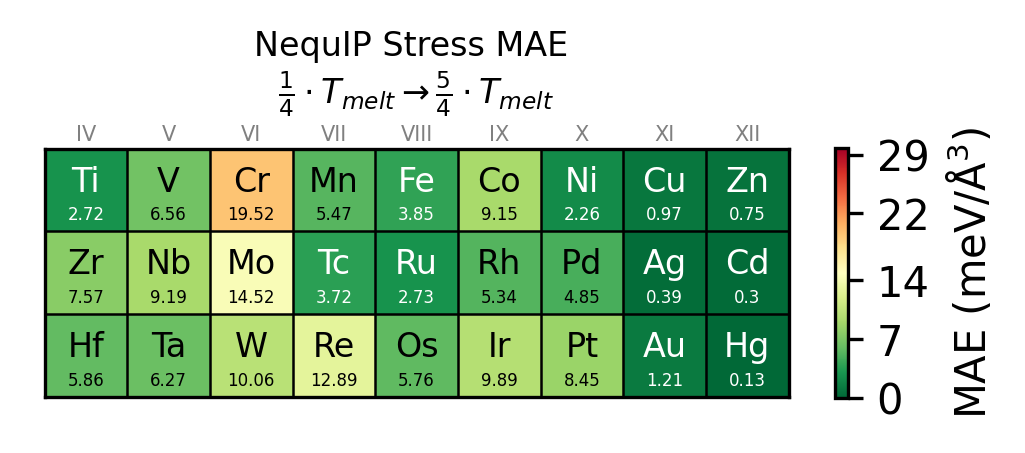}
        \caption{}
    \end{figure*}

    \begin{figure*}
        \centering
        \includegraphics[width=.8\textwidth]{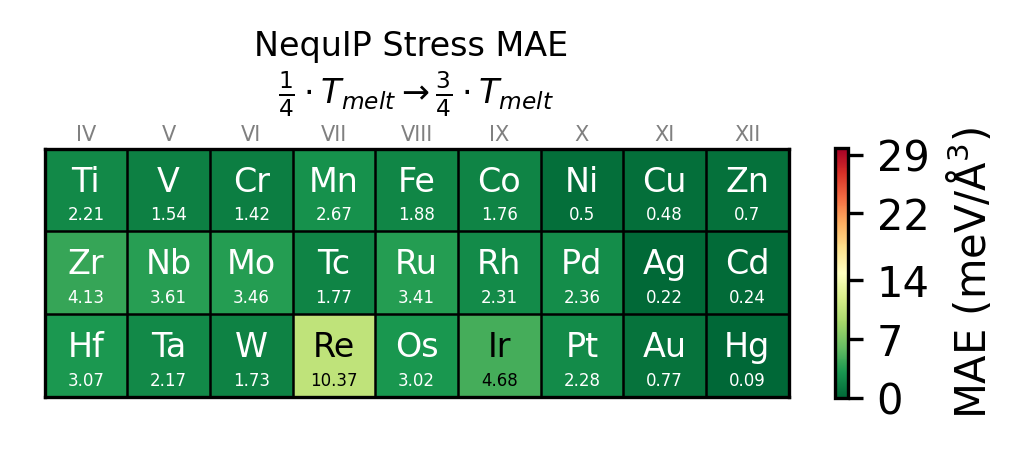}
        \caption{}
    \end{figure*}

    \begin{figure*}
        \centering
        \includegraphics[width=.8\textwidth]{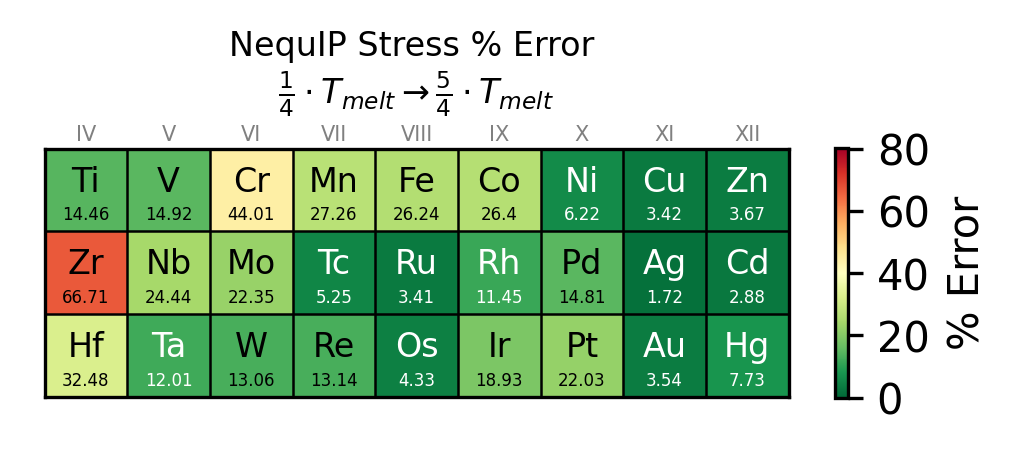}
        \caption{}
    \end{figure*}

    \begin{figure*}
        \centering
        \includegraphics[width=.8\textwidth]{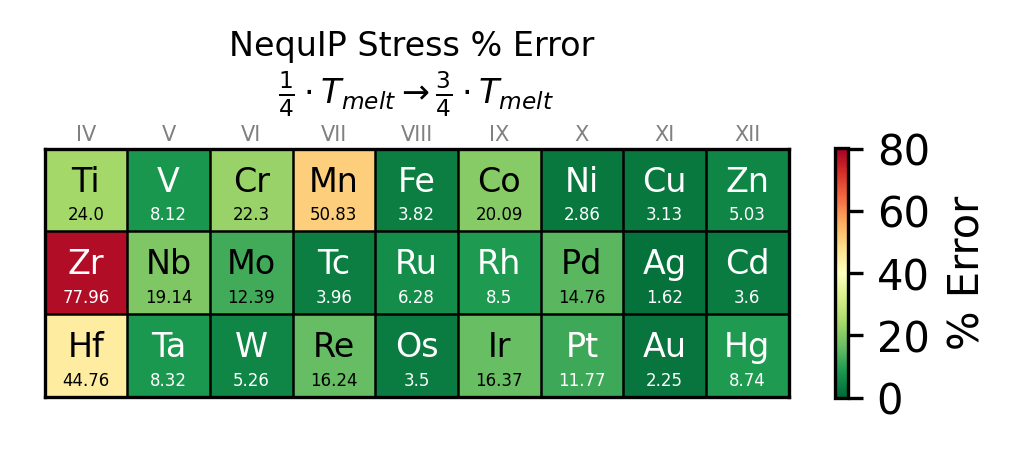}
        \caption{}
        \label{S-fig:lasterror}
    \end{figure*}


\begin{figure*}
        \centering
        \includegraphics[width=.65\textwidth]{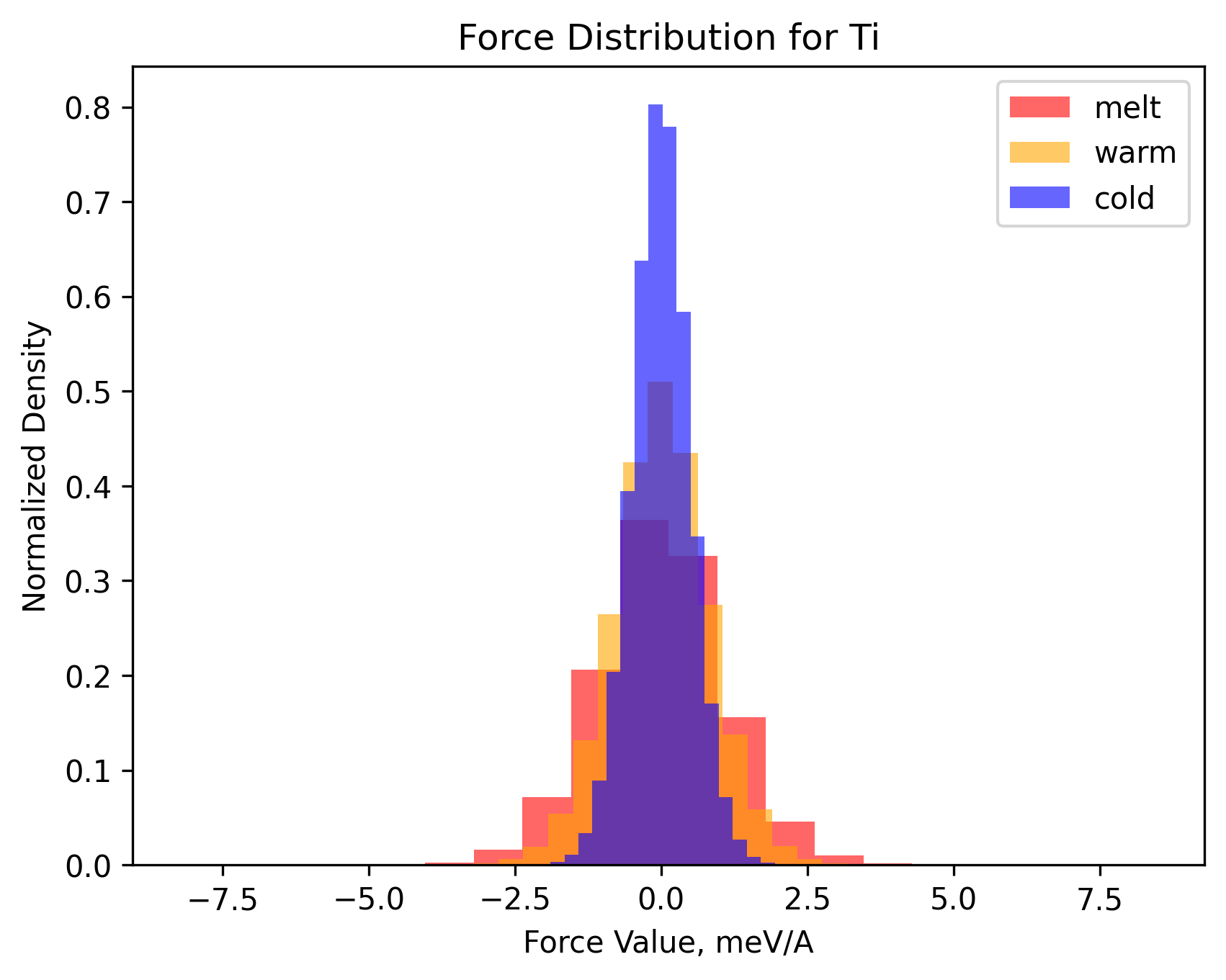}
        \caption{}
        \label{S-fig:firsterrordist}
    \end{figure*}
    
\begin{figure*}
        \centering
        \includegraphics[width=.65\textwidth]{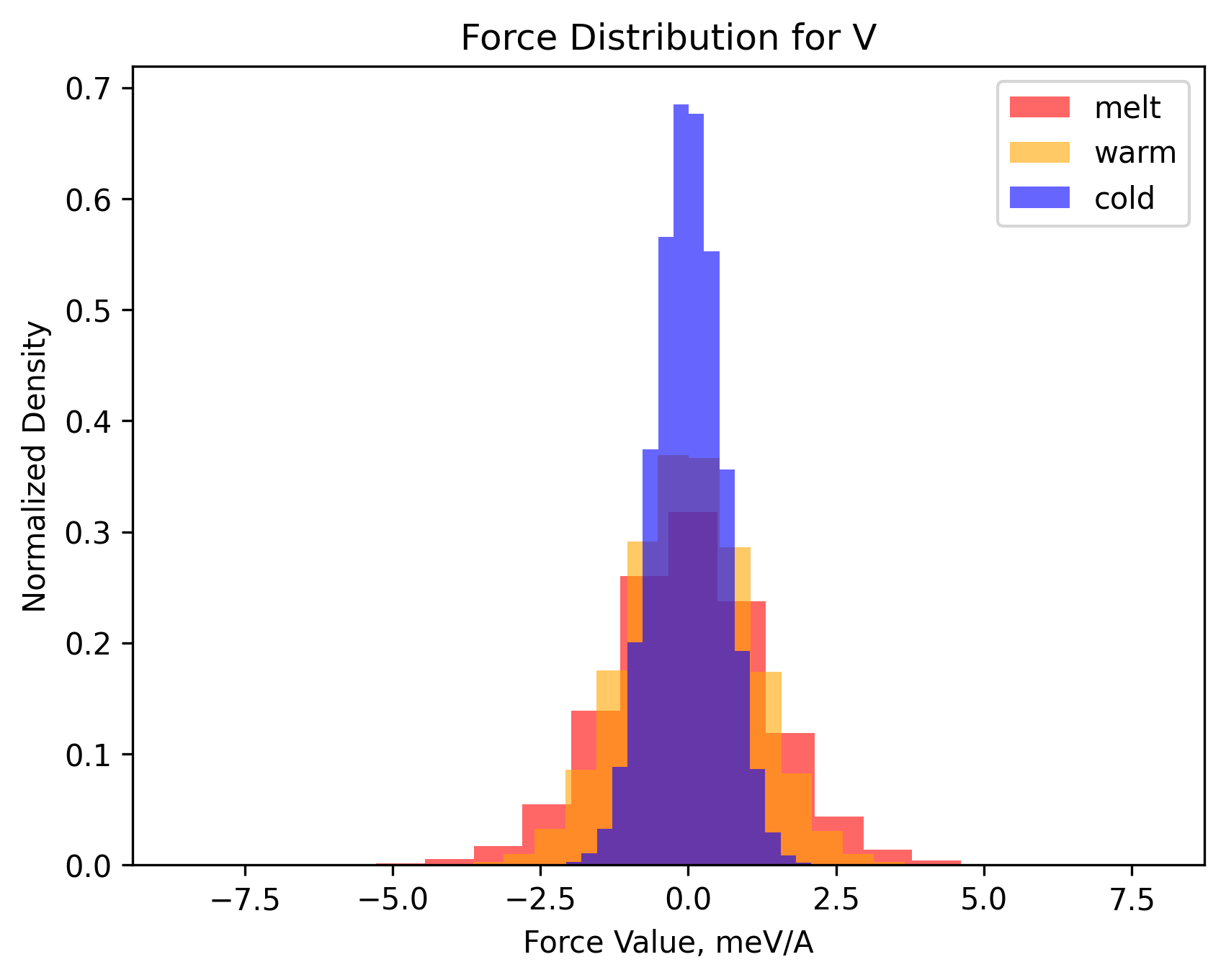}
        \caption{}
    \end{figure*}
    
\begin{figure*}
        \centering
        \includegraphics[width=.65\textwidth]{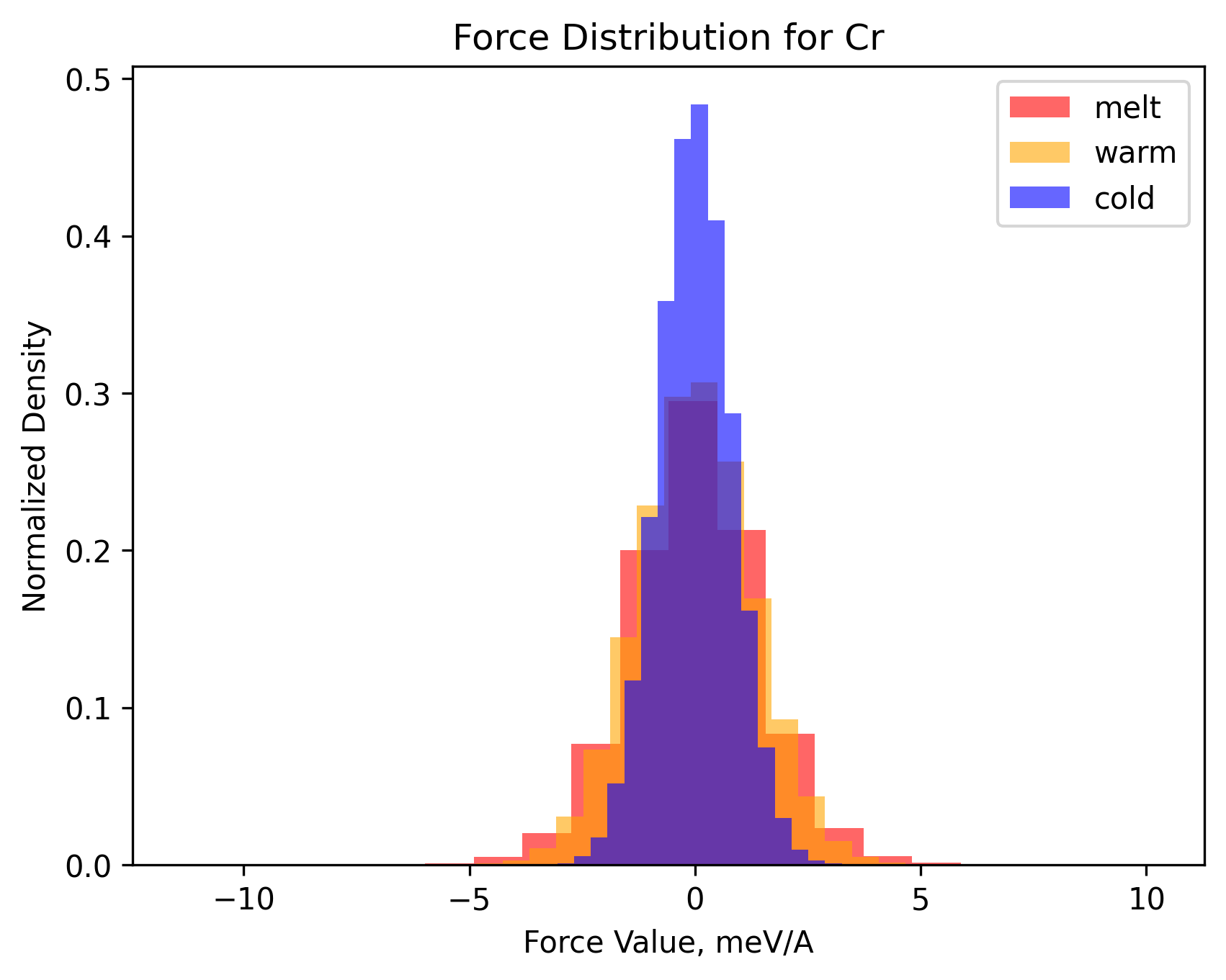}
        \caption{}
    \end{figure*}
    
\begin{figure*}
        \centering
        \includegraphics[width=.65\textwidth]{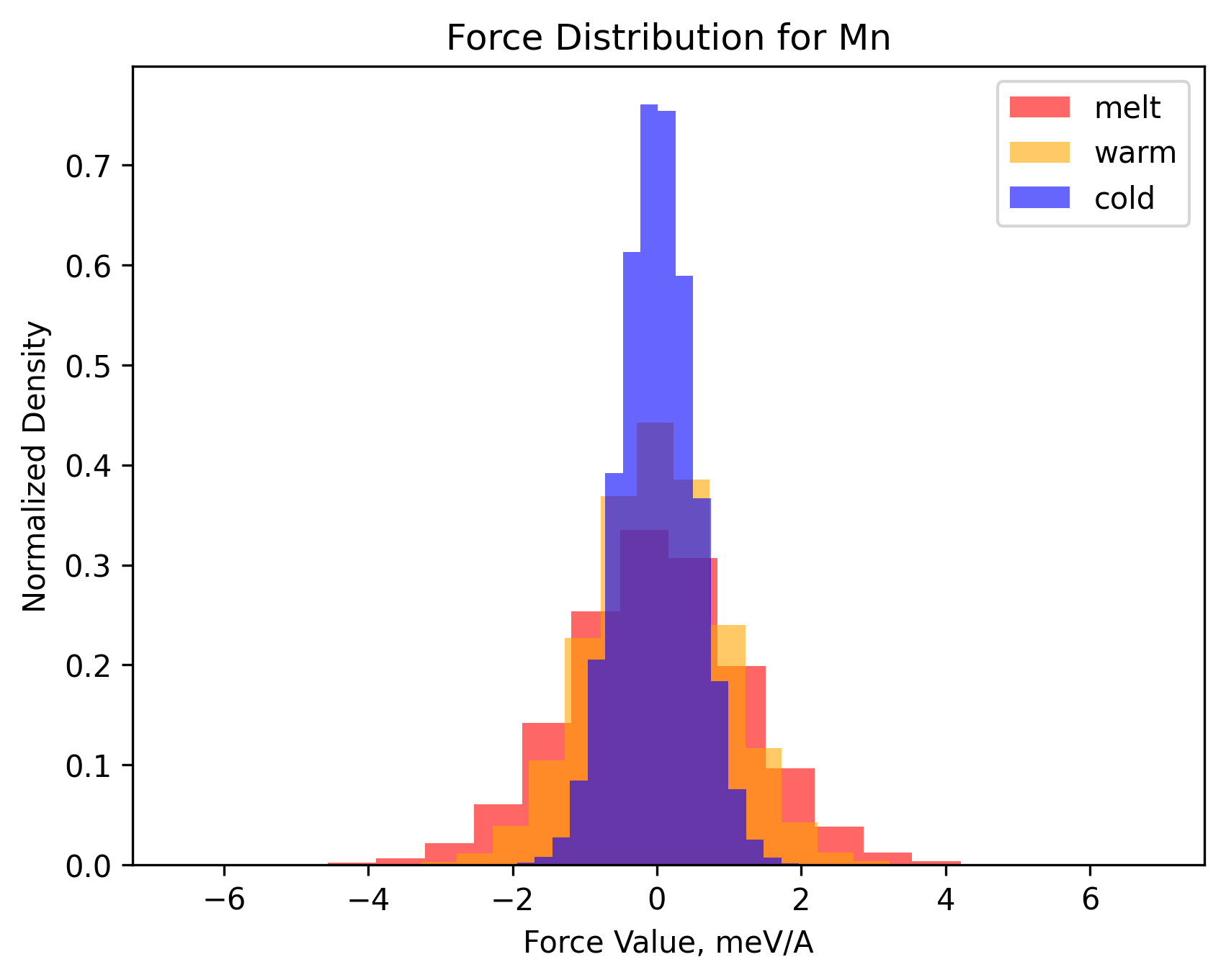}
        \caption{}
    \end{figure*}
    
\begin{figure*}
        \centering
        \includegraphics[width=.65\textwidth]{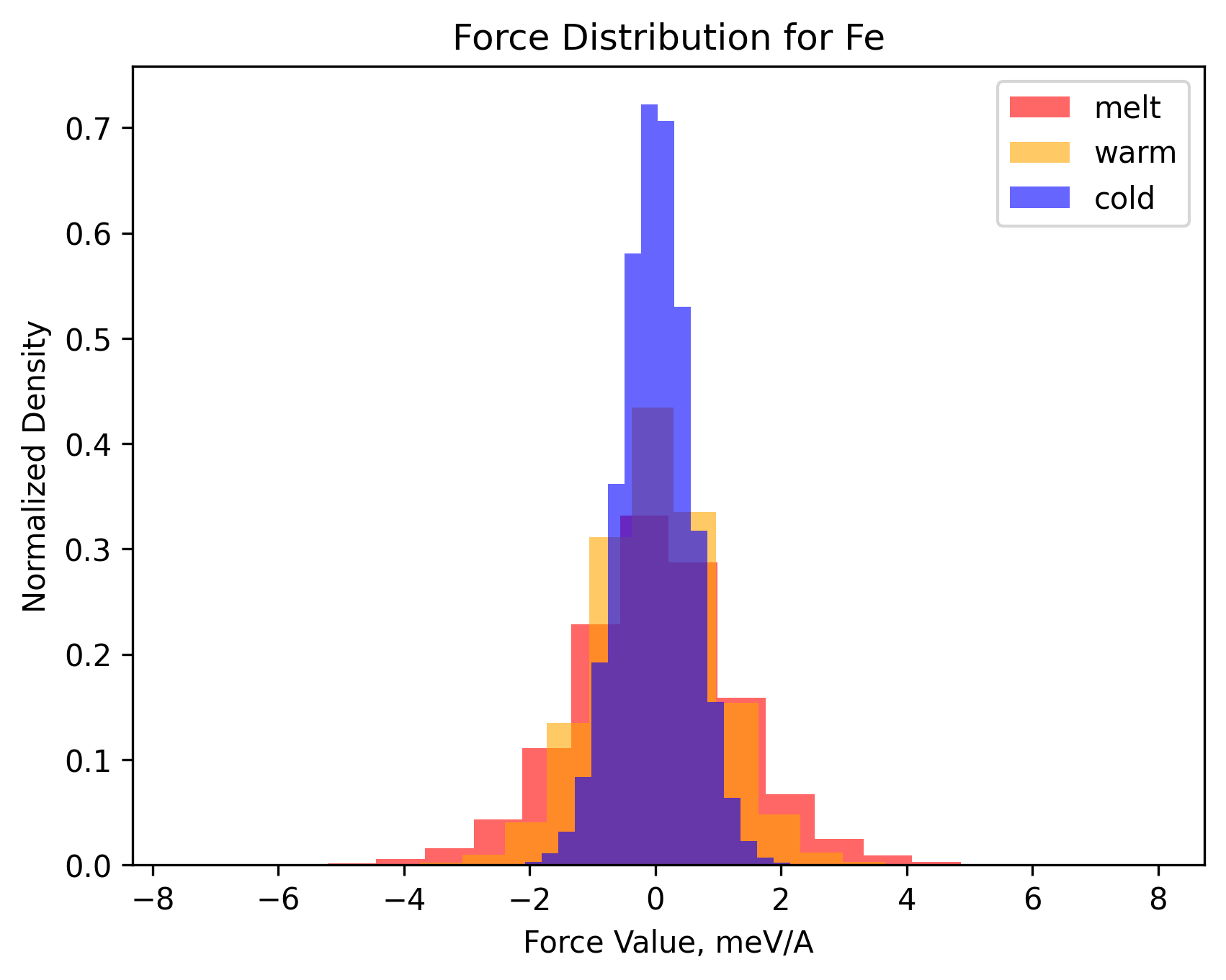}
        \caption{}
    \end{figure*}
    
\begin{figure*}
        \centering
        \includegraphics[width=.65\textwidth]{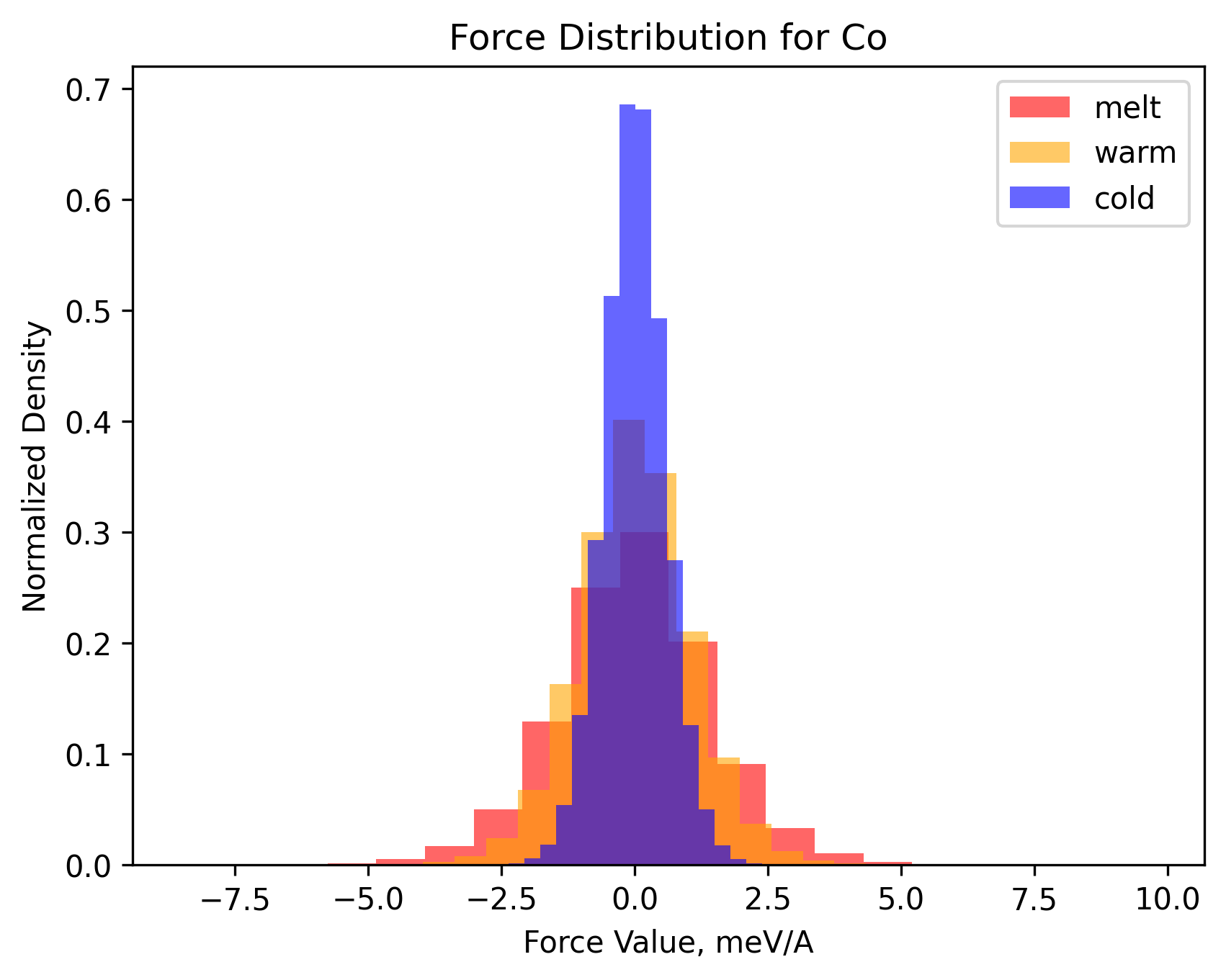}
        \caption{}
    \end{figure*}
    
\begin{figure*}
        \centering
        \includegraphics[width=.65\textwidth]{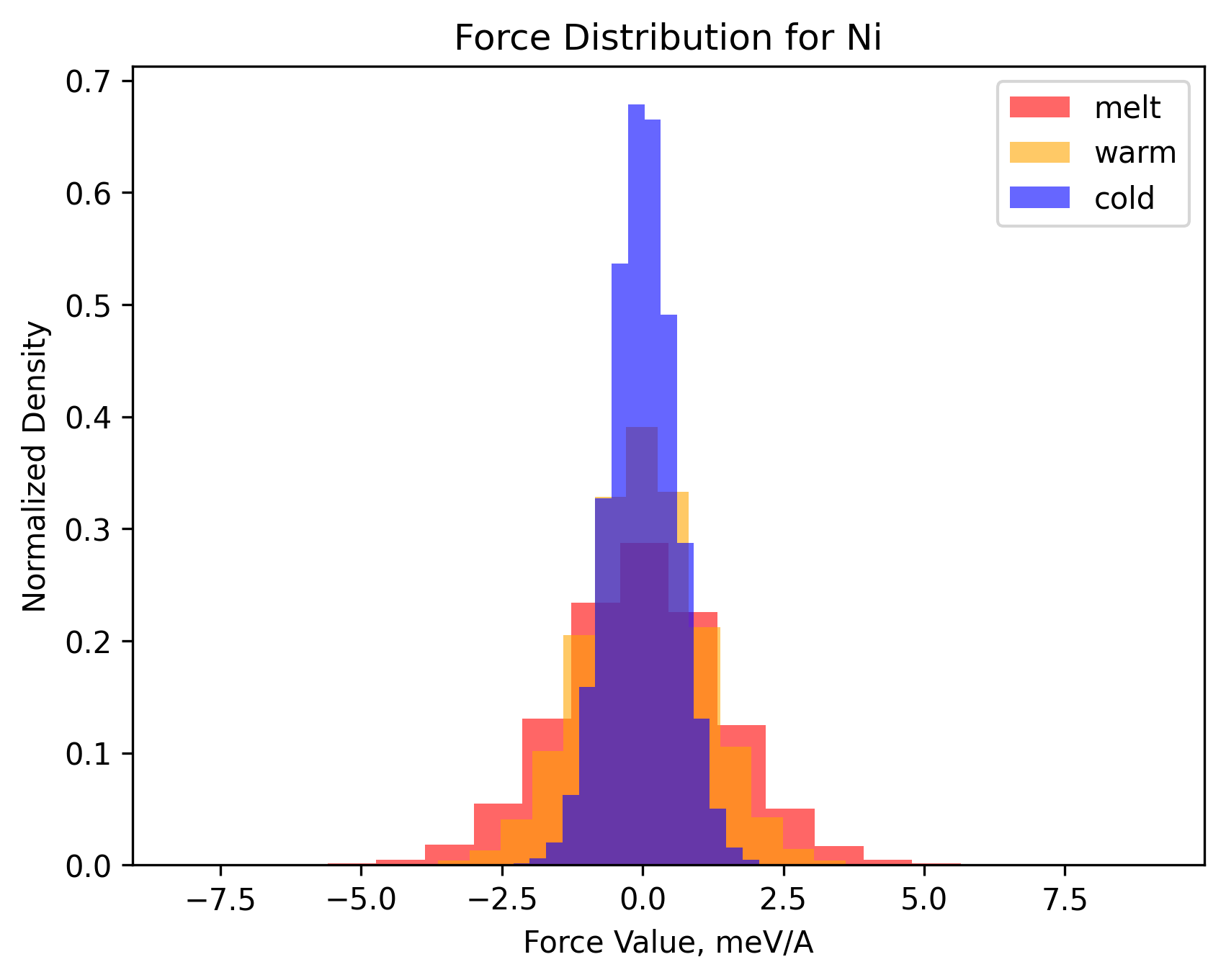}
        \caption{}
    \end{figure*}
    
\begin{figure*}
        \centering
        \includegraphics[width=.65\textwidth]{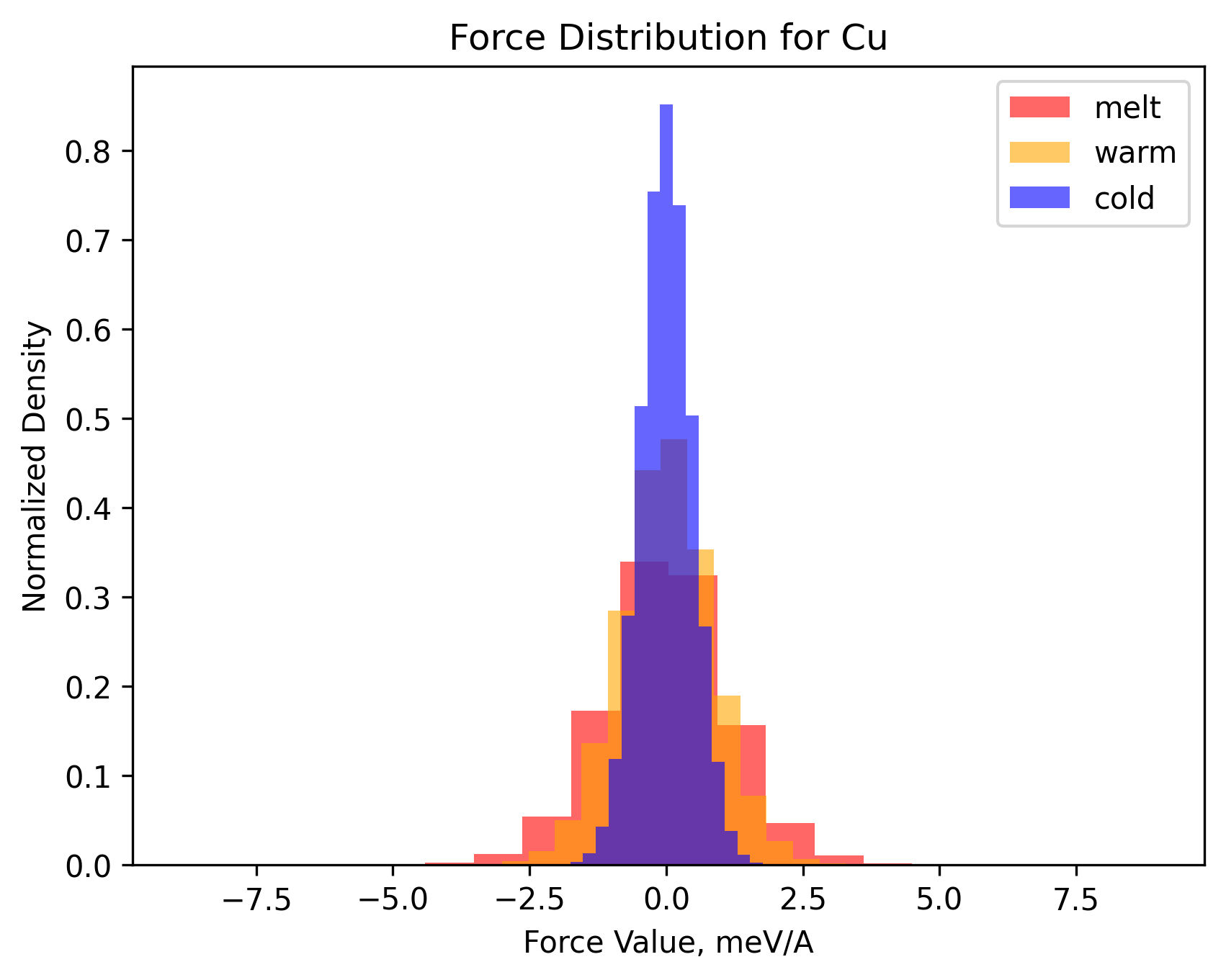}
        \caption{}
    \end{figure*}
    
\begin{figure*}
        \centering
        \includegraphics[width=.65\textwidth]{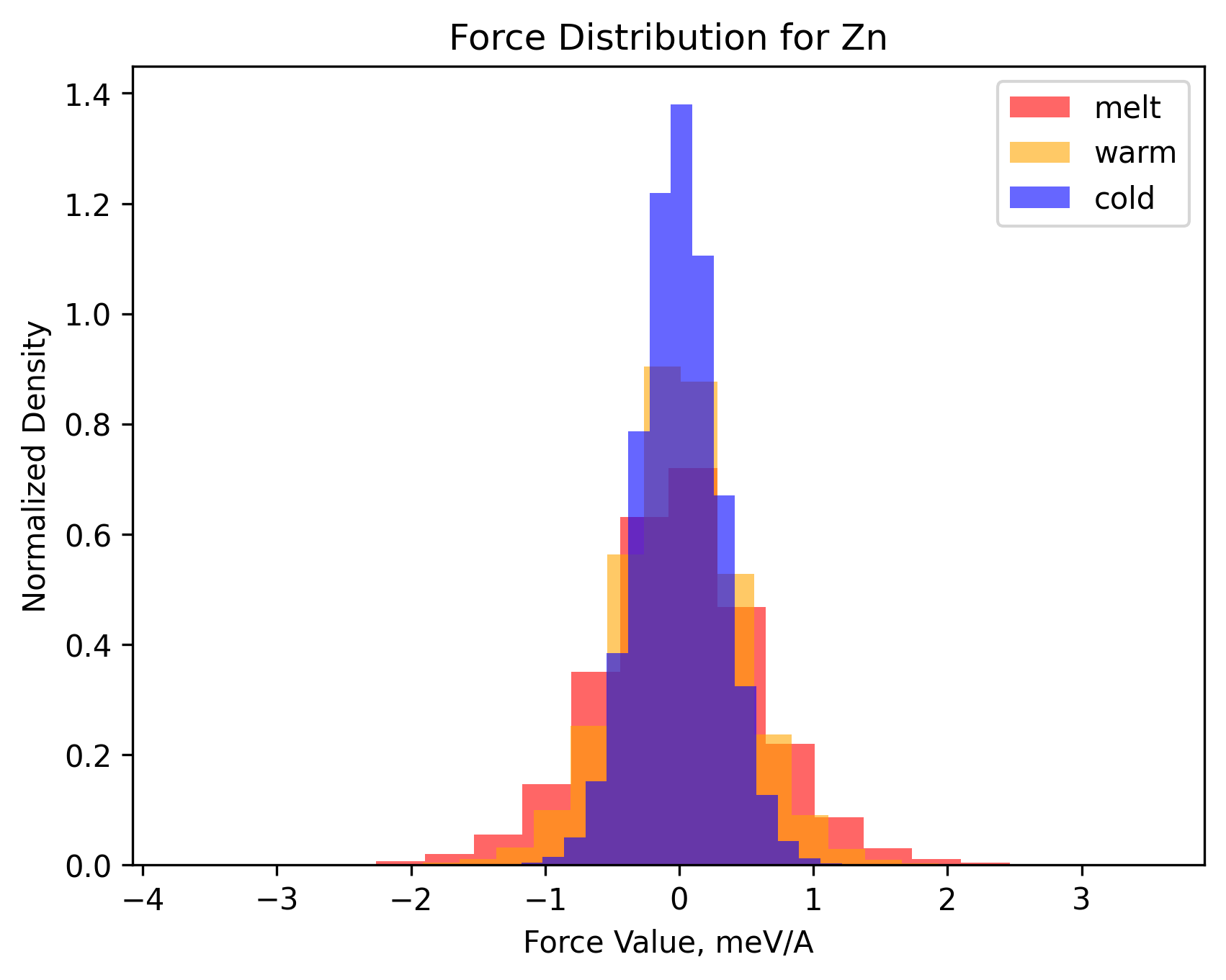}
        \caption{}
    \end{figure*}
    
\begin{figure*}
        \centering
        \includegraphics[width=.65\textwidth]{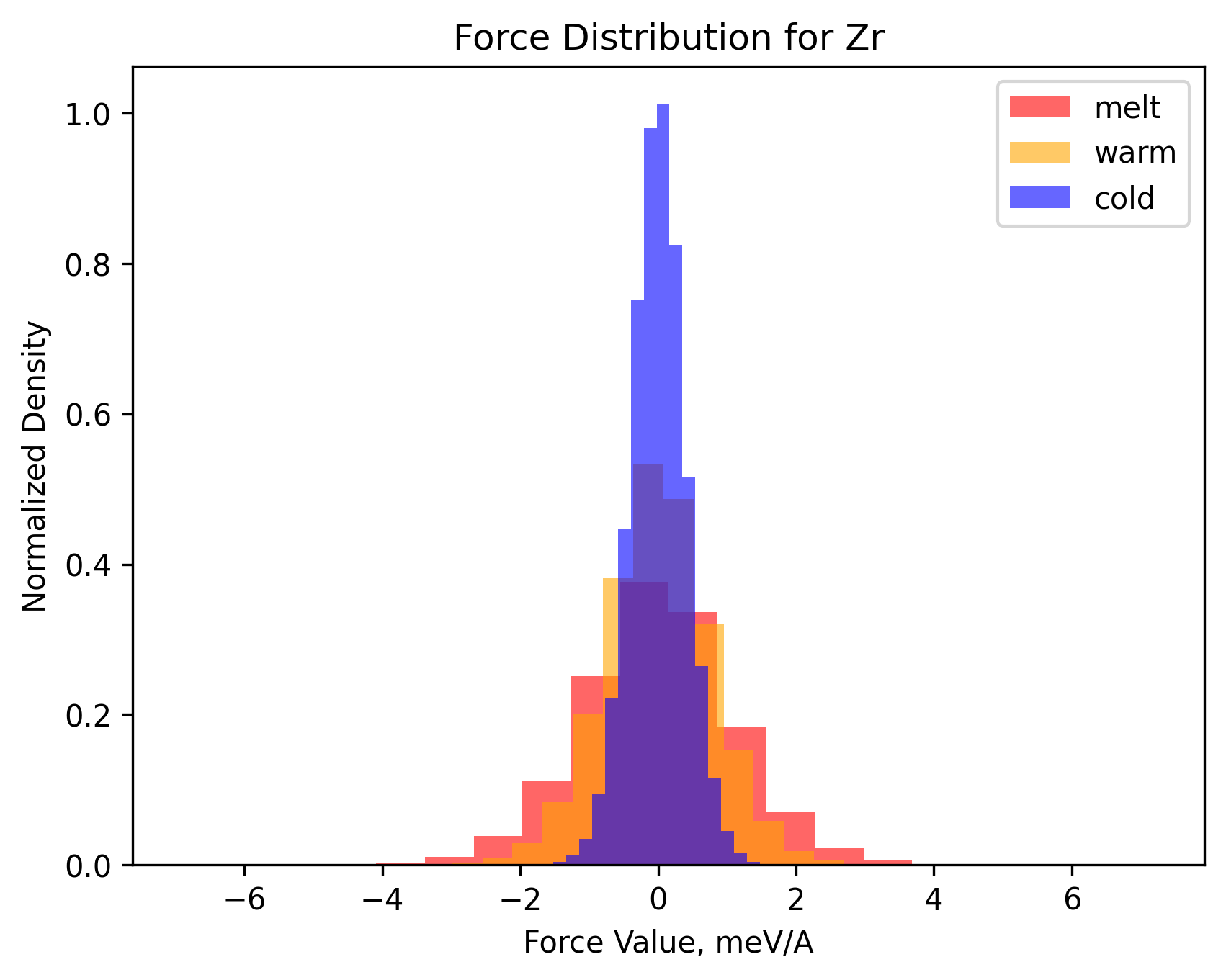}
        \caption{}
    \end{figure*}
    
\begin{figure*}
        \centering
        \includegraphics[width=.65\textwidth]{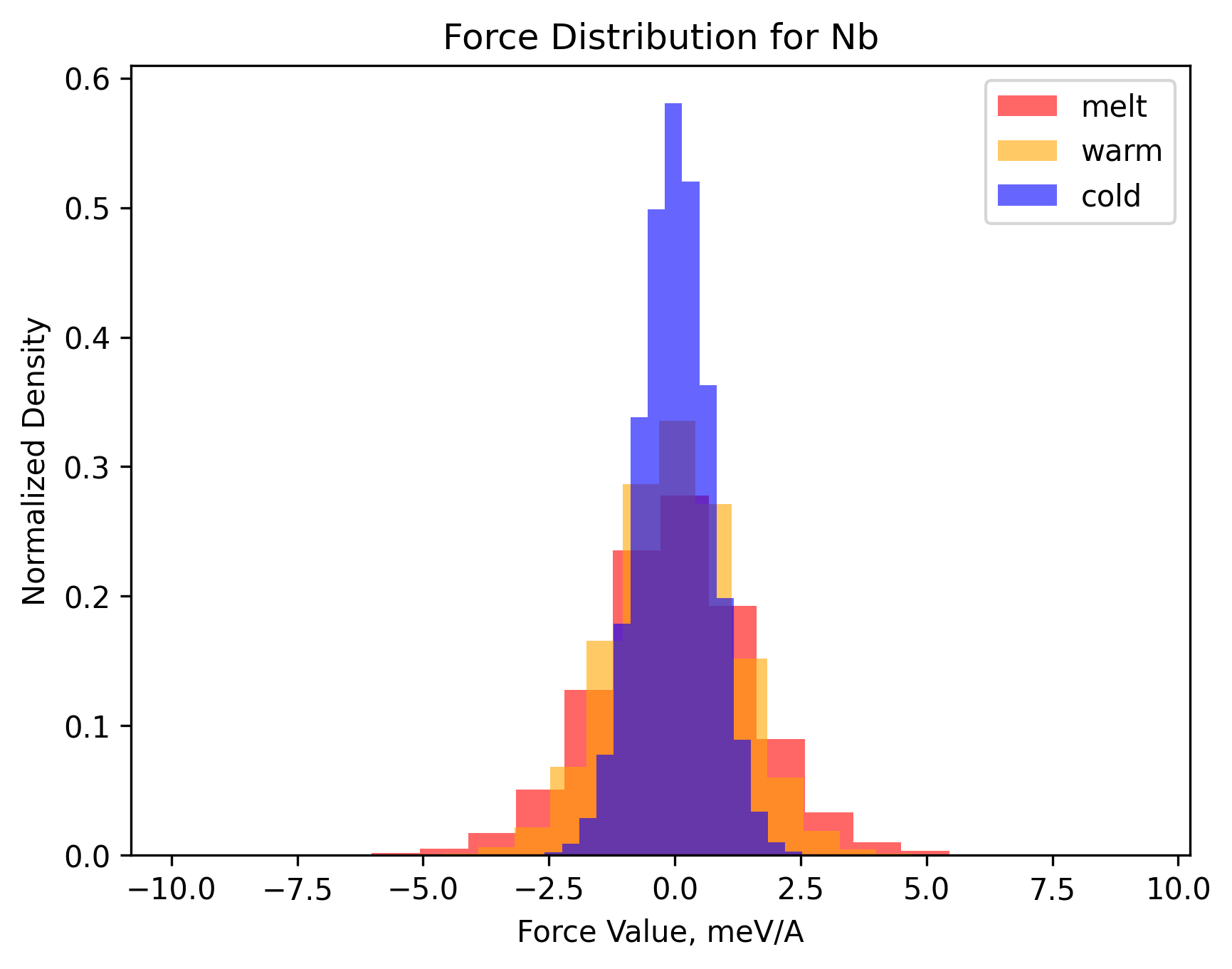}
        \caption{}
    \end{figure*}
    
\begin{figure*}
        \centering
        \includegraphics[width=.65\textwidth]{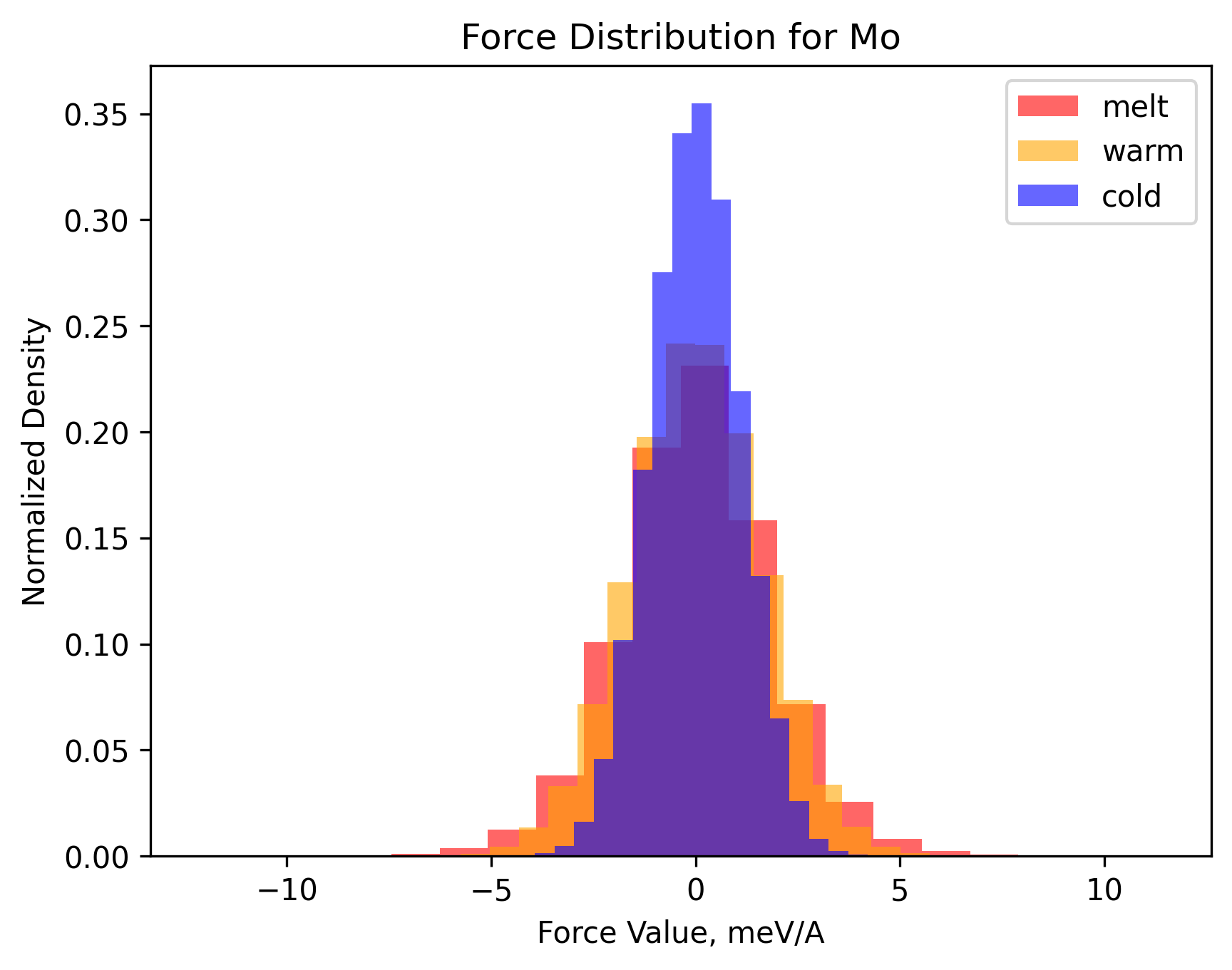}
        \caption{}
    \end{figure*}
    
\begin{figure*}
        \centering
        \includegraphics[width=.65\textwidth]{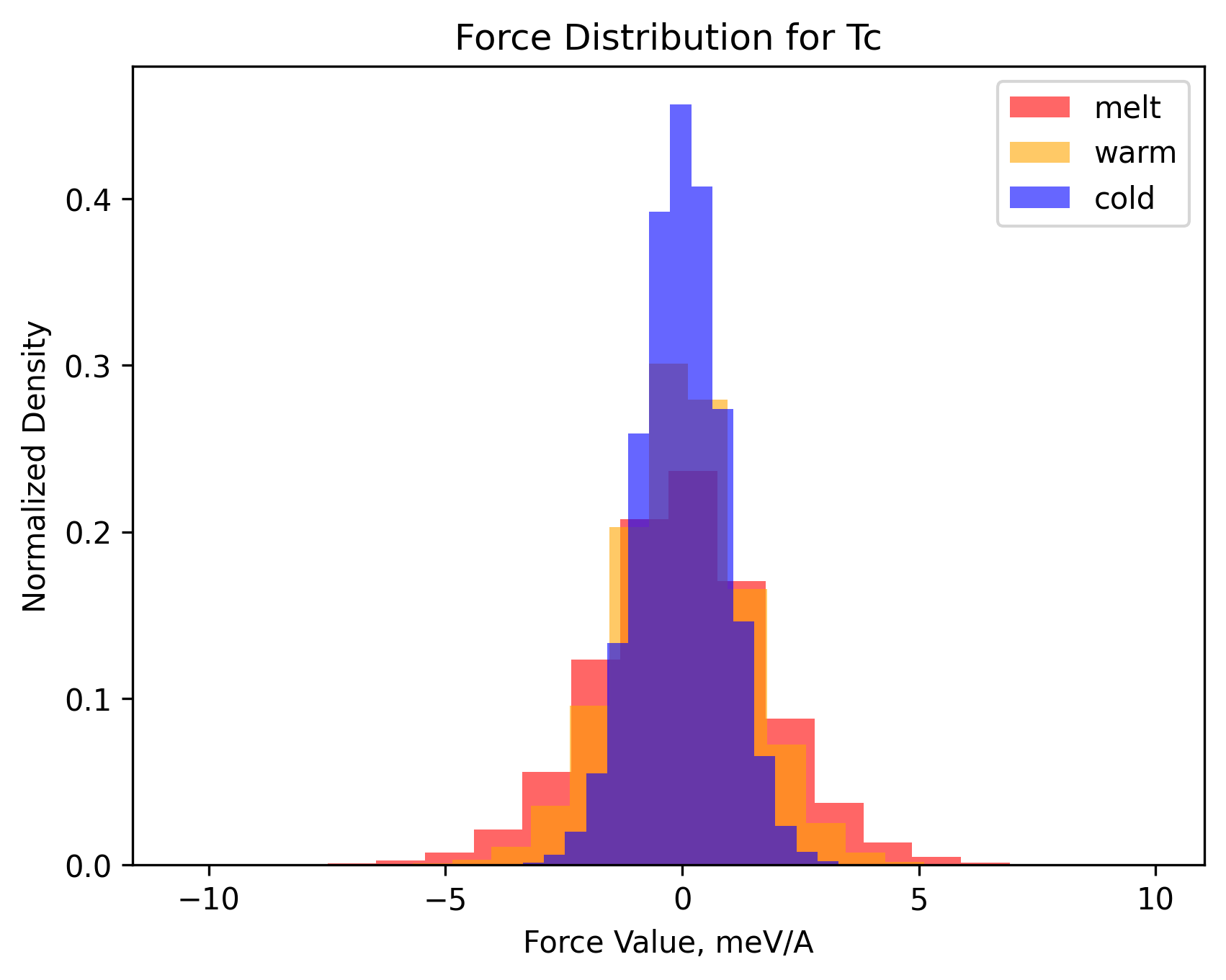}
        \caption{}
    \end{figure*}
    
\begin{figure*}
        \centering
        \includegraphics[width=.65\textwidth]{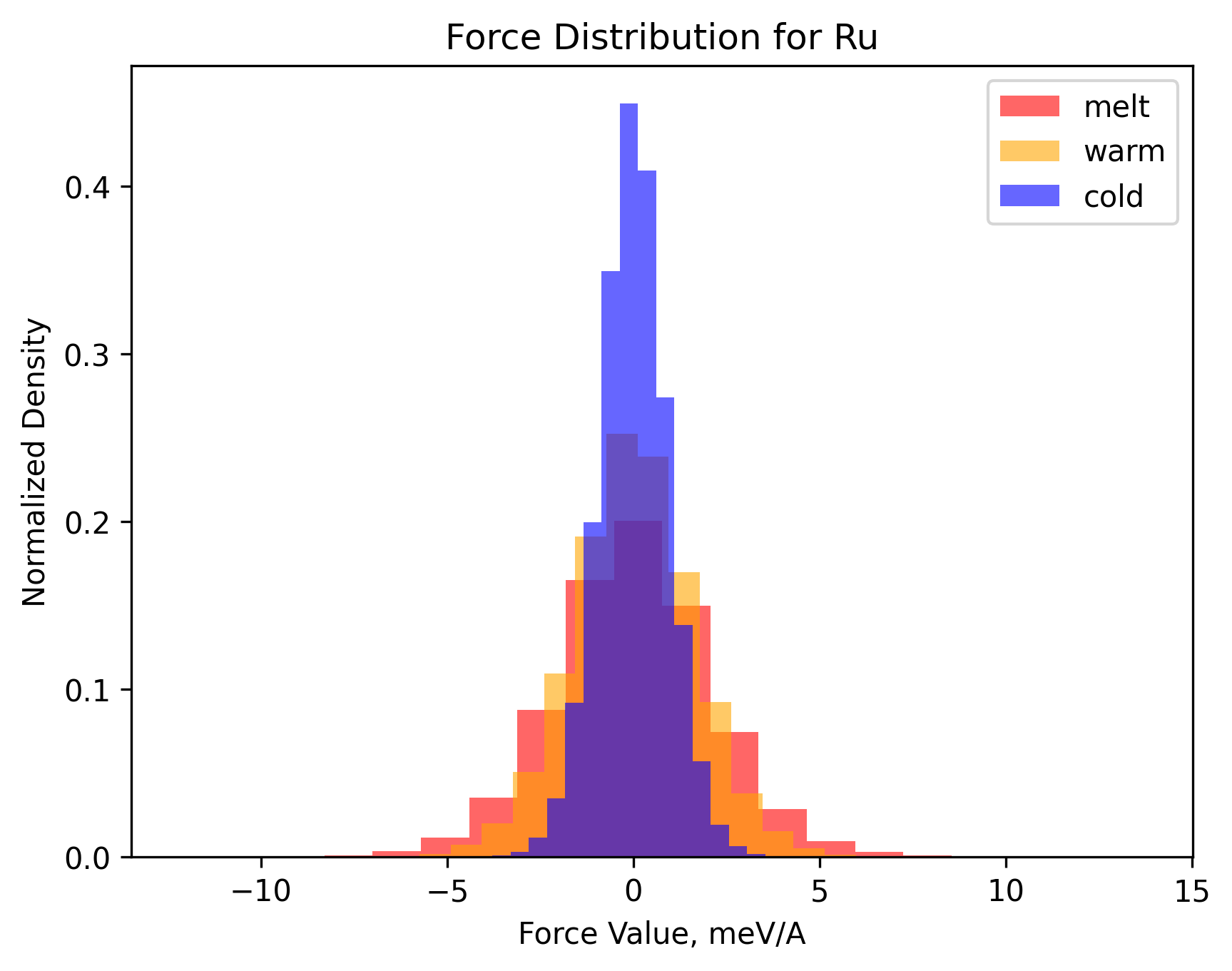}
        \caption{}
    \end{figure*}
    
\begin{figure*}
        \centering
        \includegraphics[width=.65\textwidth]{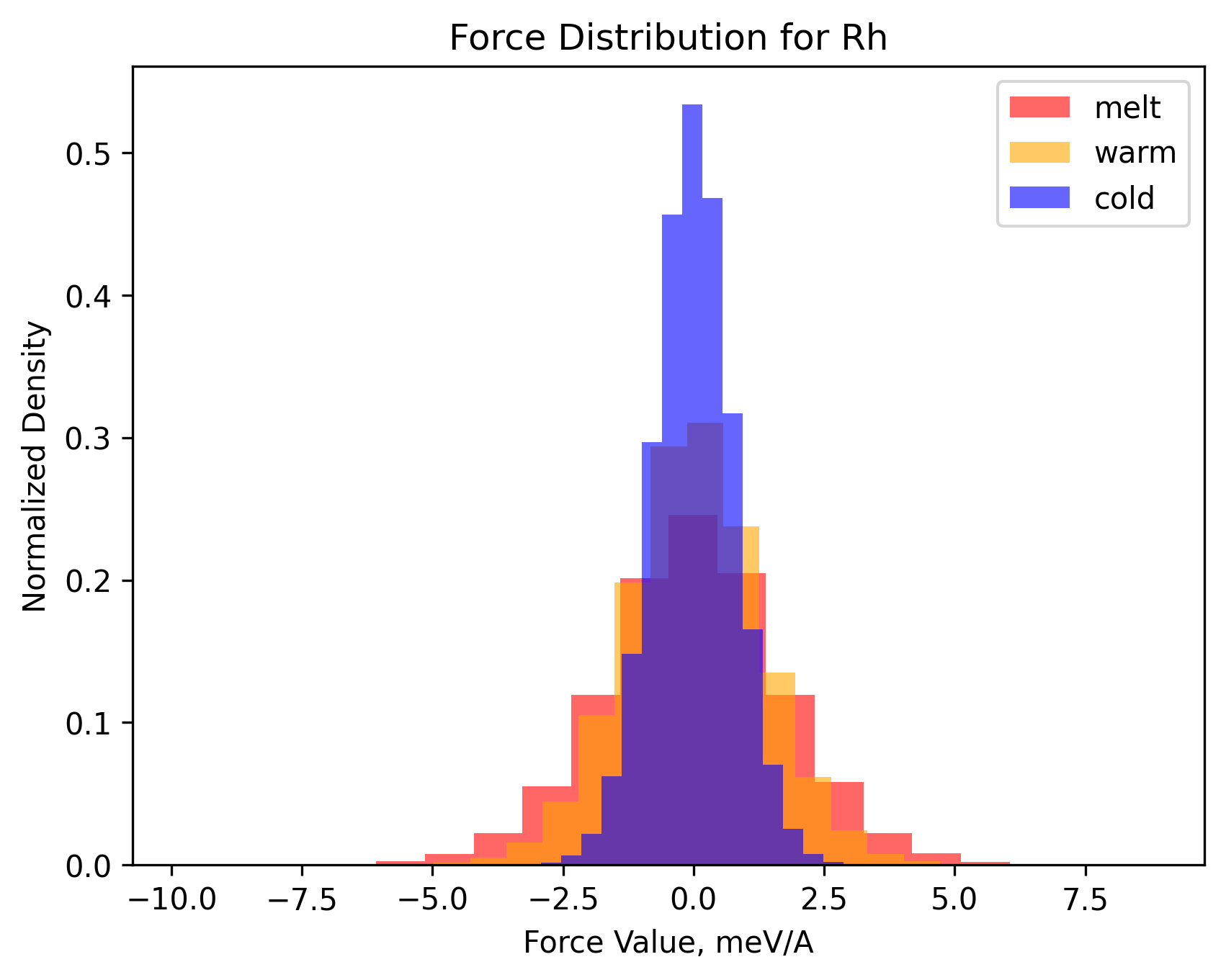}
        \caption{}
    \end{figure*}
    
\begin{figure*}
        \centering
        \includegraphics[width=.65\textwidth]{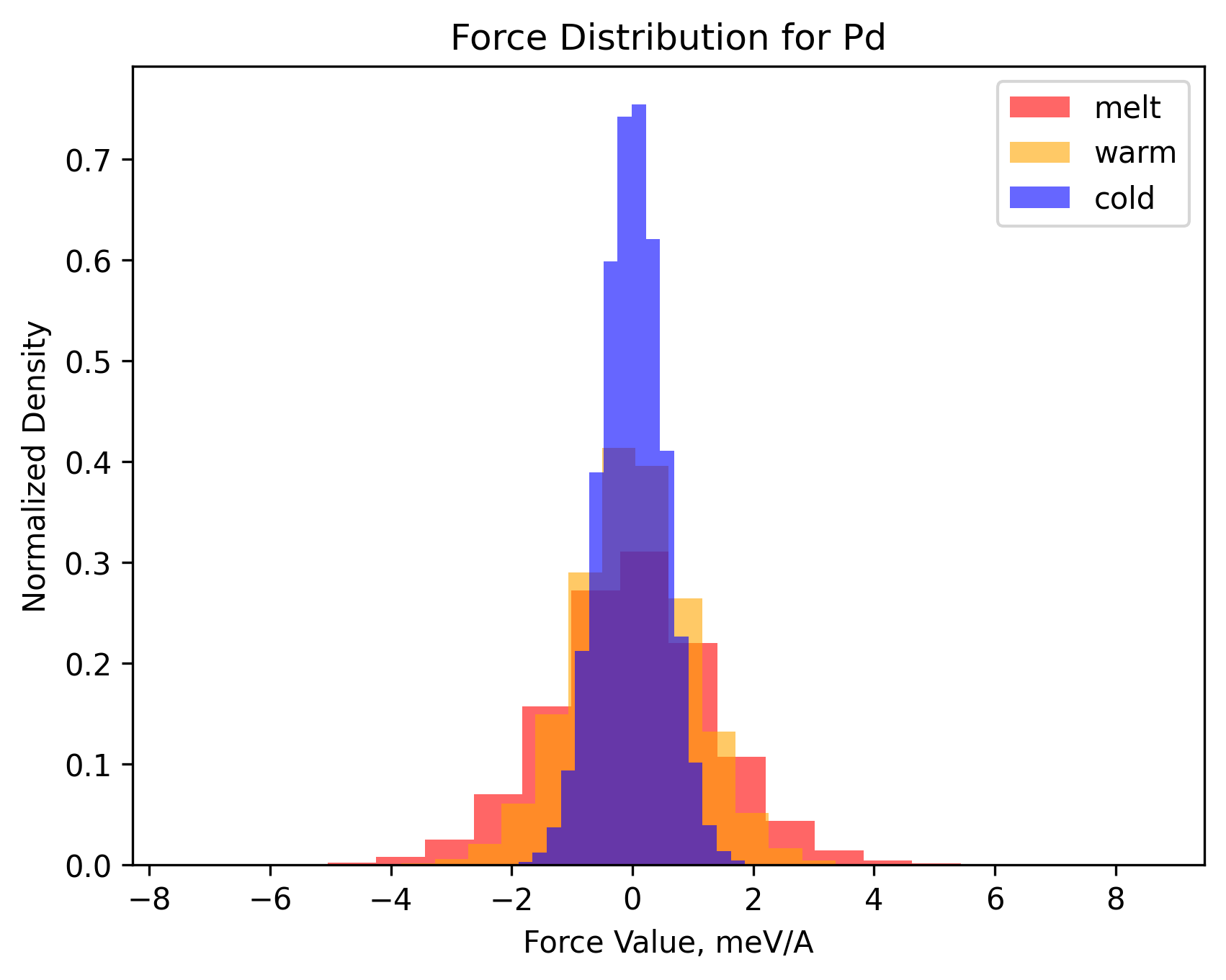}
        \caption{}
    \end{figure*}
    
\begin{figure*}
        \centering
        \includegraphics[width=.65\textwidth]{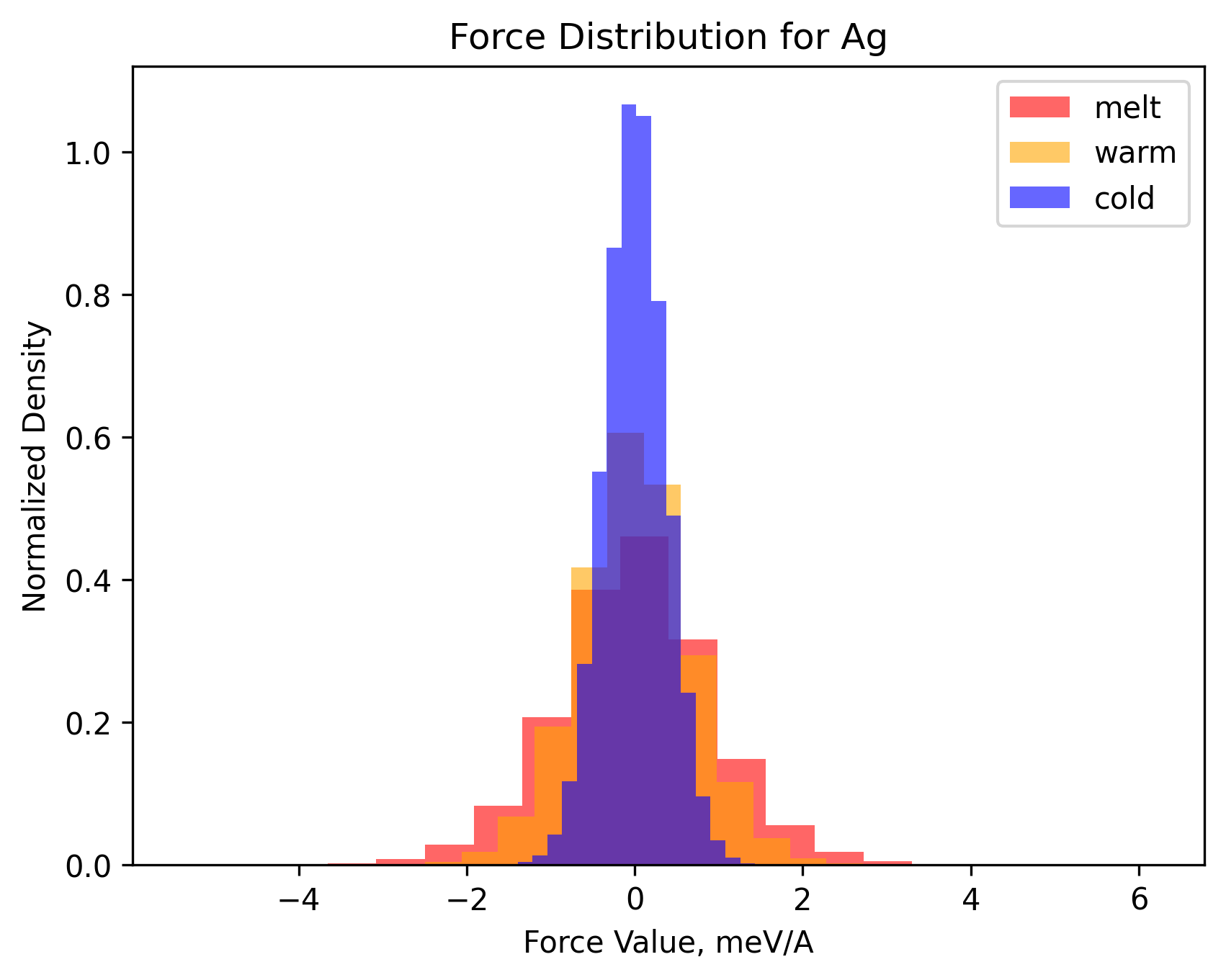}
        \caption{}
    \end{figure*}
    
\begin{figure*}
        \centering
        \includegraphics[width=.65\textwidth]{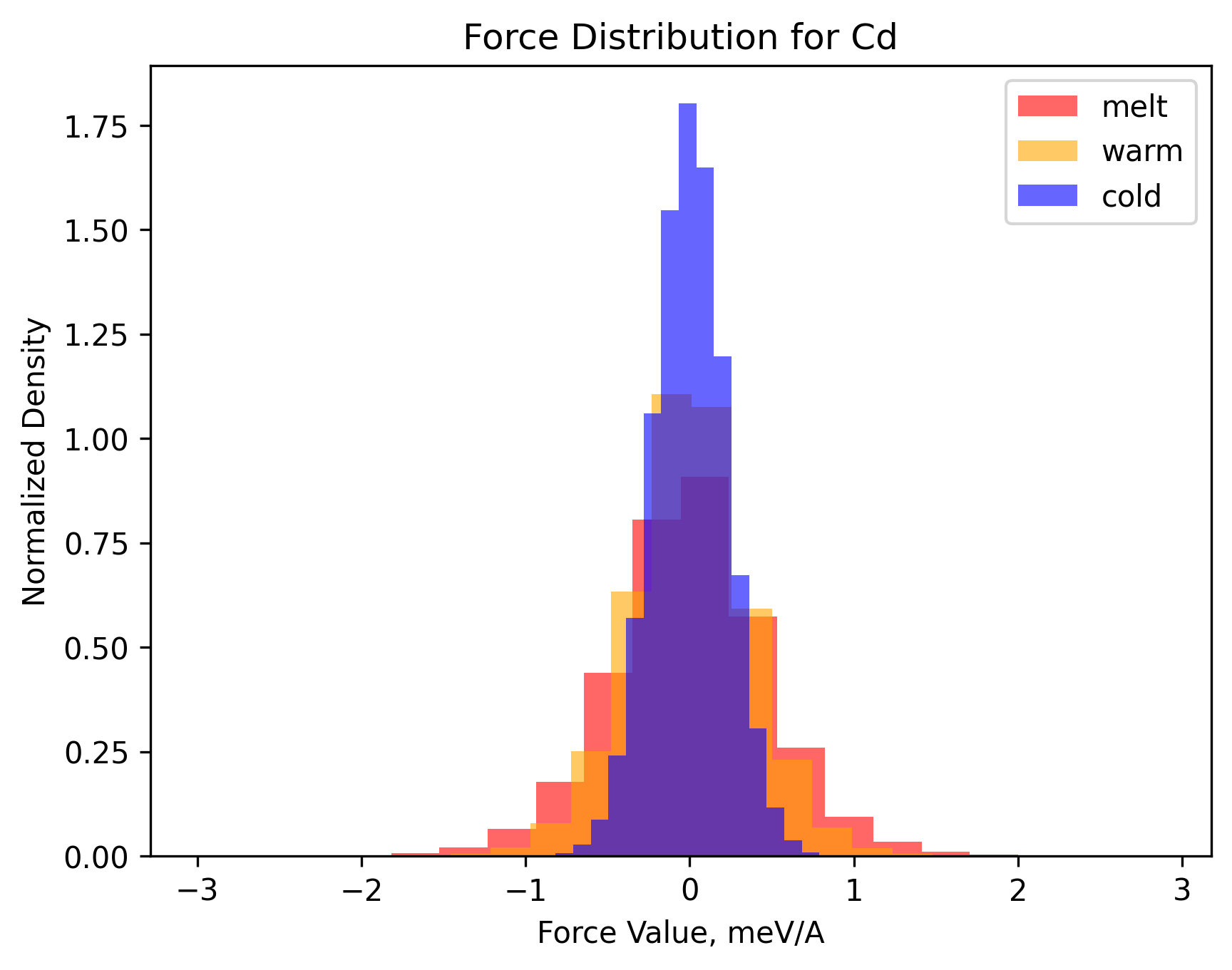}
        \caption{}
    \end{figure*}
    
\begin{figure*}
        \centering
        \includegraphics[width=.65\textwidth]{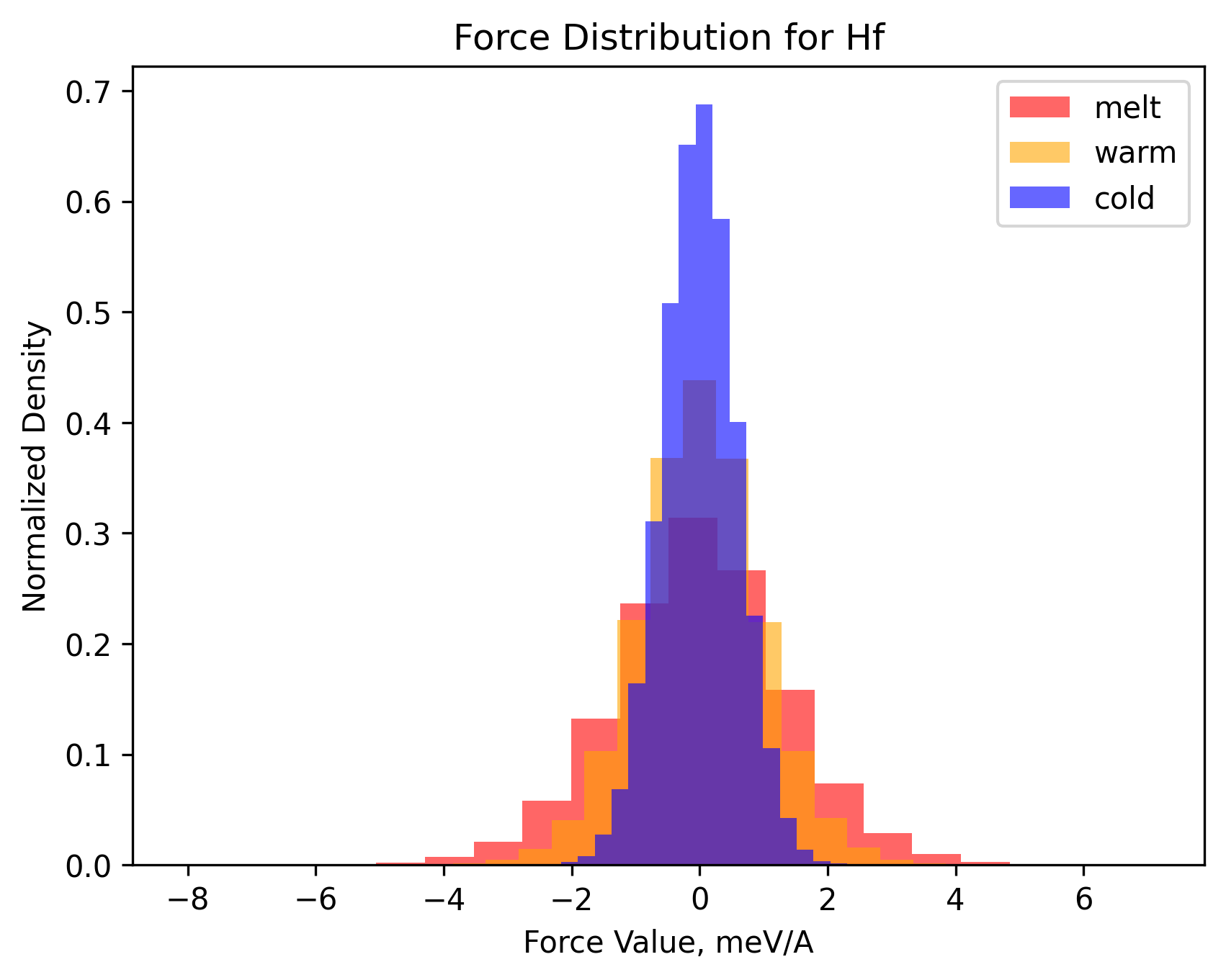}
        \caption{}
    \end{figure*}
    
\begin{figure*}
        \centering
        \includegraphics[width=.65\textwidth]{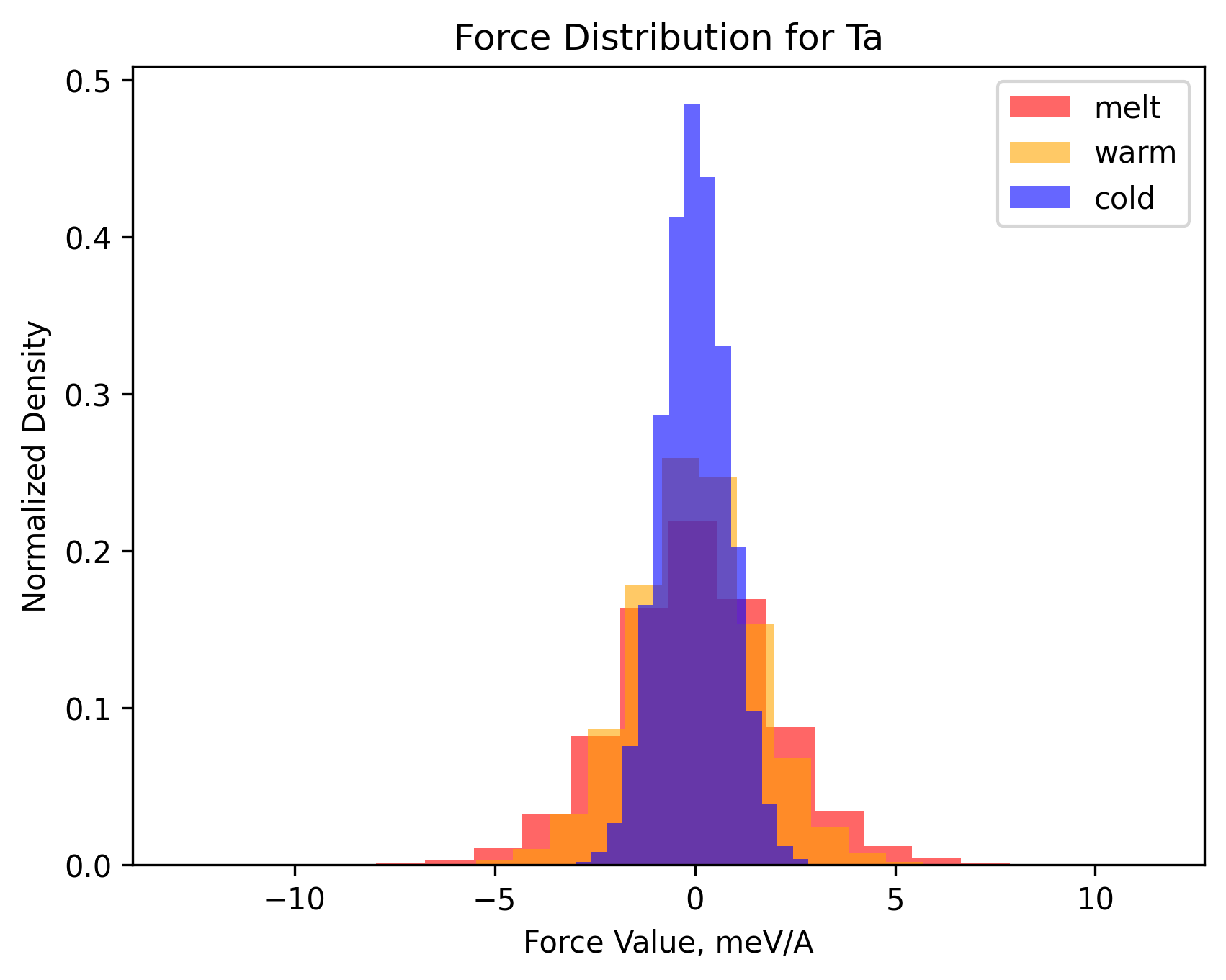}
        \caption{}
    \end{figure*}
    
\begin{figure*}
        \centering
        \includegraphics[width=.65\textwidth]{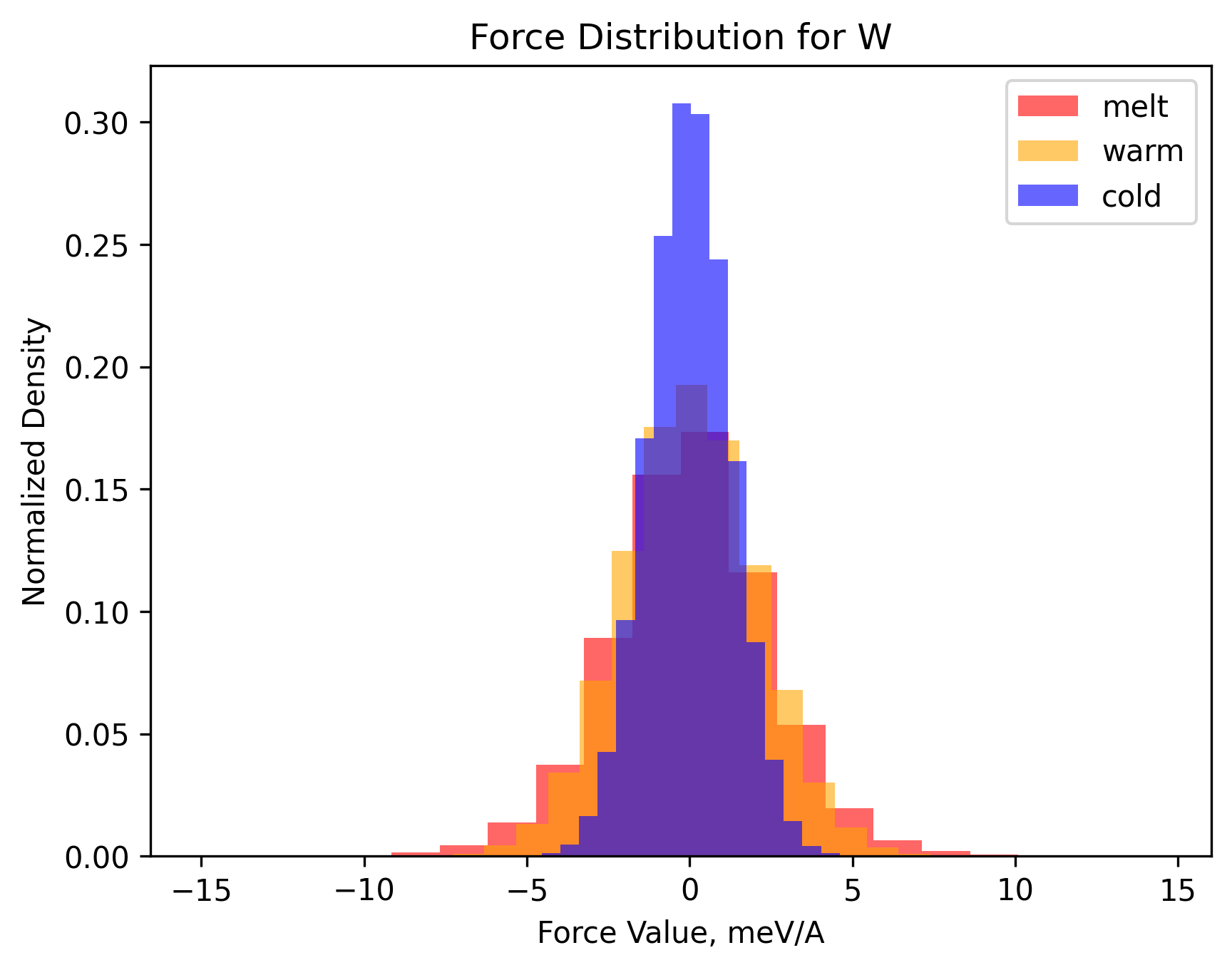}
        \caption{}
    \end{figure*}
    
\begin{figure*}
        \centering
        \includegraphics[width=.65\textwidth]{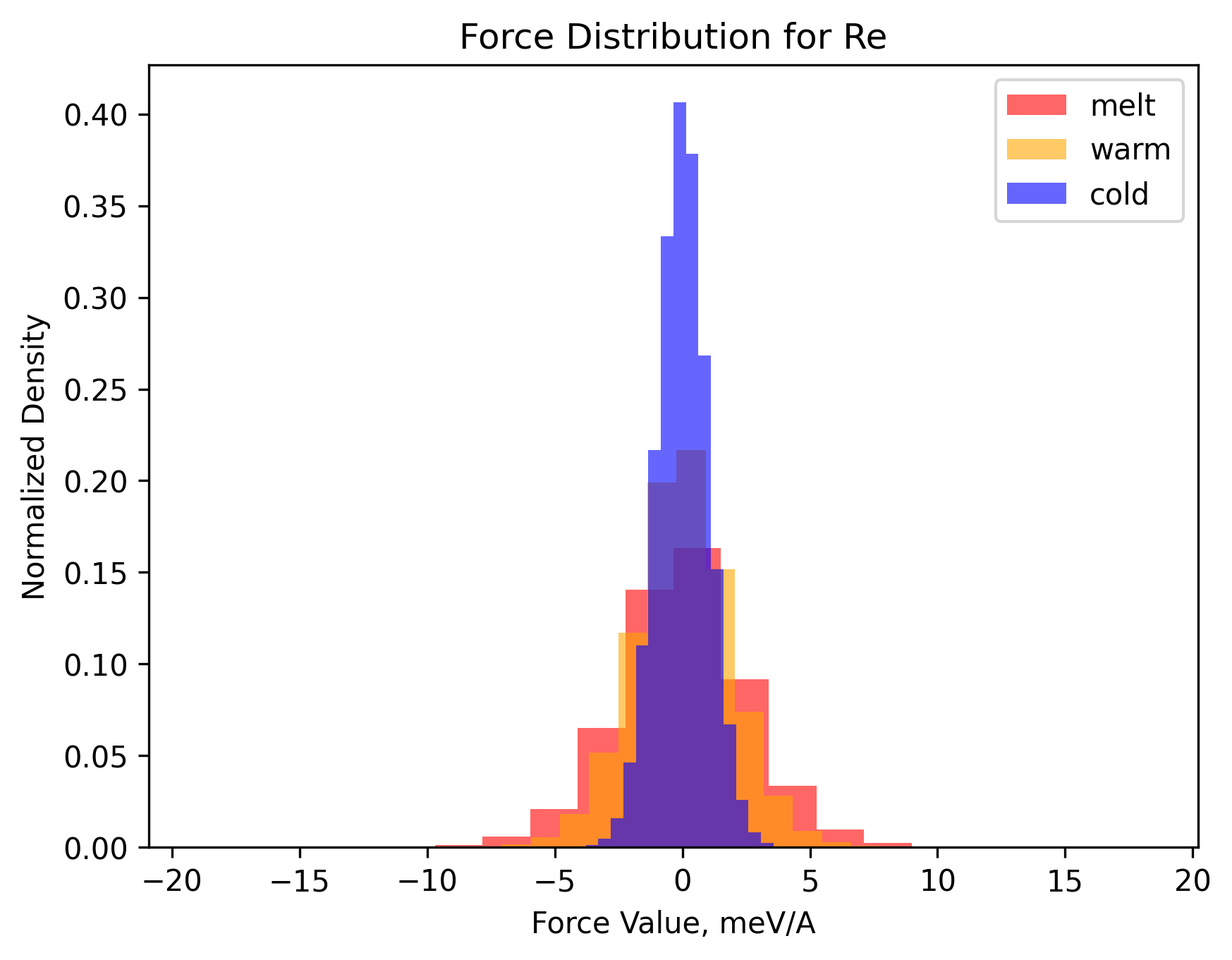}
        \caption{}
    \end{figure*}
    
\begin{figure*}
        \centering
        \includegraphics[width=.65\textwidth]{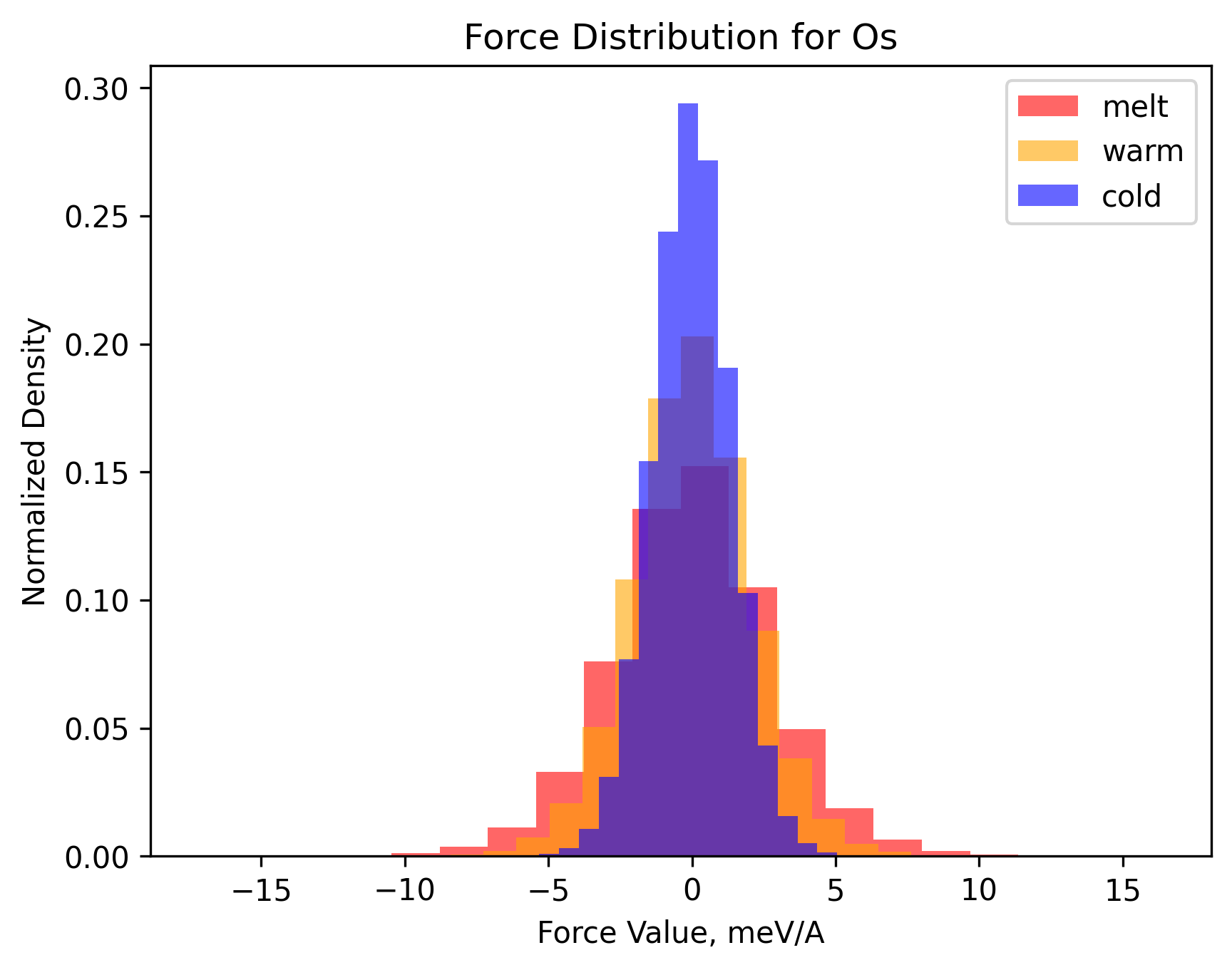}
        \caption{}
    \end{figure*}
    
\begin{figure*}
        \centering
        \includegraphics[width=.65\textwidth]{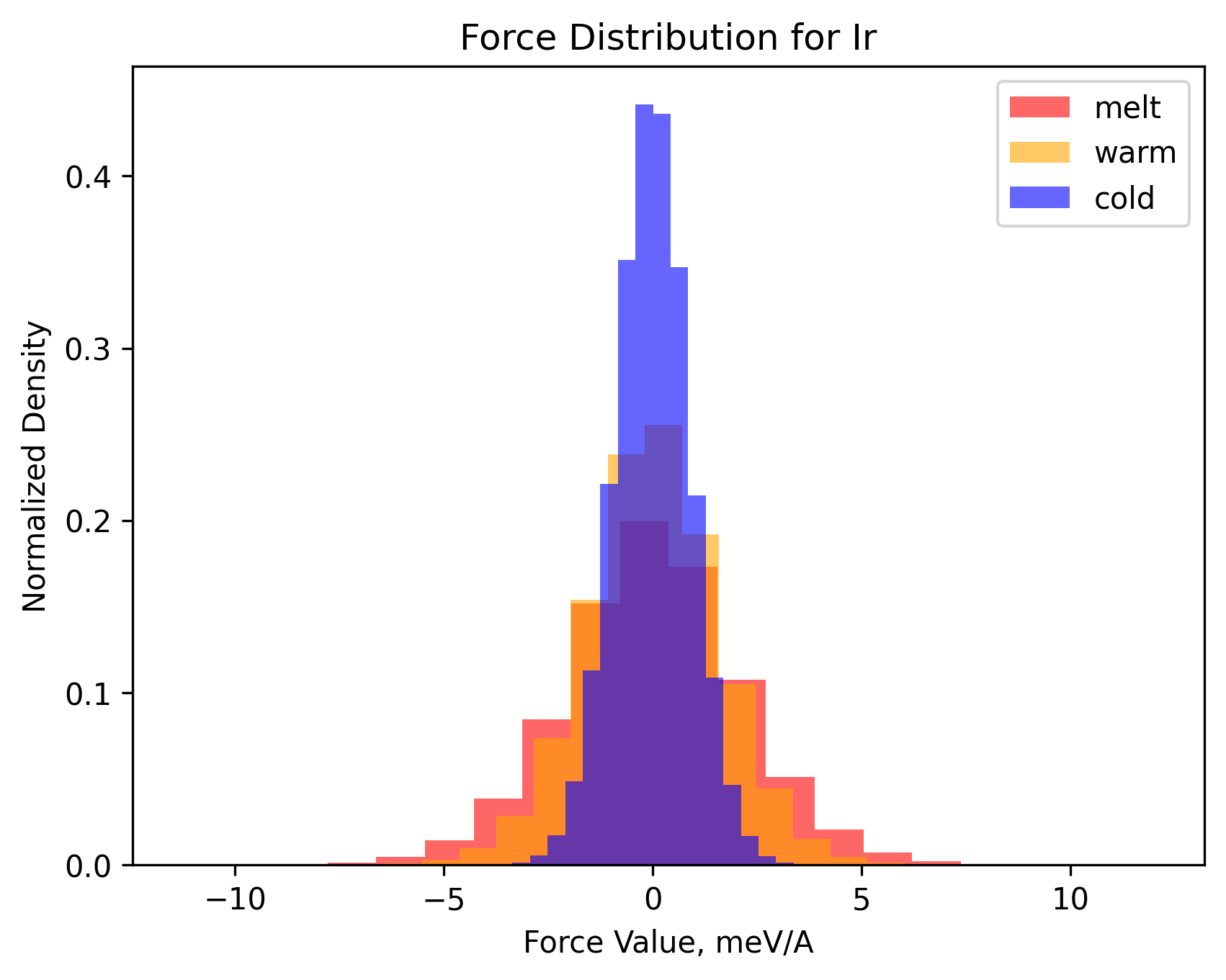}
        \caption{}
    \end{figure*}
    
\begin{figure*}
        \centering
        \includegraphics[width=.65\textwidth]{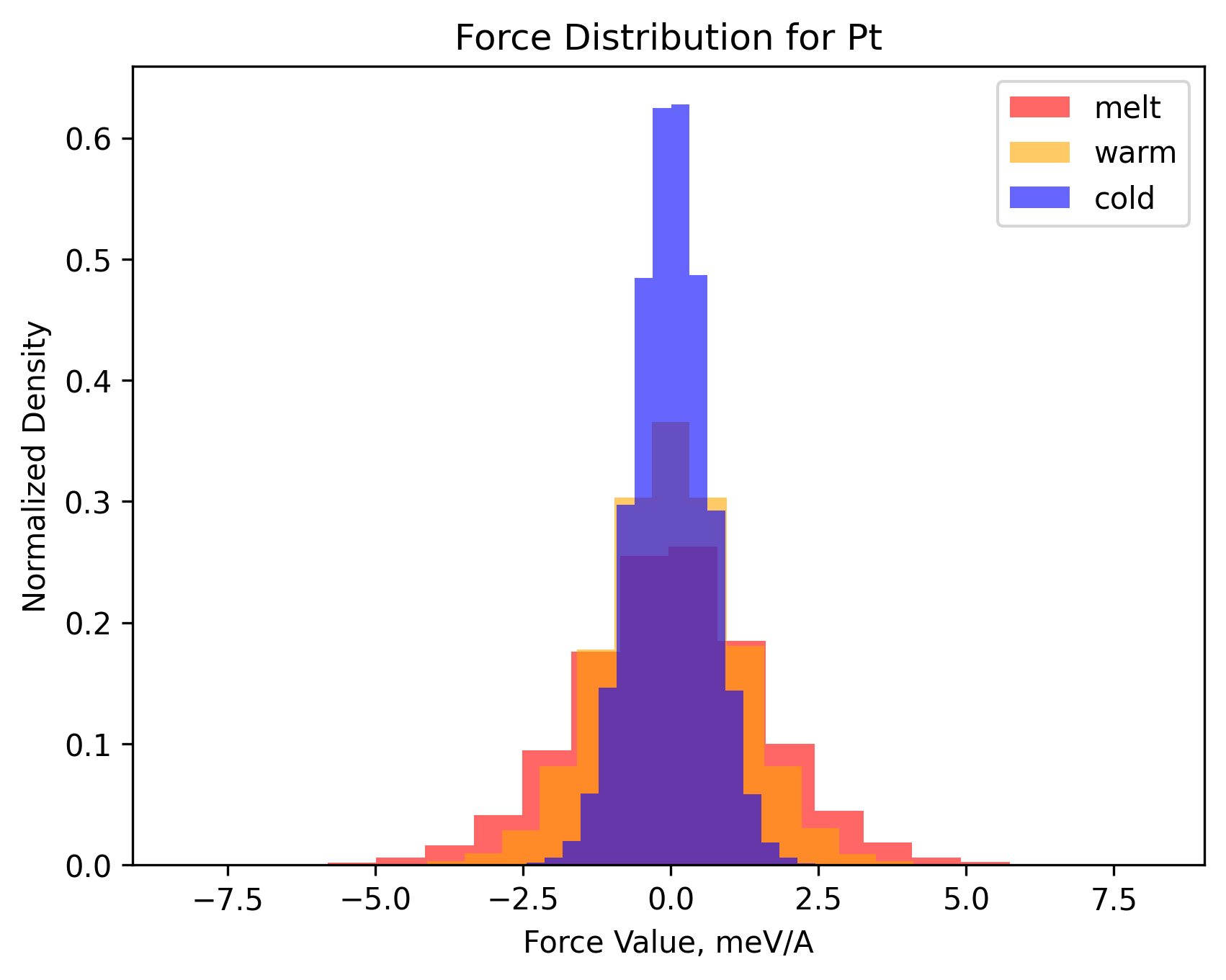}
        \caption{}
    \end{figure*}
    
\begin{figure*}
        \centering
        \includegraphics[width=.65\textwidth]{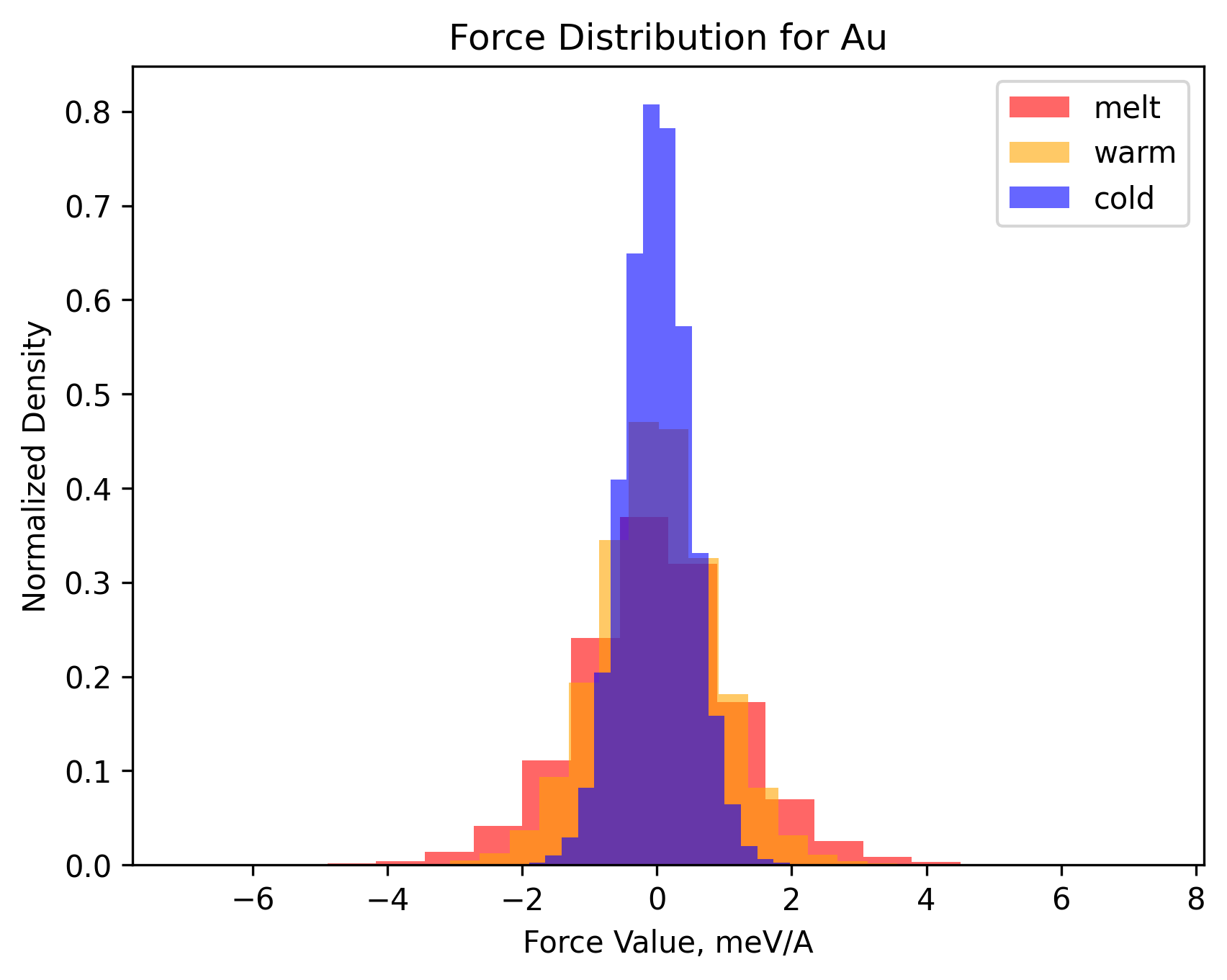}
        \caption{}
    \end{figure*}
    
\begin{figure*}
        \centering
        \includegraphics[width=.65\textwidth]{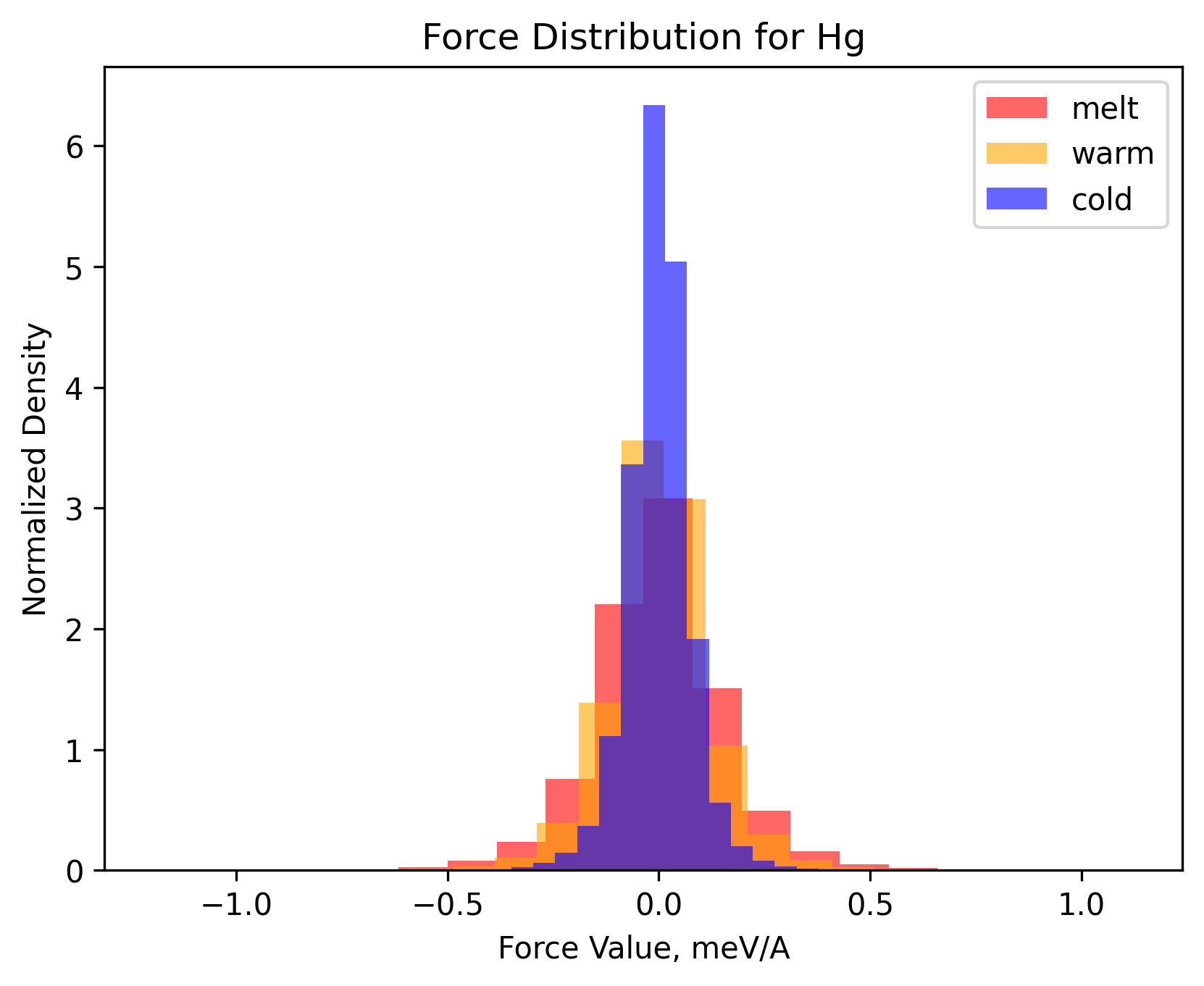}
        \caption{}
        \label{S-fig:lasterrordist}

    \end{figure*}

\clearpage


\begin{figure*}
        \centering
        \includegraphics[width=.65\textwidth]{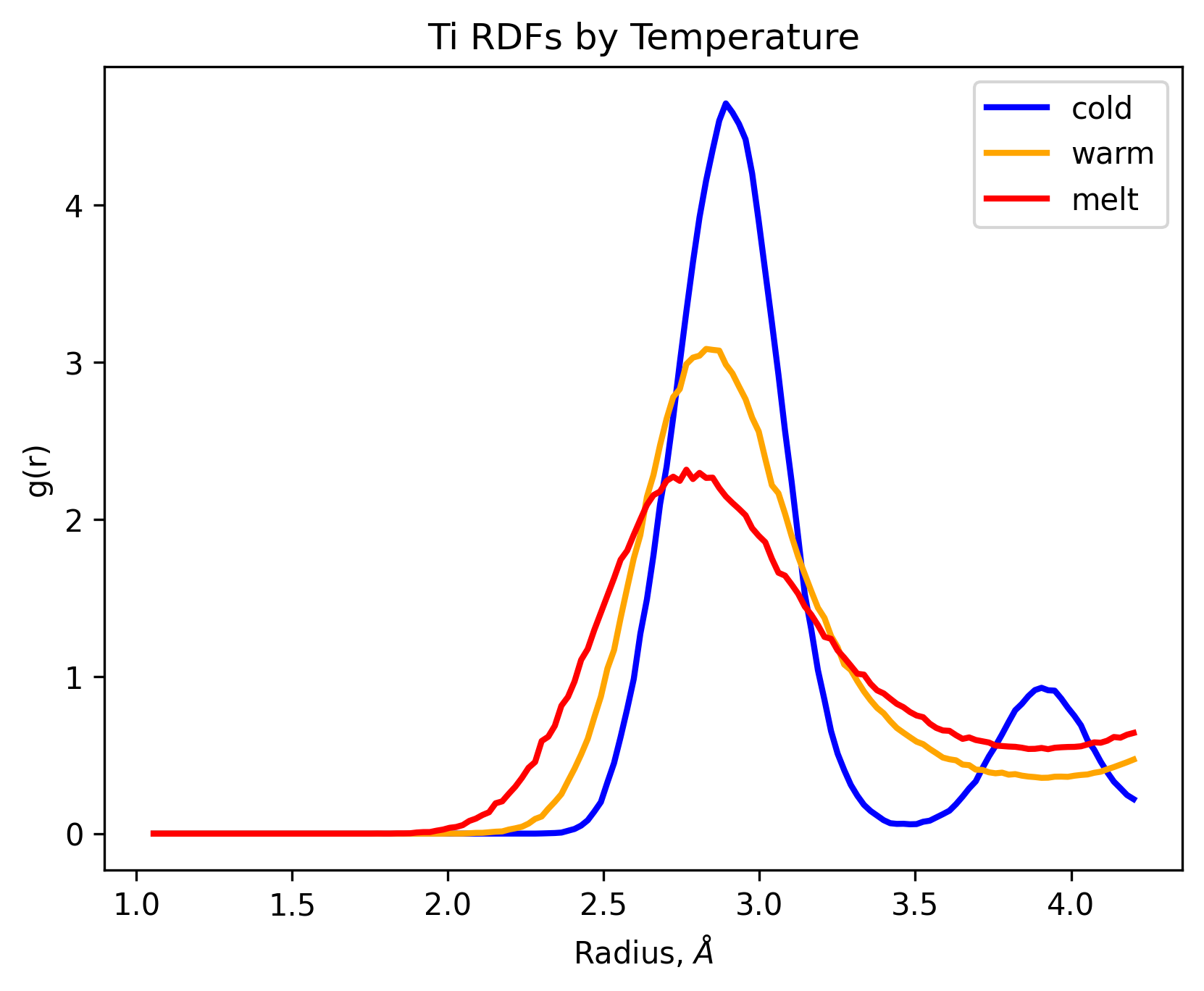}
        \caption{}
        \label{S-fig:firstrdf}
    \end{figure*}
    
\begin{figure*}
        \centering
        \includegraphics[width=.65\textwidth]{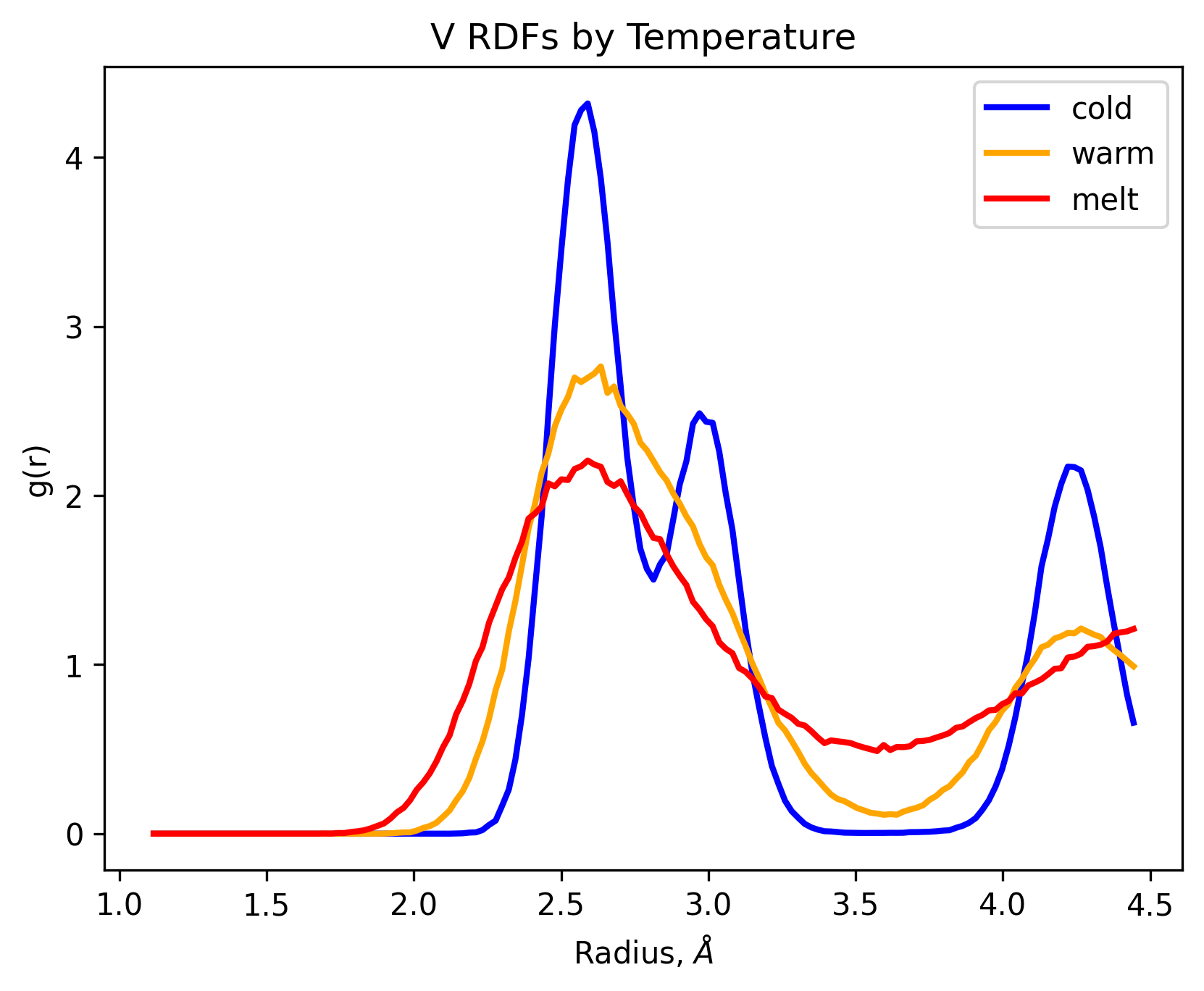}
        \caption{}
    \end{figure*}
    
\begin{figure*}
        \centering
        \includegraphics[width=.65\textwidth]{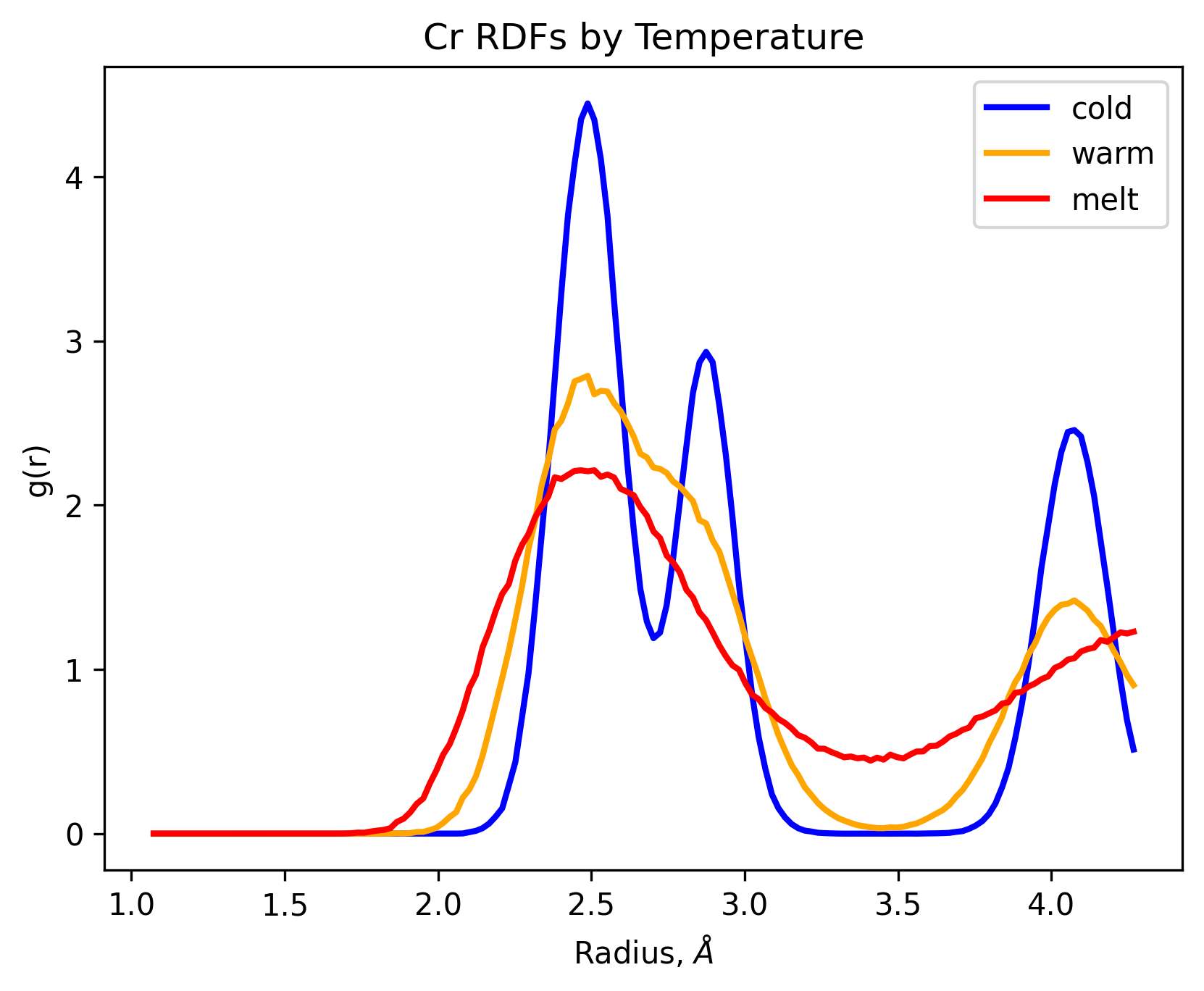}
        \caption{}
    \end{figure*}
    
\begin{figure*}
        \centering
        \includegraphics[width=.65\textwidth]{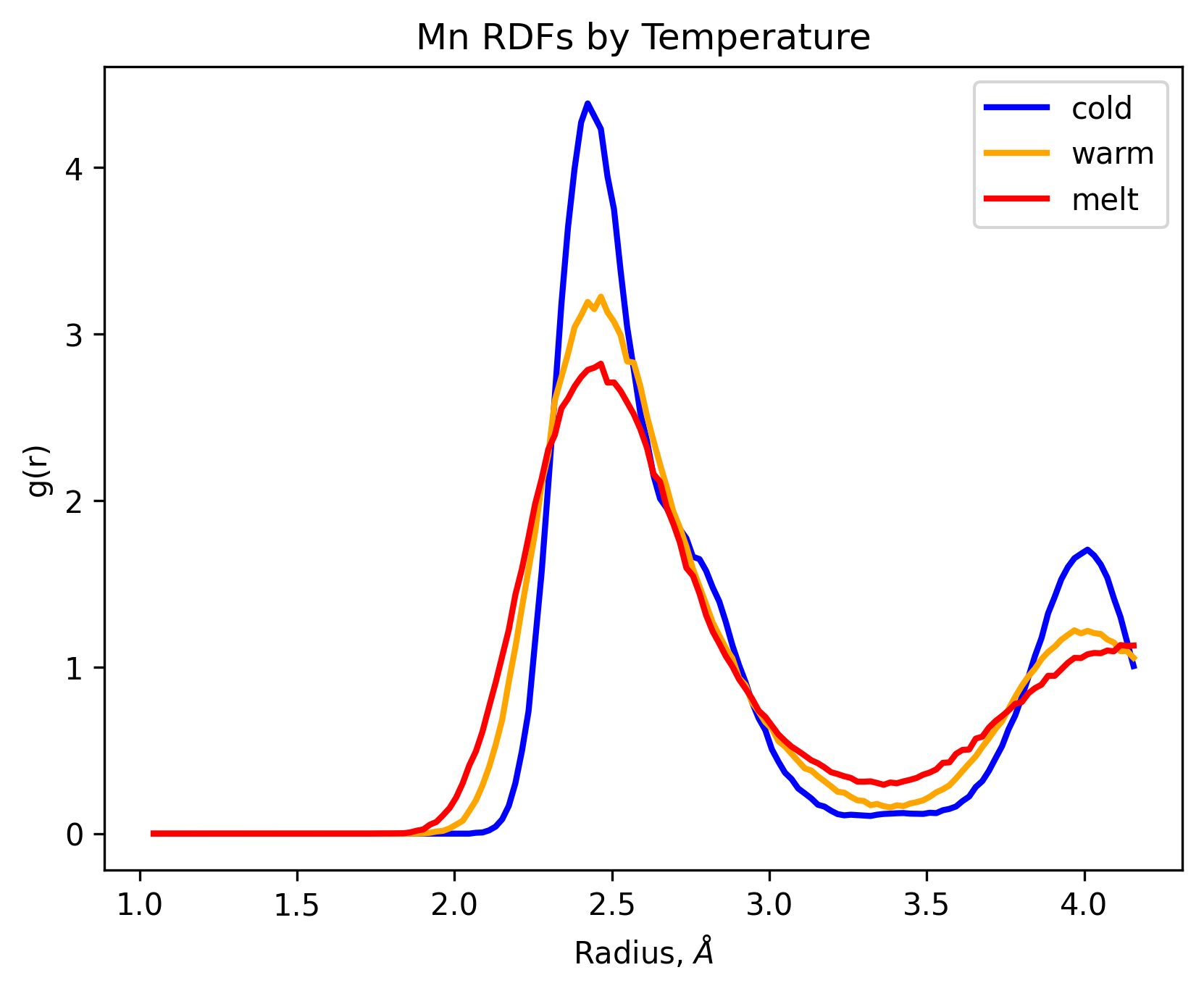}
        \caption{}
    \end{figure*}
    
\begin{figure*}
        \centering
        \includegraphics[width=.65\textwidth]{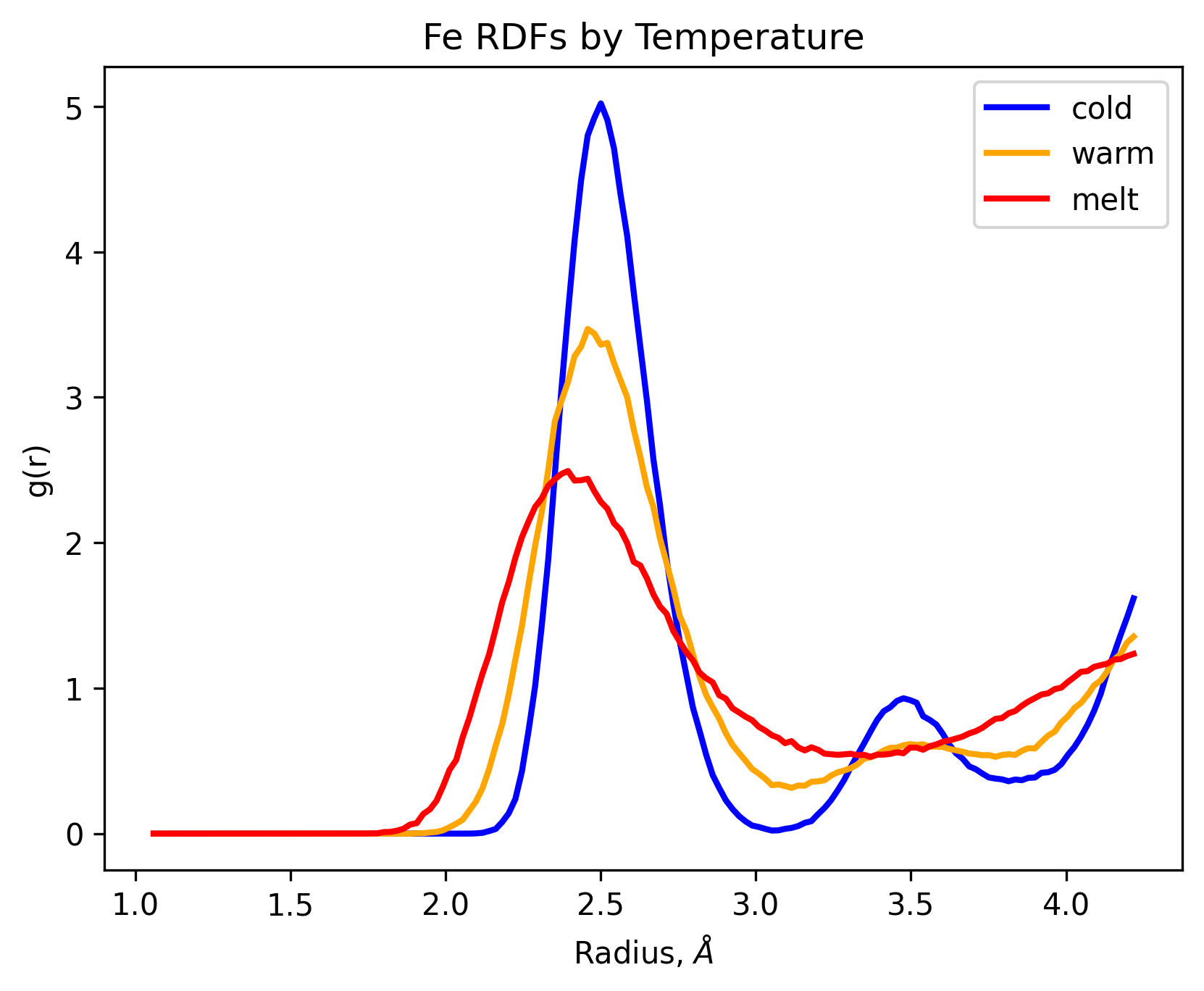}
        \caption{}
    \end{figure*}
    
\begin{figure*}
        \centering
        \includegraphics[width=.65\textwidth]{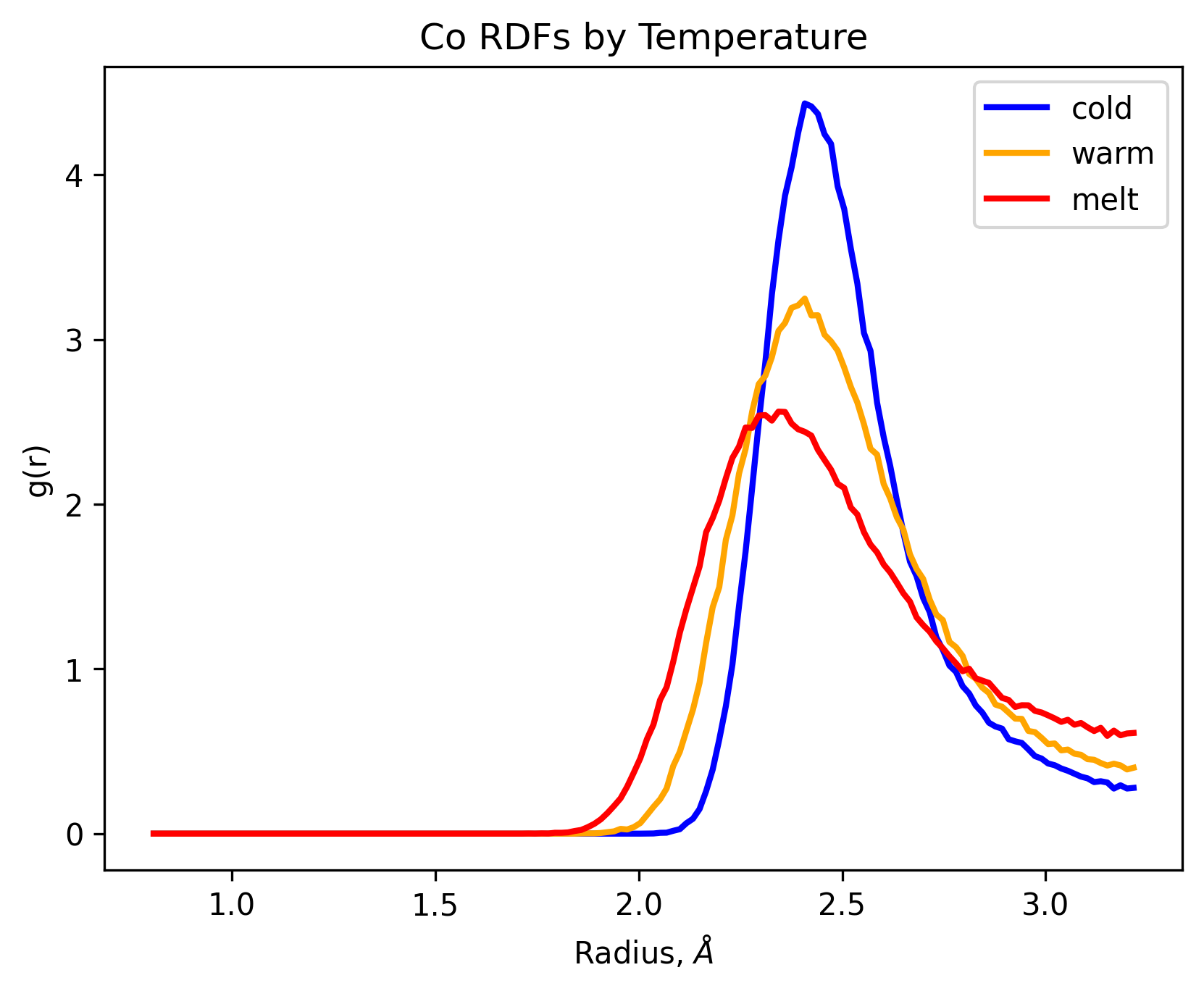}
        \caption{}
    \end{figure*}
    
\begin{figure*}
        \centering
        \includegraphics[width=.65\textwidth]{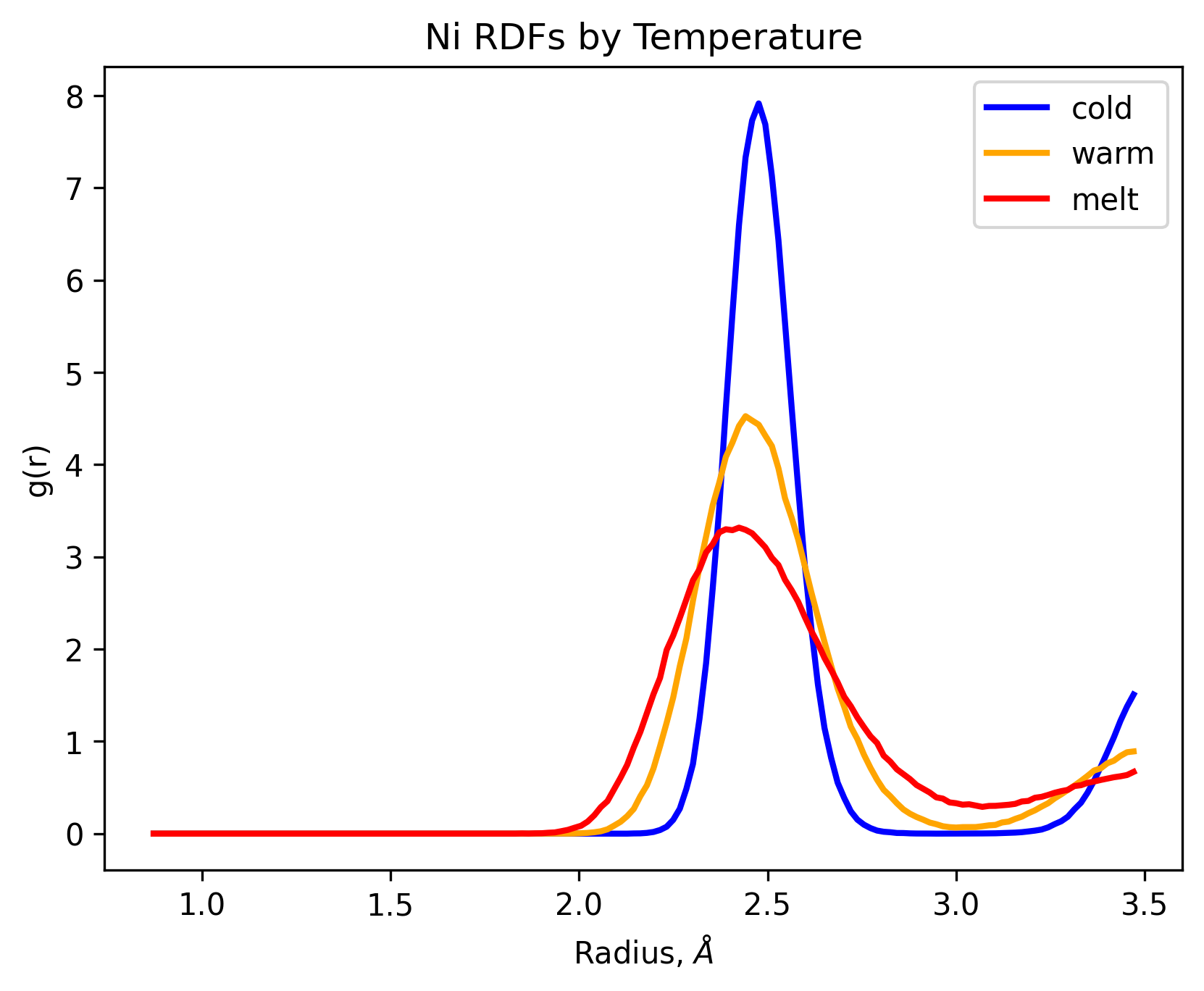}
        \caption{}
    \end{figure*}
    
\begin{figure*}
        \centering
        \includegraphics[width=.65\textwidth]{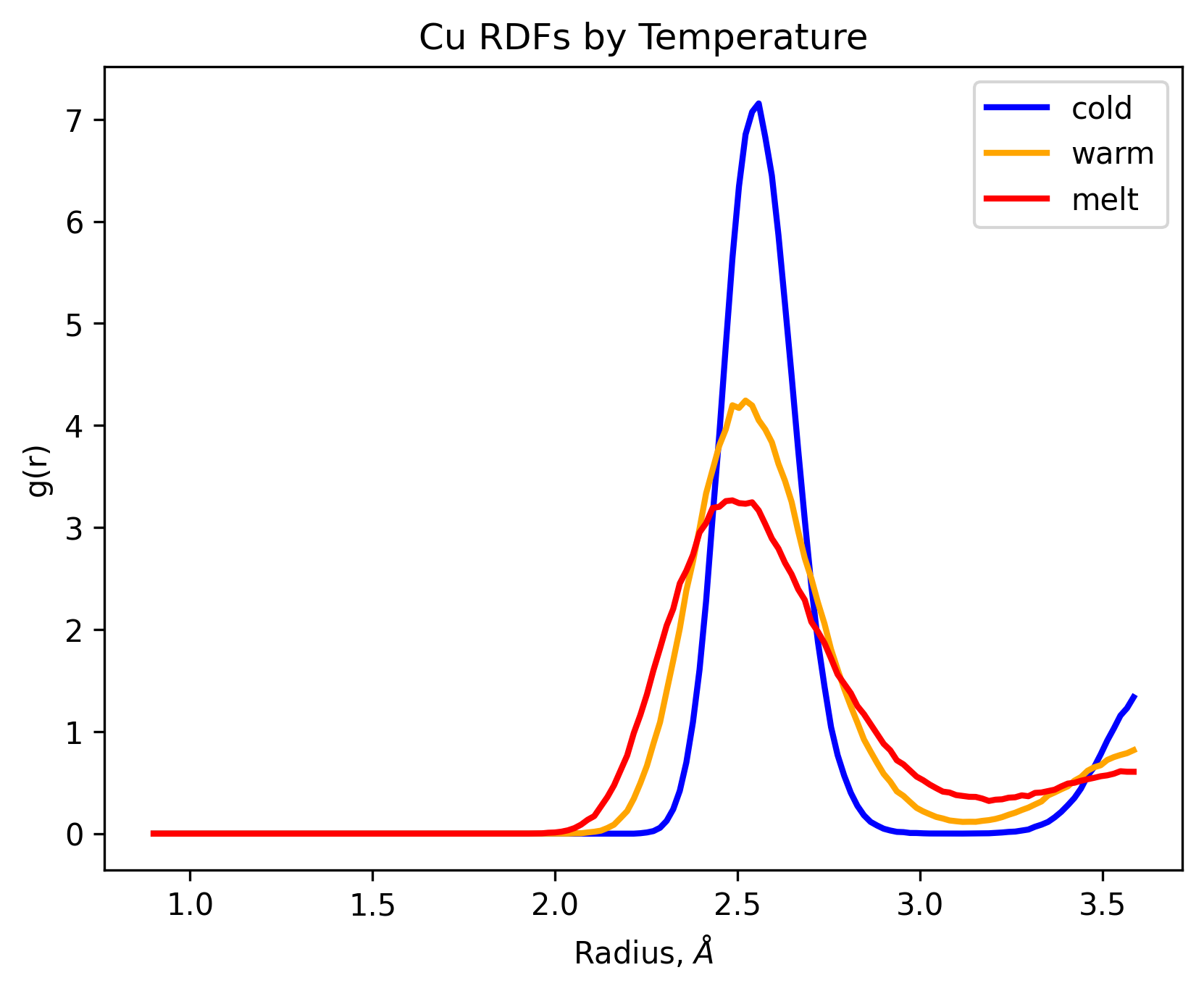}
        \caption{}
    \end{figure*}
    
\begin{figure*}
        \centering
        \includegraphics[width=.65\textwidth]{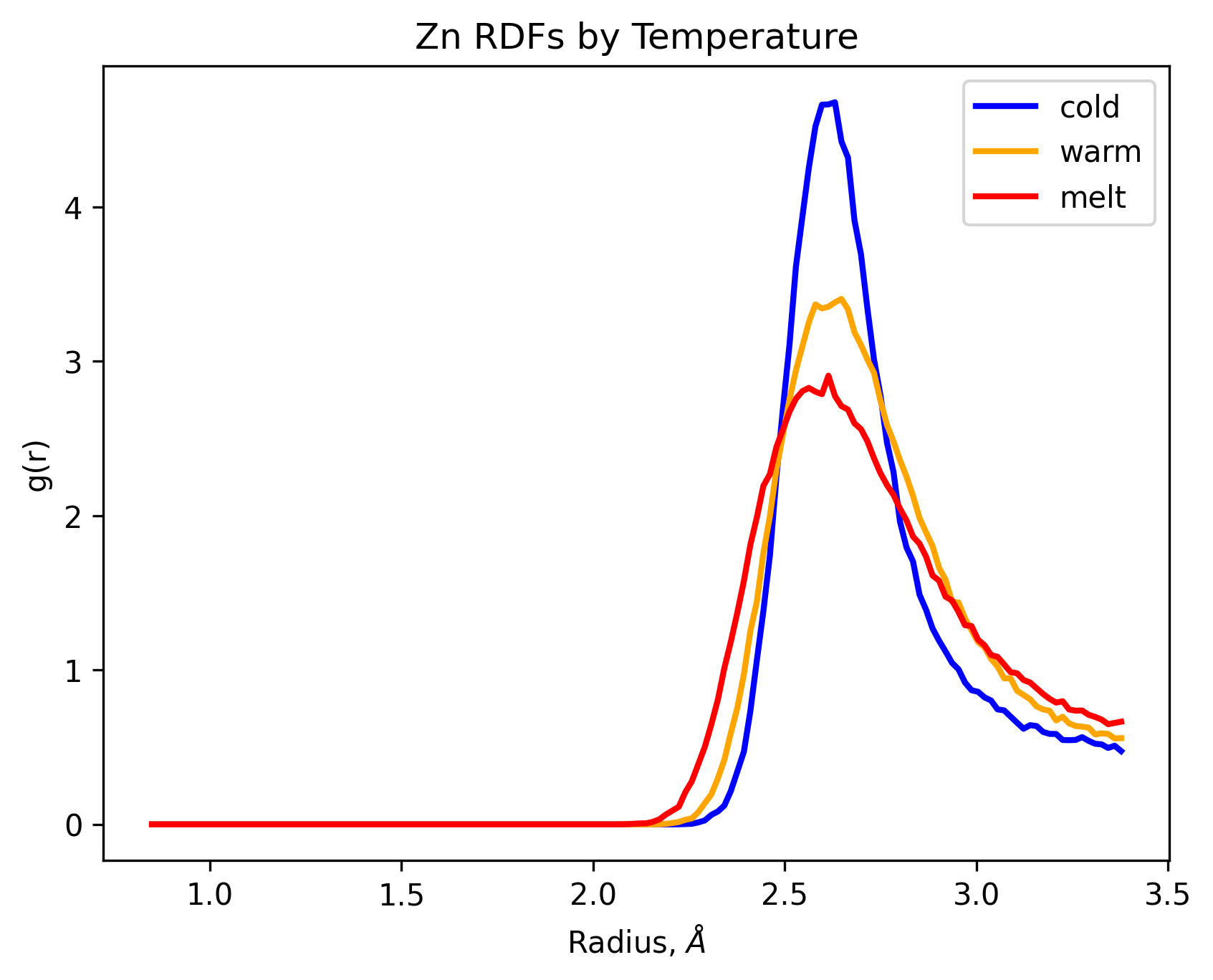}
        \caption{}
    \end{figure*}
    
\begin{figure*}
        \centering
        \includegraphics[width=.65\textwidth]{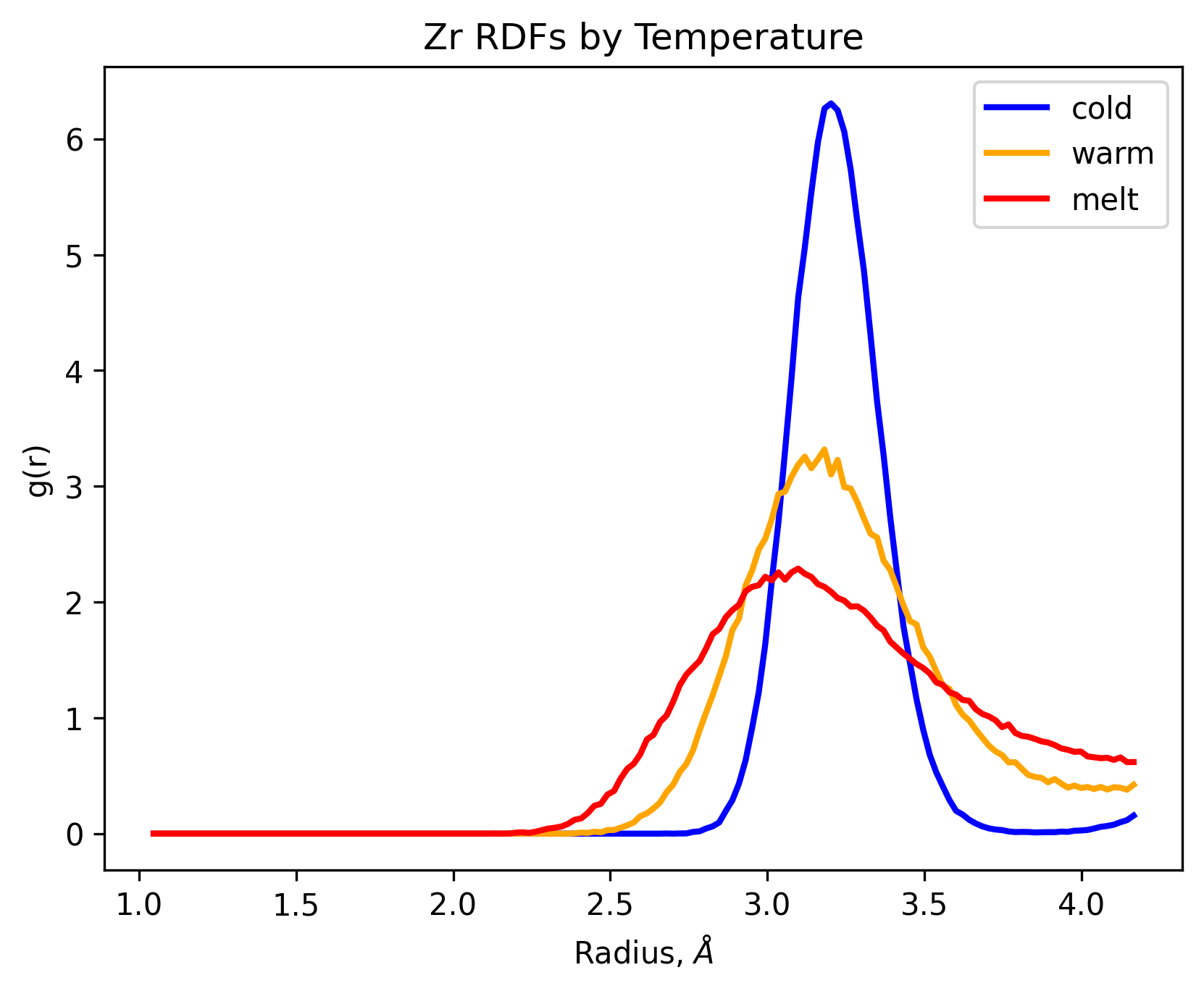}
        \caption{}
    \end{figure*}
    
\begin{figure*}
        \centering
        \includegraphics[width=.65\textwidth]{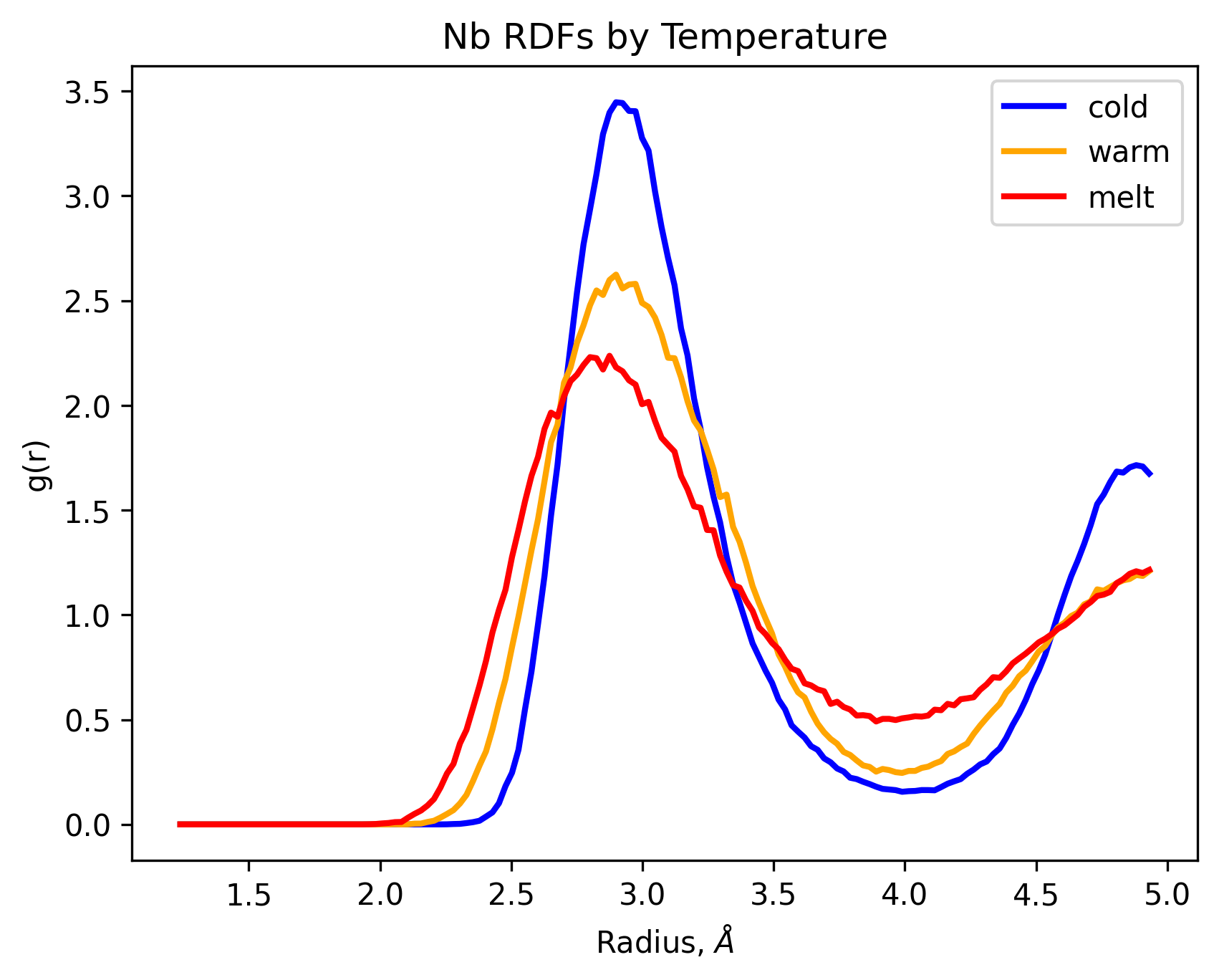}
        \caption{}
    \end{figure*}
    
\begin{figure*}
        \centering
        \includegraphics[width=.65\textwidth]{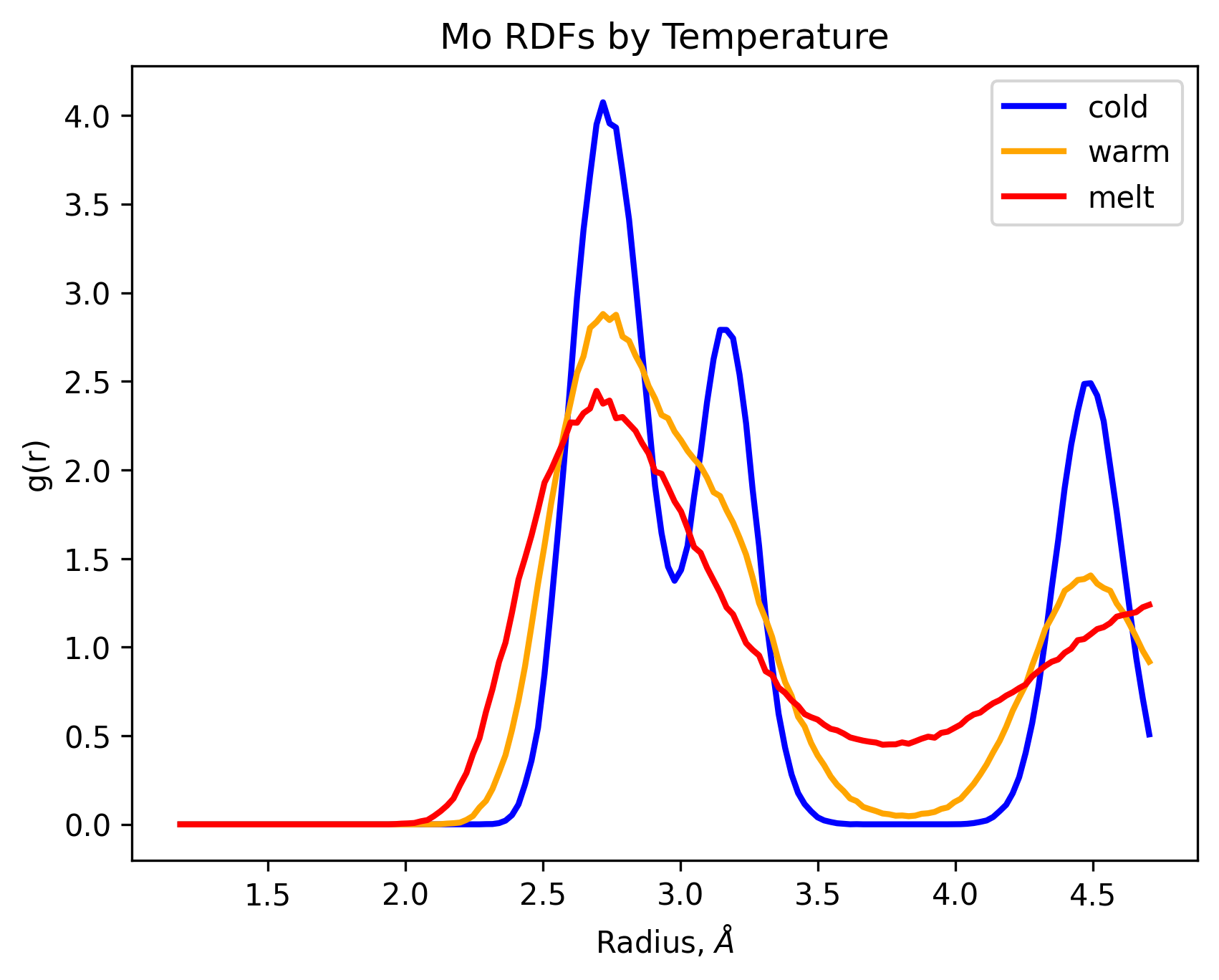}
        \caption{}
    \end{figure*}
    
\begin{figure*}
        \centering
        \includegraphics[width=.65\textwidth]{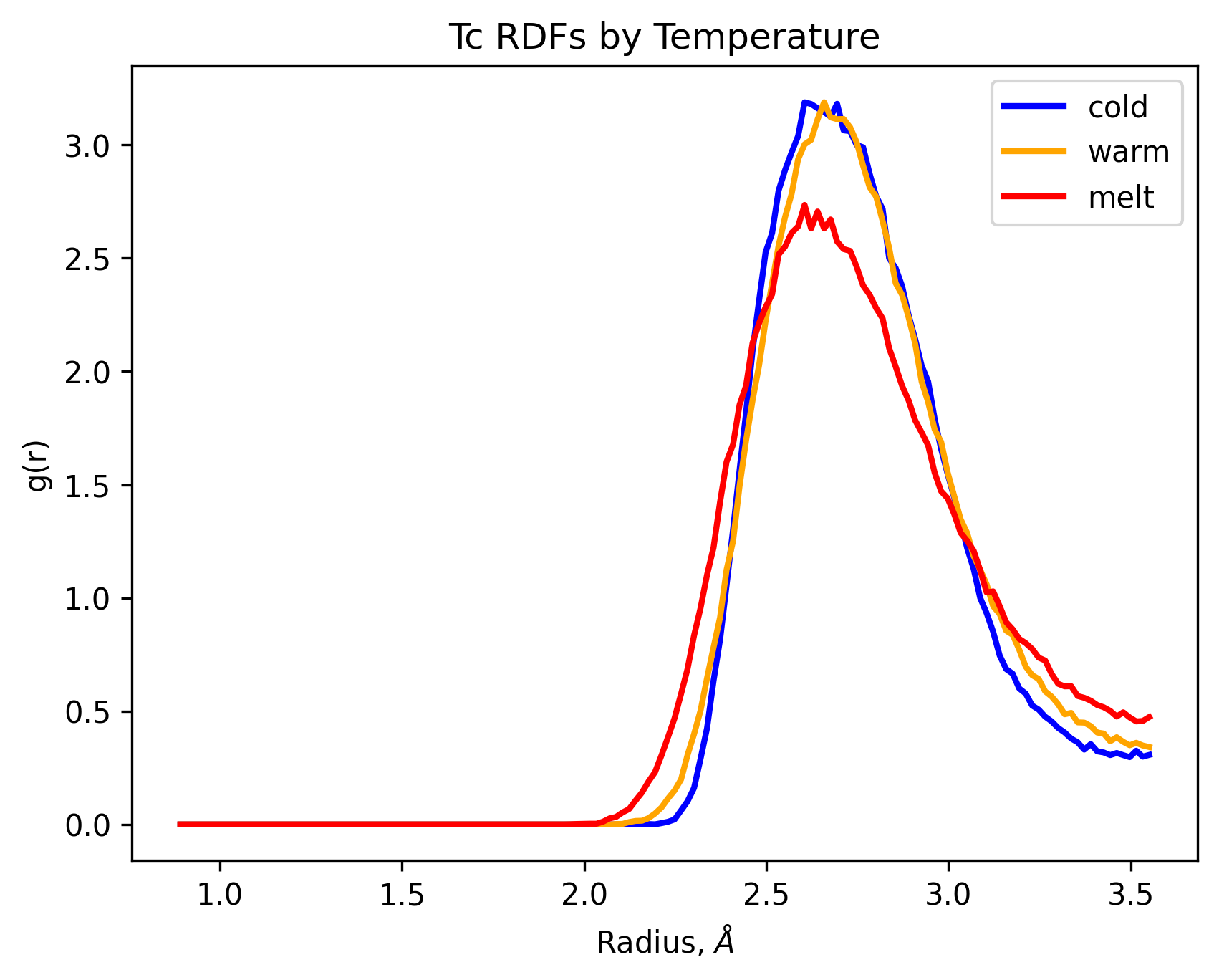}
        \caption{}
    \end{figure*}
    
\begin{figure*}
        \centering
        \includegraphics[width=.65\textwidth]{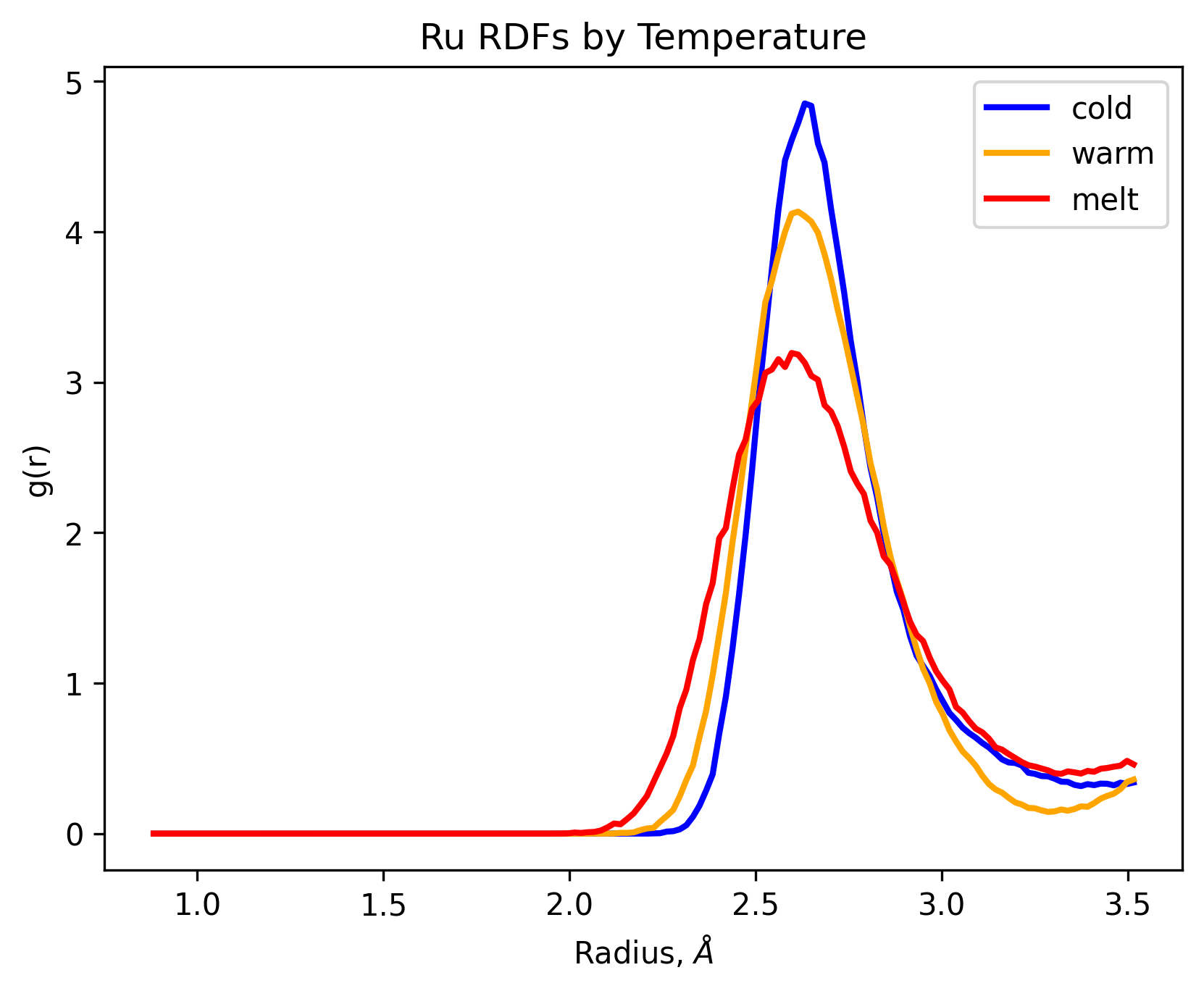}
        \caption{}
    \end{figure*}
    
\begin{figure*}
        \centering
        \includegraphics[width=.65\textwidth]{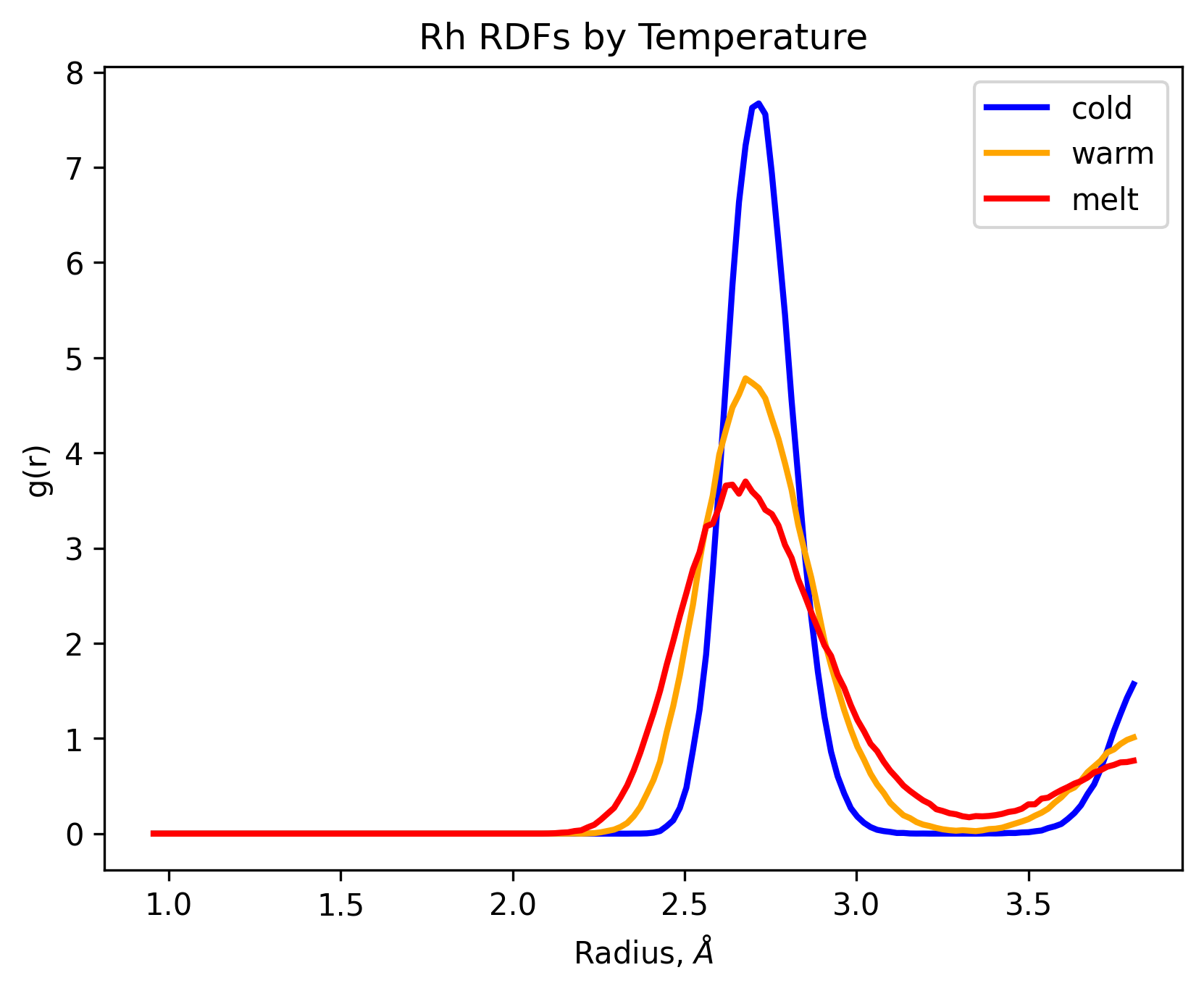}
        \caption{}
    \end{figure*}
    
\begin{figure*}
        \centering
        \includegraphics[width=.65\textwidth]{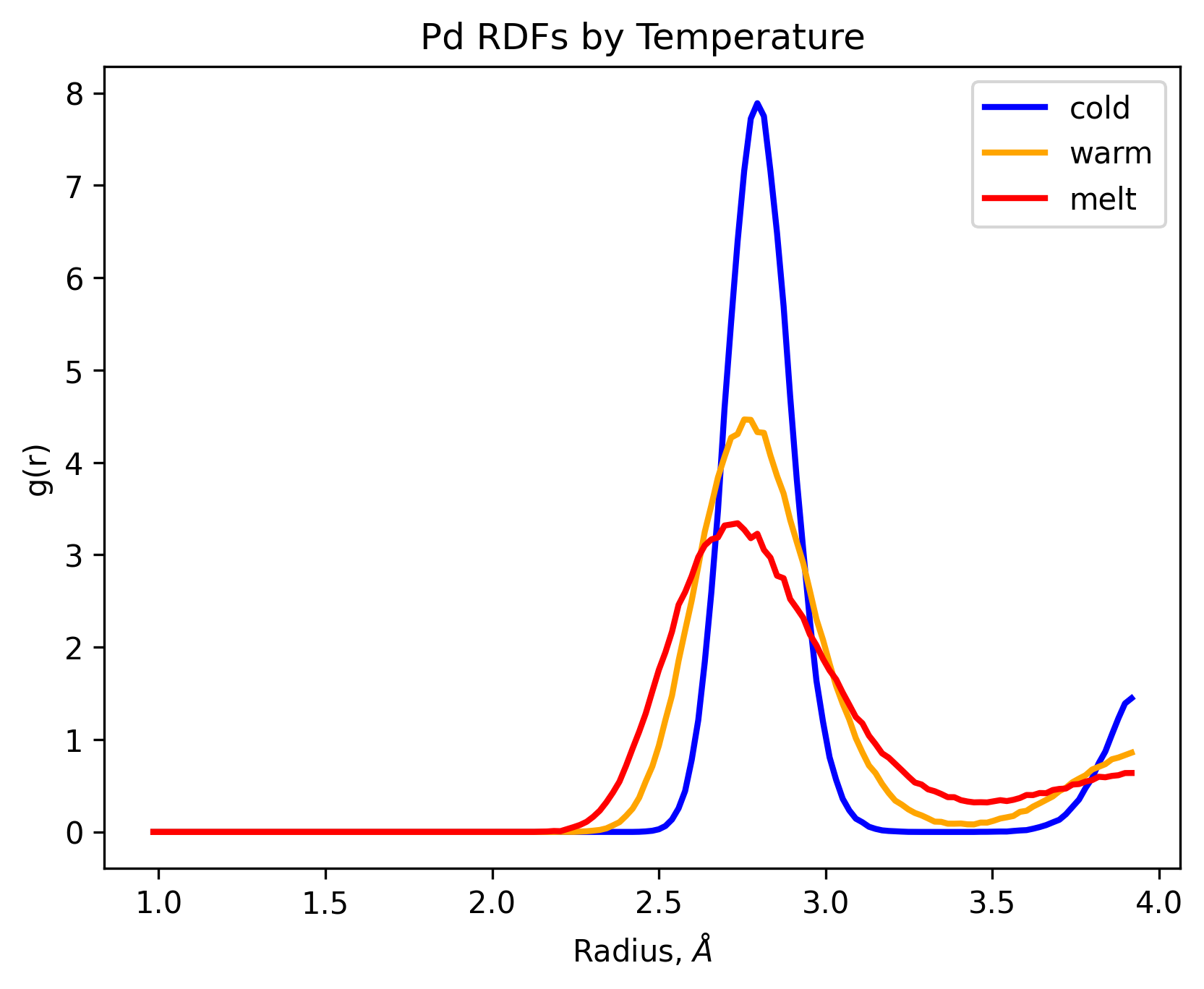}
        \caption{}
    \end{figure*}
    
\begin{figure*}
        \centering
        \includegraphics[width=.65\textwidth]{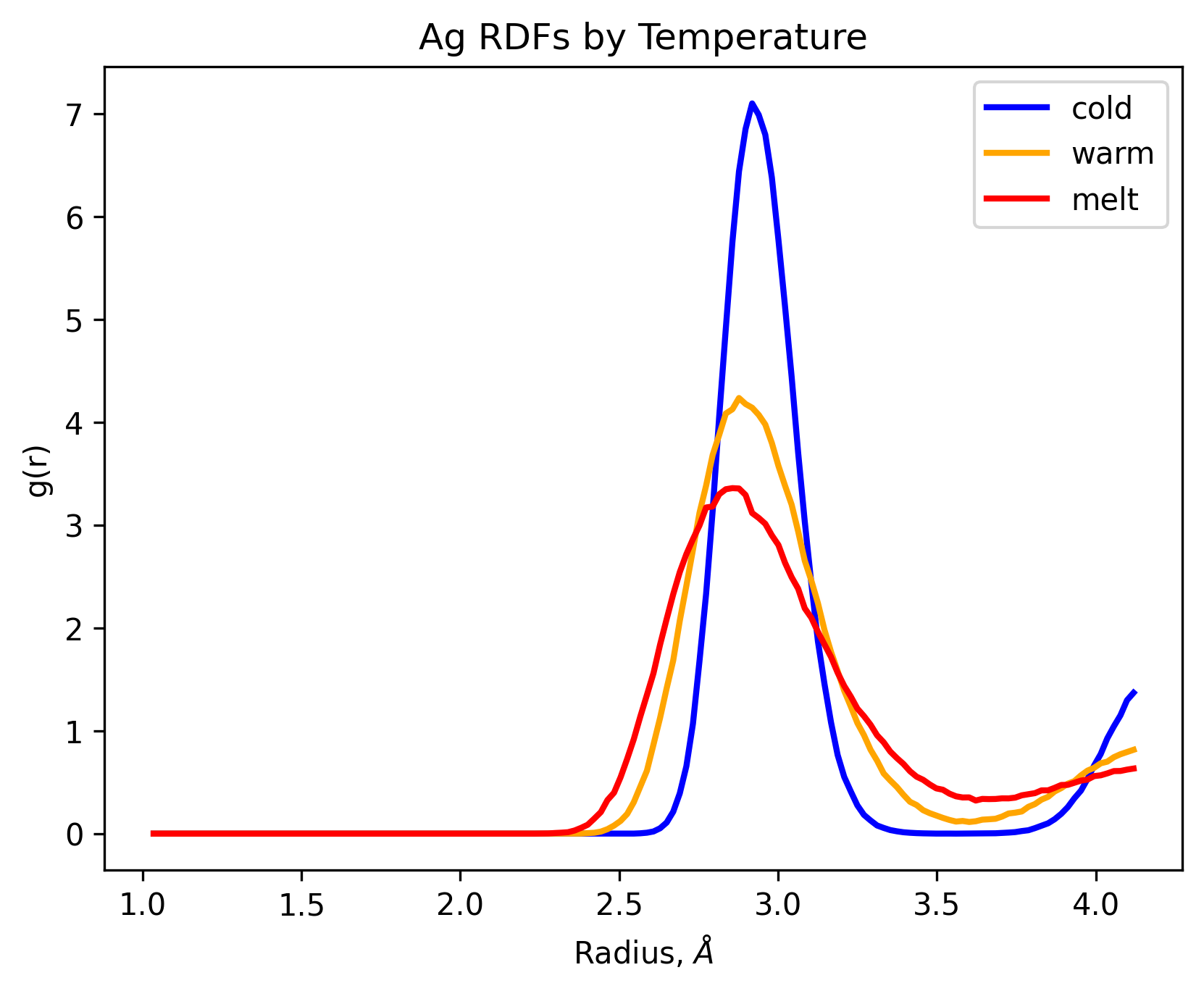}
        \caption{}
    \end{figure*}
    
\begin{figure*}
        \centering
        \includegraphics[width=.65\textwidth]{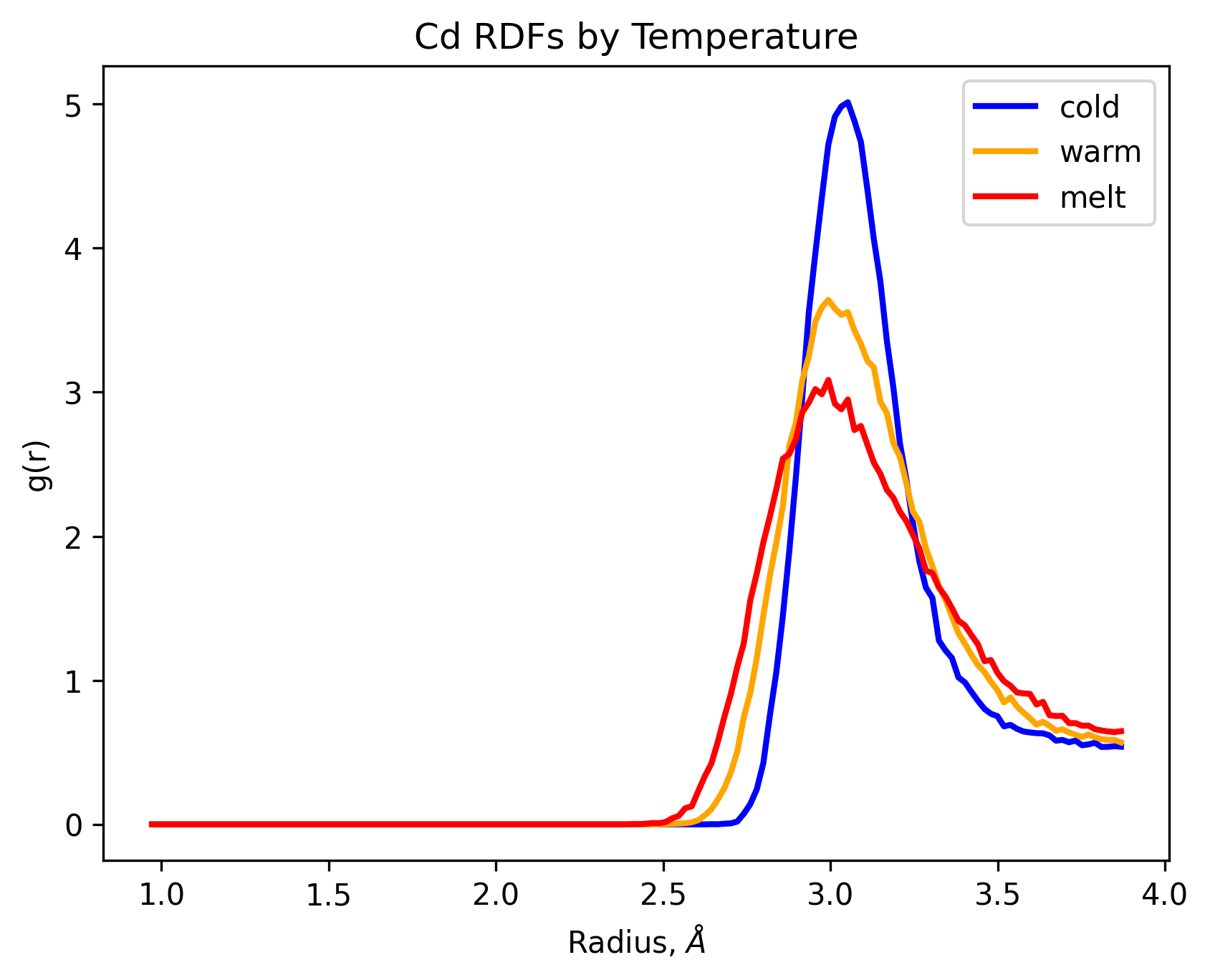}
        \caption{}
    \end{figure*}
    
\begin{figure*}
        \centering
        \includegraphics[width=.65\textwidth]{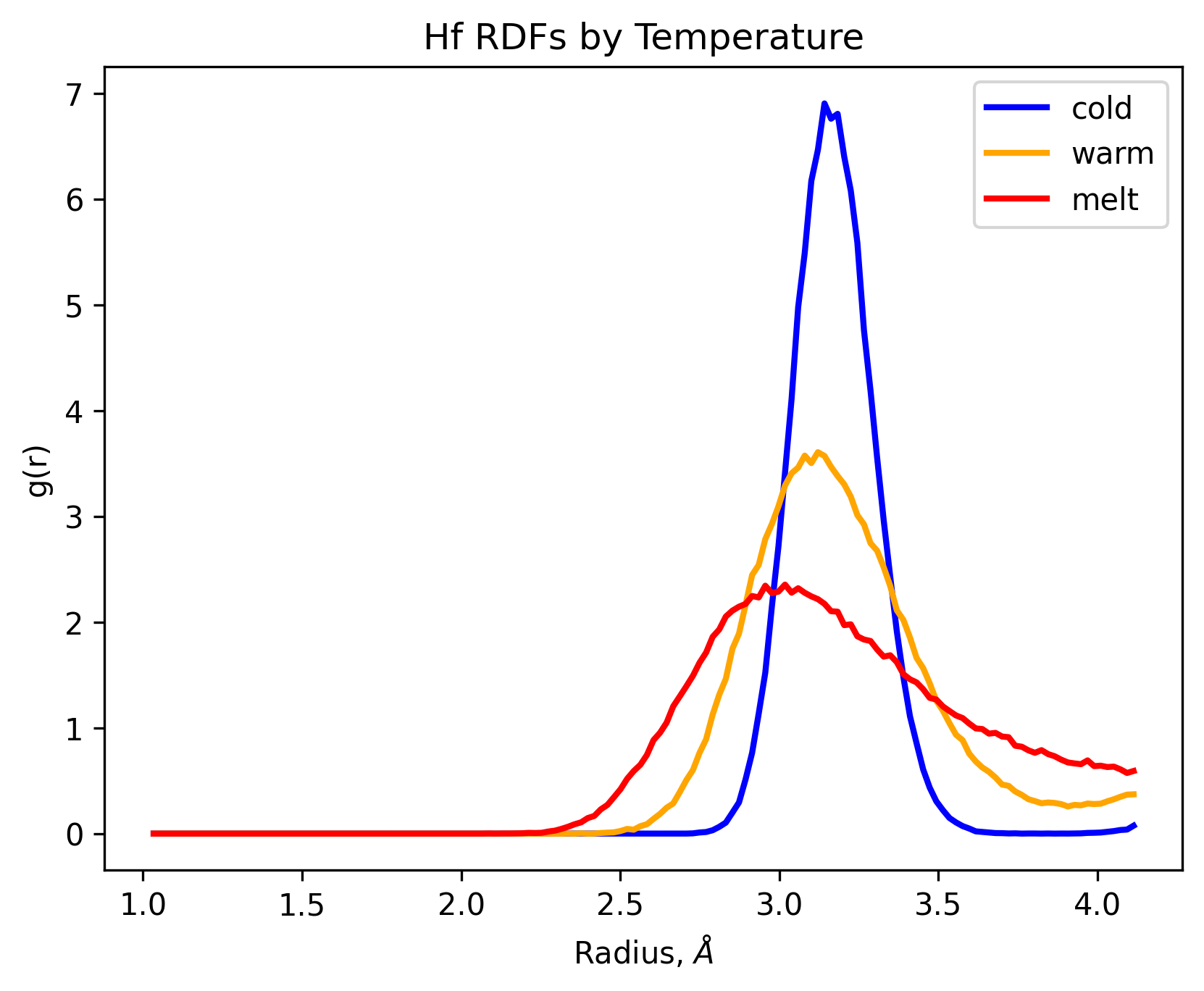}
        \caption{}
    \end{figure*}
    
\begin{figure*}
        \centering
        \includegraphics[width=.65\textwidth]{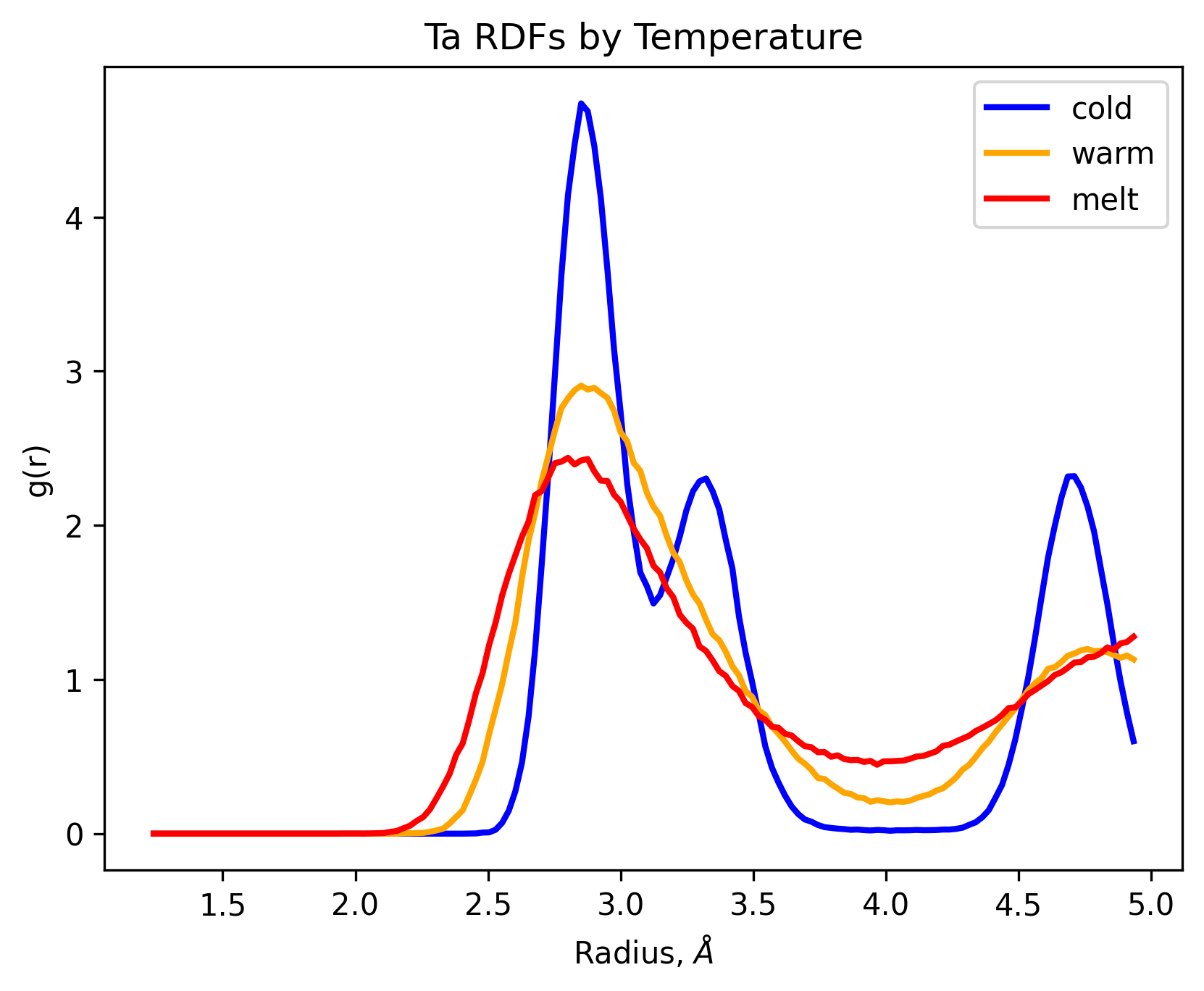}
        \caption{}
    \end{figure*}
    
\begin{figure*}
        \centering
        \includegraphics[width=.65\textwidth]{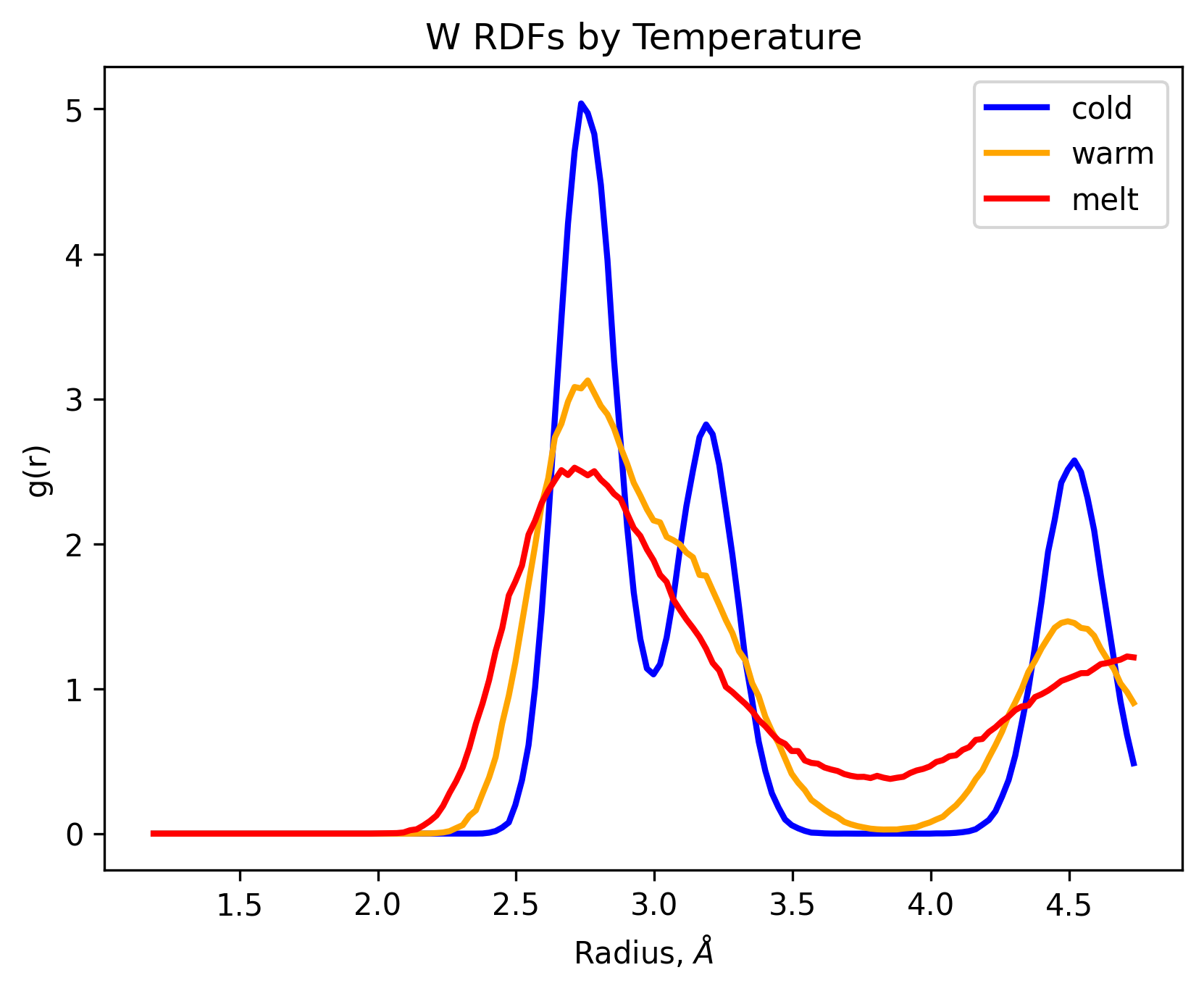}
        \caption{}
    \end{figure*}
    
\begin{figure*}
        \centering
        \includegraphics[width=.65\textwidth]{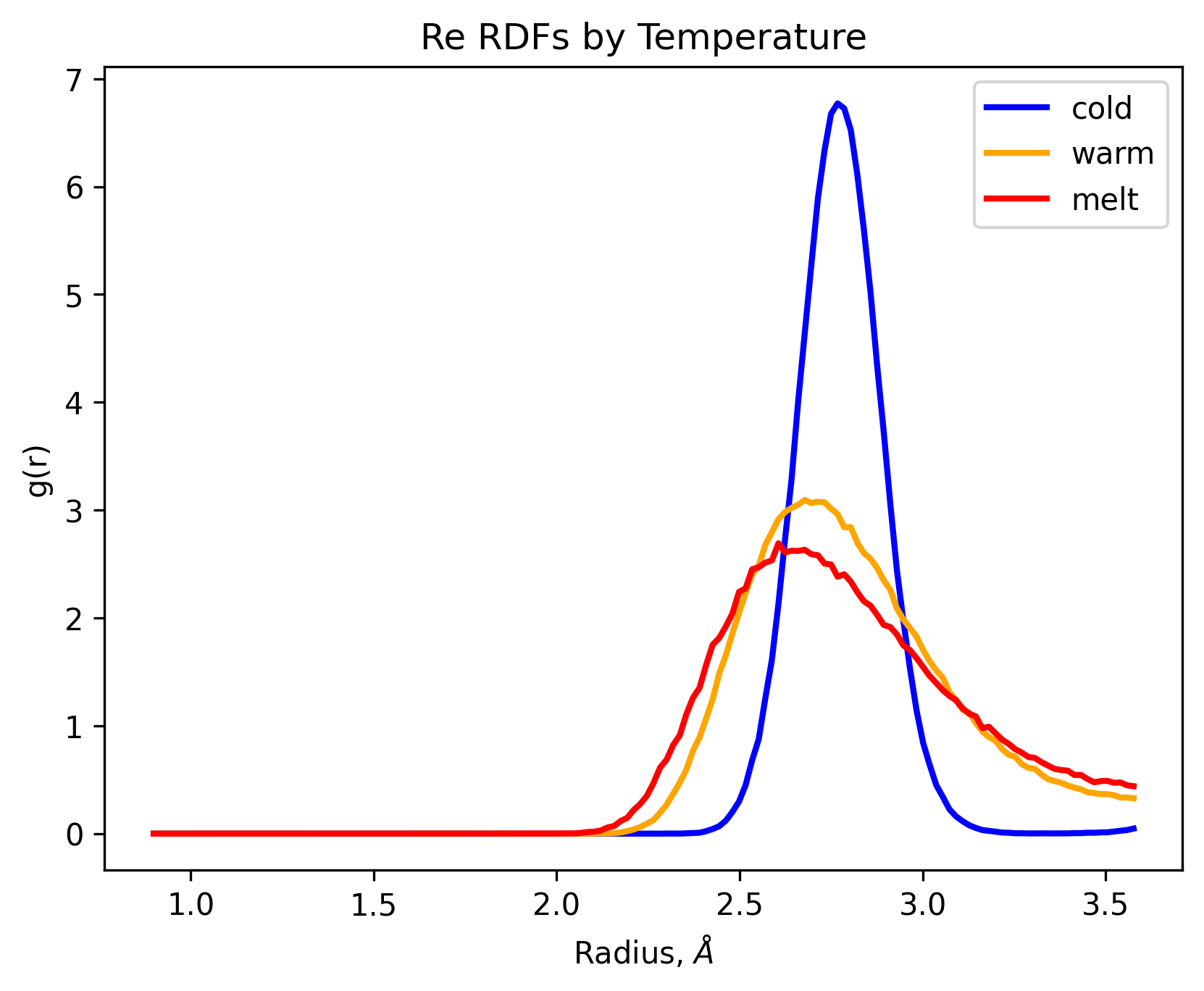}
        \caption{}
    \end{figure*}
    
\begin{figure*}
        \centering
        \includegraphics[width=.65\textwidth]{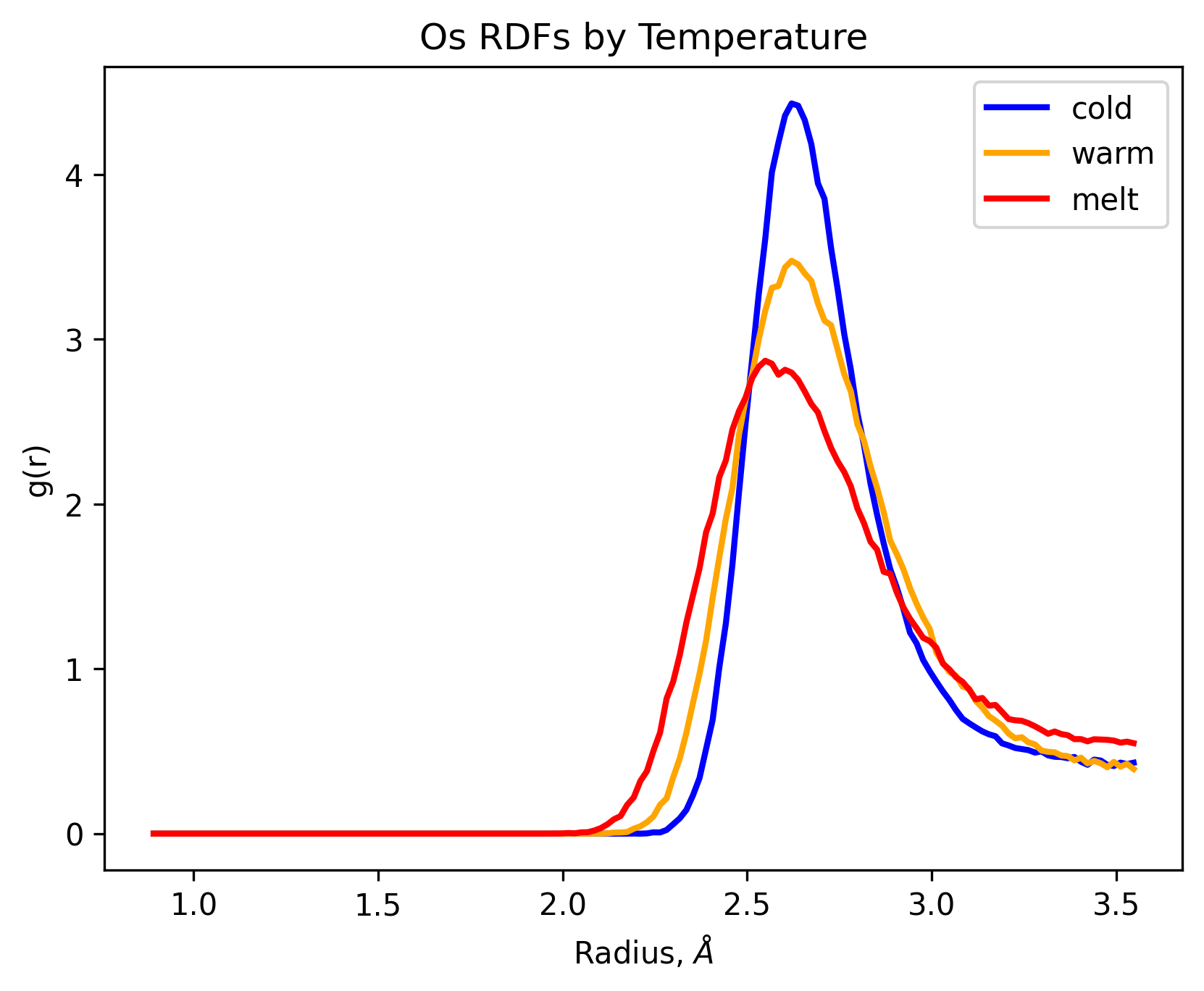}
        \caption{}
    \end{figure*}
    
\begin{figure*}
        \centering
        \includegraphics[width=.65\textwidth]{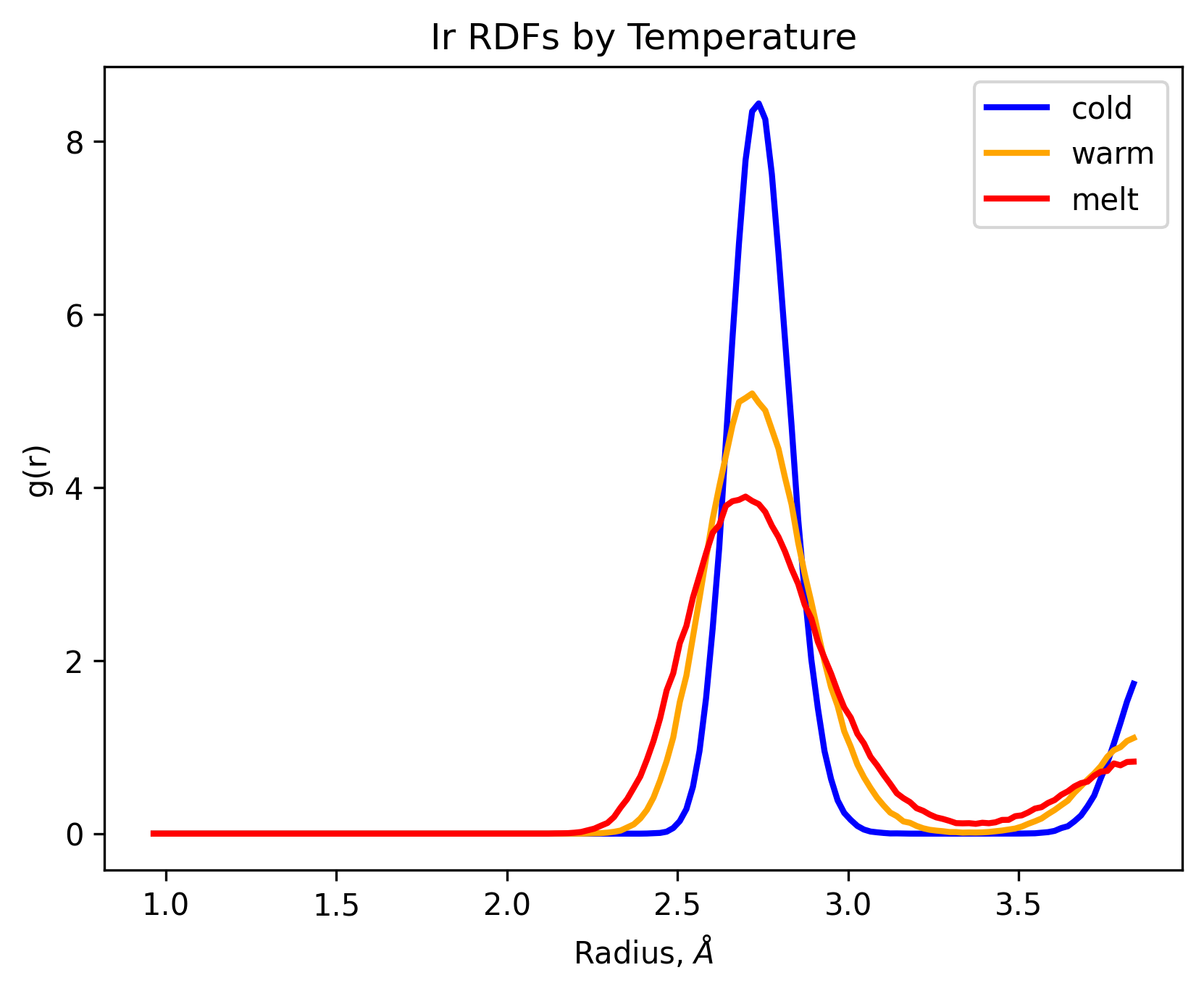}
        \caption{}
    \end{figure*}
    
\begin{figure*}
        \centering
        \includegraphics[width=.65\textwidth]{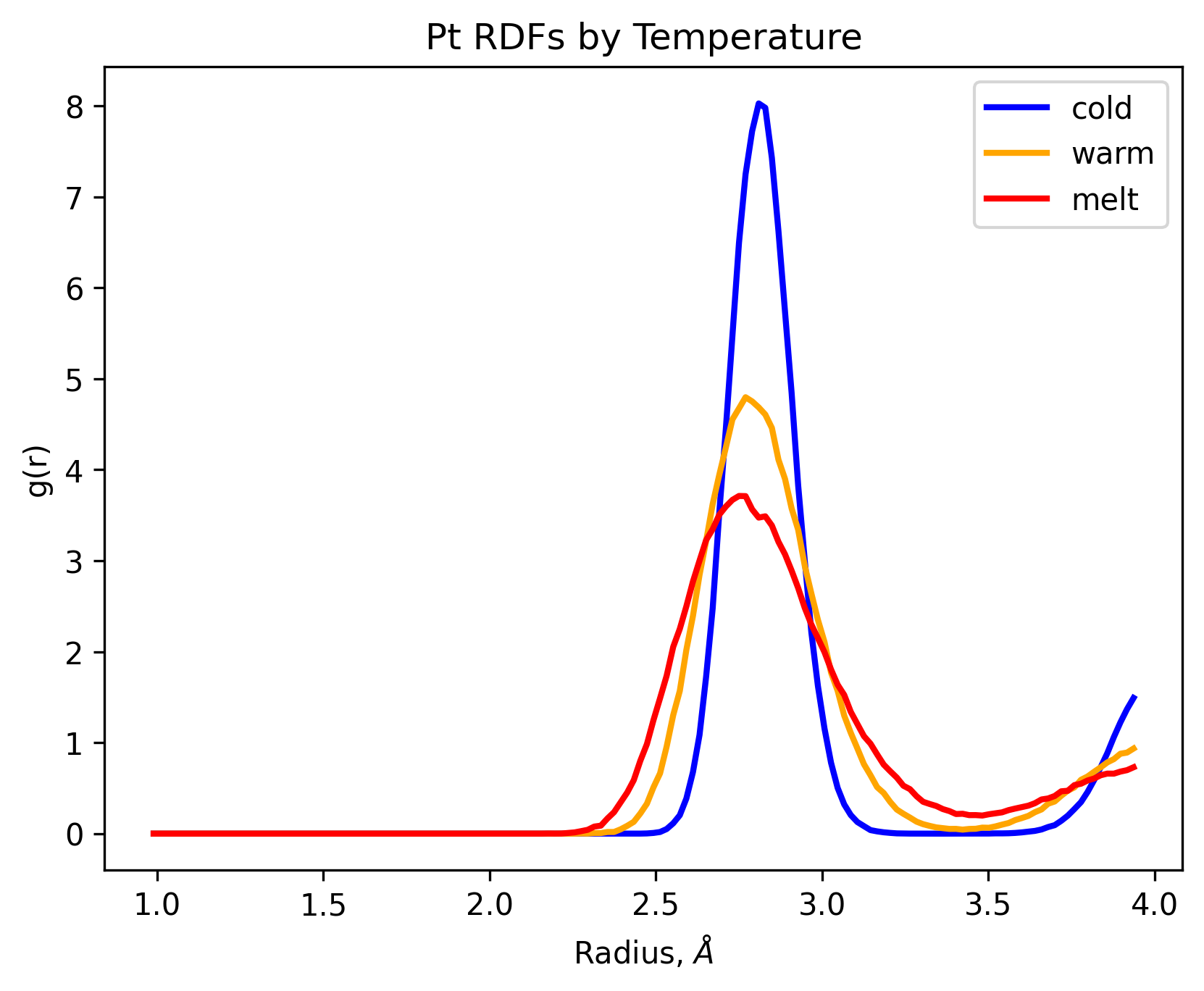}
        \caption{}
    \end{figure*}
    
\begin{figure*}
        \centering
        \includegraphics[width=.65\textwidth]{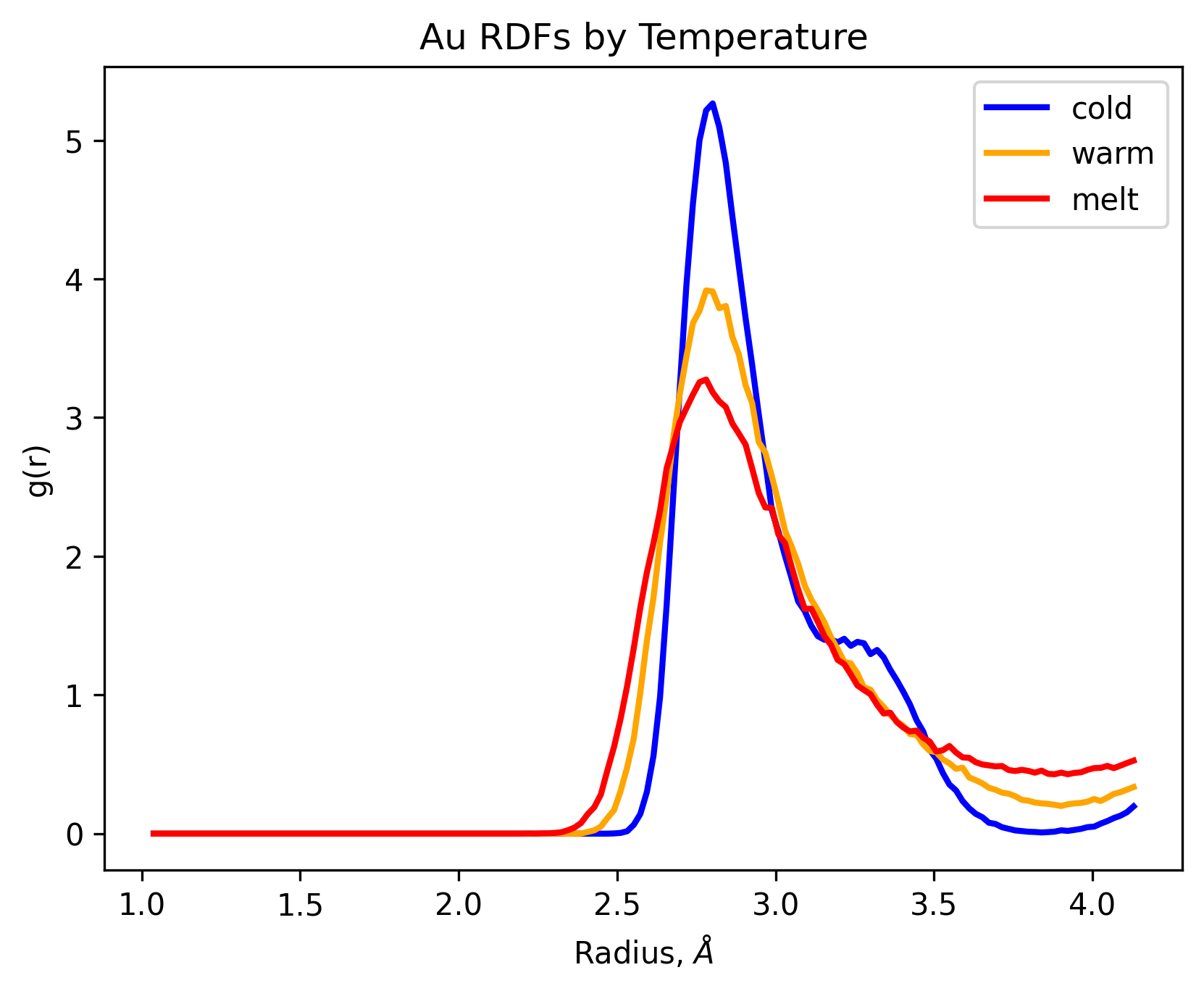}
        \caption{}
    \end{figure*}
    
\begin{figure*}
        \centering
        \includegraphics[width=.65\textwidth]{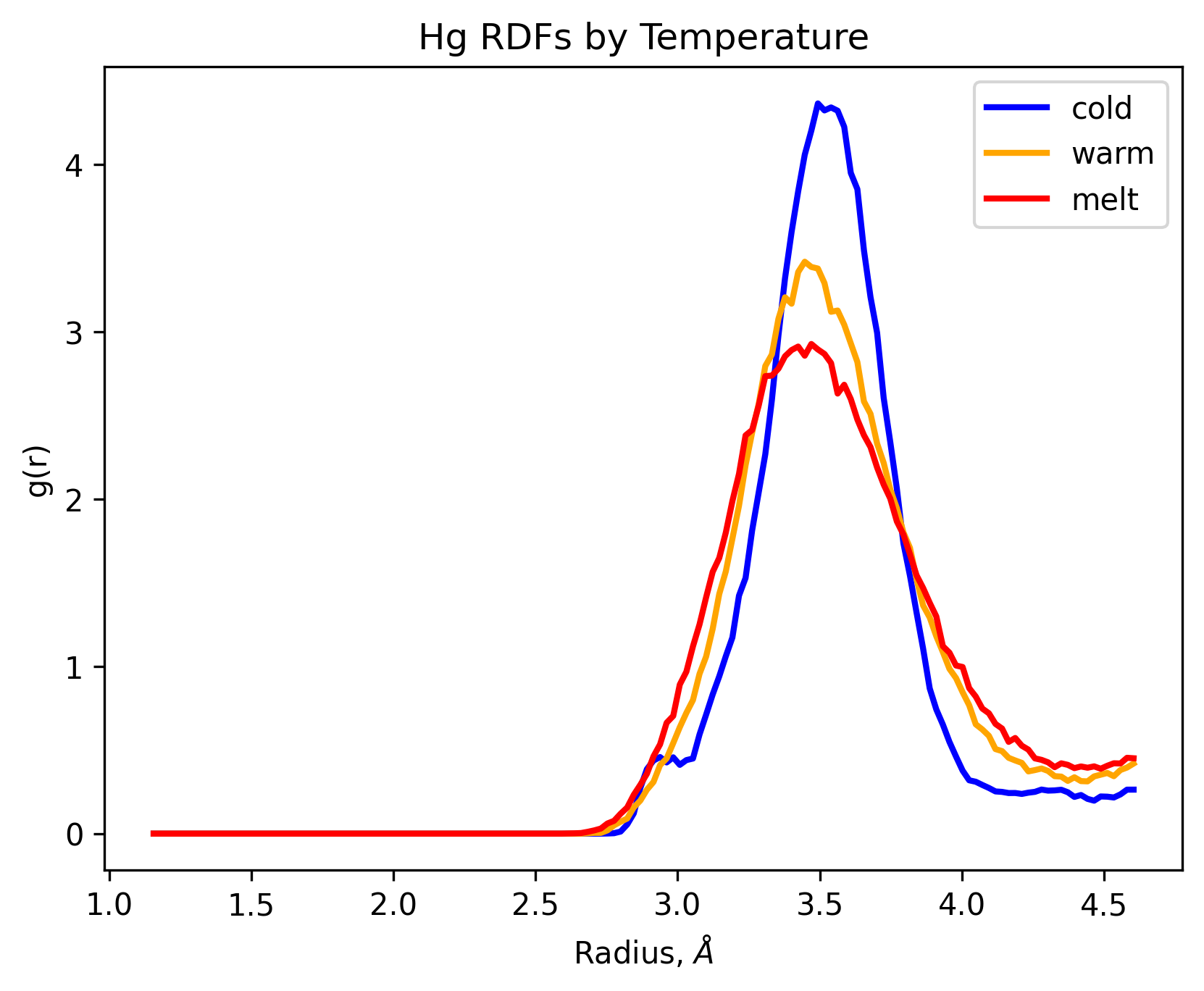}
        \caption{}
        \label{S-fig:lastrdf}
    \end{figure*}